\documentclass[aps,prd,preprint,superscriptaddress,tightenlines,nofootinbib,floatfix]{revtex4}
\usepackage{bm}
\usepackage{graphicx}
\usepackage{amsfonts}
\usepackage[mathscr]{eucal}
\usepackage{latexsym}
\usepackage{epsfig}
\DeclareGraphicsExtensions{eps,ps}

\newcommand{\postscript}[2]{\setlength{\epsfxsize}{#2\hsize}
   \centerline{\epsfbox{#1}}}
\newcommand{\beq}{\begin{equation}}
\newcommand{\eeq}{\end{equation}}

\newcommand{\mb}{\text{mb}}
\newcommand{\mpmin}{M_p^{\text{min}}}
\newcommand{\lstar}{L_{*}}

\newcommand{\mplanck}{M_{\text{Pl}}}
\newcommand{\mstar}{M_{\ast}}
\newcommand{\md}{M_D}
\newcommand{\mbh}{M_{\text{BH}}}
\newcommand{\mbhmin}{M_{\text{BH}}^{\text{min}}}

\newcommand{\xmin}{x_{\text{min}}}

\newcommand{\ev}{\text{eV}}

\newcommand{\gev}{\text{GeV}}
\newcommand{\tev}{\text{TeV}}

\newcommand{\cm}{\text{cm}}
\newcommand{\km}{\text{km}}
\newcommand{\g}{\text{g}}

\newcommand{\xmax}{X_{\text{max}}}
\newcommand{\etal}{{\em et al.}}

\newcommand{\eqref}[1]{Eq.~(\ref{#1})}

\def\snp{\sigma^{tot}_{\nu p}}
\def\emin{E_{\rm min}}
\def\emax{E_{\rm max}}
\def\deg{^{\circ}}

\def\deg{\ifmmode{^{\circ}}\else ${^{\circ}}$\fi}
\def\bi{\begin{itemize}}
\def\ei{\end{itemize}}
\def\ed{\end{document}}

\def\pri{^{\, \prime}}
\def\prii{^{\, \prime\prime}}
\def\cf#1{\ifmmode{\cal #1}\else${\cal #1}$\fi}

\def\be{\begin{equation}}
\def\ee{\end{equation}}
\def\beas{\begin{eqnarray*}}
\def\eea{\end{eqnarray}}
\def\bea{\begin{eqnarray}}
\def\eeas{\end{eqnarray*}}
\def\tfrac#1#2{{\textstyle\frac{#1}{#2}}}
\def\thalf{\tfrac{1}{2}}
\def\gev{\ifmmode{\mbox{GeV}}\else GeV\fi}
\def\es{{\rm erg\ s}^{-1}}

\def\evmss{\ \mbox{eV}^2\mbox{m}^{-2}\mbox{s}^{-1}\mbox{sr}^{-1}}
\def\lf{L_{41}}
\def\ton{T_{\rm on}}
\def\vr{\vec{r}}

\begin{document}
\baselineskip=15.5pt
\pagestyle{plain}
\setcounter{page}{1}

\vskip -1.3cm

\rightline{\small{\tt NUB-3228/Th-02}}
\rightline{\tt hep-ph/0206072}

\vspace{.5cm}

\begin{center}

{\Large {\bf Ultrahigh Energy Cosmic Rays:}}\\
{\Large {\bf  The state of the art before the Auger Observatory}}

\vskip 1.5cm
{\large Luis Anchordoqui, Thomas Paul, 
Stephen Reucroft, and John Swain}

\medskip
\medskip
{\large {\it Department of Physics, Northeastern University\\
Boston, MA 02115, USA}}
\vspace{.5cm}

{\tt l.anchordoqui@neu.edu, tom.paul@cern.ch,\\ stephen.reucroft@cern.ch, 
john.swain@cern.ch}  
\vspace{.5cm}

\end{center}

\setcounter{footnote}{0}

\begin{center}
{\large {\bf Abstract}}
\end{center}

\noindent  In this review we discuss the important progress made in recent 
years towards understanding the experimental data on cosmic rays with 
energies $\agt 10^{19}$~eV. We begin with a brief survey 
of the available data, including a description of the energy spectrum, 
mass composition, and arrival directions. At this point we also give a short 
overview of experimental techniques. After that, we 
introduce the fundamentals of acceleration and propagation in order 
to discuss the conjectured nearby cosmic ray sources. We then turn to 
theoretical notions of physics beyond the 
Standard Model where we consider both exotic primaries and exotic physical 
laws. Particular attention is given to the role that TeV-scale gravity 
could play
in addressing the origin of the highest energy cosmic rays. 
In the final part of the review we discuss the potential of future 
cosmic ray experiments for the discovery of tiny black holes that should be 
produced in the Earth's atmosphere if TeV-scale gravity is realized in Nature.

\newpage

\tableofcontents

\newpage

\section{Appetizer}

Cosmic ray (CR) and accelerator based particle physics
share common roots, and in fact many of the key discoveries
early in the history of particle physics came from the study of
CRs. After a period of divergence between the two fields,
both in methodology and in the key areas of interest, a confluence 
is now underway, driven in no small part by the mystery of the highest 
energy CRs. Much of the interest in these CRs is rooted in 
recent developments in both experiment and theory. The most recent 
data reported at the 27th International Cosmic Ray Conference (ICRC) 
suggest that some CRs arrive at the outer limits of the Earth's atmosphere
with energies above $10^{20}$~eV. This is in 
itself remarkable, as it seems difficult  to explain
how such high energies can be attained without invoking new physics.
Moreover, the center-of-mass (c.m.) energies achieved when 
these CRs impinge on a stationary nucleon of the air molecules are higher than 
any that can be reached in present-day terrestrial experiments, or indeed 
any experiment in 
the foreseeable future. Adding to the puzzle, the HiRes and AGASA experiments 
continue to revise their estimates of the fluxes and energies, 
making the subject more confusing.

Although current theoretical and experimental uncertainties make it
impossible to determine the origin and nature of the highest energy events 
with any certainty, the mere presence of such events suggests the exciting 
possibility that CRs will, once again, provide us with glimpses of physics 
beyond our present theories. Speculations ranging from supersymmetry to 
magnetic monopoles, large extra dimensions, and captivating tiny black holes 
produced in CR collisions  fill the Los Alamos preprint archives at an 
ever-increasing rate. There is clearly great hope that we are at the verge 
of learning something revolutionary.

In this spirit then, it seems opportune to review our present understanding
of the highest energy CRs: what has been measured and with what confidence,
what do we know about the composition 
and possible acceleration mechanisms for these particles, and how much do our
measurements depend on uncertainties in Standard Model calculations? 
Throughout the first part of this
review we  focus tightly on the interplay between experiment and
phenomenology assuming no physics beyond the Standard Model. This provides 
something of a
sanity check on where we are now and where we think we are going. 
Afterwards, we relax our grip on sanity and summarize some of the new, and 
sometimes fantastic, ideas that may be probed by the next generation of 
experiments. To paraphrase Neils Bohr, the hope is that none of 
these ideas is crazy enough, and that in a few years we can look forward 
to genuine surprises.

\section{Observation of the highest energy cosmic rays}

\subsection{The energy spectrum}

In 1912 Victor Hess carried out a series of pioneering balloon flights
during which he measured the levels of ionizing radiation as high as 5 km
above the Earth's surface~\cite{Hess}.  His discovery of increased radiation 
at high altitude revealed that we are bombarded by ionizing particles from 
above. These CR  particles are now known to consist primarily of
protons, helium, carbon, nitrogen and other heavy ions up to iron.     

Below $10^{14}$ eV the flux of particles is sufficiently large that 
individual nuclei can be studied by detectors carried aloft in balloons or 
satellites. From such direct experiments we know the 
relative abundances and the energy spectra of a variety of atomic 
nuclei, protons, electrons and positrons as well as the intensity, 
energy and spatial distribution of $X$-rays and $\gamma$-rays. 
Measurements of energy and isotropy showed conclusively that one obvious 
source, the Sun, is not the main source. Only below 100~MeV  
kinetic energy or so, where the solar wind shields protons 
coming from outside the solar system, does the Sun dominate the observed 
proton flux. Spacecraft missions far out into the solar 
system, well away from the confusing effects of the Earth's atmosphere 
and magnetosphere, confirm  that the abundances around 1~GeV are 
strikingly similar to those found in the ordinary material of the solar 
system. Exceptions are the overabundance of elements like 
lithium, beryllium, and boron, originating from the spallation of heavier 
nuclei 
in the interstellar medium.

Above $10^{14}$ eV, the flux becomes so low that only ground-based 
experiments with large apertures and long exposure times can hope to acquire 
a significant number of events. Such experiments exploit the atmosphere as a 
giant calorimeter. The incident cosmic radiation interacts with the atomic 
nuclei of air molecules and produces extensive air
showers (EASs) which spread out over large areas. Already in 1938, 
Pierre Auger concluded from the size of EASs that the spectrum extends up 
to and perhaps beyond $10^{15}$~eV~\cite{Auger:1938,Auger:1939}. Nowadays 
substantial progress has been made in measuring the extraordinarily low 
flux ($\sim 1$ event km$^{-2}$ yr$^{-1}$) above $10^{19}$ eV. Continuously running 
experiments using both arrays of particle detectors on the 
ground and fluorescence detectors which track the cascade through the 
atmosphere, have detected events with primary particle energies higher 
than $10^{20}$~eV~\cite{Linsley,Suga:jh,Garmston,Winn:un,Lawrence:cc,Glushkov:sn,Bird:yi,Bird:wp,Bird:1994uy,Hayashida:1994hb,Takeda:1998ps,Antonov:kn,Jui:1999xz,Sakaki}, with no evidence that the 
highest energy recorded thus far is Nature's upper limit.

\begin{figure}
\postscript{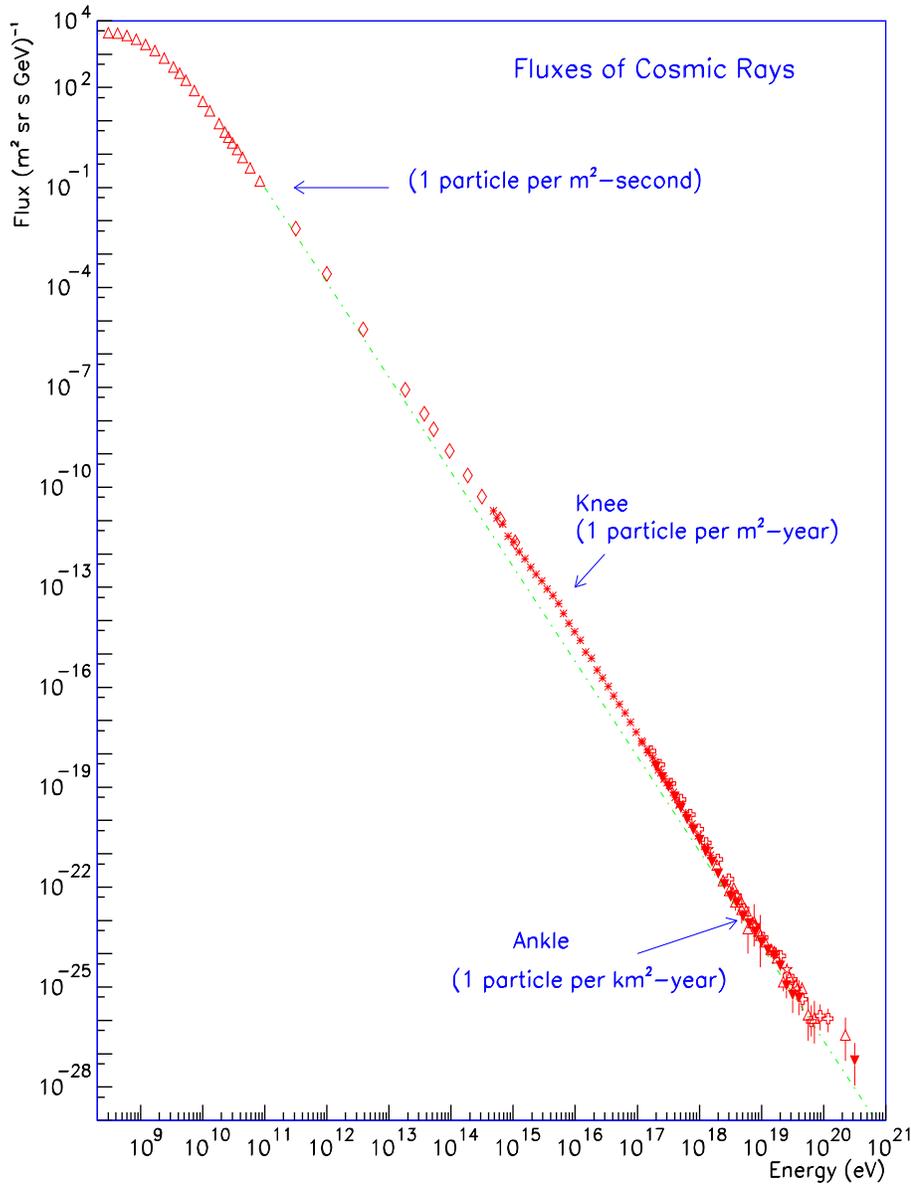}{0.80}
\caption{Compilation of measurements of the differential energy spectrum of 
CRs. The dotted line shows an $E^{-3}$ power-law for comparison. 
Approximate integral fluxes (per steradian) are also shown~\cite{Cronin}.}
\label{f1}
\end{figure}

In contrast to the irregular shape of the isotropic electromagnetic 
background spectrum from, say,
$10^{8} - 10^{20}$ Hz, the CR energy spectrum above $10^{9}$ eV can be 
described by a series of power laws, with the flux falling about 3 orders 
of magnitude for each decade
increase in energy (see Fig.~\ref{f1}). In the decade centered at 
$\sim 10^{15.5}$~eV (the knee) the spectrum steepens from  $E^{-2.7}$ 
to $E^{-3.0}$. This feature, discovered around 40 years ago \cite{Kulikov}, 
is still not 
consistently explained. The spectrum steepens further to $E^{-3.3}$ above 
$\sim 10^{17.7}$~eV (the dip) and then flattens to  $E^{-2.7}$ at 
$\sim 10^{18.5}$~eV (the ankle). Within the statistical uncertainty of the 
data collected by AGASA~\cite{Hayashida:2000zr}, which is large 
above $10^{20}$ eV, the tail of the spectrum is 
consistent with a simple extrapolation at that slope to the highest 
energies, possibly with a hint of a slight accumulation 
around $10^{19.5}$ eV. A very widely held 
interpretation of the ankle is that above $10^{18.5}$~eV a new population of 
CRs with extragalactic origin begins to dominate the more steeply falling 
Galactic population~\cite{Cocconi:et}. This hypothesis is supported by the data recorded at 
AGASA, which shows that 
around $10^{18}$ eV the angular 
distribution correlates with the Galactic center (anisotropy $\sim 4\%$) 
and is consistent with a Galactic origin, whereas at higher energy the 
anisotropy disappears~\cite{Hayashida:1998qb,Teshima:2001}.

\begin{figure}
\postscript{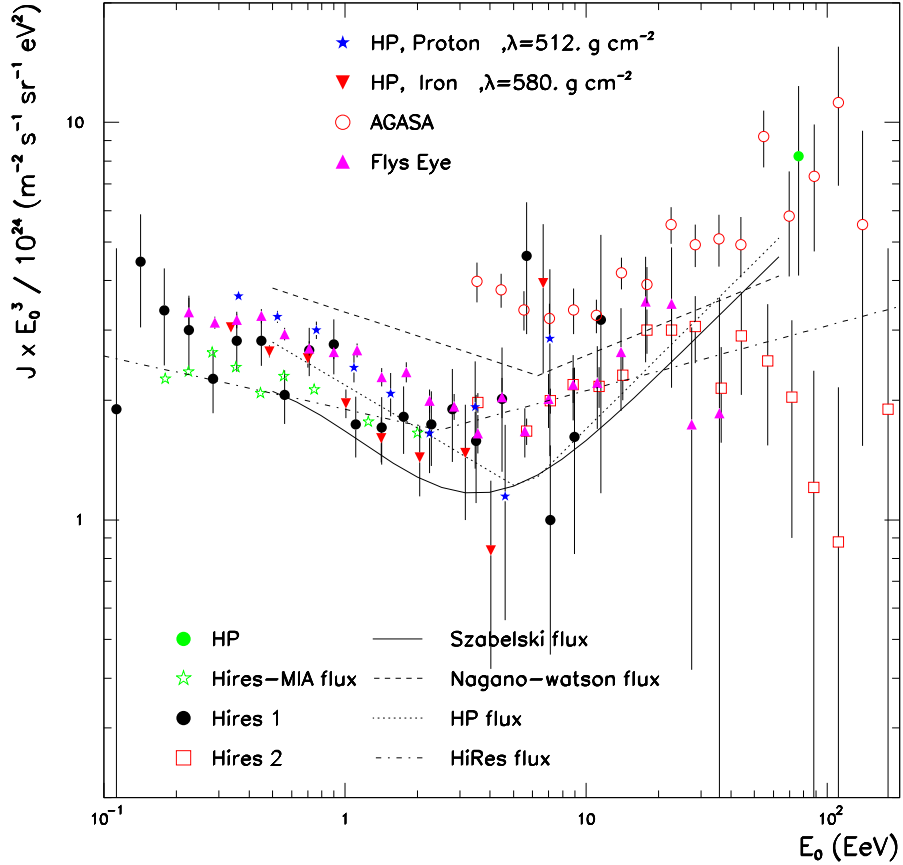}{0.80}
\caption{A composite energy spectrum including recently reanalysed Haverah 
Park data assuming proton and iron primaries (the parameter $\lambda$ 
measures the attenuation length of the density of charged particles at 600~m 
from the shower core), stereo Fly's Eye data, monocular HiRes data from 
both eyes up to $60^\circ$, and hybrid HiRes--MIA data. The measurements are 
compared to spectrum parameterizations given by different authors as 
described in the text. Published in \protect{Ref.~\cite{Watson:2001et}}.}
\label{f2}
\end{figure}

At the very high end of the spectrum the flux now appears more uncertain than was 
thought as recently as a 
year ago. At that time the event rate at $10^{19}$ eV reported by 
different experiments was in agreement at the 10 - 15\% level. 
In particular, preliminary data reported by the HiRes group~\cite{Jui:1999xz} 
showed 7 events above $10^{20}$~eV, in good accord with the number 
anticipated from the flux observed by the AGASA 
experiment~\cite{Hayashida:2000zr}. Very recently, the AGASA Collaboration 
reported a total of 17 events above $10^{20}$ eV~\cite{Sakaki}, quite 
consistent with their previous work~\cite{Hayashida:2000zr}.
However, one of the HiRes cameras, which has an exposure slightly greater 
than that of the AGASA experiment, recorded only 2 events above $10^{20}$~eV 
compared to the 20 expected for a spectrum similar to that reported earlier 
by AGASA \cite{Sommers:01}. The HiRes group also reported data from their 
stereo system, which has 20\% of the monocular exposure. They observed  
one event with an energy estimated to be close to 
$10^{20.5}$ eV. These unexpected discrepancies are not yet 
understood. The ``disappearance'' of the events 
reported as being above $10^{20}$~eV in 1999~\cite{Jui:1999xz} is attributed 
to a better understanding of the atmosphere, which is now claimed to be 
clearer than had been previously supposed. Adding to the puzzle, the Haverah 
Park energy estimates have been re-assessed~\cite{Ave:2001hq}, resulting in 
a steeper reconstructed spectrum which shows differences of up 
to 30\% compared to the one derived by fitting Akeno, AGASA and 
the ``old'' Haverah Park 
data \cite{Nagano:ve}.\footnote{It is noteworthy that AGASA energies have been 
estimated under the assumption that  the primaries are protons above 
$10^{18}$ eV. Though there is no evidence as to what mass species is 
dominant at the highest energies, the results from AGASA decrease
by only about 20\% when the hypothesized chemical composition is changed.}
The average energy of the 4 events observed by the Haverah Park experiment 
that were previously above $10^{20}$~eV is shifted 30\% downwards to 
energies below $10^{20}$ eV. The energy weighted cosmic ray flux 
corresponding to these 4 events 
is $JE^3 = 8.3^{+6.5}_{-4.0} \times 10^{24}$ eV$^2$ sr$^{-1}$ m$^{-2}$ 
s$^{-1}$ at an energy of $7.62 \times 10^{19}$ eV (see Fig.~\ref{2}). The highest 
energy Haverah Park 
event is now estimated to have an energy of $8.3 \times 10^{19}$ eV. 
In the ``old'' Haverah Park data sample, this event had a reported energy of
$1.2 \times 10^{20}$~eV.
In Fig.~\ref{f2} we show a compilation of the updated data from Haverah 
Park~\cite{Ave:2001hq}, 
AGASA~\cite{Hayashida:2000zr}, monocular HiRes (HiRes 1 and HiRes 2)~\cite{Sommers:01}, 
HiRes-MIA~\cite{Abu-Zayyad:2000ay} and Fly's Eye stereo~\cite{Bird:wp}. 
For comparison, the figure also shows 4 recent parameterizations of the 
energy spectrum including the fit to the Akeno, AGASA and Haverah Park 
data sets  
by Nagano and Watson~\cite{Nagano:ve}, the one obtained by Szabelski, 
Wibig and Wolfendale~\cite{Szabelski}, the parameterization given by the HiRes 
Collaboration~\cite{Sommers:01} and 
a parameterization based on a recent analysis of 
Haverah Park data~\cite{Ave:2001hq}. The recalibrated Haverah Park spectrum 
is in very good agreement with that 
obtained by the fluorescence detectors of Fly's Eye and HiRes. 
Reconciling the observations of the various experiments depends on a detailed
understanding of detector behavior and atmospheric properties as well
as reliable models to predict the evolution of cosmic air showers.  These
issues are discussed in the following sections.  

\subsection{Properties of extensive air showers and mass composition}  \label{sec:properties}

We begin this section with a brief discussion of the principal 
observables characterizing an EAS.  Next, we discuss 
how the shower development differs for different primary species and
how these differences are manifest in the shower observables.  

\begin{figure}
\postscript{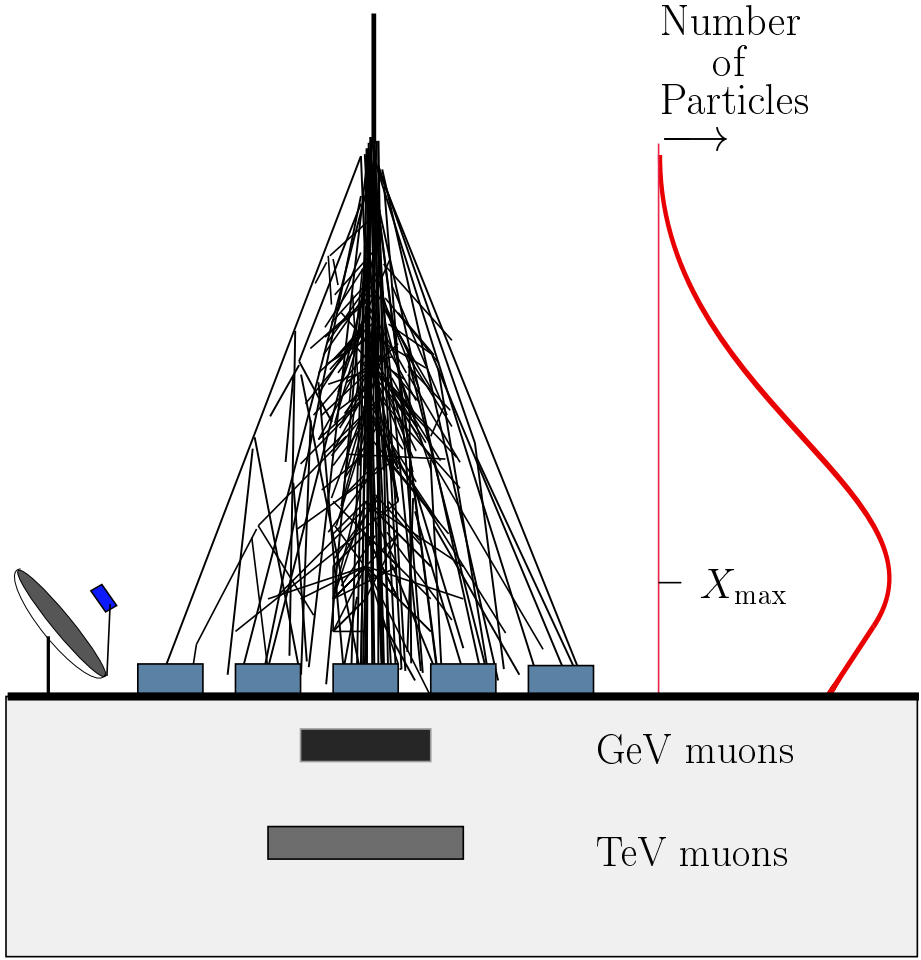}{0.90}
\caption{Particles interacting near the top of the atmosphere initiate 
an electromagnetic and hadronic cascade. Its profile is shown on the right. 
The different detection methods are illustrated. Mirrors collect the 
\v{C}erenkov and nitrogen fluorescent light, arrays of detectors sample the 
shower reaching the ground, and underground detectors identify the muon 
component of the shower.}
\label{f3}
\end{figure}

The incidence of a single high energy particle on the upper atmosphere 
gives rise to a roughly conical cascade of particles which 
reaches the Earth in the form of a giant ``saucer'' traveling
at nearly the speed of light. As the cascade develops in the atmosphere, 
the number of particles in the shower increases until the energy of the
secondary particles is degraded to the level where ionization
losses dominate. At this point the density of particles starts to decline.
The number of particles as a function of the amount of atmosphere
penetrated by the cascade (in g~cm$^{-2}$) is known as
the ``longitudinal profile'' shown schematically in Fig.~\ref{f3}. 
The atmospheric depth at which the number of particles in the showers 
reaches its maximum, 
$X_{\rm max}$, is often regarded as the most useful observable
of the shower, as it strongly depends on the primary energy and composition. 
For example,
$X_{\rm max}$  increases with primary energy since more cascade generations
are required in the cooling of secondary products. The way  
the average depth of maximum $\langle X_{\rm max} \rangle$ changes with 
energy depends on the primary composition and particle interactions 
according to 
\begin{equation}
\langle X_{\rm max} \rangle = D_e \ln \left( \frac{E}{E_0} \right), 
\end{equation}
where $D_e$ is the so-called ``elongation rate'' and $E_0$ is a 
characteristic energy that depends on the primary 
composition~\cite{Linsley:gh}. Therefore, 
since 
$\langle X_{\rm max} \rangle$ and $D_e$ can be determined directly from the 
longitudinal shower profile measured with a fluorescence 
detector,  
$E_0$ and thus the composition, can be extracted after estimating $E$ from 
the total fluorescence yield. Indeed, the parameter often 
measured is $D_{10}$, the rate of change of $\langle X_{\rm max} \rangle$ per 
{\it decade} of energy.  We note that one can discern changes in 
the primary composition from breaks in the elongation rate, and such 
breaks are relatively insensitive to certain systematic uncertainties.

If the primary particle is a nucleon or a nucleus the shower begins with 
a hadronic interaction. The number of hadrons increases through subsequent 
generations of particle interactions. However, in each generation about 30\% of the energy is transferred to an electromagnetic cascade by the prompt 
decay of 
neutral pions. Ultimately, the electromagnetic cascade dissipates around 90\%
of the primary particle's energy, and hence the total number of 
electromagnetic particles is very nearly proportional to the shower energy.
The remaining energy is carried by muons and neutrinos from $\pi^{\pm}$ decays.
The number of muons does not increase linearly with energy, since at
higher energy more generations are required to cool the pions to the point
where they are likely to decay before interaction.   
Production of extra generations results in a larger fraction of the energy 
being lost to the electromagnetic cascade, and hence a smaller fraction 
of the original energy 
being delivered to the $\pi^{\pm}$. 
With this in mind, one can evaluate the muon 
production in a heavy nucleus shower relative to that in a proton shower 
by viewing a nucleus with atomic number $A$ and energy $E$ as a collection 
of individual nucleons each with energy $E/A$. The muon production 
in a proton shower increases with energy as $E^{0.85}$~\cite{Gaisser:vg}, 
and thus the total number of muons produced by 
a nucleus of mass $A$ is $N_\mu^A \propto A (E/A)^{0.85}$, or, comparing to 
proton showers, $N_\mu^A = A^{0.15} N_\mu^p$. Therefore, an iron nucleus 
produces a shower with around 80\% more muons than a proton shower of the 
same energy. Around $10^{20}$~eV, the hadronic mean free path is roughly 
40~g/cm$^2$.

On the other hand, if the primary particle is a $\gamma$-ray the basic 
interactions that cool the cascade development are pair production, 
Bremsstrahlung, ionization losses, and Compton scattering.
Above $10^{19}$~eV the Landau-Pomeranchuk-Migdal (LPM) effect leads to 
significant suppression of the Bethe-Heitler cross section 
for pair production and Bremsstrahlung, and so the growth 
of $N_\mu$ with energy gradually disappears~\cite{Cillis:1998hf}. However, 
the muon suppression in electromagnetic showers caused by the LPM effect 
become largely compensated by the interactions of $\gamma$-rays with the 
geomagnetic field~\cite{Cillis:ij}. Above the ankle ($\sim 10^{18.5}$~eV), 
a $\gamma$-shower produces  
less than 20\% as many muons as a proton shower of the same 
energy. The energy dependence of this effect 
has a rather complicated (non power-law) form~\cite{Plyasheshnikov:2001xw}.
Gamma-rays of energy $\sim 3 \times 10^{20}$~eV have interaction lengths
of approximately 60 g/cm$^2$ with a standard deviation of 80 g/cm$^2$ 
if the LPM effect is taken into account, but these figures reduce
to 46 g/cm$^2$ and 45 g/cm$^2$, respectively, if the LPM effect 
is ignored~\cite{Cillis:1998hf}.

The experimental observables for inclined showers (incident zenith
angle $\theta > 60^\circ$) are complimentary to those for vertical 
showers, since
inclined showers are mostly comprised of muons which were produced
far from the detection zone~\cite{Dova:2001jy}. In contrast to vertical 
showers the shower front is extremely flat (radius of curvature 
$> 100$~km) and the 
particle time spread is very narrow ($\Delta t < 50$~ns). 
For $\theta > 70^\circ$ the electromagnetic component arising from the 
hadronic channel through $\pi^0$ decays is totally extinguished, as more 
than 3 equivalent vertical atmospheres are traversed. There remains only 
a residual 
electromagnetic component in the shower front which is produced by 
the muons themselves, mostly through muon decay. 

Extracting information on the nature of the primaries from
$D_e$ and $\langle X_{\rm max} \rangle$, or from the variation of the muon content has proved to be 
exceedingly difficult for a number of reasons. The most fundamental drawback is that the first few 
cascade steps are subject to large inherent fluctuations and consequently 
this limits the event-by-event mass resolution of the experiments. 
In addition, the c.m. energy of the first interactions is 
well beyond any reached in collider experiments. Thus, one needs to rely on 
hadronic interaction models that attempt to extrapolate, using different 
mixtures of theory and phenomenology, our understanding of particle 
physics.\footnote{A brief summary of the current status of hadronic 
models is provided in Appendix A.} 

The analysis of the elongation rate and the spread 
in $X_{\rm max}$ at a given energy reported by the Fly's Eye 
Collaboration suggests a change from an iron dominated composition at 
$10^{17.5}$~eV to a proton dominated composition near 
$10^{19}$~eV~\cite{Bird:yi}. Such behavior of $D_e$ is in agreement with 
an earlier analysis from Haverah 
Park~\cite{Walker:sa}. However, the variation of the density of muons
with energy reported by the Akeno Collaboration favours a composition that 
remains mixed over the $10^{18} - 10^{19}$~eV 
decade \cite{Hayashida:95}. At this stage, it is worthwhile to recall that 
any inferred mass composition would be model-dependent (see Appendix A). 
Recently, Wibig and Wolfendale~\cite{Wibig} 
reanalyzed the Fly's Eye data considering not only  proton and iron 
components (as in~\cite{Bird:yi}) but a larger number of atomic mass 
hypotheses. Additionally, they adopted a different hadronic model 
that shifts the prediction of $X_{\rm max}$ for primary protons of 
$10^{18}$ eV from 730 g cm$^{-2}$~\cite{Bird:yi} to 751 g cm$^{-2}$. The 
difference, although apparently small, has a significant effect on the mass 
composition inferred from the data. The study indicates that at the highest 
energies ($10^{18.5} - 10^{19}$~eV and somewhat above) there is a 
significant fraction of nuclei with charge greater than unity. This result 
is more in accord with the conclusions of the Akeno group than those of 
the Fly's Eye group.  Deepening the problem, there is a difference between 
the conclusion reached 
from data on $X_{\rm max}$ from the Fly's Eye experiment~\cite{Bird:yi} 
and those reached from  the HiRes prototype, operated with the MIA 
detector in the range $10^{17}$ to $10^{18}$~eV~\cite{Abu-Zayyad:2000np}. 

Nowadays, the most powerful tool for extracting the primary mass spectrum
relies on comparing the flux of vertical showers to that of inclined
showers.  For example, if all primaries were photons, one would expect
to observe a great asymmetry in the number of vertical and inclined
events. Ave {\it et al.}~\cite{Ave:2000nd,Ave:2001xn} have used the 
observed Fly's Eye spectrum to predict
the rate of inclined events.  Comparing the predicted rate to the 
rate observed by Haverah Park for showers in the range 
$60^\circ< \theta < 80^\circ$, they
conclude that above $10^{19}$ eV, less 
than 48\% of the primary cosmic rays can be photons  and 
above $4 \times 10^{19}$ eV less than 50\%  can be 
photons. Both of these statements are made at the 95\% CL. This 
analysis is consistent with the search 
for showers which have significantly fewer muons than normal as reported
by the AGASA Collaboration~\cite{Shinozaki}. In addition, 
according to the predictions of {\sc qgsjet} for high energy 
interactions (see Appendix A),
the difference in abundance of muons with respect to photons and electrons
in the Haverah Park inclined shower measurements seems to favour a 
light composition above $10^{19}$~eV~\cite{Ave:2001sd}. A more robust 
statement awaits a better understanding of the 
sensitivity of this kind of analysis to different models of particle 
interactions. 

Finally, if the primary is a neutrino the first 
interaction occurs deep in the atmosphere, triggering showers in the 
volume of air immediately above the detector. The shower would thus present 
a very 
curved 
front (radius of curvature of a few km), with particles arriving 
over ${\cal O}(\mu$s). If the primaries are $\nu_e$ and $\nu_\mu$, 
as expected from $\pi^{\pm}$ decays, one expects a different type of
shower for each neutrino species:
``mixed'' (with full energy) for $\nu_{e}$ or 
``pure  hadronic'' (with reduced energy) for $\nu_\mu$. In the charged current 
interaction of a $\nu_e$, an ultrahigh energy electron is produced which 
initiates a large electromagnetic cascade parallel to the hadronic cascade. 
In contrast, the charged current interaction of a $\nu_\mu$ produces an 
ultrarelativistic  muon 
that is not detectable by the experiments. In the presence of maximal 
$\nu_\mu/\nu_\tau$ mixing, $\nu_\tau$ showers must also be considered. However,
since the $\tau$ mean flight distance is $\sim 50E$ km/EeV, only $\tau$'s 
with energy $\alt 8 \times 10^{17}$~eV 
will decay. Thus, $\nu_\tau$ showers above this energy will be 
indistinguishable from $\nu_\mu$ showers. The characteristics of 
these anomalous showers can be easily identified by fluorescence eyes 
and surface arrays. 

\subsection{The arrival direction distribution}

The investigation of anisotropy, when taken together with analyses of the 
spectral shape and particle species, can yield very important clues to 
reveal the CR origin. For example, if CRs mostly come from the direction 
of the Galactic plane and are largely protonic in composition, one expects 
to see a dipole anisotropy 
favoring the direction of the Galactic center. Moreover, 
since magnetic rigidity increases with energy, the angular 
width of the Galactic plane as seen in CRs would shrink slowly 
with rising energy. Of course, a heavy 
composition dominated by iron nuclei would 
show much smaller anisotropy at any given energy because of the smaller 
Larmor radius. The deviation from isotropy in galactic latitude is generally 
expressed in terms of the Wdowczyk-Wolfendale~\cite{Wdowczyk:xj} plane 
enhancement function,
\begin{equation}
\frac{I_{\rm obs}(b)}{I_{\rm exp}(b)} = (1 - f_E) + 1.437 \,\,f_E \,\,
\exp \left[ - b^2 \right]\,,
\label{fg}
\end{equation}
where $b$ is the galactic latitude in radians, $I_{\rm obs}$ 
and $I_{\rm exp}$ are observed and expected (for isotropy) 
intensities at latitude $b$, 
and $f_E$ is an energy dependent galactic latitude enhancement factor. 
A galactic origin for most of the particles would be expected to result in 
a positive value of $f_E$ that increases steadily with energy. A 
negative $f_E$ shows depression around the plane and $f_E =0$ indicates 
the arrival direction distribution is isotropic. Much of the data suggests 
that the magnitude of the Galactic plane enhancement increases 
systematically with energy until a little more than $10^{19}$~eV, above which 
it disappears and, indeed, there is evidence for a deficit in the 
direction of the Galactic plane 
above this energy~\cite{Szabelski:rx}. The most significant 
Galactic plane enhancement factor reported by Fly's Eye, $f_E = 0.104 
\pm 0.036$, is in the energy range 
$(0.4 - 1.0) \times 10^{18}$~eV~\cite{Bird:1998nu}. In a 
similar energy range, the AGASA Collaboration reported a strong anisotropy 
with first harmonic amplitude of $\sim 4\%$, corresponding to a chance 
probability of 0.2\% after taking the number of independent trials into 
account~\cite{Hayashida:1998qb}. A recent analysis of SUGAR data confirms the 
existence of an excess flux of CRs from a direction near the Galactic 
center~\cite{Bellido:2000tr}. The signal is consistent with that from a 
point source~\cite{Bednarek:2002nm}.

\begin{figure}[htpb]
\postscript{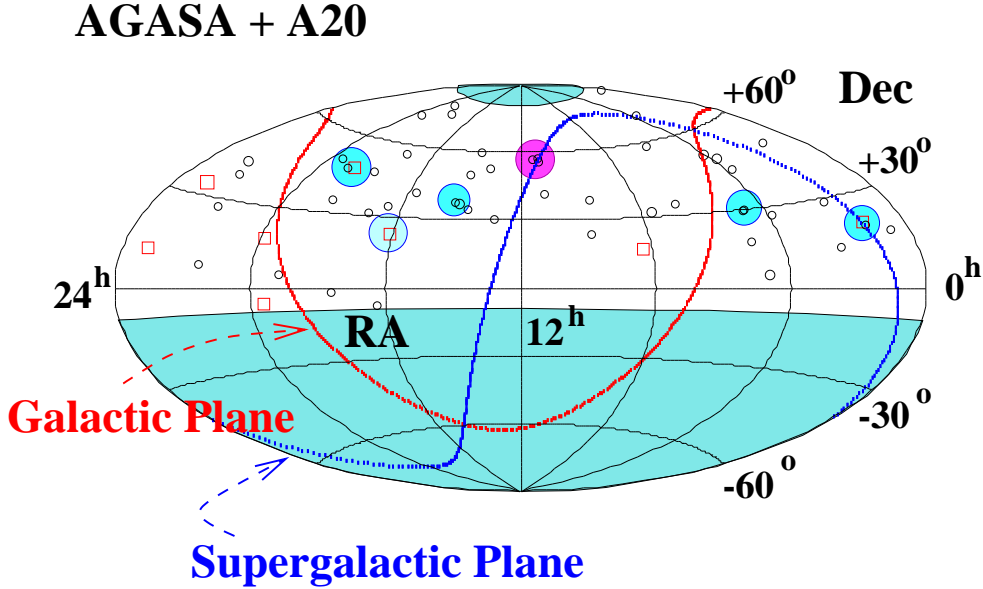}{0.90}
\caption{Arrival directions of cosmic rays detected by the AGASA and 
Akeno (A20)
experiments in  equatorial coordinates. Open circles and open squares 
represent cosmic rays with energies $(4 - 10) \times 10^{19}$~eV, 
and $\geq 10^{20}$~eV, respectively. The Galactic and super-Galactic planes 
are shown. Large shaded circles indicate event clusters 
within $2.5^{\circ}$. The shaded regions indicate the celestial regions 
excluded by a zenith angle cut of $\leq 45^\circ$. Published 
in Ref.~\cite{Hayashida:2000zr}.}
\label{f4}
\end{figure}

In a much higher energy range, the analysis of Haverah Park data 
seems to indicate that events with primary energy $> 4 \times 10^{19}$~eV 
reach the Earth preferentially from the direction of the super-Galactic 
plane, a swath 
in the sky along which radio galaxies are clustered~\cite{Stanev:1995my}. 
The magnitude of the 
observed excess is found to be 2.5 - 2.8 $\sigma$ in terms of Gaussian 
probabilities. However, such large-scale correlation with the super-Galactic 
plane was not observed in the data sets of the AGASA~\cite{Hayashida:bc}, 
SUGAR~\cite{Winn:up,Kewley:zt} and Fly's Eye~\cite{Bird:1998nu} experiments.

The arrival direction of the events with energy $> 4 \times 10^{19}$~eV 
registered by the AGASA experiment is shown in Fig.~\ref{f4}. The data show no 
significant super-Galactic plane enhancement. However, some excess of 
particles from the general direction of the super-Galactic plane is observed  
in the energy bin $\log(E[{\rm eV}]) = [19.1, 19.2]$, 
with $f_E^{\rm SG} = 0.36 \pm 0.15$~\cite{Takeda:1999sg}.\footnote{The super-Galactic plane enhancement factor is given by
$I_{\rm obs}(b)/I_{\rm exp}(b) = (1 - f^{\rm SG}_E) + 1.402 \,\,f^{\rm SG}_E \,\,\exp \left[ - b^2 \right]$. Compare this with the Fly's Eye parameterization for the Galactic plane given in Eq.~(\ref{fg}).}
The highest energy events are distributed widely over the sky, without 
any apparent counterparts, such as sources in the Galactic plane or in the 
local supercluster. Moreover, the data are consistent with an isotropic 
distribution of sources, in sharp contrast to the anisotropic distribution 
of light in the local supercluster~\cite{Waxman:1996hp}.

\begin{figure}
\epsfxsize=6.0in
\postscript{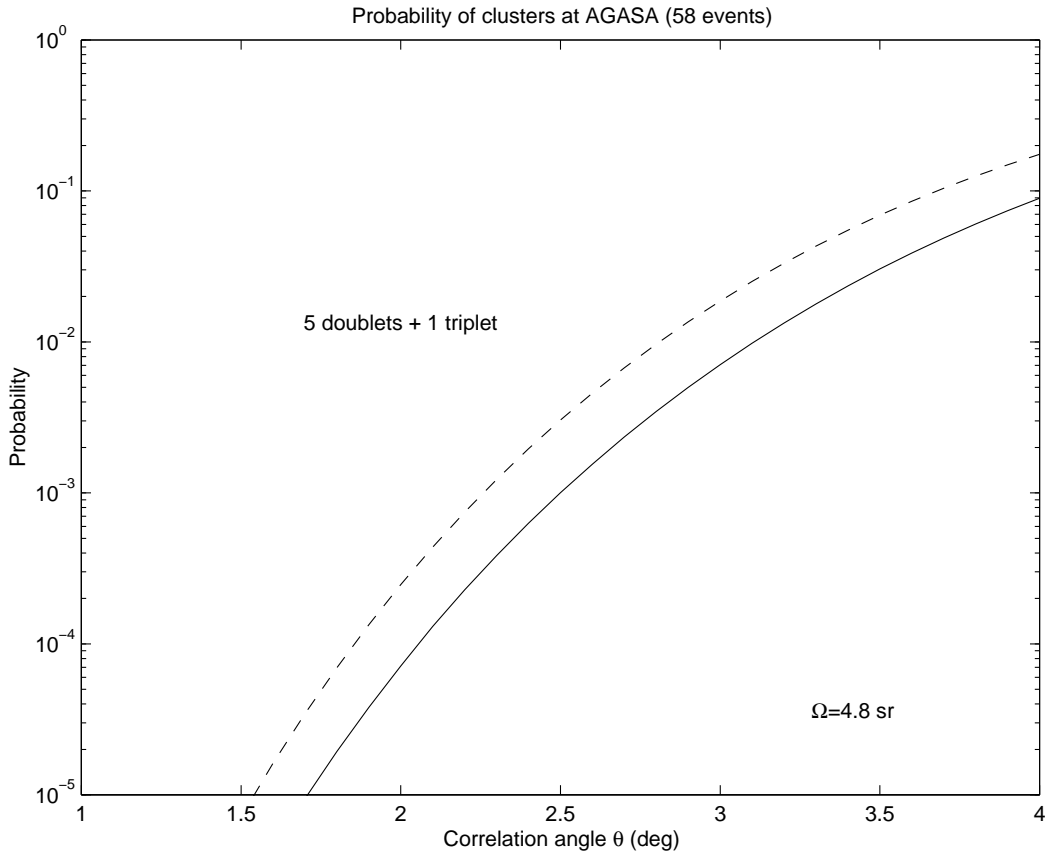}{0.90}
\caption{Inclusive probabilities for five doublets and one triplet in a 58-event sample at AGASA. Solid (exact), dashed (Poisson). From Ref.~\cite{Anchordoqui:2001qk}.}
\label{f5}
\end{figure}

Adding to the puzzle, the AGASA experiment draws attention to the fact 
that the pairing of events on the celestial sky could be occurring at 
higher than chance coincidence~\cite{Hayashida:2000zr}. Specifically,
when showers with separation angle less 
than the angular resolution $\theta_{\rm min}= 2.5^\circ$ are 
paired up, AGASA finds five doublets and one triplet
among the 58 events reported with mean energy above $4 \times 10^{19}$ eV 
(see Fig.~\ref{f4}). The chance probability of observing such topology in an 
isotropic distribution can be estimated by means of the Goldberg-Weiler 
formalism~\cite{Goldberg:2000zq}. To do so,
consider the solid angle $\Omega \sim 4.8$~sr on the celestial sphere 
covered by AGASA to be divided into $N$ equal angular bins, each with 
solid angle $\omega \simeq \pi \theta^2$. 
Then, by tossing $n$ events randomly into
\begin{equation}
N \,\simeq \,\frac{\Omega}{\pi\, \theta^2} = 1045 \,\,\frac{\Omega}
{1 \,\, {\rm sr}} \,\,
\left(\frac{\theta}{1^{^\circ}} \right)^{-2}
\label{N}
\end{equation}
bins, one is left with a random distribution. Now, identify each event 
distribution by specifying the partition of the total sample of AGASA  
into a number $m_0$ of empty bins, a number $m_1$ of single hits, a 
number $m_2$ of double hits, etc., among the $N$ angular bins that constitute 
the whole exposure. The probability to obtain 
a given event topology is~\cite{Goldberg:2000zq}
\begin{equation}
P = \frac{N!}{N^N}\,\,\frac{n!}{n^n} \,\, \prod_{j=0}
\,\,\frac{(\overline{m_j})^{m_j}}{m_j\,!}\,\,\,,
\label{exact}
\end{equation}
where 
\begin{equation}
\overline{m_j} \equiv N \, \left(\frac{n}{N}\right)^j\, \frac{1}{j\,!}.
\end{equation}
Using Stirling's
approximation for the factorials with the further assumption $N\gg n\gg 1$, 
Eq. (\ref{exact}) can be re-written in a quasi-Poisson form
\begin{equation}
P \approx {\cal P}
\left[ \prod_{j=2} \frac{(\overline{m_j})^{m_j}}{m_j!}
e^{-\overline{m_j}\,r^j (j-2)!} \right]\,,
\label{largeN}
\end{equation}
 where $r\equiv (N-m_0)/n\approx 1$, and the prefactor ${\cal P}$ is given by
\begin{equation}
{\cal P}=e^{-(n-m_1)}\,\left(\frac{n}{m_1}\right)^{m_1 +\frac{1}{2}}\,.
\end{equation}
For ``sparse events'', where $N\gg n$, one expects the number of singlets 
$m_1$ to approximate the number of events $n$. In such a case the prefactor
is near unity. Figure~\ref{f5} shows 
the inclusive probabilities  for observing  5 doublets and one triplet 
given 58 events 
at AGASA as a function of the angular resolution.\footnote{The specified 
number of $j$-plets plus any other cluster, counts as all the $j$-plets 
+ extra-clusters.} The inclusive probability is 
extremely sensitive to the angular binning.
The chance probability within the experimental angular resolution of AGASA 
is less than $10^{-3}$, in agreement with the estimates of the AGASA 
Collaboration~\cite{Takeda:1999sg}.

The ``world'' data set has 
also been studied~\cite{Uchihori:1999gu}. Six doublets and two triplets out 
of 92 events with energies $ > 4 \times 10^{19}$~eV were found, with the 
chance 
probability being less than 1\% in the restricted region within $\pm 10^\circ$ of 
the super-Galactic plane. Very recently, the angular two-point correlation 
function of a 
combined data sample of AGASA ($E > 4.8 \times 10^{19}$~eV) and Yakutsk 
($E> 2.4 \times 10^{19}$~eV) was analyzed~\cite{Tinyakov:2001ic}. For a 
uniform 
distribution of sources, the probability of chance clustering is reported to 
be as small as $4 \times 10^{-6}$. The implication of such clusters would be profound, but the case for
them is not yet proven.  To calculate a meaningful statistical
significance in
such an analysis, it is important to define the search procedure {\it a
priori} in order to
ensure it is not (inadvertently) devised especially to suite the
particular data set after
having studied it~\cite{Clay:01}. In the above mentioned analysis, 
for instance, the angular bin size was not defined
ahead of time.

\subsection{Measurement techniques}

There are several techniques which can be employed in detecting ultrahigh 
energy CRs, ranging from direct sampling of particles in the shower
to measurements of associated fluorescence, \v{C}erenkov or radio emissions,
or possibly even radar detection of the air shower.
Direct detection of shower particles is the most commonly used method, and
involves constructing an array of sensors spread over a large area to
sample particle densities as the shower arrives at the Earth's surface.
Another well-established method involves measurement of the longitudinal
development of the EAS by sensing the fluorescence light produced
via interactions of the charged particles in the atmosphere.  A more recently
proposed technique uses radar echos from the column of ionized air
produced by the shower.  In the rest of this section,
we give an overview of the main features of these experimental techniques, and
afterwards outline the status of existing and pending ultrahigh energy CR 
experiments. More detailed and rigorous treatments of the current 
experimental situation of ultrahigh energy CRs are given in several review 
articles~\cite{Nagano:ve,Sokolsky:rz,Yoshida:1998jr,Watson:kk,Bertou:2000ip}.

\subsubsection{Surface arrays}

A surface array is comprised of particle detectors, such as plastic
scintillators or \v{C}erenkov radiators, distributed with approximately 
regular spacing. Such detectors measure the energy deposited by particles 
in the EAS as a function of time, and from energy density measured at the 
ground and the relative timing of hits
in the different detectors one can estimate the energy and direction of the
primary CR.\footnote{An assumption of axial symmetry is generally made; this
assumption is only valid for zenith angles $\theta < 60 \deg$, because at
larger angles the low energy secondaries are deflected by 
geomagnetic fields.}

\begin{figure}
\postscript{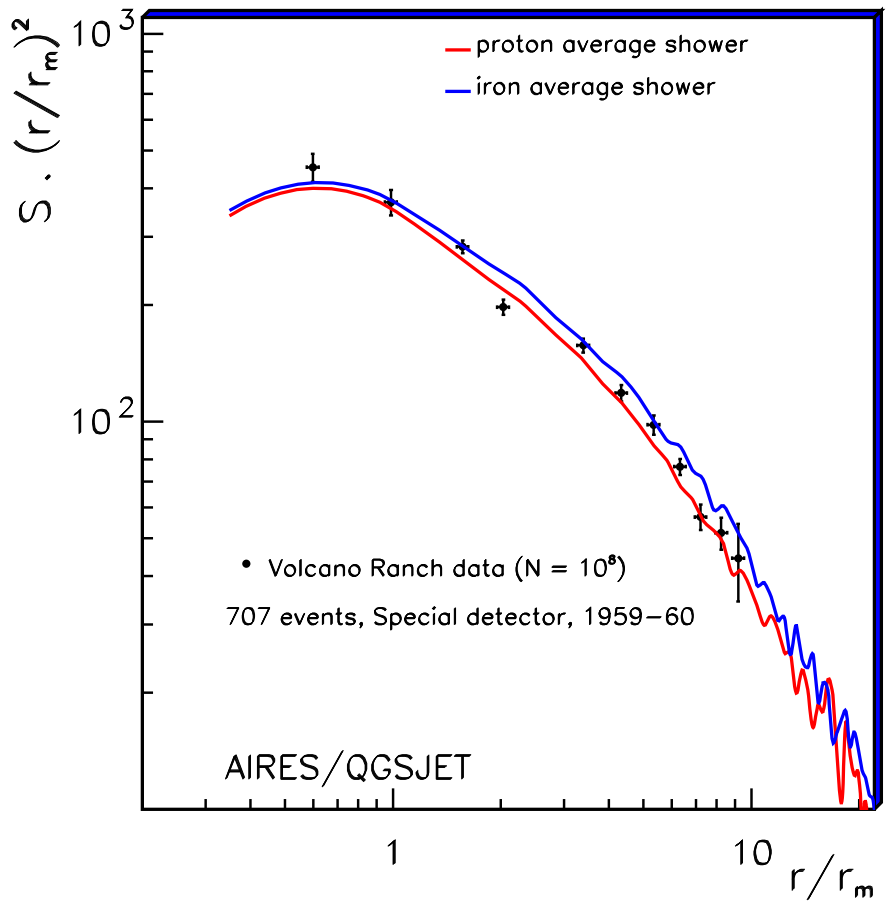}{0.80}
\caption{Average lateral distribution of simulated showers (with the program 
{\sc aires}/{\sc qgsjet}~\cite{Sciutto:1999rr}) compared to average measurements from Volcano 
Ranch~\cite{Linsley:icrc}. Here $r/r_{\rm M}$ is the distance to the shower axis in units of the Moliere 
radius (at the Volcano Ranch elevation $r_{\rm M} \approx 100$~m) and $S$ is the lateral distribution of 
particles at ground in units of minimum ionizing particles per square meter (mips/m$^2$). 
The estimated size of the showers is $10^8$ particles, which corresponds to an energy of 
about $10^{18}$~eV. Taken fron Ref.~\cite{Tom}.}
\label{LD}
\end{figure}

\begin{figure}
\postscript{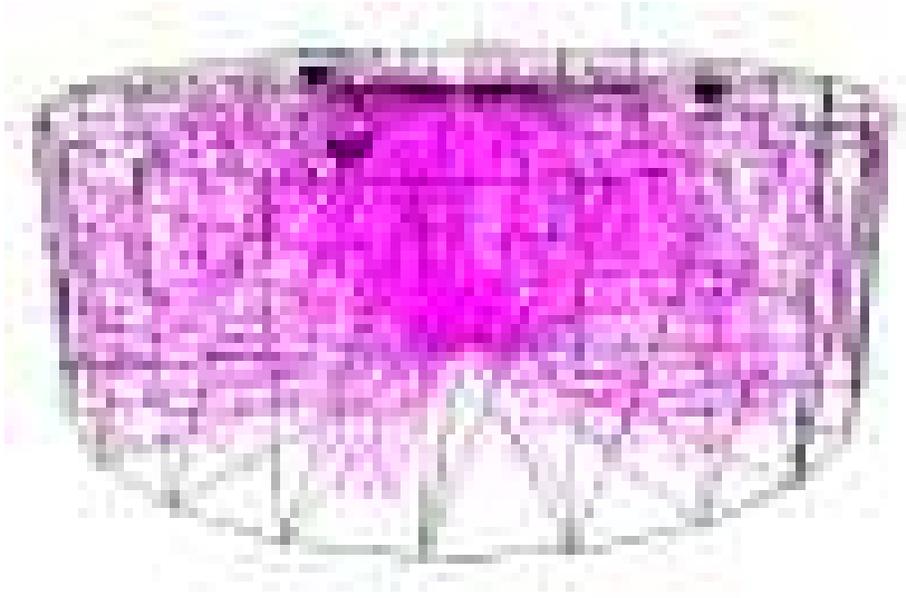}{0.80}
\caption{Simulation of a muon entering a tank of water from above and 
radiating \v{C}erenkov light. Housings for three photomultiplier tubes 
are visible at the top of the tank. This is a simulation of a Pierre 
Auger Observatory water tank~\cite{Anchordoqui:2000rb}.}
\label{tank}
\end{figure}

Reconstruction of air showers 
involves fitting the lateral distribution (LD) function of particle 
densities at the ground (see for example Fig.~\ref{LD}).
The exact form of the LD depends on how the experimental
apparatus responds to the shower particles.
The AGASA experiment, for example, employs 5~cm
thick plastic scintillators.  For energies which are typical of particles 
at about 1~km from the shower axis 
(usually called the ``core'')
these scintillators generate similar signals for the electron and muon components 
of the shower, while producing much smaller signals for the photons.
The Haverah Park and Auger experiments, on the other hand, employ 
tanks of water about a meter deep and a meter and a half in radius, as illustrated in Fig.~\ref{tank}.
In this case, \v{C}erenkov radiation produced by charged particles 
passing through the water bounces off the reflective tank walls and is picked
up by photomultiplier tubes mounted inside.
The signal size is thus proportional to the track length of the charged
particle in the water.  For energies typical of particles about 1~km from the core,
muons tend to penetrate the full depth of the tank and thus produce large signals,
whereas electrons and photons are quickly absorbed and produce signals an order of magnitude 
smaller.

Furthermore, the fraction of the energy deposited by the muon and electromagnetic components 
changes with distance from the core and has been determined experimentally. 
For the Haverah Park experiment, for instance, the 
lateral density distribution of the water \v{C}erenkov signal in 
units of fully penetrating vertical equivalent muons per m$^2$, can be fitted with a modified 
power-law~\cite{Coy}   
\begin{equation}
\rho (r) = k \, r^{-(\eta + r/4000)+\beta}\, \left(\frac{r}{800}\right)^\beta,
\end{equation}
where the distance to the shower core, $r$, is in meters, 
$k$ is a normalization parameter, $\beta = 1.03 \pm 0.05$, and $\eta$ is 
given by
\begin{equation}
\eta = 3.49 - 1.29 \sec \theta + 0.165 \log \left(\frac{E}{10^{17}\, {\rm eV}}\right)\,,
\end{equation}
with $\theta$ the shower zenith angle, and $E \agt 10^{19}$~eV the shower 
energy.

Since the response of plastic scintillator slabs to electrons, muons and
photons is somewhat different from that of deep water \v{C}erenkov radiators, a
different form of the LD function is appropriate.  For instance, in the
case of AGASA, the empirical formula for the LD of charged particles 
with $\sec \theta \leq 1.7$ reads~\cite{Yoshida:1994jf}
\begin{equation}
S(r) = C \, \left(\frac{r}{r_{\rm M}} \right)^{-\alpha}\, 
\left(1 + \frac{r}{r_{\rm M}} \right)^{-\eta + \alpha} \, 
\left[ 1 + \left( \frac{r}{1~{\rm km}} \right)^2 \right]^{-\delta}\,,  
\end{equation}
where $r$ is the distance in meters from the core, $r_{\rm M}$ is the Moliere 
radius at two radiation lengths above the observation level 
(91.6~m at Akeno), $C$ is a normalization factor, $\alpha$ and $\delta$ are 1.2 and 0.6, respectively. $\eta$ is the parameter which indicates the slope of the LD\footnote{The selection criterion  of deeply developing showers is $\eta$ of 4 to 6, which indicates a steeper lateral distribution than that of the average EAS (more on this below)~\cite{Inoue:cn}.} 
for $r \geq r_{\rm M}$ as a function of the zenith angle of the 
arrival direction, and is given by
\begin{equation}
\eta = 3.97 - 1.79 \,(\sec \theta - 1.0)\,.
\end{equation}
The primary CR energy is determined using a formula derived from simulations, 
\begin{equation}
E = 2.0 \times 10^{17} \, S_0(600)\,\,\,\,{\rm eV}\,, 
\end{equation}
where $S_0(600)$ is the charged particle density in 1/m$^2$ at 600~m from the core for a vertical equivalent shower. $S_0(600)$ is evaluated from the observed local density at 600~m,  
\begin{equation}
S_0(600) = S(600) \,\,\exp \left[ -\frac{X_0}{\Lambda_1}\, (\sec \theta - 1) -
\frac{X_0}{\Lambda_2} \, (\sec \theta -1)^2 \right]\,,
\end{equation}
where $X_0 = 920$~g/cm$^2$, $\Lambda_1 = 500$~g/cm$^2$, and 
$\Lambda_2 = 594^{+ 268}_{-120}$~g/cm$^2$. The first order approximation 
agrees with measurements for $\sec \theta < 1.4$. The error quoted in the 
second term is at the 68\% CL.

As discussed previously, the muon content of the EAS at ground level depends on
the composition of the primary CR.  Thus, surface arrays with some ability 
to distinguish muons from
electrons and photons can determine something about the primary CR species.
Water \v{C}erenkov detectors and scintillator-absorber combinations have such
distinguishing
power, for example.  It is also possible to gauge the muon content of an EAS
from the signal rise time, as the muonic component tends to be compressed in 
time compared to the electromagnetic component. Typically experiments measure 
something like the time interval between the arrival of the 10\% and 50\% 
points of the integrated signal $(t_{10}/t_{50})$.

\subsubsection{Fluorescence eyes}

As an EAS develops in the atmosphere it dissipates much of its energy by
exciting and ionizing air molecules along its path.  Excited nitrogen
molecules fluoresce producing ultraviolet radiation.
The shower development appears as a rapidly moving spot of light
describing a great circle path across a night-sky background of starlight,
atmospheric airglow, and man-made light pollution. The angular motion of
the
spot depends on both the distance and the orientation of the shower axis.
The apparent brightness of the spot depends on the instantaneous number of
charged particles present in the shower, but it is also affected
by \v{C}erenkov contamination and atmospheric scattering.

The fluorescence trail is emitted isotropically with an intensity that is
proportional to the number of charged particles in the shower,
$N_e$.\footnote{We stress that $N_e$ denotes the charged multiplicity, but
in practice it is often used to express the number of electrons and
positrons
since they completely dominate the total number of charged particles.}
The ratio of the energy emitted as fluorescence light to the total energy
deposited is  less than 1\%, hence low energy ($< 10^{17}$~eV)
showers can hardly be observed.  Furthermore,
observations can only be done on clear moon-less nights, resulting in
an average 10\% duty cycle.
The emitted light is typically in the 300 - 400 nm ultraviolet
range to which the atmosphere is quite transparent. Under favorable
atmospheric conditions EASs can be detected at distances as large as
20~km,
about 2 attenuation lengths in a standard desert atmosphere at ground
level.

A fluorescence eye consists of several large light collectors (or
telescopes)
which image regions of the sky onto clusters of light sensing and
amplification devices. The basic elements of a telescope are the
diaphragm,
which defines the telescope aperture, the spherical mirror that must be
dimensioned to collect all the light entering the diaphragm in the
acceptance
angular range, and the camera which consists of an array of
photomultiplier
tubes (PMTs) positioned approximately on the mirror focal
surface~\cite{Matthiae}.\footnote{A potential future development could
involve the use of avalanche photodiodes (APDs) instead of PMTs. APDs have
several
advantages~\cite{Musienko:jj}.}  The PMTs effectively pixelize the region
of the
sky covered by the telescope.  The shower
development is detected as a long, rather narrow sequence of hit PMTs.
As the
point-like image of the shower proceeds through an individual PMT, the
signal
rises, levels off, and falls again. The collection of vectors pointing
from the hit PMTs defines the shower detector plane (SDP).

The sensitivity of the detector depends primarily on the signal ($S$) to 
noise ($N$) ratio. The signal is proportional to the diaphragm area, whereas 
the background ($B$) is proportional to the pixel (the area read out by a 
single PMT) solid angle times the diaphragm area, thus
\begin{equation}
\frac{S}{N} = \frac{S}{\sqrt{B}} \propto \frac{d_{\rm dph}}{\alpha_{\rm pix}}\,
\end{equation}
where $d_{\rm dph}$ is the diaphragm diameter, and $\alpha_{\rm pix}$ is the 
angular diameter of a pixel.

\begin{figure}
\postscript{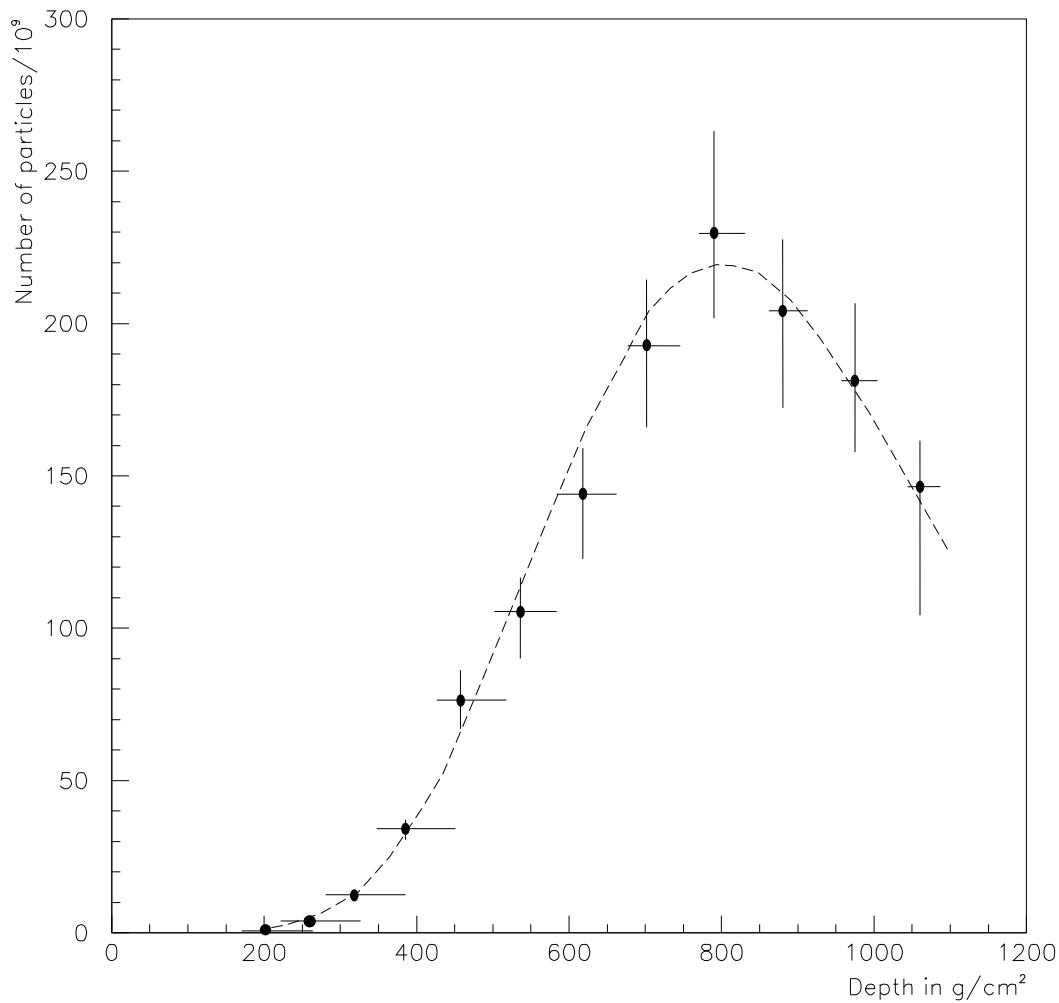}{0.90}
\caption{The 3-parameter best fit shower profile for the highest energy
event
observed by the Fly's Eye experiment. Published in~\cite{Bird:1994uy}.}
\label{fe}
\end{figure}

Shower reconstruction involves first determining the geometry
of the shower, and then reconstructing the shower's longitudinal
profile.  The first step of the geometrical reconstruction is the
SDP determination from the PMT hit pattern.  Next, the tube hit times
are used to find the shower impact parameter and incident angle
in the SDP; the resolution obtained for these parameters depends
on the length of the observed track~\cite{Bird:icrc}. Once the geometry 
is known, the track can be chopped
into angular bins and the longitudinal profile can be determined.
Extracting the profile usually involves 3-parameter fits to the
Gaisser-Hillas
function~\cite{Gaisser:icrc},
\begin{equation}
N_e (X) = N_{e,{\rm max}} \left(
\frac{X - X_0}{X_{\rm max}-X_0}\right)^{[(X_{\rm max} - X_0)/\lambda]}\,\,
\exp\left\{\frac{X_{\rm max} - X}{\lambda}\right\},
\,\,\, X \geq X_0,
\end{equation}
where
$N_{e,{\rm max}}$ is the size at the maximum, $X_0$ is the depth of the
first observed interaction, and $\lambda = $70 g/cm$^2$.  The integral of
the
longitudinal profile is a calorimetric measure of the total
electromagnetic
shower energy. A charged particle in the cascade
deposits an average of 2.2 MeV into the atmosphere in each depth interval
of 1 g/cm$^2$~\cite{Longair:jc}, so the total electromagnetic energy (in
MeV)
is given by
\begin{equation}
E_{\rm em} = 2.2 \int N_e(X)\,\, dX\,\,.
\end{equation}
The largest cosmic ray air shower ever recorded has an estimated energy of
$(3.2 \pm 0.9) \times 10^{20}$~eV, reaching the maximum size near a depth
of 815 g/cm$^2$~\cite{Bird:1994uy}. The longitudinal development is shown
in
Fig.~\ref{fe}. The size at maximum is greater than 200 billion
particles!

\subsubsection{Radio detection techniques}
 
In addition to the ultraviolet emission exploited by fluorescence eyes,
it may also be possible to call into service the radio frequency (RF) energy
generated by air-showers.  CR showers may induce radio pulses through
several mechanisms, though it is thought that from about 20-100 MHz,
the dominant process can be described as coherent synchrotron emission
by the electron and positron pairs propagating in the Earth's magnetic
field~\cite{Kahn:1966,Wilson:1967}.
 
RF pulses coincident with EASs were first measured in
the 1960's~\cite{Jelley:1965}, and these early results were followed up
with additional work over another decade or so.  Subsequently the technique
was mostly abandoned in light of the promising results from surface    
arrays and fluorescence eyes,
though quite recently a few groups have undertaken a new examination of the
method.
For example, a prototype RF detection
system using a
single antenna in conjunction with the buried muon detectors of the
CASA-MIA array has recently been evaluated, with the goal of 
demonstrating the feasibility of
incorporating such an RF receiver into
an existing experiment like the Pierre Auger Observatory~\cite{Green:2002ad}.
Though no significant EAS-induced pulses were observed, a number of
technical issues relevant to future development were uncovered.
 
More ambitious possibilities have been recently
described~\cite{Falcke:2002tp} in the context of next-generation digital
telescopes.  In particular, the planned Low-Frequency Array
(LOFAR)~\footnote{{\tt http://www.lofar.org}}, which could begin  
construction as early as 2004, would be well-suited to study RF
emissions from EASs.  This device
will comprise around 100 stations of 100 dipoles distributed over a
region some 400 km in radius.
A cluster of computers will correlate data from the antennae to
synthesize ``virtual telescopes'' which
look in any desired direction.  The CR energy range for which one
achieves both a reasonable signal-to-noise
ratio and a reasonable event rate depends on the number of dipoles
actually employed.
A single LOFAR-type station of about 100 dipoles, for example, would be
useful for measurements in
the range $10^{15}$--$10^{17}$~eV; indeed just such a prototype station,
known as LOPES~\cite{Huege:2002bh}, 
is scheduled to be operational in 2003 at the KASCADE experiment.
Falcke and Gorham~\cite{Falcke:2002tp} estimate that LOFAR should be able
to observe events up to $10^{20}$~eV   
at a rate of order 1 per year.
 
In addition to such passive radio techniques, it may also be possible to
detect radar
reflections off the ionization columns produced by EASs.  This idea was
suggested already in 1940~\cite{Blackett}, and recently the issue
has been re-explored~\cite{Gorham:2000da,Gorham:2000fp} as either an
independent method to study air showers, or as
a compliment to existing fluorescence and surface detectors.
For example, one could use a fluorescence detector to trigger
a radar system to interrogate in the approximate direction of the
shower.  It has been estimated that showers from
primaries of order $10^{18}$~eV should be detectable~\cite{Gorham:2000da}, 
and that range estimates
with precision around 20~m may be possible.  The technique would
work best for distant or horizontal showers, the latter being of
particular interest in searches for deeply penetrating showers induced
by neutrino primaries.  One important uncertainty is the lifetime of
the ionization trail, which decays away due to electron
recombination and attachment and can thus limit the time available
for radar interrogation; these effects depend on altitude and
atmospheric conditions.  A proposal has recently been put forth
to evaluate the method using the Jicamarca radar system near Lima,
Peru~\cite{Vinogradova:2000fr}.

\subsection{CR-observatories: past and present}

The first measurements of ultrahigh energy CRs 
were carried out by Linsley at Volcano Ranch 
($35^\circ 09'$N, $106^\circ47'$W) in
the late 1950's~\cite{Linsley:prl,Linsley:pr}. The detector comprised an array of 19 scintillation counters each of 
3.3~m$^2$, spaced on a 884~m hexagonal grid. More recently, experiments 
have been 
performed at Haverah Park in England ($53^\circ 58'$N, $1^\circ38'$W), 
Yakutsk in Russia ($62^\circ$N, $130^\circ$E), Sydney in 
Australia ($30^\circ 32'$S, $149^\circ 43'$E), Dugway in the 
Utah desert ($40^\circ$N, $112^\circ$W), and near the 
village of Akeno, about 100~km west of Tokyo ($38^\circ47'$N, 
$138^\circ 30'$E).

The Haverah Park detector~\cite{Lawrence:cc} ceased operation in July 1987 
after a life of 19 
years. The detector array consisted of deep (1.2~m) water \v{C}erenkov 
detectors with stations ranging in area from 1 to 54 m$^2$ 
(typically 34 m$^2$) distributed over an area of approximately 12 km$^2$. 
Restrictions on land access made it impossible to position the detectors on a 
uniform grid. Therefore, a 4 detector array with 500~m spacing was deployed at 
at the center and 6 sub-arrays of 50~m and 150~m spacing were located at about
2~km from the center. In the ongoing analyses 
of the data,  primary energies are 
derived from the water \v{C}erenkov detector density at 600~m from the shower 
axis, $\rho(600)$. Random errors in the determination of $\rho(600)$ are less 
than 30\%~\cite{Cunningham}. The energy threshold was approximately 
$6 \times 10^{16}$~eV. 
The angular resolution of the array was measured to be: for zenith 
angle ($\theta$), $2.5^\circ \times \sec \theta$ with $\theta 
\in (0^\circ, 75^\circ)$, whereas for azimuth angle ($\phi$), $2.5 \times {\rm cosec}\, \theta$ 
with $\theta \in (15^\circ, 90^\circ)$~\cite{Edge:rr}.

The Yakutsk array, in operation since 1972, was expanded in 1974 for 
sensitivity to ultrahigh energy CRs, and spread over a ground area of 
approximately 
18~km$^2$. Most of the detectors are plastic scintillators arranged on a 
triangular grid. The spacing of the detectors was around 1 km, with the center 
of the array filled with 7 detectors on a 500~m triangular 
grid~\cite{Khristiansen}. 
Underground stations measured the muon flux, with an energy  threshold 
of 1~GeV. The primary energy is estimated from the particle density 
at 600~m from the shower core. In 1995, the array was 
re-arranged into a smaller area ($\sim 10$~km$^2$) in order to investigate in 
detail the spectrum around $10^{19}$~eV~\cite{Glushkov:au}.

The Sydney University Giant Air-shower Recorder (SUGAR)~\cite{Winn:un}
operated between 1968 and 1979, and is the largest array to date in the 
Southern 
hemisphere. 47 detector stations were arranged to cover an area of 
approximately 70~km$^2$. Each of these stations consisted of two 6~m$^2$ conical liquid scintillator tanks separated by 50~m. The angular 
resolution of the array is quoted~\cite{Winn:un} to be 
$3^\circ \times \sec \theta$ for showers detected by more than three 
stations. However, most of the events were only viewed by 3 stations, and 
consequently the resolution is closer to $6^\circ$ for vertical 
showers~\cite{Clay:aa}. 

The Akeno Giant Air Shower Array (AGASA)~\cite{Chiba:1991nf,Ohoka:ww} 
consists of 111 scintillation detectors each of area 2.2 m$^2$,
spread over an area of $100$~km$^2$ with 1 km spacing. 
This gives an acceptance of $125$~km$^2$ sr for CRs above 
$10^{19}$~eV.\footnote{The array has been 
divided into 4 branches for the purpose of trigger and data acquisition.  
These branches are called the ``Akeno Branch'', the ``Sudama 
Branch'', the ``Takane Branch'', and the ``Nagasaka Branch''.
The 20 km$^2$ array~\cite{Teshima:nim}, operational since the end of 1984, 
forms a part of the ``Akeno Branch''.}
The array detectors are connected and
controlled through a sophisticated optical fiber network. The array
also contains a number of shielded scintillation detectors which
provide information about the muon content of the showers. The full
AGASA experiment has been running since 1992.  

The fluorescence technique has so far been implemented only in the Dugway 
desert by the Fly's Eye group from the University of Utah. Following a 
successful trial at Volcano Ranch~\cite{Bergeson:nw} the Utah group built 
a device containing 
two separated Fly's Eyes made up of 880 and 460 PMTs, 3.3~km 
apart~\cite{Baltrusaitis:mx,Baltrusaitis:ce}. The two-eye configuration 
monitored the sky from 1986 until 1993. 
The first detector, Fly's Eye I, had full operation since 1981. It was made 
of 67 telescopes of $d_{\rm dph}= 1.5$~m and spherical curvature, each with 12 
or 14 PMTs at the focus. The mirrors were arranged so that the 
entire  night sky was imaged, with each PMT viewing a hexagonal 
region of the sky with $\alpha_{\rm pix}=  5^\circ$. Fly's Eye II 
was made of of 
36 mirrors of the same design. This detector viewed only half of the night 
sky. 

As an up-scaled version of Fly's Eye, the High Resolution (HiRes) Fly's Eye 
detector has recently begun operations~\cite{Corbato:fq}. 
It uses  14 spherical telescopes of $d_{\rm dph}= 2.0$~m  to collect the light 
from a 0.95 sr portion of the sky. The image plane of each telescope is 
populated with an array of 256 hexagonal PMTs, 
yielding $\alpha_{\rm pix} = 1^\circ$. 
In monocular mode, the effective 
acceptance of this instrument is $\sim 350 (1000)$~km$^2$~sr at 
$10^{19}\, (10^{20})$~eV, on average about 6 times the Fly's Eye acceptance, 
and the threshold energy is $10^{17}$~eV. This takes into account a duty 
cycle of 
about 10\% typical of the fluorescence technique. 
The field of view of 
the telescopes is centered on the Chicago Air Shower Array (CASA) and the 
Michigan Muon Array 
(MIA)~\cite{Borione:iy}, situated 3.3 km to the northeast. The combination 
of these instruments has been used as a prototype hybrid detector in which 
HiRes 
records the development profile, CASA records the ground particle density 
and MIA detects the muonic component of a common shower.

\subsection{The potential of the Pierre Auger Observatory}

The Pierre Auger Observatory (PAO)~\cite{Zavrtanik:zi,Beatty:ff} is 
designed to work in 
a hybrid mode, employing fluorescence detectors overlooking 
a ground array of deep water \v{C}erenkov radiators. 
During clear, dark nights,  events 
will be simultaneously observed by 
fluorescence light and particle detectors at ground level.
The PAO is expected to measure the energy, arrival direction and 
primary species with unprecedented statistical precision. It will eventually 
consist of two sites, one in the Northern hemisphere and one in the Southern, 
each covering an area of $3000~\km^2$ and consisting of 1600 particle detectors
overlooked by 4 fluorescence detectors. The overall acceptance (2 sites) is
14000~km$^2$ sr. A prototype engineering array, 1/40th-scale, is (at the time of writing) already taking data 
in Malarg\"ue, Argentina ($35^\circ 12'$S, $69^\circ 12'$W)~\cite{Allekote}.

The surface array stations are cylindrical water \v{C}erenkov detectors 
with area 10 m$^2$, spaced 1.5 km from each other in a hexagonal grid.  
\v{C}erenkov radiation emitted by charged particles penetrating the detector 
is read out by 3 PMTs. The output signal is digitised 
by flash ADCs, with the aim of separating the 
electromagnetic signal (low energy electrons and photons) from the muons 
crossing the tank. Event
timing is made possible via global positioning system (GPS)
satellites with a precision of a few tens of ns. Communication between the 
stations is achieved using radio signals by methods similar to 
cellular telephone techniques. The stations are powered by solar panels 
and batteries which allow autonomous operation.

The configuration of the fluorescence detectors is arranged to optimize the 
hybrid detector performance. The number of eyes and their location 
on the site are chosen so that all showers of energy $> 10^{19}$~eV that hit 
the surface array are seen by at least one eye. A further constraint comes 
from the need to limit the systematic error in the measurements deriving 
from  uncertainty in the attenuation length of the atmosphere traversed by 
the light in its path from shower to detector. The optimal configuration 
was determined by Monte Carlo simulation, guided by the orographic 
constraints on the site. The base-line design of the detector 
includes 4 fluorescence eyes, each comprised of six telescopes
with $d_{\rm dph} = 1.7$~m and $\alpha_{\rm pix} = 1.5^\circ$~\cite{Cester}. 

The angular and energy resolution of the ground array (without coincident 
fluorescence data) are typically less than $1.5^\circ$ and less than 
20\%, respectively. ``Golden events,'' events detected by
both methods simultaneously, will have a directional reconstruction 
resolution of about $0.3^\circ$ for energies near $10^{20}$~eV. If an 
event trigger is 
assumed to require 5 detectors above threshold, the array is fully efficient 
at $10^{19}$~eV. In three years of running, the surface arrays in both 
hemispheres, operating 24 hours per day, will collect more than 1000 showers 
above $4 \times 10^{19}$~eV with approximately uniform sky 
exposure. This will enable a straightforward search for correlations with discrete 
sources and also a sensitive large scale anisotropy analysis.

In addition, PAO offers a window for neutrino astronomy 
above $10^{17}$~eV. For standard neutrino interactions in the atmosphere, 
each site of PAO reaches $\sim 15$~km$^{3}$~w.e.~sr of target mass
around $10^{19}$~eV~\cite{Capelle:1998zz}, which is comparable to other 
neutrino detectors being planned.\footnote{w.e. $\equiv$ water equivalent.} 
An even greater acceptance~\cite{Letessier-Selvon:2000kk} should be 
achievable for the case of 
Earth-skimming neutrinos which produce a $\tau$, as we discuss in 
Sec.II-I.

\subsection{The Telescope Array Project}

The Telescope Array will comprise a collection of fluorescence detectors
dispersed over a large area near Salt Lake City, Utah~\cite{Teshima:2000xz}.
Ten observational stations separated from each other by $30-40$~km are
planned, each station containing 40 telescopes 
of $d_{\rm dph} = 3$~m. An imaging camera with 256 PMTs 
($\alpha_{\rm pix} = 1^\circ$) will be installed on the focal 
plane of each telescope. The effective aperture of the array will be 
approximately 5000~km$^2$~sr for $10^{20}$~eV particles 
assuming a 10\% observation duty factor.  This aperture is around 
30 times larger than 
the existing AGASA ground array. The energy, arrival direction, and 
shower maximum  will be determined with an accuracy of 6\%, 0.6$^\circ$, and 
20~g/cm$^2$, respectively. In addition, the Telescope Array will observe 
high energy gamma rays from point sources in the sub-TeV energy region. 
Gamma rays will be distinguished from the large hadronic background using the 
imaging patterns observed with many telescopes.

\subsection{200?: A space odyssey} 

Recently, NASA initiated a concept study for space-based detectors which will 
stereoscopically image, from equatorial orbit, the nitrogen 
fluorescence light generated by EASs induced by ultrahigh energy ($>\, {\rm 
few}\, \times 10^{19}$~eV) CRs. The Orbiting Wide-angle 
Light-collectors (OWL)~\cite{Krizmanic:qn} mission will involve photodetectors 
mounted on 2 satellites orbiting at 640~km above the Earth's surface. The 
eyes of the OWL will monitor a large atmospheric volume and record the 
ultraviolet fluorescent trails with a time resolution of $\sim 1~\mu$s or 
less in segmented focal plane arrays. The segmentation of the arrays will be 
such as to sample the atmosphere near the Earth's surface in $\sim 1$~km$^2$ 
pixels. The use of a space-based platform enables 
an extremely large event acceptance, allowing a high statistics 
measurement. OWL is set for possible implementation after 2007. 

Two different baseline instruments have been proposed to achieve wide-field of views. The first is a refractive design using two Fresnel lenses 
which focus onto a large focal plane array~\cite{Lamb:aip}. The second design 
uses Fresnel correcting optics which focus onto a spherical reflector in a Maksutov design that in turn focuses the light onto a focal plane 
array~\cite{Lamb,Pitalo}. 
For a satellite separation of 500 (2000)~km, the instantaneous acceptance is 
$1.5  \times 10^{6}$ $(3.75  \times 10^{6})$~km$^2$~sr. 
Using a 10\% duty factor, the 500 (2000)~km configuration leads to an 
effective acceptance of  
$1.5 \times 10^{5}$ $(3.75 \times 10^{5}$)~km$^2$~sr. Therefore, assuming 
a continuation of the CR-spectrum  $\propto E^{-2.75}$, one expects 
rates of $1500$ events/yr (500 km) and $3750$ 
events/yr (2000~km) for $E\agt 10^{20}$~eV~\cite{Krizmanic}.
The effective acceptance for the Maksutov baseline assuming 
1000~km orbits and 500 (2000)~km is $2.0 \times 10^5$ ($4.0 \times 
10^5$)~km$^2$~sr which leads to an event rate of 2000 (4000) events per year
for $E \agt 10^{20}$, again assuming an $E^{-2.75}$ spectrum 
continuation~\cite{Krizmanic}.  

For the near future ($\approx 2006$), the European Space Agency is studying 
the feasibility of placing a single eye on the International Space Station, 
which will serve as a pathfinder mission to develop the required technology 
to observe the fluorescent trails of EASs. The smaller 
viewing volume of the Extreme Universe Space Observatory 
(EUSO)~\cite{Catalano:mm} will result in a smaller event rate by a 
factor of $\sim 5$ compared to OWL. 

\subsection{Ultrahigh energy neutrino experiments} \label{sec:neutrinos}
Up to now, we have discussed experiments designed primarily to 
study air showers initiated by hadrons or photons.  As pointed out
in Sec.II-B, however, neutrinos may also
induce extensive air showers, so current and future air shower
experiments might also function as neutrino detectors. 
In this section, we briefly review dedicated neutrino detectors 
as well as neutrino searches carried out by existing air shower experiments.

The traditional technique to observe high energy neutrinos 
involves detecting the optical \v{C}erenkov light emitted by muons produced in 
charged current interactions of neutrinos with nucleons either in ice or 
water. The largest pilot experiments ($\sim 0.1$~km in size) are: 
the now defunct DUMAND (Deep Underwater Muon and Neutrino Detector)  
experiment~\cite{Roberts:re} in the deep sea near Hawaii, the
underwater experiment in Lake Baikal~\cite{Balkanov:1999xh},
and AMANDA (Antarctic Muon And Neutrino Detector Array )~\cite{Kowalski:2001th} in the South Pole ice. Next generation neutrino telescopes 
aim towards an active volume in the range of 1~km$^{3}$ of water.
Projects under construction or in the proposal stage are: two deep sea 
experiments in the Mediterranean, the French ANTARES (Astronomy with a 
Neutrino telescope Abyss environment RESearch)~\cite{Aslanides:1999vq} 
and NESTOR (Neutrino Experiment SouthwesT Of 
GReece~\cite{Grieder:ha}), and ICECUBE~\cite{Alvarez-Muniz:2001gb}, 
a scaled up version of the AMANDA detector.

Above $10^{17}$~eV the neutrino interaction length is below 2000~km w.e. in 
rock, and so upward going neutrinos are typically blocked by the Earth. 
This shadowing severely restricts the high energy event rates in underground detectors. 
Current limits on high energy 
($10^{15} - 10^{17}$~eV) neutrino fluxes come from measurements of the 
Extensive Air Shower array on the TOP of the 
underground Gran Sasso Laboratory in central Italy 
(EAS-TOP)~\cite{Aglietta:ex,Aglietta:ny}, and the Fr\'ejus 
detector~\cite{Berger:1987ke,Rhode:es}, located in an underground 
laboratory near the middle of the road tunnel 
connecting Modane (France) and Bardonecchia (Italy) in the Alps. 

Large ground arrays or fluorescence eyes, for which the interaction 
medium is not the Earth but the atmosphere, are complementary to neutrino 
telescopes in the analysis of the energy spectrum. Specifically, 
the neutrino cross section at ultrahigh energies is non-negligible 
(about $10^{-32}$~cm$^2$ at $10^{18}$~eV~\cite{Gandhi:1998ri}), 
and so quasi-horizontal ($\theta > 75^\circ$) neutrinos that traverse 
an atmospheric depth of up to 360~m~w.e. can initiate a shower. As  
discussed earlier, hadronic showers at large
zenith angles have their electromagnetic component 
extinguished, and only high energy muons created in the first stages of the 
shower development survive. Therefore, the shape of the shower front is 
relatively flat. The Fly's Eye Collaboration searched for deeply developing 
air showers (DDASs) and upward-moving showers above $10^{17}$~eV and no 
unusual 
events were found in $10^{6}$~s of running time~\cite{Baltrusaitis:mt}. 
Additionally, the AGASA  
group searched for giant deeply developing air showers~\cite{Inoue:cn}. 
During the observation live time of $9.7 \times 10^{7}$~s no candidate 
DDAS was found. From this, the AGASA Collaboration concluded that the 
90\% CL upper limit on the flux of DDASs with  energy  
$\agt 10^{19.5}$~eV is $1.9 \times 10^{-16}$ m$^{-2}$ s$^{-1}$ sr$^{-1}$, 
a factor of 10 lower than the flux of cosmic rays above $10^{19.5}$~eV. The 
current upper bound ($E_{\nu} \sim 10^{19.5}$~eV) on the total cosmic 
neutrino flux, obtained by combining the exposures of the AGASA and 
Fly's Eye experiments (the latter integrated over all its operating epochs), 
is found to be $5.1 \times 10^{-12}$ m$^{-2}$ s$^{-1}$ sr$^{-1}$,  
at the 95\% CL~\cite{Anchordoqui:2002vb}. The sensitivity of surface arrays 
and fluorescence eyes 
could be significantly enhanced by triggering on neutrinos that skim the 
Earth, traveling at low 
angles along chords with lengths of order their interaction 
length~\cite{Domokos:1997ve,Domokos:1998hz,Fargion:2000iz,Fargion:2000rn,Bertou:2001vm,Feng:2001ue,Kusenko:2001gj}. Some of 
these Earth-skimming neutrinos may be converted into tau leptons in the 
Earth's crust. Unlike electrons, which do not escape from rocks, or muons, 
that do not produce any visible signal in the atmosphere, taus can escape 
even from inside the rock and produce a clear signal if they decay above the 
detector, increasing the $\nu$-event rate. 

The study of radio pulses from electromagnetic 
showers created by neutrino interactions in ice would provide an increase in 
the effective area up to $10^{4}$~km$^2$. A prototype of this technique is 
the Radio Ice \v{C}erenkov Experiment (RICE)~\cite{Kravchenko:2001id}. 
Similar concepts are  used by  the Goldstone Lunar ultrahigh energy 
neutrino Experiment (GLUE) to set an upper bound 
on the ultrahigh energy neutrino flux~\cite{Gorham:1999cr,Gorham:2001aj}. 
In this experiment, the non-observation of microwave \v{C}erenkov pulses 
from electromagnetic showers induced by neutrinos 
interacting in the moon's rim leads to flux upper limits: 
$\log (E^2 J)  = -3.03,$ $-2.66,$ and 
$-2.30$~GeV cm$^{-2}$~s$^{-1}$~sr$^{-1}$, for $1.0 \times 10^{22},$ 
$3.0 \times 10^{22},$ and $1.0 \times 10^{23}$~eV, respectively. 

For a more extensive discussion on the status of ultrahigh energy neutrino 
experiments the reader is referred to~\cite{Halzen:nu}.

\section{Cosmic ray acceleration}

It is most likely that the bulk of the cosmic radiation is a 
result of some very general 
magneto-hydrodynamic (MHD) phenomenon in space which transfers kinetic or 
magnetic energy into cosmic ray energy. The details of the acceleration 
process and the maximum attainable energy depend on the particular physical 
situation under consideration. There are basically two types of mechanism 
that one might invoke for bottom-up CR production.\footnote{In bottom-up 
models the CR particles start with low energy and are accelerated, as 
opposed to top-down models where usually exotic particles start initially 
with very high energy and cascade decay to the CR particles.} 
The first type assumes the particles are accelerated directly to high energy 
by an extended 
electric field~\cite{Hillas:1984}. This idea can be traced back to the 
early 1930's when 
Swann~\cite{Swann} pointed out that betatron acceleration may take place
in the increasing magnetic field of a sunspot. These so-called ``one-shot'' 
mechanisms have been worked out in greatest detail, and the electric field 
in question is now generally associated with the rapid rotation of small, 
highly magnetized objects such as neutron stars (pulsars) or active 
galactic nuclei (AGN). Electric field acceleration has the advantage of 
being fast, but 
suffers from the circumstance that the acceleration occurs in 
astrophysical sites of very high energy density, where new opportunities for 
energy loss exist. Moreover, it is usually not obvious how to obtain the 
observed power law spectrum in a natural way, and so this kind of 
mechanism is not widely favored these days. The second type of process is 
often referred to as statistical acceleration, because 
particles gain energy gradually by numerous encounters with moving magnetized 
plasmas. These kinds of models were mostly pioneered
by Fermi~\cite{Fermi}. In this case the $E^{-2}$ spectrum very convincingly 
emerges.\footnote{A point worth noting at this juncture: A power law spectrum 
does not necessarily point to the Fermi mechanism~\cite{Anchordoqui:2002xu}.} 
However, the process of acceleration is slow, and it is hard to 
keep the particles confined within the Fermi engine. 
In this section we first provide a summary of statistical acceleration  
based on the simplified version given in Ref~\cite{Gaisser:vg} 
(See also~\cite{Protheroe:1998hp}). For a more detailed and rigorous 
discussion the reader is referred to~\cite{Blandford:pw}. 
After reviewing statistical acceleration, we turn to 
the issue of the maximum achievable energy within diffuse 
shock acceleration and explore the viability of some proposed ultrahigh energy 
CR-sources.

\subsection{The Fermi mechanism}

In his original analysis, Fermi~\cite{Fermi} considered the scattering of CRs 
on moving 
magnetized clouds. In this case, a CR entering into a single cloud 
with energy $E_i$ and incident angle $\theta_i$ with the cloud's direction 
undergoes diffuse scattering on the 
irregularities in the magnetic field. 
After diffusing inside the cloud,
the particle's average motion coincides 
with that of the gas cloud. The energy gain 
by the particle, which emerges at an angle $\theta_f$ with energy $E_f$, 
can be obtained by applying Lorentz transformations
between the laboratory frame (unprimed) and the cloud frame (primed).
In the rest frame of the moving cloud, the CR particle has a total initial 
energy
\begin{equation}
E_i' = \Gamma \,E_i\, (1 - \beta\, \cos \theta_i)\,, 
\end{equation}
where $\Gamma$ and $\beta = V/c$ are the Lorentz factor and velocity
of the cloud in units of the speed of light, respectively.   
In the frame of the cloud we expect no change in energy ($E_i' = E_f'$), 
because all the scatterings inside the cloud are due only to motion in the 
magnetic field (so-called collisionless scattering). There is elastic 
scattering between the CR and the cloud as a whole, which is much more 
massive than the CR. Transforming to the laboratory frame
we find that the energy of the particle after its encounter with the cloud is
\begin{equation}
E_f = \Gamma \,E_f'\, (1 + \beta \cos \theta_f)\,.
\end{equation}
The fractional energy change in the 
laboratory frame is then
\begin{equation}
\frac{\Delta E}{E} = \frac{E_f-E_i}{E_i} = \frac{1 - \beta \cos \theta_i + \beta \cos \theta_f' - \beta^2 \cos \theta_i \cos \theta_f'}{1 - \beta^2} - 1 \,.
\label{flash0}
\end{equation}
Inside the cloud the CR direction becomes randomized and so 
$\langle \cos \theta_f'\rangle = 0.$ The average value of 
$\cos \theta_i$ depends on the relative 
velocity between the cloud and the particle. The 
probability $P$ per unit solid angle $\Omega$ of having a collision at angle 
$\theta_i$ is proportional to $(v - V \cos \theta_i)$, where $v$ is the CR 
speed. In the 
ultrarelativistic limit, i.e., $v \sim c$ (as seen in the laboratory frame),
\begin{equation}
\frac{dP}{d \Omega_i} \propto (1 - \beta \cos\theta_i)\,,
\end{equation}
so
\begin{equation}
\langle \cos \theta_i \rangle  = -\frac{\beta}{3}\,.
\label{flash}
\end{equation}
Now, inserting Eq.~(\ref{flash}) into Eq.(\ref{flash0}) one obtains for 
$\beta \ll 1$, 
\begin{equation}
\frac{\langle\Delta E\rangle}{E} = \frac{1 +\beta^2/3}{1-\beta^2} - 1 \approx \frac{4}{3} \,\beta^2\,.
\label{ja}
\end{equation}
Note that $\langle \Delta E \rangle/E \propto \beta^2$, so even though the 
average magnetic 
field may vanish, there can still be a net transfer of the macroscopic kinetic 
energy 
from the moving cloud to the particle. However, the average energy gain is very small, because $\beta^2 \ll 1$.

\begin{figure}
\postscript{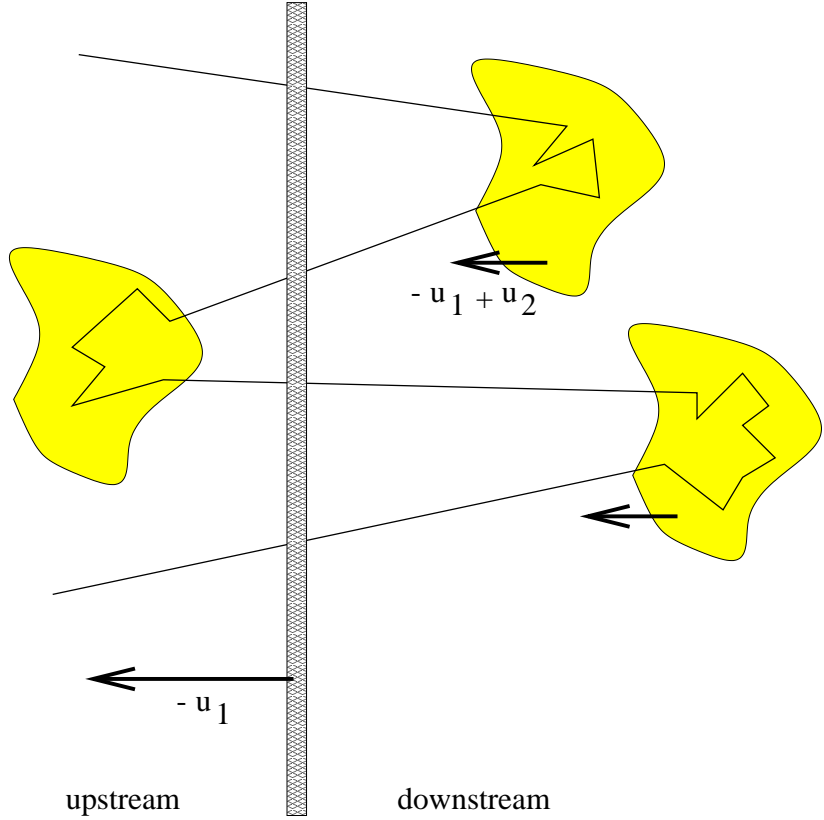}{0.70}
\caption{CR acceleration at a shock front.  A
planar shock wave is moving with velocity $-u_1$.  A
CR particle is repeatedly crossing the front and scattering
in magnetic irregularities.}
\label{shock}
\end{figure}

A version of Fermi's mechanism which is first order in $\beta$,
and thus a more efficient accelerator, is realized for CR encounters
with plane
shock fronts~\cite{Blandford:ky}.
In this case, a large shock wave propagates
with velocity $-\vec{u_1}$, as depicted in Fig.~\ref{shock}.
Relative to the shock front, the downstream shocked gas is receding
with velocity $\vec{u_2}$, where $\left| u_2 \right| < \left| u_1 \right|$,
and thus in the laboratory frame it is moving in the direction
of the front with velocity $\vec{V} = -\vec{u_1} + \vec{u_2}$.
In order to find the energy gain per shock crossing, we identify the
magnetic
irregularities on either side of the shock as the clouds of
magnetized plasma in the Fermi mechanism we discussed previously.      
By considering the rate at
which CRs cross the shock from downstream to upstream,
and upstream to downstream, one finds $\langle \cos \theta_{i}
\rangle = -2/3$ and $\langle \cos \theta_{f}^{\prime} \rangle =
2/3$. Hence, Eq.~(\ref{flash0}) can be re-written as
\begin{equation}
{\langle \Delta E \rangle \over E}  \simeq {4 \over 3} \beta = \frac{4}{3} 
\frac{u_1 -u_2}{c}.
\end{equation}
Note this is first order in $\beta=V/c$, and is therefore more
efficient than Fermi's original mechanism.  This is because of the
converging flow -- whichever side of the shock you are on, if you
are moving with the plasma, the plasma on the other side of the
shock is approaching you.

The rate at which CRs cross from upstream to
downstream is given by the projection of the isotropic CR flux onto 
the plane shock front
\begin{equation}
r_{\rm cross} =  \int_0^1 d(\cos \theta) \,\int_0^{2\pi} d \phi\,  
\frac{n_{\rm CR} v}{4 \pi}\, \cos \theta \,
\sim \frac{n_{\rm CR} v}{4} \,,
\label{eq:r_cross}
\end{equation}
where $n_{\rm CR}$ is the number density of particles undergoing acceleration.
On the other hand, the rate of convection downstream away from the shock 
front is
\begin{equation}
r_{{\rm loss}} = n_{\rm CR} u_2 \, ,
\label{eq:r_loss}
\end{equation}
so the probability of crossing the shock once and then escaping from
the shock (being lost downstream) is given by
\begin{equation}
{\rm Prob.(escape)} = \frac{r_{\rm loss}}{r_{\rm cross}} 
\sim 4\, \frac{u_2}{v} \, .
\end{equation}
The probability of returning to the shock
after crossing from upstream to downstream is
\begin{equation}
{\rm Prob.(return)} = 1 - {\rm Prob.(escape)},
\end{equation}
and so the probability of returning to the shock $n$ times and
also of crossing the shock at least $n$ times is
\begin{equation}
{\rm Prob.(cross} \ge n{\rm )} = [1 - {\rm Prob.(escape)}]^{n}.
\end{equation}
Therefore, the energy after $n$ shock crossings is
\begin{equation}
E = E_{0} \left( 1 + {\langle \Delta E \rangle \over E} \right)^{n}
\end{equation}
where $E_{0}$ is the overall initial energy. With this in mind, the number of 
encounters needed to reach an energy $E$ reads 
\begin{equation}
n = {\ln (E/E_{0}) \over \ln (1 + \langle \Delta E \rangle /E)}\,,
\label{klk}
\end{equation}
and the number of particles accelerated to energies greater than $E$ is
\begin{equation}
Q(>E) \propto \sum_{m=n}^\infty [1 - {\rm Prob.(escape)}]^{m} = \frac{[1 - {\rm Prob.(escape)}]^{n}}{{\rm Prob.(escape)}}\,.
\label{kl}
\end{equation}
Substitution of Eq.~(\ref{klk}) into Eq.~(\ref{kl}) leads to
\begin{equation}
Q(>E) \propto \frac{1}{{\rm Prob.(escape)}}\, \left(\frac{E}{E_0}\right)^{-\gamma}
\label{Ae}
\end{equation}
with
\begin{equation}
\gamma =  {\ln [1-{\rm Prob.(escape)}]^{-1} \over \ln (1 + \langle \Delta
E \rangle/E)}\,.
\label{eq:gamma}
\end{equation}
All in all, Fermi's mechanism yields the desired power law spectrum
for CR acceleration. Note that in the first order mechanism the 
spectral index, $\gamma$, is independent of the absolute magnitude of the velocity of the plasma, and depends only on the ratio of the upstream and downstream velocities.

\subsection{Desperately seeking Zevatrons}

\subsubsection{Plausible sources}

Now we focus attention on the direct identification of candidate CR-sources.
For well-researched reviews on this topic see~\cite{Biermann:tr,Ostrowski:2001ej}.
A variety of astrophysical objects have been proposed to account for the 
origin of high energy CRs. The options include:
\begin{itemize}
\item Supernovae explosions~\cite{Lagage,Voelk}.
\item Large scale Galactic wind termination shocks~\cite{Jokipii}.
\item Pulsars (neutron stars)~\cite{Venkatesan:1996jw}.
\item Active galactic nuclei (AGNs)~\cite{Protheroe:qs}.
\item BL Lacertae (BL Lac) -- a sub-class of 
AGN~\cite{Tinyakov:2001nr,Gorbunov:2002hk}.
\item Spinning supermassive black holes associated with presently 
inactive quasar remnants~\cite{Boldt:1999ge,Boldt:2000dx}.\footnote{An unexpected apparent 
correlation between the highest energy events reported by the AGASA 
Collaboration 
and dead 
quasars was recently reported~\cite{Torres:2002bb}.}
\item Large scale motions and the related shock waves resulting 
from structure formation in the Universe~\cite{Norman} such as accretion 
flow onto galaxy clusters and cluster 
mergers~\cite{Kang:1996rp,Miniati:1999ww}.\footnote{The famous AGASA triplet points towards the merger galaxies Arp 299 (NGC 3690 + IC 694), at a distance of
70~Mpc~\cite{Smialkowski}.}
\item  Relativistic jets and ``hot-spots'' produced by powerful radiogalaxies. Relativistic jets occurring 
in very powerful radiogalaxies carry large amounts of energy up to the radio hot-spots situated far 
($\sim 100$~kpc) from the central engine.  These hot-spots are believed to 
harbor strong, mildly relativistic shocks dissipating the jet bulk kinetic energy into heating plasma, thus
generating magnetic fields which efficiently accelerate CRs to ultrahigh 
energies~\cite{Rachen:1992pg,Rachen:1993gf,Anchordoqui:jn}. Additionally, a velocity shear layer at the 
relativistic jet side boundary  can play an active role
 in the acceleration process~\cite{Ostrowski:1998ic,Ostrowski:mnras}.
\item The electrostatic polarization fields 
that arise in plasmoids produced in planetoid impacts onto neutron star 
magnetospheres~\cite{Litwin:2001ku}.
\item Magnetars -- pulsars with dipole magnetic fields 
approaching 
$\sim 10^{15}$~G~\cite{Kouveliotou,Kouveliotou:apj,Kouveliotou:n}-- appear 
also as serious candidates~\cite{Bhattacharjee:1998qc,Arons:2002yj}. 
\item Starburst galaxies~\cite{Elbert:1994zv,Anchordoqui:1999cu,Anchordoqui:2001ss}.
\item Magnetohydrodynamic (MHD) winds of newly formed strongly magnetized 
neutron 
stars~\cite{Blasi:1999xm}.
\item Gamma ray burst (GRB) fireballs~\cite{Waxman:1995vg,Vietri:1995hs,Milgrom:1995wz,Miralda-Escude:1996kf}.\item Strangelets, stable lumps of quark matter, accelerated in astrophysical 
environments~\cite{Madsen:2002iw}.
\item Hostile aliens with a big CR gun~\cite{Anchordoqui:movie}. 
\end{itemize}

\begin{figure}
\postscript{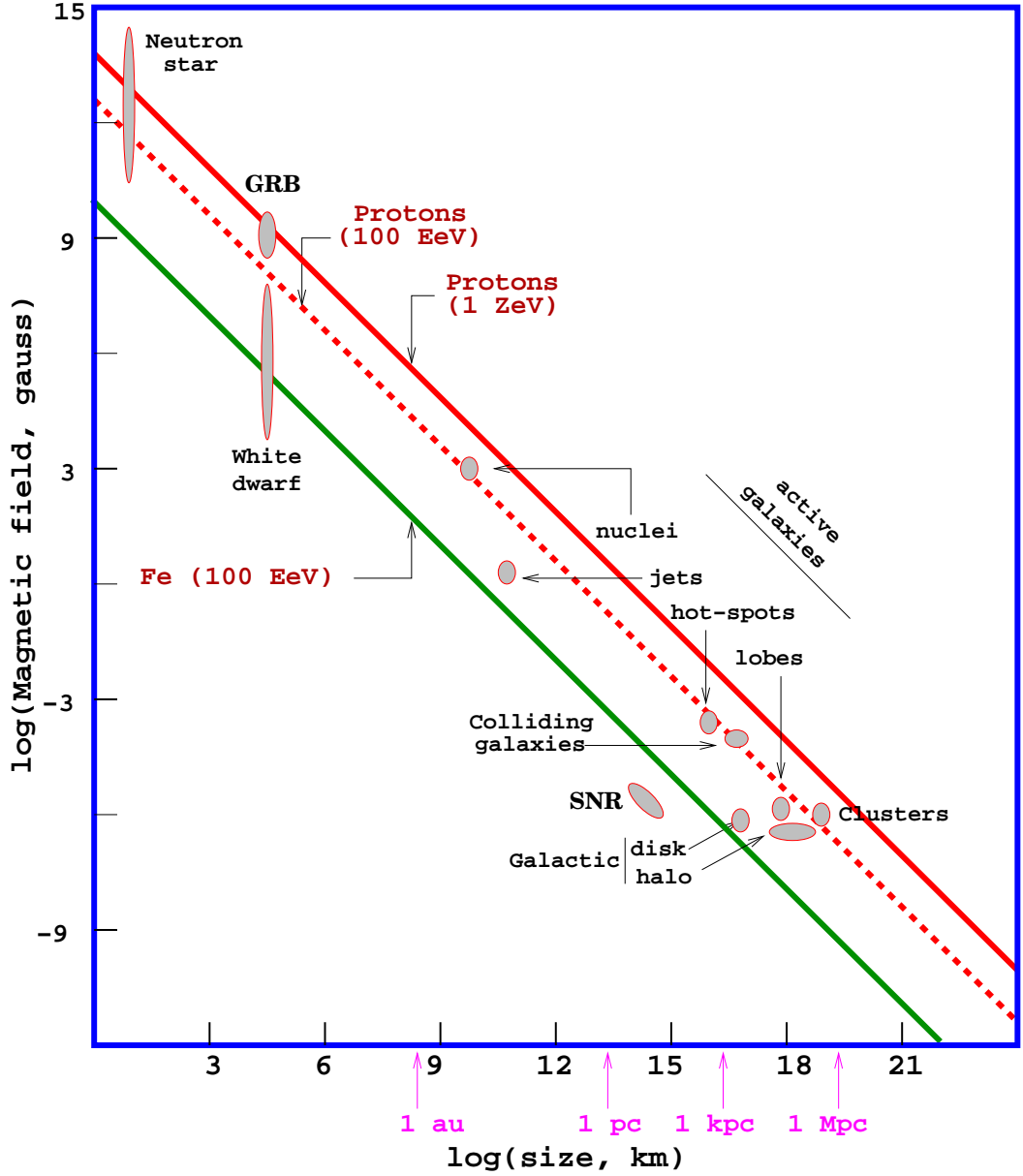}{0.90}
\caption{The Hillas diagram showing size and magnetic field strengths of 
possible sites of particle acceleration. Objects below the diagonal lines 
(from top to bottom), derived from Eq.~(\ref{hillas}) assuming the 
extreme value $\beta =1$, cannot accelerate 
protons above $ 10^{21}$~eV, above $ 10^{20}$~eV and iron nuclei above 
$10^{20}$~eV, respectively.
(This version of the picture is courtesy of Murat 
Boratav).}
\label{hillasf}
\end{figure}

In general, the maximum attainable energy of Fermi's mechanism is 
determined by the time scale over which particles are able to interact 
with the plasma. 
Sometimes the acceleration region itself only exists for a limited period 
of time; for example, supernovae shock waves  dissipate after about 
$10^{4}$~yr~\cite{Torres:2002af}.  In such a case, Eq.~(\ref{Ae}) would have to be modifed 
accordingly. 
Otherwise, if the plasma disturbances persist for much longer periods, the 
maximum energy may be limited by an increased likelihood of escape from the 
region. The latter case is relevant to ultrahigh energies, because when the 
Larmor radius of the particle (of charge $Ze$), 
\begin{equation} 
r_{_{\rm L}} \sim 110 \frac{E_{20}}{Z\,B_{\mu{\rm G}}}\,\, {\rm kpc}\,,
\label{rg}
\end{equation}
approaches the accelerator size it becomes very 
difficult to confine the CR magnetically to the acceleration region. 
Here, $B_{\mu{\rm G}}$ is the magnetic field in units of $\mu$G, 
and $E_{20} \equiv E/10^{20}$~eV.
If one includes the effect of the characteristic velocity $\beta c$ 
of the magnetic scattering centers, the above argument leads to the general 
condition~\cite{Hillas:1984},
\begin{equation}
E_{\rm max} \sim 2 \beta\, c\, Ze\,B\, r_{_{\rm L}}\,,
\label{hillas}
\end{equation}
for the maximum energy acquired by a particle traveling in a 
medium with magnetic field $B$.  This is sometimes called the ``Hillas criterion.''
The upper limit on the energy of one-shot acceleration scenarii 
turns out to be quite similar to the shock acceleration case of Eq.~(\ref{hillas}). For instance, a dimensional analysis suggests that 
the maximum energy that can be obtained  from a pulsar is \cite{Hillas:1984}
\begin{equation}
E_{\rm max} = \frac{\omega}{c}\, Ze\, B_s \,r_{\rm ns}^2\,,
\label{hillas2}
\end{equation}
where $\omega$ is the pulsar angular velocity, $B_s$ the surface magnetic 
field and $r_{\rm ns}$ the neutron star radius. Therefore, 
if $B_s \sim 10^{12}$~G, $r_{\rm ns} \sim 10$~km, and 
$\omega \sim 60 \pi$~s$^{-1}$ (as for the Crab pulsar), a circuit connected 
between pole and equator would see an emf $\sim 10^{18}$~V for an 
aligned or oblique dipole. When realistic 
models of acceleration are constructed, however, this ideal dimensional limit 
is not realized, because the large potential drop 
along the magnetic field lines is  significantly short-circuited 
by electron and positrons moving in the opposite directions along the field 
lines~\cite{Venkatesan:1996jw}. 

The dimensional arguments of Eqs.~(\ref{hillas}) and (\ref{hillas2}) are 
usually summarized in the form of the popular ``Hillas 
diagram''~\cite{Hillas:1984} shown in Fig.~\ref{hillasf}. Clearly, very 
few sites 
appear able to generate particle energies $> 10^{20}$~eV; either this 
occurs on highly condensed objects with huge $B$ or enormously 
extended objects.\footnote{For a comprehensive discussion on the 
electrodynamical limitations of CR sources the reader is referred to 
the recent study in Ref.~\cite{Aharonian:2002we}.} From a glance at 
Fig.~\ref{hillasf} it seems that
the structures associated 
with active galaxies, neutron stars and GRBs have sufficient size and field strength to be 
considered as potential sources. In subsequent sections we discuss these
potential sources in more detail.

\subsubsection{Radiogalaxies}

Fanaroff-Riley II (FRII) galaxies~\cite{Fanaroff} are the largest known
dissipative objects (non-thermal sources) in the Universe.
Localized regions of intense synchrotron emission, known as
``hot spots'', are observed within their lobes. These regions are
presumably produced when the bulk kinetic energy of the jets
ejected by a central active nucleus (supermassive black hole +
accretion disk) is reconverted into relativistic particles and
turbulent fields at a ``working surface'' in the head of the jets
\cite{Blandford}. Specifically, the speed $v_{\rm h}$ with which the 
head of a jet advances
into the intergalactic medium of particle density $n_{\rm e}$ can be obtained 
by balancing the momentum flux in the jet against the momentum flux of the 
surrounding medium. Measured in the frame comoving with the advancing head,
$v_{\rm h}\approx \;v_{\rm j}\,[ 1 + ( n_{\rm e} /n_{\rm j})^{1/2}]^{-1}$,
where $n_{\rm j}$ and $v_{\rm j}$ are the particle density and the velocity
of the jet flow, respectively. Clearly, $v_{\rm j}> v_{\rm h}$ for 
$n_{\rm e} \geq n_{\rm j}$, in such a way that the jet will decelerate. The 
result is the formation of a strong collisionless shock, 
which is responsible for particle reacceleration and magnetic field 
amplification~\cite{Begelman}. 
The acceleration of particles up to ultrarelativistic energies in the hot
spots is the result of repeated scattering back and forth across the shock
front, similar to that discussed in Sec.III-A. The particle deflection in 
this mechanism is produced by Alfv\'en waves in the turbulent magnetic field. 
This process has been studied in detail by Biermann and 
Strittmatter~\cite{Biermann:ep}. Dimensional arguments 
suggest that the energy density per unit of wave number of MHD turbulence 
is of the Kolmogorov type~\cite{Kolmogorov}, and so 
for strong shocks the acceleration time
for protons is~\cite{Drury}
\begin{equation}
\tau_{\rm acc} \simeq \frac{40}{\pi}\, \frac{1}{c\,\beta_{\rm jet}^2}
\,\frac{1}{u}\,\left(\frac{E}{eB}\right)^{1/3}\, R^{-2/3}
\label{acc}
\end{equation}
where
$\beta_{\rm jet}$ is the jet velocity in units of $c,$ $u$ is the ratio of
turbulent to ambient magnetic energy density in the region of the shock
(of radius $R$), and
$B$ is the total magnetic field strength. 
The acceleration process will be efficient as long as the
energy losses by synchrotron radiation and photon--proton
interactions do not become dominant. The subtleties surrounding the 
conversion of a particle kinetic
energy into radiation provide ample material for 
discussion~\cite{Biermann:ep,Mannheim:ac,Mannheim:1998fs,Mannheim:1999hf,Mannheim:jg,Aharonian:2000pv}. The most popular mechanism to date relates
$\gamma$-ray emission to the development of electromagnetic
cascades triggered by secondary photomeson products that cool
instantaneously via synchrotron radiation~\cite{Biermann:ep,Mannheim:ac,Mannheim:1998fs,Mannheim:1999hf,Mannheim:jg}.
The synchrotron loss time for protons is given by~\cite{Rybicki}
\begin{equation}
\tau_{\rm syn} \sim \frac{6\, \pi\, m_p^3\,c}{\sigma_{\rm T}\,m_e^2\,\Gamma\,B^2}\,,
\label{tausyn}
\end{equation}
where $m_e$, $m_p$, $\sigma_{\rm T}$ and $\Gamma$ are the electron mass, 
proton mass, Thomson cross section, and Lorentz factor, respectively. 
The characteristic single photon energy in synchrotron radiation
emitted by an electron is
\begin{equation}
E_\gamma = \left(\frac{3}{2}\right)^{1/2} 
\frac{h\,e\,E^2\,B}{2 \,\pi\, m_e^3 \,
c^5} \sim 5.4\times 10^{-2}\, B_{\mu{\rm G}}\, 
E_{20}^{2} \, {\rm TeV}\ \ .\label{synch}
\end{equation}
For a proton this number  is $(m_p/m_e)^3 \sim 6 \times 10^9$
times smaller. Thus, it is evident that high energy gamma ray production 
through proton synchrotron radiation requires very large (${\cal O}(100\ {\rm
G})$) magnetic fields. 
Considering an average cross
section $\bar{\sigma}_{\gamma p}$ for the three dominant
pion--producing interactions~\cite{Armstrong:1971ns}, 
\begin{equation}
\gamma p \rightarrow p  \pi^0\,, 
\end{equation}
\begin{equation}
\gamma  p \rightarrow n  \pi^+\,,
\label{neutron}
\end{equation}
\begin{equation}
\gamma  p\rightarrow p  \pi^+  \pi^- \,,
\end{equation} 
the time scale of the energy losses, including synchrotron and photon 
interaction losses, reads~\cite{Biermann:ep}
\begin{equation}
\tau_{\rm loss} \simeq \frac{6\pi\ m_p^4\ c^3}{\sigma_{\rm T}\ m_e^2\ B^2\ (1+Aa)}\ E^{-1}\ = \frac{\tau_{\rm syn}}{1 + Aa} \,,
\end{equation} 
where $a$ stands for the ratio of photon to magnetic energy densities 
and $A$ gives a measure of the relative strength of $\gamma p $
interactions versus the synchrotron emission. Note that  channel 
(\ref{neutron}) involves the creation of ultrarelativistic neutrons 
(but $\Gamma_n \alt  \Gamma_p$) with mean free path in the observer rest 
frame given by 
$\lambda_n = \Gamma_n c \tau_n$, where $\tau_n \sim 900$~s, is the neutron 
lifetime. Since $\lambda_n > \lambda_p$ for $\Gamma_n \alt \Gamma_{p \ {\rm max}}$, such neutrons can readily escape the system, thereby modifying the high end of 
the proton spectrum. Biermann and
Strittmatter~\cite{Biermann:ep} have estimated that $A \approx 200$, almost
independently of the source parameters. The most energetic protons
injected in the intergalactic medium will have an energy that can be
obtained by balancing the energy gains and losses~\cite{Anchordoqui:2001bs}
\begin{equation}
E_{20}=1.4\times 10^5\,\,B_{\mu{\rm G}}^{-5/4}\,\,\beta_{\rm jet}^{3/2}
\,\,u^{3/4}\,\,R_{\rm kpc}^{-1/2}\,\,(1+Aa)^{-3/4}\ ,
\label{ab}
\end{equation}
where $R_{\rm kpc}\equiv R/1\,{\rm kpc}$. 

As noted in the previous section, the canonical theory of diffuse shock acceleration not only assumes an infinite time, but also an infinite space for the particles to diffuse in the medium around the shock front. Therefore,
in order to ascertain the capability of FRII to accelerate
particles to ultrahigh energies, one also has to apply the Hillas
criterion~\cite{Hillas:1984} for localizing the Fermi engine in space,
namely that the Larmor radius be less than the size of the
magnetic region. For typical hot-spot conditions ($B \sim 300~\mu$G, 
$u \sim 0.5$, and $\beta_{\rm jet} \sim 0.3$) and assuming that the magnetic 
field of the hot spot is limited to the observable region, one obtains 
$E < 5 \times 10^{20}$~eV for $a<0.1$~\cite{Rachen:1992pg}. Therefore, one concludes that
protons can be accelerated to the highest observed energies in typical 
FRII hot spots. Moreover, the shock structure in hot spots is likely to be much more extended than the visible region in the nonthermal radioemission, as suggested by magnetohydrodynamical modeling~\cite{Rachen:1992pg}.

Particles can also attain ultrahigh energies ($E \agt 10^{20}$~eV)  within the jets or the AGNs 
themselves. For instance, the knot A in the M87 jet, 
with a length scale $l_{87} \sim 2 \times 10^{20}$ cm, has a 
magnetic field strength $B_{87} \sim 300~\mu$G~\cite{Stocke}.  Typical AGN
sizes are $l_{\rm AGN} \sim 10^{15}$~cm, and 
$B_{\rm AGN} \sim 1$~G~\cite{Angel}. Observational evidence suggests that in the jets 
$a\ll 1$, whereas $a \sim 1$ for AGNs~\cite{Biermann:ep}.

In Sec.V the ideas developed in this section will be applied to specific models.

\subsubsection{$\gamma$-ray burst fireballs}

Gamma ray bursts (GRBs) are flashes of high energy radiation that can be 
brighter,
during their brief existence, than any other gamma ray source in the sky. The
bursts present an amazing variety of temporal profiles, spectra, and
timescales
that have puzzled astrophysicists for almost three decades
\cite{Fishman:95}.
In recent years, our observational insight of this phenomenon has been
dramatically
increased by the huge amount of data collected by the 
Burst and Transient Source
Experiment (BATSE) on board the Compton Gamma Ray Observatory
(CGRO), a satellite launched by NASA in 1991. In 9
years of operation, BATSE has accumulated a database of more than 2000
observations.

The temporal distribution of the bursts is one
of the most striking signatures of the GRB phenomenon.
There are at least four classes of distributions,
from single-peaked bursts, including the fast rise
and exponential decaying (FREDs) and their inverse (anti-FREDs),
to chaotic structures~\cite{Link:1996ss,Romero:1999mg}.
There are well separated
episodes of emission, as well as bursts with
extremely complex profiles. Most of the bursts are time
asymmetric but some are symmetric. Burst timescales go through the 
30~ms scale to hundreds of seconds. The measurement of these timescales
is a rather complicated task, since it may depend on the intensity
of both the background and the source. At high energies ($>100$  MeV),
some extremely long bursts have been detected.
The angular distribution of these bursts is isotropic
within the statistical limits, and the paucity of comparatively faint 
bursts implies
that we are seeing to near the edge of the source population~\cite{Meegan:xg}.
 Both effects, isotropy
and non-homogeneity in the distribution, strongly suggest a
cosmological origin of the phenomenon. Moreover, the recent detection 
of ``afterglows'', delayed low energy ($X$-ray to radio) emission of 
GRBs, has confirmed the cosmological origin of the burst via a redshift 
determination of several GRB host-galaxies~\cite{Metzger:1997wp,Kulkarni}.

If the sources are so distant, the energy necessary to
produce the observed events by an intrinsic mechanism is astonishing: about
$10^{51}$ erg of gamma rays
must be released in less than 1
second.
The most popular interpretation of the GRB-phenomenology
is that the observable effects are due to the dissipation of the kinetic 
energy of a relativistic expanding plasma wind, a ``fireball'', whose primal cause 
is not yet known~\cite{Cavallo:78,Paczynski:1986px,Goodman,Meszaros:1,Meszaros:2,Piran:kx,Piran:1999bk}.
The very short timescale observed in the time
profiles indicate an extreme compactness that implies a source
initially opaque (because of $\gamma \gamma $ pair creation) to
$\gamma$-rays.  The radiation pressure on the optically thick source
drives relativistic expansion, converting internal energy into the kinetic
energy of the inflating shell. Baryonic pollution in this expanding
flow can trap the radiation until most of the initial energy has gone
into bulk motion with Lorentz factors of $\Gamma
\ge 10^2 - 10^3$~\cite{Waxman:2001tk}. The kinetic energy, however, can be 
partially
converted into heat when the shell collides with the interstellar
medium or when shocks within the expanding source collide with one
another. The randomized energy can then be radiated by
synchrotron radiation and inverse Compton scattering yielding
non-thermal bursts with timescales of seconds. 

We now consider Fermi acceleration in the fireball internal shocks. As usual,
$r_{_{\rm L}}$ should be smaller than the largest scale $l_{\rm GRB}$ over which the 
magnetic field fluctuates, since otherwise Fermi acceleration may not be 
efficient. One may estimate $l_{\rm GRB}$ as follows. The comoving 
time, i.e., the time measured in the fireball rest frame, is $t = r/\Gamma c$. 
Hence, the plasma wind properties fluctuate over comoving 
scale length up to $l_{\rm GRB} \sim r/\Gamma$, because
regions separated by a comoving distance larger than $r/\Gamma$ 
are causally disconnected. Moreover, the internal energy is decreasing 
because of the expansion and thus it is available 
for proton acceleration (as well as for $\gamma$-ray 
production) only  over a comoving time $t$. 
The typical acceleration time scale is then~\cite{Waxman:1995vg}
\begin{equation}
\tau_{\rm acc}^{\rm GRB} \sim \frac{r_{_{\rm L}}}{c \beta^2}\,,
\label{accgrb}
\end{equation} 
where $\beta c$ is the Alfv\'en velocity. In the GRB scenario $\beta \sim 1$, so Eq.~(\ref{accgrb})   
sets a lower limit on the required comoving magnetic field strength, and the 
Larmor radius $r_{_{\rm L}} = E'/eB = E/\Gamma eB$, where $E' = E / \Gamma$ 
is the proton energy measured in the 
fireball frame. The dominant energy loss process in this case is synchrotron 
cooling. Therefore, the condition that the synchrotron loss time 
of Eq.~(\ref{tausyn}) be smaller than the acceleration time sets the upper 
limit on the magnetic field strength. A dissipative ultra-relativistic wind, 
with luminosity and variability time implied by GRB observations, satisfies 
the constraints necessary to accelerate protons to energy $> 10^{20}$~eV, 
provided that $\Gamma > 100$, and the magnetic field is close to 
equipartition with electrons. We stress that the latter must be satisfied 
to account for both $\gamma$-ray emission and afterglow 
observations~\cite{Waxman:2001tk}. At this stage, it is worthwhile to point 
out that for the acceleration process at shocks with large $\Gamma$ the particle 
distributions are extremely anisotropic in shock, with the particle angular 
distribution opening angles $\sim \Gamma^{-1}$ in the upstream plasma rest frame.
Therefore, when transmitted downstream the shock particles have a limitted chance 
to be scattered efficiently to re-eneter the shock~\cite{Bednarz:1998pi}. However, 
in this particular case, the energy gain by any ``successful'' CR can be comparable 
to its original energy, i.e., $\langle \Delta E \rangle / E \sim 1$.

Now we comment on the data. It was pointed out that two of the highest energy 
CRs come from directions that are within the error boxes of two remarkable 
GRBs detected by BATSE 
with a delay of ${\cal O} (10)$ months after the 
bursts~\cite{Milgrom:1995um}. However, a rigorous analysis 
shows no correlation between the arrival direction of ultrahigh energy CRs 
and GRBs from the third BATSE catalog~\cite{Stanev:1996qc}. Moreover, no 
correlations were found between a pre-CGRO burst catalog and the Haverah 
Park shower set that covered approximately the same period of time.
These analyses, however, could have been distorted by the angular 
resolution ($\Delta \theta \sim 3^\circ$) of the GRB measurements. A 
sensitive anisotropy analysis between ultrahigh energy CRs and GRBs will 
be possible in the near future with the facilities of PAO  
and the High Energy Transient 
Explorer (HETE).\footnote{This satellite, designed to localize GRBs, is
expected to detect around 50 events per year with a 10 arc-minute accuracy
in the medium energy X-ray band, and around 10 events per year with 10
arc-second accuracy in the soft X-ray band. {\tt http://space.mit.edu/HETE/Welcome.html}.} 

Finally, it is also interesting to note that if the GRBs are uniformly 
distributed (independent of  redshift), and if in the past they
emitted the same amount of energy in ultrahigh energy 
($\sim 10^{14}$~MeV) CRs as in $\sim$ MeV photons, 
the energy input of these particles into the extragalactic space would be 
enough to account for the observed CR flux~\cite{Waxman:1995vg}. However, 
recent afterglow studies indicate that their 
redshift distribution likely follows the average star formation rate of the 
Universe and that GRBs are more numerous at high redshift. If this is the 
case, ultrahigh energy CRs coming from GRBs would produce too sharp a 
spectral energy cutoff to be consistent with the AGASA data, because of 
energy degradation {\it en route} to 
Earth~\cite{Stecker:1999wh,Scully:2000dr}.\footnote{It has been noted by  
Bahcall and Waxman~\cite{Bahcall:2002wi} that if one excludes the AGASA 
results, the remaining data from Fly's Eye, HiRes, and Yakutsk CR experiments 
show evidence for a cutoff in the CR energy spectrum below $10^{20}$~eV 
with $7\sigma$ significance. In such a case, the door is open for a common 
origin between GRBs and ultrahigh energy CRs~\cite{Loeb:2002ee}.}

\section{Interactions {\em en route} to Earth} 

In this section we review the propagation of cosmic rays. 
We start with the hadronic component, continue with a discussion of 
electromagnetic cascades 
initiated by ultrahigh energy photons in extragalactic space, and then 
discuss how propagation can be influenced by cosmic magnetic fields. 
For alternative discussions see, for 
instance, Refs.~\cite{Bhattacharjee:1998qc,Sigl:2000vf}.

\subsection{The GZK cutoff}

Shortly after the discovery of the microwave echo of the big 
bang~\cite{Penzias:wn}, Greisen~\cite{Greisen:1966jv} and independently
Zatsepin and Kuzmin~\cite{Zatsepin:1966jv} (GZK) pointed out 
that this radiation 
would make the Universe opaque to protons of sufficiently high energy. 
At energies above a few $10^{19}$~eV, the thermal photons are seen 
highly blue-shifted by the protons in their rest frames. The energy of the 
relic photons is sufficient to excite  baryon resonances thus 
draining the energy of the proton via pion production and, coincidentally, 
producing a source of ultrahigh energy gamma rays and neutrinos. 
Since the implications of this process have been revisited in many 
forms~\cite{Stecker:68,Hill:1983mk,Berezinsky:wi,Stecker:1989ti,Aharonian:90,Yoshida:pt,Aharonian:nn,Protheroe:1995ft,Anchordoqui:gs,Anchordoqui:1996ru,Lee:1996fp,Achterberg:1999vr,Stanev:2000fb,Fodor:2000yi,Anchordoqui:2000ad,Berezinsky:2002nc,Berezinsky:2002vt}, the concept of GZK sphere is by now a somewhat fuzzy notion. 
Na\"{\i}vely, it is 
the radius of the sphere within which a source has to lie in order to 
provide us with protons of $10^{20}$~eV. In what follows we 
extensively discuss the phenomenology of proton-photon interactions 
and discuss in some  detail the GZK radius.

There are three sources of energy loss of ultrahigh energy protons. These are the 
adiabatic fractional energy loss due to the expansion of the Universe,  pair 
production $(p \gamma \rightarrow p  e^+  e^-$) 
and photopion production ($p \gamma \rightarrow \pi  N$), each successively 
dominating as the proton energy increases. The adiabatic fractional energy 
loss at the present cosmological epoch is given by
\begin{equation}
-\frac{1}{E} \left(\frac{dE}{dt}\right)_{\rm adiabatic} = H_0 \,,
\end{equation}
where $H_0 \sim 100\,h$~km~s$^{-1}$~Mpc$^{-1}$ is the Hubble constant, 
with $h \sim (0.71 \pm 0.07) \times ^{1.15}_{0.95}$ the normalized Hubble 
expansion rate~\cite{Groom:in} (see Appendix B). 
The fractional  energy loss due to interactions
with the cosmic background radiation at a redshift
$z=0$ is determined by the integral of the nucleon 
energy loss per collision multiplied by
the probability per unit time for a nucleon collision in an isotropic gas of photons \cite{Stecker:68}. This integral can be
explicitly written as follows,
\begin{equation}
-\frac{1}{E} \frac{dE}{dt} =\frac{c}{2 \Gamma^2}\,\sum_j\, \int_0^{w_m} dw_r \,\, 
K_j  \, \,  
\sigma_j (w_r)\, \, w_r \, \int_{w_r/2 \Gamma}^{w_m} dw \, 
\frac{n(w)}{w^2}   
\label{conventions}
\end{equation} 
where $w_r$ is the photon energy in the rest frame of the nucleon, and
$K_j$ is the inelasticity, i.e., the average fraction of the energy lost by
the photon to the nucleon in the laboratory 
frame for the $j$th reaction channel.\footnote{Here the laboratory frame is the one
in which the cosmic microwave background is at a temperature $\approx 2.7$ K.} The
sum is carried out over all channels, 
$n(w)dw$ stands for the number density
of photons with energy between $w$ and $dw$,  $\sigma_j(w_r)$ is 
the total cross section of the $j$th interaction channel, 
$\Gamma$ is the usual  Lorentz factor of the nucleon, and $w_m$ is the 
maximum energy of the photon in the photon gas.

Pair production and photopion production processes are only of importance 
for interactions with the 2.7 K  blackbody background radiation. Collisions 
with optical and infrared photons give a negligible contribution. Therefore, 
for interactions with a blackbody field of temperature $T$, the photon 
density is that of a Planck spectrum~\cite{Mather:pc}, so the 
fractional energy loss is given by
\begin{equation}
-\frac{1}{E} \frac{dE}{dt} = - \frac{ckT}{2 \pi^2 \Gamma^2 (c \hbar)^3}
\sum_j \int_{w_{0_j}}^{\infty}  dw_r \,
\sigma_j (w_r) \,K_j \, w_r \, \ln ( 1 -  e^{-w_r / 2 \Gamma kT})
\label{phds!}
\end{equation}
where $k$ and $\hbar$ are Boltzmann's and Planck's constants
respectively, and $w_{0_j}$ is the threshold energy for the $j$th reaction 
in the rest frame of the nucleon.

Berezinsky and Grigor'eva have examined in detail the pair production 
process~\cite{Berezinsky:wi}.
At energies $E\ll m_e\,m_p/kT = 2.1 \times 10^{18}$~eV 
(i.e., $w_r/m_e -2 \ll 1$), when the reaction takes place on the photons 
from the high energy tail of the 
Planck distribution, the fraction of energy lost in one collision and the 
cross section can be approximated by the threshold values
\begin{equation}
K_{e^+e^-} = 2\,\frac{m_e}{m_p}\,,
\end{equation}
and
\begin{equation}
\sigma_{e^+e^-}(w_r) = \frac{\pi}{12}\, \alpha\,\, r_0^2 \,\,
\left(\frac{w_r}{m_e} - 2\right)^3\,,
\end{equation}
where $\alpha$ is the fine structure constant  
and $r_0$ is the classical radius of the electron.
The fractional energy loss due to pair production for $E \alt 10^{18}$~eV is 
then, 
\begin{equation}
-\frac{1}{E}\, \left(\frac{dE}{dt}\right)_{e^+e^-} = 
\frac{16 c}{\pi}\, \frac{m_e}{m_p}\, 
\alpha\, r_0^2\,
\left(\frac{kT}{hc}\right)^3 \, \left(\frac{\Gamma k T}{m_e}\right)^2\, 
\exp \left(-\frac{m_e}{\Gamma kT}\right).
\end{equation}
At higher energies ($E > 10^{19}$ eV) the characteristic time for the energy 
loss due to pair production is $t \approx 5\times 10^9$ 
yr~\cite{Blumenthal:nn}.  In this energy regime, the photopion 
reactions $p \gamma \rightarrow p \pi^0$ and
$p \gamma \rightarrow \pi^+  n$ on the tail of the Planck distribution give 
the main contribution to proton energy loss. The cross sections of these 
reactions are well known and the kinematics is simple.

Photopion production turns on at a photon energy in the proton rest frame of 
145~MeV with a strongly increasing cross section at the $\Delta (1232)$ 
resonance, which decays into the one pion channels $\pi^+ n$ and $\pi^0 p$.  
With increasing energy, heavier baryon resonances occur and the proton 
might reappear only after successive decays of resonances. The most important 
channel of this kind is $p\gamma \rightarrow \Delta^{++} \pi^-$ with 
intermediate $\Delta^{++}$ states leading finally to 
$\Delta^{++} \rightarrow p\pi^+$. $\Delta^{++}$ examples in this category are
the $\Delta (1620)$ and $\Delta(1700)$ resonances. The cross section in this 
region can be described by either a sum or a product of Breit-Wigner 
distributions over the main 
resonances produced in $N \gamma$ collisions considering $\pi N$, 
$\pi \pi N$ and $K\Lambda$ ($\Lambda \rightarrow N \pi$) final 
states~\cite{Barnett:1996hr}. At high energies, $3.0\, {\rm GeV} 
< w_r < 183 \, {\rm GeV}$, the CERN-HERA and COMPAS Groups have made 
a fit to the $p\gamma$ cross section~\cite{Montanet:1994xu}. 
The parameterization is
\begin{equation}
\sigma_\pi(w_r) = A + B \,\,\ln^2\left(\frac{w_r}{{\rm GeV}}\right) + C \,\,
\ln \left(\frac{w_r}{{\rm GeV}}\right) \,\,{\rm mb}\,,
\end{equation} 
where $A = 0.147 \pm 0.001$, $B = 0.0022\pm 0.0001$, 
and $C = -0.0170 \pm 0.0007$. In this energy range, the 
$\sigma_{{\rm total}} (n\gamma)$ is 
to a good approximation identical to $\sigma_{{\rm total}} (p\gamma)$. 

We turn now to the kinematics of photon-nucleon interactions. 
Assuming that reactions mediated 
by baryon resonances have spherically symmetric decay angular distributions, 
the average energy loss of the nucleon after $n$ resonant 
decays is given by
\begin{equation}
K_\pi(m_{R_0}) = 1 - \frac{1}{2^n} \prod_{i=1}^{n} \left( 1 + \frac{m_{R_{_i}}^2 -
m_M^2}{m_{R_{_{i-1}}}^2} \right)\,,
\label{kj}
\end{equation}
where $m_{R_{_i}}$ denotes the mass of the $i^{\rm th}$
resonant system of the decay chain,  $m_M$ the mass of the associated
meson, $m_{R_{_0}} = \sqrt{s}$ is the total energy of the reaction 
in the c.m., and $m_{R_{_n}}$ the mass of the nucleon. See 
Appendix C for details. For multi-pion production the case is much more complicated because of the 
non-trivial final state kinematics. However,
it is well established experimentally~\cite{Golyak:cz} that, at very 
high energies ($\sqrt{s} \agt 3$ GeV), the incoming particles 
lose only one-half their energy via pion photoproduction independently of the
number of pions produced. This is the ``leading particle effect''.

\begin{figure}
\postscript{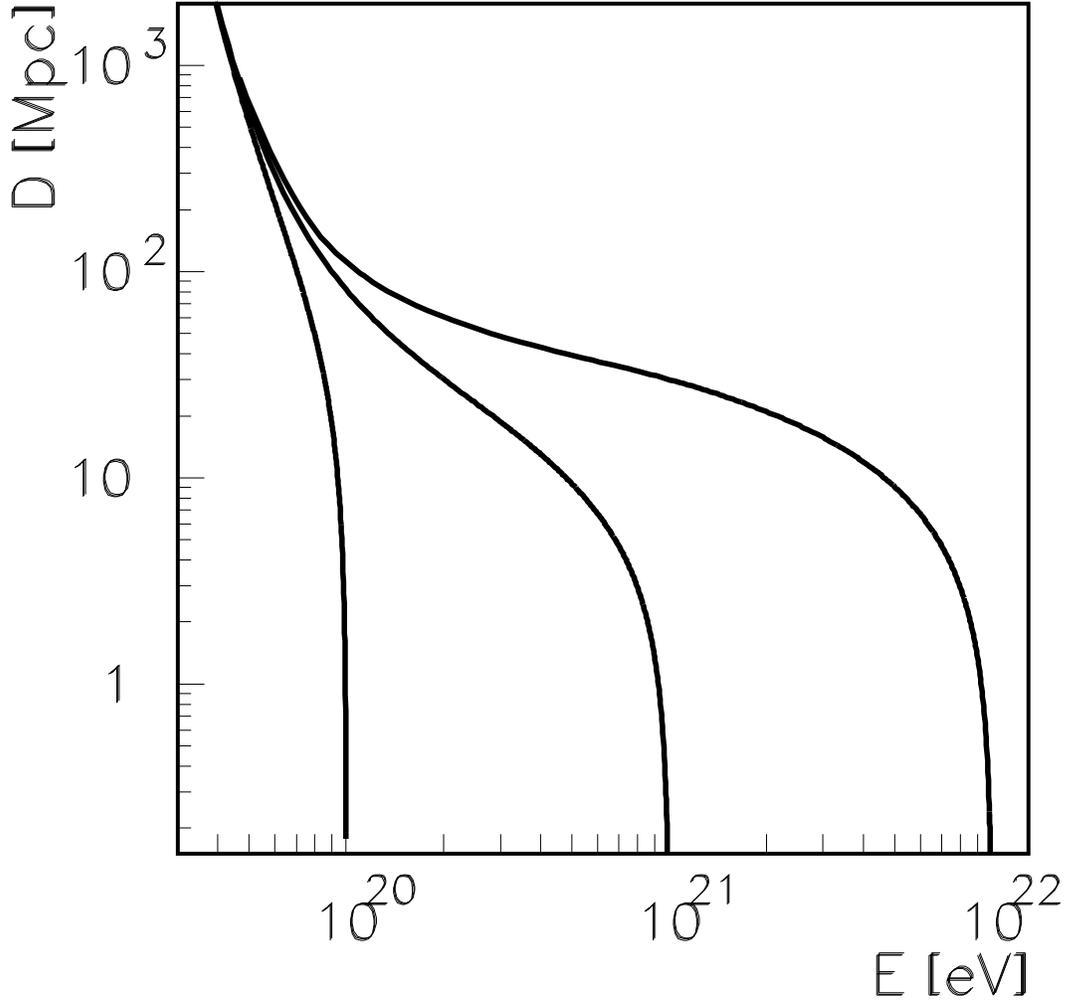}{0.95}
\caption{Energy attenuation length of nucleons in the intergalactic
medium. Note that after a distance of $\sim 100$~Mpc, or propagation time 
$\sim 3 \times 10^{8}$~yr, the mean energy is essentially independent of the 
initial energy of the protons, with a critical energy around $10^{20}$~eV. 
Published in Ref.~\cite{Anchordoqui:1996ru}.} 
\label{gzk}
\end{figure}

For $\sqrt{s} < 2$ GeV, the best maximum likelihood fit to Eq.~(\ref{phds!}) 
with the exponential behavior
\begin{equation} 
- \frac{1}{E}\,\left(\frac{dE}{dt}\right)_\pi = A \, {\rm exp} [ - B / E ]\,,
\end{equation}
derived from the values of cross section 
and fractional energy loss at threshold, gives~\cite{Anchordoqui:1996ru} 
\begin{equation}
A = ( 3.66  \pm 0.08 )
\times 10^{-8} \, {\rm yr}^{-1}, \,\,\,\,\, B = (2.87 \pm 0.03 )\times
10^{20}\, {\rm eV}\,.
\label{parame}
\end{equation} 
The fractional energy loss due to production of multipion final
states at higher c.m. energies ($\sqrt{s} \agt 3$ GeV) 
is roughly a constant, 
\begin{equation}
-\frac{1}{E}\,\left(\frac{dE}{dt}\right)_\pi = C = ( 2.42 \pm 0.03 ) 
\times 10^{-8}  \,\, {\rm yr}^{-1} \,.
\label{ce}
\end{equation}
From the values determined for the fractional energy loss, it is 
straightforward to compute the energy degradation of ultrahigh energy 
cosmic rays in terms of their flight time. This is given by,
\begin{equation}
A \, t \, - \,  {\rm Ei}\,(B/E) 
+ \, {\rm Ei}\, (B/E_0) 
= 0\,,\,\,\,\,\,\,  {\rm for} \,\,10^{19}\,{\rm eV} \alt E \alt 10^{21} \,{\rm eV} \,,
\label{degradacion}
\end{equation}
and
\begin{equation}
E (t) = E_0 \exp[- \,C \ t \,]\,,\,\,\,\,\,\, {\rm for}\,\, E \agt 10^{21} \, 
{\rm eV}\,,
\end{equation}
where Ei is the exponential integral~\cite{Abramowitz}. Figure~\ref{gzk} 
shows the 
proton energy degradation as a function of the mean flight distance. Notice 
that, independent of the initial energy of the nucleon, the mean energy 
values approach  $10^{20}$~eV after a distance of $\approx$ 100~Mpc.

Nevertheless, the definition of the GZK radius is subject to possible 
magnetic field deflections. To obtain quantitative estimates, one generally 
defines the 50\% horizon $R_{50}$ as the light propagation 
distance from the source at which $1/e$ of all injected protons have retained 
50\% or more of their energy, i.e., $R_{50}$ is achieved when
\begin{equation}
\int_{E_0/2}^{E_0} \, \frac{dN}{dE} \, dE = N_0 \exp(-1)\,,
\end{equation}
where $N_0$ is the number of particles injected with energy 
$E_0$~\cite{Stanev:2000fb}. For 
injection energies $> 10^{20}$~eV the protons are not affected much by 
magnetic field strengths in the range $1 - 10$~nG (more on this below) 
since their 
scattering angles are small, and so the horizon energy dependence is similar 
to that of the energy loss distance. At $E=10^{20}$~eV $R_{50}$ is about 
100~Mpc, whereas at $2 \times 10^{20}$~eV it decreases to 20~Mpc and becomes 
smaller than 10~Mpc for energies above 
$3 \times 10^{20}$~eV~\cite{Stanev:2000fb} . Below $10^{20}$~eV the 
scattering in the magnetic field increases the propagation 
time and thus causes additional energy loss and an increase of the ratio 
between the mean energy loss length and $R_{50}$.
In summary, the GZK-cutoff implies that if the primary ultrahigh energy 
CRs are protons energetic sources should be very close to the Earth, 
say within about 20 Mpc.

On a different track, the GZK-cutoff guarantees a cosmogenic flux  
of neutrinos, caused by the decay of charged 
pions produced in the photon-nucleon interactions~\cite{Beresinsky,Stecker:1979ah}.  The resulting neutrino flux
depends critically on the cosmological evolution of the cosmic ray
sources and on their proton injection
spectra~\cite{Yoshida:pt,Protheroe:1995ft,Engel:2001hd}. 
Of course, the neutrino intensity will also depend on the homogeneity of
sources.  For example, semi-local objects, such as the Virgo
cluster~\cite{Hill:1983mk}, could countribute to the high
energy tail of the neutrino spectrum.  Additionally, there is a weak dependence
on the details of the cosmological expansion of the Universe.  For
example, a small cosmological constant tends to increase the
contribution to neutrino fluxes from higher
redshifts~\cite{Engel:2001hd}.

\begin{figure}[tbp]
\postscript{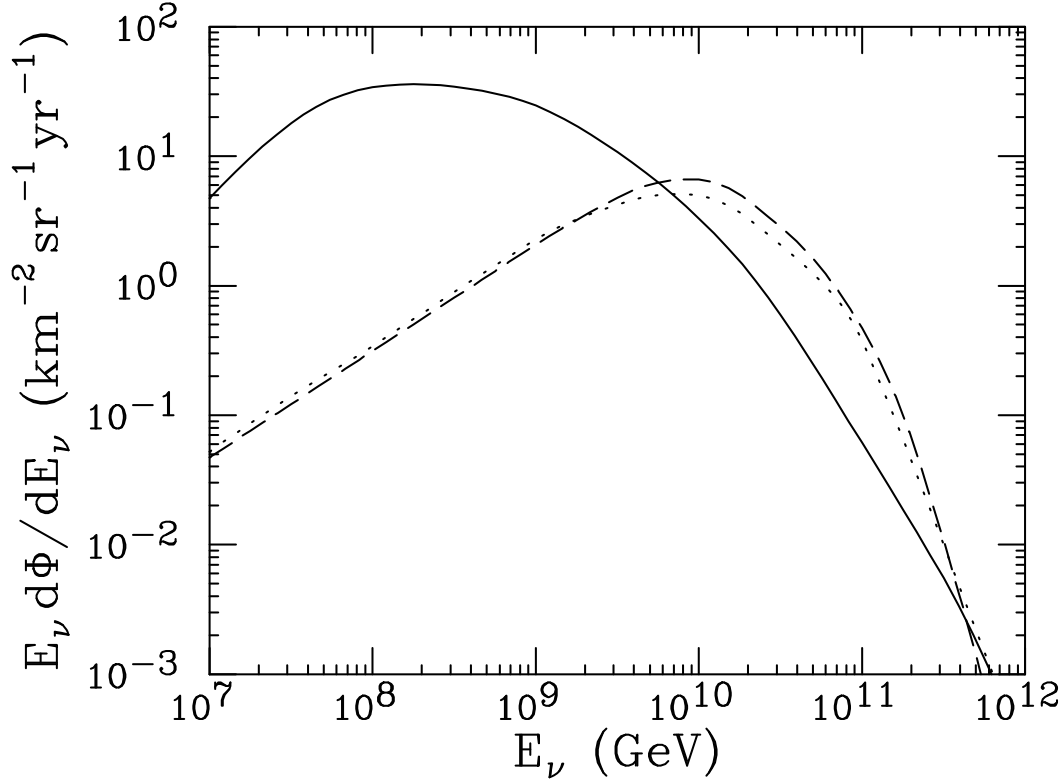}{0.90}
\caption{Cosmogenic $\nu_{\mu} + \bar{\nu}_{\mu} + \nu_e$ fluxes from
Protheroe and Johnson with energy cutoff of $3 \times 10^{21}~\ev$
(solid)~\protect\cite{Protheroe:ei}, Hill and Schramm
(dashed)~\protect\cite{Hill:1983mk}, and previous estimate by Stecker
without source evolution (dotted)~\protect\cite{Stecker:1979ah}. 
Published in Ref.~\cite{Anchordoqui:2001cg}.}
\label{nuflux}
\end{figure}

In Fig.~\ref{nuflux} we show the cosmogenic neutrino flux 
obtained by Protheroe and 
Johnson~\cite{Protheroe:1995ft} using Monte Carlo simulation 
techniques for a proton injection 
spectrum with
$E_{\rm cutoff} = 3 \times 10^{21}~\ev$. 
This study incorporates the cosmological evolution of the sources from
estimates~\cite{Rachen:1992pg} of the power per comoving volume
emitted as protons by powerful radiogalaxies, taking into account
the radio luminosity functions given in Ref.~\cite{Peacock}. The flux 
peaks around $E \approx 2 \times
10^{17}$~eV, which is roughly the same energy suggested by other
analyses following a source evolution scaling like
$(1+z)^4$~\cite{Yoshida:pt,Engel:2001hd}. The figure also shows 
the flux-estimates by Hill and Schramm~\cite{Hill:1983mk}
considering contributions from semi-local nucleon sources, and an 
earlier estimate by Stecker which does not take into account the source 
evolution~\cite{Stecker:1979ah}.

In summary, cosmic microwave background (CMB) renders the Universe opaque
to protons above about $5 \times 10^{19}$~eV.  If proton sources are
are at cosmological distances ($\agt 100$ Mpc), the observed proton spectrum
should display the GZK cutoff.  

\subsection{Photonuclear interactions}

The relevant mechanisms for the energy loss that extremely high energy nuclei 
suffer during their trip to Earth are: Compton interactions, 
pair production in the field of the nucleus, photodisintegration, 
and hadron photoproduction. The Compton interactions have no threshold energy.
In the nucleus rest-frame, pair production has a threshold at 
$\sim 1$ MeV, photodisintegration is particularly important at the peak of the 
giant dipole resonance (15 to 25~MeV), and photomeson production has a 
threshold energy of $\sim 145$~MeV. 

Compton interactions result in only a negligibly small energy loss for 
the nucleus given by~\cite{Puget:nz}
\begin{equation}
-\frac{dE}{dt} = \frac{Z^4}{A^2} \rho_\gamma \left( \frac{E}{A m_p c^2}
\right)^2 \, \,\, {\rm eV\, s^{-1}} \,
\end{equation}
where $\rho_\gamma$ is the energy density of the ambient photon field in 
eV cm$^{-3}$, $E$ is the total energy of the nucleus in eV,
and $Z$ and $A$ are the atomic number and weight of the nucleus. 
The energy loss rate due to photopair production is $Z^2/A$ 
times higher than for a proton of the same Lorentz factor~\cite{Chodorowski}, 
whereas the 
energy loss rate due to photomeson production remains roughly the same. 
The latter is true because the cross section for photomeson production by 
nuclei is proportional to the mass number $A$~\cite{Michalowski:eg}, 
while the inelasticity is 
proportional to $1/A$. However, it is photodisintegration rather than 
photopair and photomeson production that determines the energetics of 
ultrahigh energy cosmic ray nuclei. During this process some fragments of 
the nuclei are released, mostly single neutrons and protons.  
Experimental data of photonuclear interactions are consistent with a 
two-step process: photoabsorption by the nucleus to form a compound state, 
followed by a statistical decay process involving the emission of one or more 
nucleons.

Following the conventions of Eq.~(\ref{conventions}), the 
disintegration rate with production of $i$ nucleons is given 
by~\cite{Stecker:fw}
\begin{equation}
R_{Ai} = \frac{1}{2 \Gamma^2} \int_0^{\infty} dw \,
\frac{n(w)}{w^2} \, \int_0^{2\Gamma w} dw_r
 \, w_r \sigma_{Ai}(w_r)
\label{phdsrate}
\end{equation}
with $\sigma_{Ai}$ the cross section for the interaction. 
Using the expressions for the cross section fitted by 
Puget, Stecker and Bredekamp~\cite{Puget:nz}, it is possible to work out
an analytical solution for the nuclear disintegration rates (see Appendix D). 
Summing over all the possible channels for a given
number of nucleons, one obtains the effective nucleon loss rate 
$R = \sum_i i R_{Ai}$. The effective nucleon loss rate for 
light elements, as well as for those in  the carbon, silicon 
and iron groups
can be scaled as~\cite{Puget:nz}
\begin{equation}
\left. \frac{dA}{dt}\right|_A \sim \left. \frac{dA}{dt}\right|_{\rm Fe} \left(\frac{A}{56}\right) = \left. R\right|_{_{\rm Fe}} \left(\frac{A}{56}\right)\,,
\end{equation}
with the photodisintegration rate parametrized by~\cite{Anchordoqui:1997rn}
\begin{equation}
\left. R(\Gamma)\right|_{_{\rm Fe}} =3.25 \times 10^{-6}\, 
\Gamma^{-0.643}                                          
\exp (-2.15 \times 10^{10}/\Gamma)\,\, {\rm s}^{-1} 
\label{oop}
\end{equation}
for $\Gamma \,\in \, [1.0 \times 10^{9}, 36.8 \times 10^{9}]$, and 
\begin{equation}
\left. R(\Gamma)\right|_{_{\rm Fe}} =1.59 \times 10^{-12} \, 
\Gamma^{-0.0698}\,\, {\rm s}^{-1}   
\end{equation}
for $ \Gamma\, 
\in\, 
[3.68 \times 10^{10}, 10.0 \times 10^{10}]$. 

For photodisintegration, the averaged fractional
energy loss results equal the fractional loss in mass number of the 
nucleus, because the nucleon emission is isotropic in the
rest frame of the nucleus. During the photodisintegration process the 
Lorentz factor of the nucleus is conserved, unlike the cases of pair 
production and photomeson production processes which involve the creation 
of new 
particles that carry off energy. The total fractional energy loss 
 is then 
\begin{equation}
-\frac{1}{E} \frac{dE}{dt} = \frac{1}{\Gamma} \frac{d\Gamma}{dt} + \frac{R}{A} \,.
\end{equation}
For $w_r \alt 145$~MeV the reduction in $\Gamma$ comes from the nuclear 
energy loss due to pair production. The $\gamma$-ray momentum absorbed by the 
nucleus during the formation of the excited compound nuclear state that 
precedes nucleon emission is ${\cal O} (10^{-2})$ times the energy loss by 
nucleon emission~\cite{Stecker:1998ib}.
For $\Gamma > 10^{10}$ the  energy loss due to photopair production is 
negligible, and thus
\begin{equation}
E (t)  \sim  938\,\, A(t)\,\, \Gamma\,\,\, {\rm MeV}  \nonumber \\
   \sim  E_0\, \exp\left[\frac{- \left. R(\Gamma)\right|_{_{\rm Fe}}\,t}
{56}\right]\,.
\label{pats}
\end{equation}
\begin{figure}
\postscript{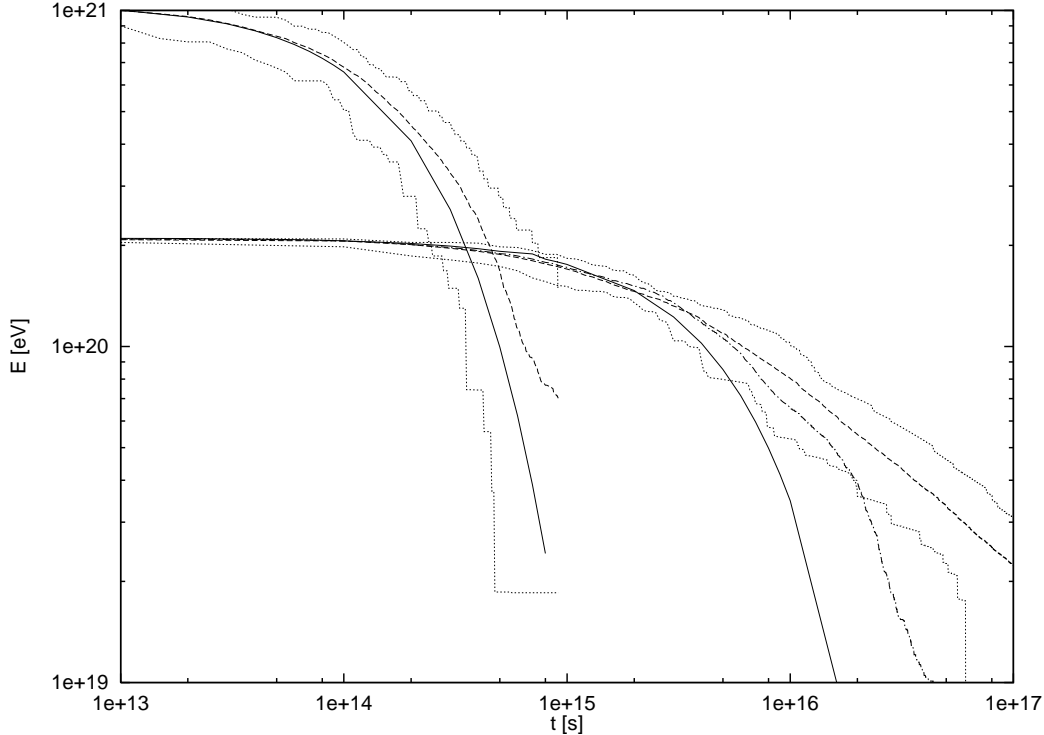}{0.90}
\caption{The energy of the surviving fragment ($\Gamma_0 = 4 \times
10^{9}$, $\Gamma_0 = 2 \times 10^{10}$) vs. propagation time
obtained using
Eq.~(\ref{pats}) is indicated with a solid line.
Also included is the energy attenuation
length obtained from Monte Carlo simulations with (dashed) 
and without (dotted-dashed) pair creation production, for comparison.
The region between the two dotted lines includes 95\% of the 
simulations. This
gives a clear idea of the range of values which can result from
fluctuations from the average behaviour. It is important to keep in mind that
a light propagation distance of $1.03 \times 10^{14}$~s corresponds to 1~Mpc.
From Ref.~\cite{Anchordoqui:1998ig}.} 
\label{gzk2}
\end{figure}

Figure~\ref{gzk2} shows the energy of the heaviest surviving nuclear
fragment as a function of the propagation time, for initial iron nuclei. 
The solid curves are obtained using Eq.~(\ref{pats}), whereas
the dashed and dotted-dashed curves are obtained by means of
Monte Carlo simulations~\cite{Epele:1998ia}. One can see that nuclei with 
Lorentz factors above $10^{10}$ cannot survive 
for more than 10 Mpc. For these distances, the approximation given in 
Eq.~(\ref{pats}) always lies in the region which includes 95\% of the 
Monte Carlo simulations. When the nucleus is emitted with a Lorentz factor 
$\Gamma_0 < 5 \times 10^9$, pair production
losses start to be relevant, significantly reducing the value of $\Gamma$ 
as the nucleus propagates distances of ${\cal O}$(100~Mpc). 
The effect has a maximum for $\Gamma_0 \approx 4 \times 10^{9}$
but becomes small again for $\Gamma_0 \leq 10^{9}$, for which
appreciable effects only appear for cosmological distances ($> 1000$
Mpc), see for instance~\cite{Epele:1998ia}. 

Note that Eq.~(\ref{pats}) imposes a  strong 
constraint on the location of nucleus-sources:  less than 1\%
of iron nuclei (or any surviving fragment of their spallations)
can survive more than $3 \times 10^{14}$~s with an energy $>10^{20.5}$ eV.
For straight line propagation, this represents a  maximum distance of 
$\sim 3$ Mpc.

Because of the position of $^{56}$Fe on the binding energy curve, it is 
generally considered to be a significant end product of stellar evolution, 
and indeed higher mass nuclei are found to be much rarer in the cosmic 
radiation. 
Specifically, the atomic abundance of nuclei heavier than iron (in the 
cosmic radiation) is expected to be smaller by 3 or 5 orders of magnitude 
relative to the lighter ones~\cite{Burbidge}. However, for super-heavy nuclei
the GZK sphere is substantially increased, because for $\Gamma \alt 10^{9}$ 
the nucleus interacts only with the tail of the Planck spectrum and infrared 
backgrounds~\cite{Anchordoqui:1999aj,Anchordoqui:2000sm}. Indeed, because 
of the paucity of data, there is 
a great uncertainty in the chemical composition at the end of the spectrum, so 
the situation is ripe for speculation on super-heavy nuclei as 
primaries~\cite{Anchordoqui:2000sk}.

\subsection{Relativistic specks of dust}

Dust is a very widespread component of diffuse matter in the galaxy and,
apparently, in intergalactic space. In particular, specks expelled by
radiation pressure from cool stars may be injected into the interstellar
medium with speeds $\sim 10^{8}$ cm s$^{-1}$~\cite{Wickramasinghe:72},
generally carrying a net electric charge. Therefore, these small solid
particles can be re-accelerated very effectively to ultra-high energies at
shock-wave fronts~\cite{Spitzer:1949,Wickramasinghe:74,Epstein:mc}.
However, the situation is unusual, because the  Lorentz factors involved
can be small even at very high energies. 

We now discuss the survival
probability of such specks of dust. The break up of dust grains is mainly
due to electrical stresses induced by the accumulation of charge as a
result of the photoelectric effect in the galactic optical radiation and
also as a result of ionization by the hydrogen atoms of interstellar gas.
The critical electric charge on the surface of a dust grain of radius $R$,
for which repulsion forces break up the particle is given by~\cite{Berezinsky:73}
\begin{equation} 
q_c = \sqrt{4\,\pi\,f_0}\,R^2\,, 
\end{equation} 
where $f_0$ denotes the breaking stress. Subrelativistic ($v \alt 3 \times
10^9$~cm/s) dust grains disintegrate in collisions with the nuclei and
electrons of the interstellar medium. Ionization takes place along the
path of the nucleus in the dust grain, heating the matter up to $10^{5}$ K
in a narrow channel of radius $\sim 10^{-7}$~cm. Close to the surface of
the channel the local pressure reaches $10^{4}$~kg mm$^{-2}$, which results in
microfractures. Because of the small size of the grain, microfractures are
unstable and increase until the complete break-up of the particle occurs.
The charge accumulation rate on a small grain $(R<\overline{R})$ is given
by~\cite{Ginzburg:sk} 
\begin{equation} 
\frac{dq}{dt} = 0.6\, A^{-1}\, v\,
n_{\rm H}\, \pi\, R^2\, \left(\frac{c}{v}\right)^2 \,(\rho R)^{0.41}
\end{equation} 
and on a large one $(R>\overline{R})$ 
\begin{equation}
\frac{dq}{dt} = 2\, A^{-1}\, v\, n_{\rm H}\, \pi\, R^2\, \sigma_T \,N_A\,
\rho \,\left(\frac{c}{v}\right)^2 \,\frac{m_ec^2}{e\,\sqrt{4\pi f_0}}\,,
\end{equation} 
with 
\begin{equation} \overline{R} = 1.1 \times 10^2
\,\rho^{1.44}\, f_0^{-1.22}\,\, \mu{\rm m}\,. 
\end{equation} 
Here, $v$ is
the velocity of the dust grain, $\rho$ its density, 
$n_{\rm H}$ is the density of hydrogen
atoms in the surrounding medium, $A$ is the mass number, $N_A$ is
Avogadro's number, and $\sigma_T$ is the Thomson cross section.
Significant accumulation of charge occurs during the trip through the
galaxy, where the density of hydrogen is $n_{\rm H} \sim 0.6$
atoms/cm$^3$. The path length up to the first break up in the galactic
gas, 
\begin{equation} 
D = v q_c \left( \frac{dq}{dt} \right)^{-1}\,,
\end{equation} 
turns out to be $D \alt 10^{15}$~cm, 
for graphite grains ($f_0 \sim 1 $~kg mm$^{-2}$), and $D \alt 10^{16}$~cm, for 
iron  grains ($f_0 \sim 10 $~kg mm$^{-2}$)~\cite{Ginzburg:sk}.

\begin{figure}
\postscript{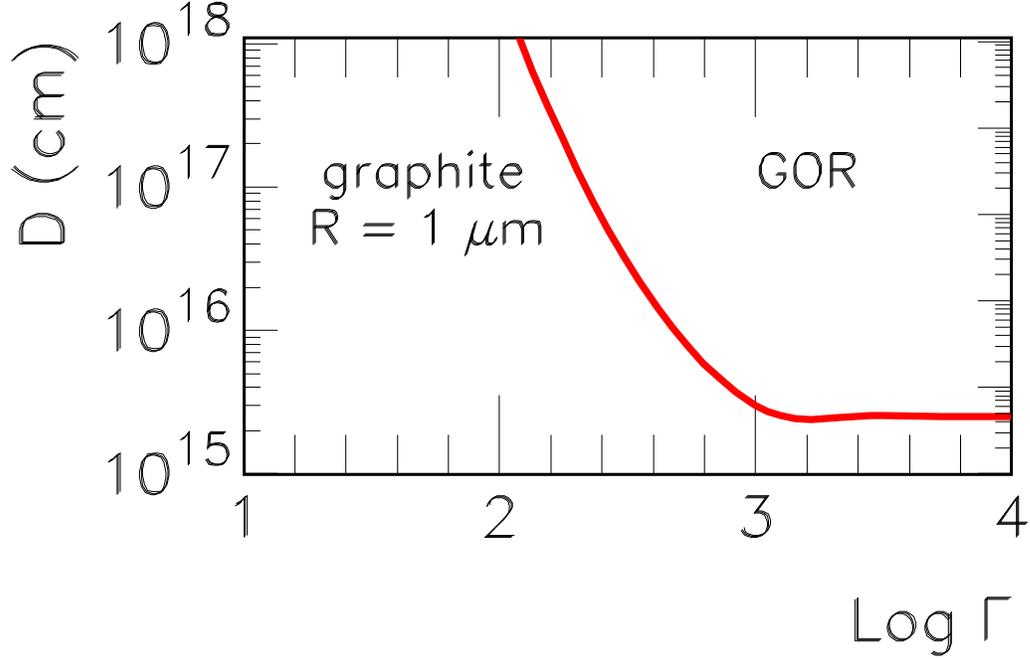}{0.90}
\caption{The path length of a graphite grain of radius $R =1\,\mu$m up to 
the first break-up, in the field of the galactic optical radiation (GOR). 
Adapted from Ref.~\cite{Ginzburg:sk}.} 
\label{gzkdust}
\end{figure}

The threshold value of the Lorentz factor of a relativistic grain 
with energy $E$ and 
density $\rho$ (measured in g cm$^{-3}$), under the influence of the 
photoelectric effect with photons of energy $w$, is given 
by~\cite{Berezinsky:73}
\begin{equation}
\Gamma \sim \left(\frac{3 E}{4 \pi \rho}\right)^{1/4} 
\left(\frac{{\cal E}}{wc} \right)^{3/4} = 9.0 \times 10^{-2} \,\left(\frac{f_0^{3/2} \,E_{_{\rm GeV}}}{w_{_{\rm eV}}^3 \rho}\right)^{1/4}\,,
\end{equation} 
where $E_{_{\rm GeV}} \equiv E/{\rm GeV}$, 
$w_{_{\rm eV}} \equiv w/{\rm eV}$, and $f_0$ is measured in kg mm$^{-2}$. 
Thus, a carbon dust grain with 
$E \sim 10^{20}$ eV and $\Gamma > 25$ is split up by  starlight photons 
($w \alt 2$~eV). Relativistic dust grain break up is particularly effective 
within the solar system. The distance the grain can approach the Sun without 
splitting up is given by~\cite{Berezinsky:73}
\begin{equation}
D \sim  \frac{L(w) \,\chi\, e \,R^2}{4\, c \,q_c} = 2.3 \times 10^{-25}\, 
\frac{\chi\, L(w)}{\sqrt{f_0}}\,
\,{\rm cm} \,,
\label{stephan}
\end{equation}
where $L(w)$ is the number of photons with energy above 
threshold emitted by the Sun measured in s$^{-1}$, and $\chi \sim 0.01 - 0.1$ 
is the number of electrons stripped from the speck of dust by one photon. 
For carbon dust grains with 
$\Gamma > 25$ ($\omega \alt 2$~eV and $L(w) \sim 3.5 \times 10^{44}$~s$^{-1}$), 
Eq.~(\ref{stephan}) leads to $D \sim 10$ pc. Figure~\ref{gzkdust} shows the 
path length of a typical graphite grain in the field of the galactic optical 
radiation.

Doubts have also been expressed about the
prospects of surviving against the heating  from photoionization within
the solar radiation field~\cite{McBreen}. Consider a grain moving
(radially) towards the Sun with constant velocity $v\equiv \beta c$ and
subtending an effective solid angle $d\Omega$, with $\alpha$ the angle
between the position vector $\hat{r}$ of the grain and the positive
$x$-axis defined by a line parallel to the direction of motion. In the
rest system of the grain, the average energy of the photons is blue-shifted 
to $\Gamma w (1 - \beta \cos \alpha)$. The Sun no longer appears
to emit isotropically; namely, the number of incident photons per second
is $\Gamma \,L(w)\, (1 - \beta \cos \alpha) d\Omega /4\pi$. To estimate
the surface temperature $T$ of the grain we assume that it radiates as a
black body,
 \begin{equation} \epsilon \,\sigma_B \,T^4\, 4 \pi R^2 =
\Gamma^2\, L(w)\, w\, (1- \beta \cos \alpha)^2 \,\frac{d \Omega}{4\pi}\,,
\label{p} \end{equation} where $\sigma_B$ is the Stefan-Boltzman constant
and $\epsilon$ the emissivity of the grain. The solid angle is determined
by the cross section for absorption by the grain. For a particle
traveling in the $r$-direction towards the Sun, \begin{equation} d\Omega
= {4\, N\, \sigma \,\pi \,R^3}{3\, r^2}\,, \label{q} \end{equation} where
$N$ is the volume density of grain atoms and $\sigma$ the ionization cross
section per atom. For iron, $\sigma \sim 3 \times
10^{-16}\,\Gamma^{-1}$~cm$^2$~\cite{Veigele:73}. Now, setting $T$ equal to
the melting point $T_c$ of the grain, Eqs.~(\ref{p}) and (\ref{q}) combine
to yield the following expression for the position $r_\alpha$ at which the
grain melts \begin{equation} r_\alpha = \tilde{r}_\alpha (1-\beta \cos
\alpha)\,, \end{equation} where \begin{equation} \tilde{r}_\alpha =
\sqrt{\frac{\Gamma^2\,R\,L(w)\,w\,N\,\sigma}{12 \pi\, \epsilon\,
\sigma_B\, T^4_c}}\,. \label{samy} \end{equation} 
For an iron grain with
$T_c \sim 1800$ K, $\epsilon \sim 0.5$ and $N \sim 8.6 \times 10^{22}$
atoms cm$^{-3}$, Eq.~(\ref{samy}) leads to \begin{equation}
\tilde{r}_\alpha \sim 3 \times 10^{15} \, \sqrt{\Gamma\, R} \,\,\,{\rm
cm}. \end{equation} Now, since \begin{equation} E = \frac{4}{3} \pi \,R^3
\,c^3 \,\Gamma\, \rho \,. \end{equation} for an iron grain of $E\sim
10^{19}$~eV, $\tilde{r}_\alpha = 1.0 \times 10^{13}\,\,\Gamma^{1/3}$~cm.
Hence, relativistic ($\log \Gamma < 2$) iron grains would melt before
reaching the Earth's orbit around the Sun, i.e., $r_\alpha \sim 1$ au =
$1.5 \times 10^{13}$~cm. There is no substantial change for an iron grain
traveling non-radially, it would melt before reaching the 
Earth~\cite{McBreen}. However,
note that dust grains composed of carbon
($T_c \sim 4000$~K) may reach the Earth's orbit with $E\sim 10^{20}$~eV if
$\Gamma \sim 20$.  

In summary, subrelativistic dust grains have no chance to survive the trip
to Earth.  However, relativistic carbon specks could survive a trip of $\approx$~100~pc if one
assumes the most optimistic parameters.  Since the closest supernova remnant is about 
200~pc away~\cite{Combi:rx}, astrophysical models cannot rule out relativistic dust grains as 
ultrahigh energy CR primaries.

\subsection{Electromagnetic cascades}

The dominant absorption process for high energy $\gamma$-rays (energy $E$) 
traversing cosmic distances is pair creation through collisions with the 
various radiation fields (energy $w$) permeating the 
Universe~\cite{Nikishov,Goldreich,Jelley,Gould:67b,Stecker:g69,Fazio:70,Berezinsky:g}. The 
fundamental process is well understood and its amplitude can be calculated 
accurately by general 
perturbation methods developed in quantum electrodynamics 
(QED)~\cite{Gould:67a,Brown:cc,Bjorken:dk}.
The lowest-order single pair creation cross section~\cite{Nikishov}
\begin{equation} 
\sigma = \frac{1}{2} \pi r_0^2 (1 - \beta^2) \left[ (3-\beta^4) \ln \left(
\frac{1 +\beta}{1-\beta}\right) - 2 \beta (2 - \beta^2) \right]\,\,,
\end{equation}
peaks near the threshold $E_{\rm th} = m_e^2/w 
\sim 2.6 \times 10^2 (w/{\rm eV})^{-1}$~GeV, and falls off 
asymptotically as $s \rightarrow \infty$~\cite{Brown:cc}, 
\begin{equation}
\sigma(s) \sim \frac{4 \pi \alpha^2}{s} 
\left[\ln \left(\frac{s}{m_e^2}\right) - 1 \right]\,.
\end{equation}
Here, $\beta = (1 - 1/s)^{1/2}$ is the electron (positron) velocity in 
the c.m. system and $r_0$ is the classical electron radius.
Higher order QED processes with more than two final state particles 
become important with rising energy.  
For example, the fourth-order two-pair creation cross section is a sharply 
rising function of $s$ near the threshold that quickly approaches its 
asymptotic value~\cite{Brown:cc}\footnote{$\zeta(3) \approx 1.202$ is 
the zeta function with argument 3.}
\begin{equation}
\sigma(s) \sim \frac{\alpha^4}{36 \pi m_e^2}\,[175 \zeta(3) - 38] \sim 6.45\,\,\mu{\rm b}\,,
\end{equation}
and becomes bigger than the single
pair creation cross section when $s\agt 1$~GeV$^2$. 
With this in mind, the most efficient targets for 
$\gamma$-rays of 
energy $E$ are background photons of energy $w \sim m_e^2/E$.
This implies that for $\gamma$-ray energies $\agt 10^{20}$~eV, 
interactions with the radio background ($w \alt 10^{-6}$~eV $\sim 100$~MHz) 
become more important than those with the CMB.

We have no direct knowledge of the cosmic radio background, mostly because it 
is difficult to disentangle the galactic and extragalactic component.
For a temperature of the extragalactic component 
of 15~K at 178 MHz one finds
\begin{equation}
n(w) \sim K w^{-2}\,,
\end{equation}
with $K = 1.09 \times 10^{20}$~erg/cm$^3$~\cite{Turtle}. 
At frequencies somewhere below 1 MHz the intensity is expected to fall off 
exponentially; the principal absorption process would be
inverse Bremsstrahlung or ``free-free'' absorption by the intergalactic 
plasma. Observational estimates are consistent with a cutoff in the spectrum 
at 
$w \alt 10^{-8}$~eV~\cite{Clark}. However, that part of the spectrum where the 
Universe is optically thick to radio emission depends on the abundance and 
clustering of electrons in the extragalactic medium and/or the radio source, and is 
uncertain between 0.1 - 2 MHz. A theoretical estimate~\cite{Protheroe:1996si} of 
the intensity down to kHz frequencies, based on the observed luminosity 
functions and radio spectra of normal galaxies and radio-galaxies, tends to 
give higher estimates than those of Ref.~\cite{Clark}.

\begin{figure}
\postscript{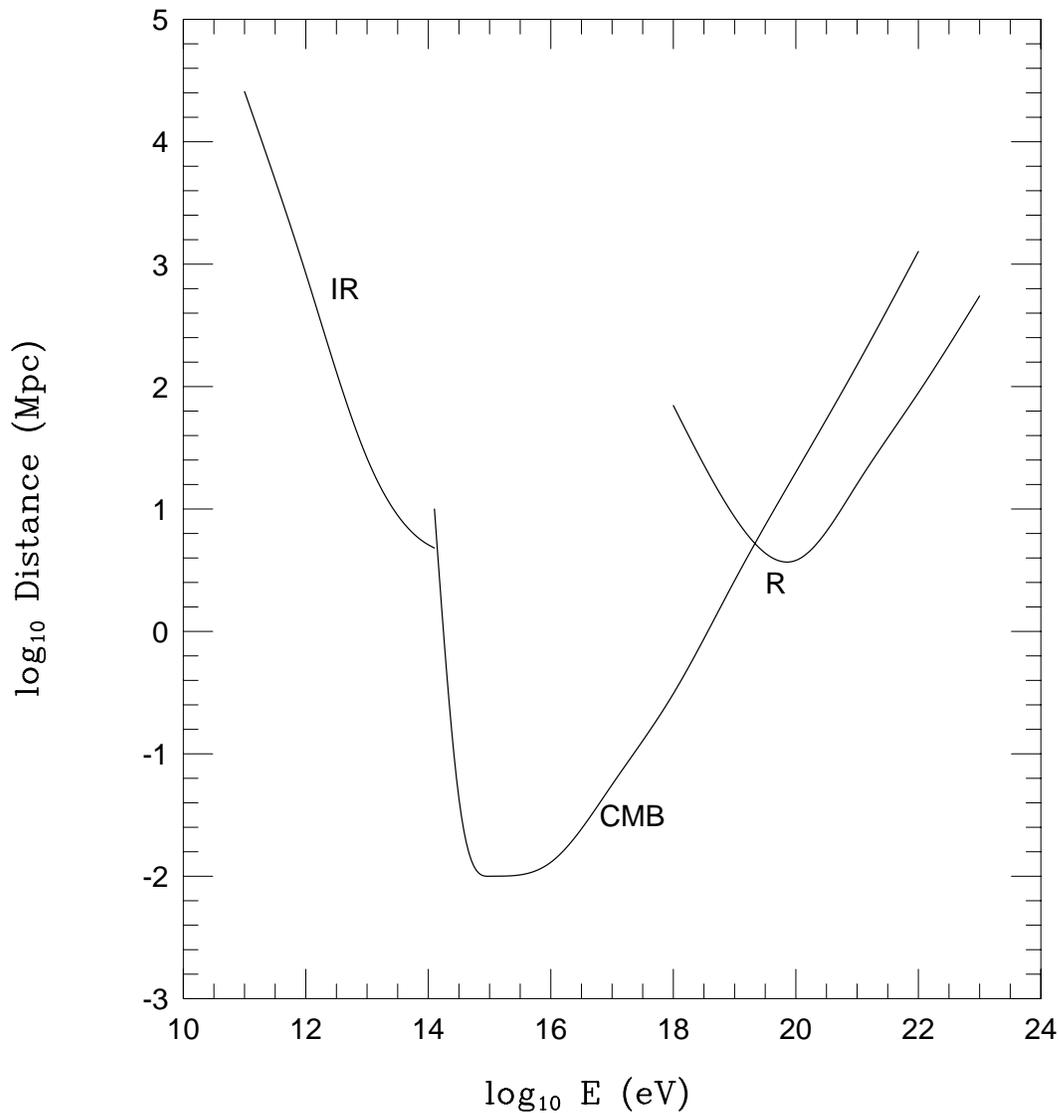}{0.90}
\caption{Mean pair production interaction length for $\gamma$-rays 
in the infrared and optical 
backgrounds (IR), radio background (R), and relic photons (CMB). Adapted 
from Ref.~\cite{Protheroe:1995ft}.}
\label{gzkgamma}
\end{figure}

The probability per unit path length  
that a $\gamma$-ray is converted into $e^+ e^-$ pairs in a 
collision with a low-energy photon is given 
by~\cite{Nikishov}
\begin{equation}
\frac{dP}{dx} = 2 \left(\frac{m_e^2}{E}\right)^2 
\int_0^\infty \frac{n(w)}{w^2}\,dw\,
\int_1^{s_0} s \, \sigma(s)\, ds\,,
\end{equation}
where $s_0 = Ew/m_e$.
The mean interaction length of pair production, showing the contribution of 
the various radiation 
fields, is given in Fig.~\ref{gzkgamma}. For the radio background, we assume
the conservative estimate of Ref.~\cite{Clark}. Using the radio spectrum 
derived by Protheroe and Biermann leads to a photon mean free path 
(at $E \sim 10^{21}$~eV) a factor of 3 -- 10 smaller~\cite{Protheroe:1996si}.

In the Klein-Nishina limit, $s\gg m_e^2$, one of the pair-produced particles
carries most of the $\gamma$-ray energy. In the absence of 
magnetic fields, this leading electron (positron) can deliver a substantial 
fraction of the original $\gamma$-ray's energy to a new $\gamma$-ray via 
inverse Compton scattering (ICS), with cross section
\begin{equation}
\sigma_{\rm ICS} (s) \sim \frac{2 \pi \alpha^2}{s} 
\left[\ln \left(\frac{s}{m_e^2}\right) - 1 \right]\,.
\end{equation}
The scattered photon, which is now the leading 
particle, can initiate a fresh cycle of pair production and inverse Compton 
scattering interactions, yielding an electromagnetic cascade.

For $s\ll m_e^2$, $\sigma_{\rm ICS} \sim \sigma_{\rm T}$, and the fractional 
energy loss is given by~\cite{Felten}
\begin{equation}
-\frac{1}{E} \left(\frac{dE}{dt}\right)_{\rm ICS} \sim \frac{4}{3}\, \sigma_{\rm T} \,
c\, \Gamma \,\frac{w\,n(w)}{m_e} \,.
\end{equation} 
Corrections to the lowest order ICS cross section from processes involving 
additional photons in the final state are less than 10\% in the ultrahigh 
energy range~\cite{Gould:pw}. Furthermore, various other possible processes, 
such as those involving production of one 
or more muons, taus, pion pairs, and Bethe-Heitler pairs ($\gamma X \rightarrow X e^+ e^-$, where $X$ stands for an atom, ion or free electron), 
can be safely neglected~\cite{Brown:cc}. For example, the total cross 
section for the lowest-order single muon pair production is around an order 
of magnitude smaller than that for electron pair production.

The development of  electromagnetic cascades depends sensitively on the 
strength of the extragalactic magnetic field $B$, which is rather uncertain. 
For $B \agt 10^{-11}$~G the 
electrons will lose much of their energy through synchrotron radiation 
before they have the opportunity to undergo ICS, and the electromagnetic 
cascade will be virtually terminated~\cite{Wdowczyk}. Specifically, an electron of energy 
$E$ moving in a field having a perpendicular component of magnitude 
$B_\perp$ radiates energy at a rate
\begin{equation}
-\frac{dE}{dx} = 3.3 \times 10^{-14} \, 
\left(\frac{B_\perp}{1\,{\rm G}}\right)^2
\left( \frac{E}{m_e c^2} \right)^2\,{\rm eV\, cm}^{-1}\,.
\end{equation}
Hence, if the radiated energy is less than say 10\% of the 
electron energy in an interaction length $\lambda$, the electron 
synchrotron cools before it can undergo ICS, and thus the cascade 
development stops. The above cooling time scale holds provided that
\begin{equation}
B_\perp^2 < 7.9 \times 10^{23} \frac{1}{E_{_{\rm eV}}\,\lambda_{_{\rm cm}}}\,
{\rm G}^2\,,
\end{equation}
where $E_{_{\rm eV}} \equiv E/1 \,{\rm eV}$ and $\lambda_{_{\rm cm}} 
\equiv \lambda/1\,{\rm cm}$.
Then, for $3 \times 10^{-12}$~G, 
$2 \times 10^{-11}$G, and $1\times 10^{-10}$~G, the intergalactic space 
becomes opaque to the propagation of $\gamma$-rays of 
$E = 10^{21}$~eV, $E=10^{20}$~eV, and  $E= 10^{19}$~eV, 
respectively.

Putting all this together, the GZK radius of the photon channel strongly 
depends on the 
strength of extragalactic magnetic fields. In principle, distant 
sources with $z > 0.03$ can contribute to the observed rays above 
$5 \times 10^{19}$~eV if the extragalactic magnetic field does not exceed 
$10^{-12}$~G~\cite{Kalashev:2001qp}. However, comparison of the 
expected and observed numbers of CRs from the direction of 
the nearby source Centaurus A allows one to place bounds on 
intergalactic magnetic fields 
of $\gg 10^{-11}$~G~\cite{Anchordoqui:2001bs}. In such a 
case, the survival probability for $3\times 10^{20}$~eV 
$\gamma$-rays to  distance $D$,  
\begin{equation}
P(>D) = \exp[-D/6.6\, {\rm Mpc}]\,,
\end{equation}
becomes less than $10^{-4}$ after traversing a 
distance of 50 Mpc~\cite{Elbert:1994zv}.

\subsection{Deflection and delay due to magnetic fields}

Cosmic rays {\em en route} to Earth suffer deflection and
delay in magnetic fields, effects which can camouflage
their origins.  In this section, we discuss bending of
cosmic ray orbits in both Galactic and extragalactic
fields.  Various combinations of field strength and
coherence length are considered, and we discuss
solutions to the diffusion equations as well as
results from Monte Carlo simulation.      

The magnetic deflection of
protons in the Galactic disk has been studied in detail 
by Stanev~\cite{Stanev:290}. This analysis includes two extreme options 
for the behaviour of the field, reflecting the different symmetries with
respect to field reversals in the $r$- and $z$-directions. The
$B$-field has a $1/r$ behaviour, with deviations calculated out to
20 kpc from the Galactic center. The r.m.s. deviation averaged
over arrival direction, for an energy of $2 \times 10^{19}$~eV 
($4\times 10^{19}$ eV), varies from $17.7^\circ$ to $23.7^\circ$ 
($7.9\deg$ to $10.5\deg$) in going between the two models. This deviation 
shows an approximately linear decrease with increasing energy. 
In the case of heavy nuclei, the arrival direction of the particles is 
strongly dependent on the coordinates of the 
source~\cite{MedinaTanco:1997rt,Harari:1999it,Harari:2000az}. The Larmor 
radius of a nucleus with 
$E/Z = 10^{19}$~eV in a uniform magnetic field of strength $B \sim 3~\mu$G 
is slightly larger than 3 kpc. The typical large scale field intensity of 
the Milky Way is a few $\mu$G, and is approximately uniform over scales of 
the order of a few kpc. Therefore, the propagation in the Galactic magnetic 
field of a nucleus with  $E/Z > 10^{19}$~eV should in general be not very 
different from a quasi-rectilinear trajectory, with deflections away from the 
straight path becoming smaller with increasing energy. However, for $10^{18}\,{\rm eV} 
< E/Z < 10^{19} \, {\rm eV},$
the deflection may be high enough ($>20^\circ$)  to make any 
identification of the sources extremely difficult.\footnote{At smaller 
energies 
the drift and diffusive regimes dominate~\cite{Lee:ki}.} Moreover, 
lensing effects 
yield (de)magnification of the CR-fluxes, and not every incoming 
particle direction is allowed between a given source and the detector. This 
generates sky patches which are virtually unobservable from 
Earth~\cite{MedinaTanco:1997rt,Harari:1999it}. On the 
other hand, the enhancement of the probability to detect events from a given 
source in a narrow energy range near the caustic implies a concentration 
of events around the location where the new pair of images forms. Therefore, 
magnetic lensing around caustics becomes a potential source of clustering 
in the angular distribution of arrival directions~\cite{Harari:2000az}. 
Precise predictions depend upon the detailed structure of the magnetic field 
configuration, which is not so well known. However, new experiments such as 
PAO may make it possible to use CRs as probes of the magnetic field 
structure~\cite{Harari:2002fj,Anchordoqui:2002dj}.

At present, surprisingly little is known about extragalactic 
magnetic fields (EGMFs). There are some measurements
of diffuse radio emission from the bridge area between the Coma
and Abell superclusters~\cite{Kim}, which, under
assumptions of equipartition allows an estimate of ${\cal
O}(0.2-0.6)\,\mu$G for the magnetic field in this 
region.\footnote{Fields of ${\cal O}(\mu{\rm G})$ are also indicated in
a more extensive study of 16 low redshift clusters~\cite{Clarke}.}
Such a large field may possibly be understood if the bridge region lies
along a filament or sheet~\cite{Ryu}. In addition, the
absence of a positive signal in Faraday rotation measurements on
distant quasars~\cite{Kronberg:1993vk} provides upper limits on magnetic fields
of any origin as a function of reversal scale~\cite{Vallee}. 
These bounds depend significantly on assumptions about the electron density
profile as a function of the redshift. When electron densities
follow that of the Lyman-$\alpha$ forest, the average magnitude
of the magnetic field receives an upper limit of $B \sim 10^{-9}$~G for
reversals on the scale of the horizon, and $B \sim 10^{-8}$~G for
reversal scales on the order of 1~Mpc~\cite{Blasi:1999hu}. The latter
upper bound is roughly coincident with the lower limit in the 
Galactic neighborhood~\cite{Anchordoqui:2001bs}.

If the Larmor radius of a particle is sizably larger than the coherence 
length $\ell$ of the extragalactic magnetic field, the typical deflection 
angle from the direction of the source, located at a distance $d$, can be 
estimated assuming that the particle makes a random walk in the magnetic
field~\cite{Waxman:1996zn}, 
\begin{equation}
\theta(E) \simeq 0.54^\circ\,Z\, \left(\frac{d}{1\ {\rm Mpc}}\right)^{1/2}\,
\left(\frac{\ell}{1 \,\, {\rm Mpc}} \right)^{1/2} \,
\frac{B_{{\rm nG}}}{E_{20}}\ \ ,\label{random}
\end{equation}
where $B_{{\rm nG}} \equiv B/1$~nG, and $E_{20} \equiv E/10^{20}$~eV.
Note that deflection also implies an average time 
delay 
\begin{equation}
\tau_{\rm delay}(E) \approx \frac{d\,\, \theta^2}{4\,c}
\label{tdelay}
\end{equation}
relative to rectilinear propagation with 
the speed of light. With increasing magnetic field strength, the Larmor 
radius decreases and the hypothesis of straight line propagation between 
scatterings, that Eq.~(\ref{random}) relies upon, breaks down. 
A diffusive approach is more appropriate for this situation~\cite{Wdowczyk:sg,Blasi:1998xp,Sigl:1998dd,MedinaTanco:1998ie,Lemoine:1999ys,Farrar:1999bw,Harari:2002dy}.

The Local Supercluster is a flattened overdensity of galaxies extending
for $\sim 30 - 40$~Mpc with a width of $\sim 10$~Mpc~\cite{Ginzburg:sk}. The 
Virgo cluster is approximately in the center of this distribution, 
about 17~Mpc away from the Milky Way, which is located at the edge.  
This large scale distribution defines the super-Galactic plane where it 
intersects almost perpendicular the Galactic plane. A broad $\sim 50^\circ$ 
region 
centered at the Local 
Supercluster seems to be endowed with a magnetic field with a uniform 
component of strength $\sim 1.5 \, \mu$G~\cite{Vallee:99}, and a random 
component of strength $\sim 1 \, \mu$G~\cite{Thomson}. If this is indeed 
the case, all CRs traveling to Earth will propagate diffusively.

The evolution of the nucleon spectrum is governed by the balance equation
\begin{equation}
\frac{\partial n}{\partial t} = \frac{\partial[b(E) n]}{\partial E} +
D(E) \,\nabla^2 n + Q
\label{spectrum}
\end{equation}
which takes into account the conservation of the total number of particles 
in the spectrum, i.e., $\int n \,d^3\vr = n_0$. 
In the first term on the right, $b(E) \equiv dE/dt$ is 
the mean rate at which particles lose energy, and the last term, $Q$ is
the number of nucleons per time generated by the source(s).
The second term stands 
for the diffusion in the extragalactic medium. 
Two extreme regimes can be distinguished depending on whether 
the particle remains trapped in magnetic subdomains or not, yielding 
different functional dependence on energy of the diffusion 
coefficient~\cite{Wentzel} 
\begin{equation}
D(E) = \frac{1}{3}\, r_{_{\rm L}}\,c \frac{B^2}{\int_{\frac{1}{r_{_{\rm L}}}}^\infty dk\,P(k)}\,.
\label{w}
\end{equation}
Here, $P(k)$ is the magnetic field power spectrum. 
As mentioned earlier, extragalactic magnetic field
strengths and coherence lengths are not well established, but 
for $r_{_{\rm L}} \alt \ell$, it
may be plausible to assume a Kolmogorov~\cite{Kolmogorov} form for the 
turbulent magnetic field power spectrum, $P(k) = P_0 (k/k_0)^{-5/3}$, 
with coherent directions on scales of 0.5 -
1 Mpc. Here, $k_0 \sim 1/\ell$ is the small wavenumber limit for the 
magnetic field and $3\,P_0\,k_0/2 = B^2$. One can now easily 
estimate that 
protons with
energies $E_c < 10^{21} \ell_{\rm Mpc} B_{\mu{\rm G}}$ eV remain
trapped inside magnetic subdomains of size 
$\ell_{\rm Mpc}= \ell/(1\ {\rm Mpc})$, attaining efficient
diffusion when the wave number of the associated Alfv\'en wave is equal to
the Larmor radius of the particle~\cite{Wentzel,Drury}. The replacement of 
the Kolmogorov spectrum into Eq.~(\ref{w}) leads to
\begin{equation}
D(E) \approx 0.048 \left( \frac{E_{20} \,\ell^2_{{\rm
Mpc}}}{B_{\mu{\rm G}}} \right)^{1/3} \,{\rm Mpc}^2/{\rm Myr}.
\label{dg}
\end{equation}
As the particle energy increases ($E>E_c$), there is a transition to 
Bohm diffusion when $r_{_{\rm L}} \sim \ell$, and the diffusion coefficient is of the 
order of the Larmor radius times the velocity. Numerical 
simulations~\cite{Sigl:1998dd} show that for Bohm diffusion, $D \sim 3 r_{_{\rm L}} c$, or equivalently, $D(E) \sim 0.1 E_{20}/B_{\mu{\rm G}}
\,{\rm Mpc}^2/{\rm Myr}$.\footnote{Criticisms 
of these assumptions have been raised in~\cite{Casse:2001be}.}

Now we consider the propagation of a burst from a single source.  
Idealizing the emission to be uniform with a 
rate $dN/dt = n_0/\tau$, we have 
\begin{equation}
Q = \frac{n_0}{\tau}\, \delta^3(\vr)\,\, [\Theta(t'- t_{{\rm on}}) - 
\Theta(t' - t_{{\rm off}})],
\end{equation}
where  $\int Q \,d^3\vr \,dt' = N_{\rm tot}$, and $\Theta$ is the 
Heaviside step function. Here $t_{{\rm on}}$ ($t_{{\rm off}}$) is the time 
since the engine turned on (off) its CR production. 
If the energy loss term  is neglected, 
the solution to Eq. (\ref{spectrum}) reads,
\begin{equation}
n(\vr, t) = \int dt' \int d^3\vr \,\,G(\vr-\vr', t-t') \,\, Q(\vr', t'),
\label{ndif}
\end{equation}
where 
\begin{equation}
G(\vr - \vr', t - t')  =  [4 \,\pi \,D \,(t-t')]^{-3/2}\,\, \Theta (t - t') 
  \exp\{-(\vr - \vr')^2/ 4D(t-t')\}\,,
\end{equation}
is the Green function~\cite{Duff-Naylor}.
Before integrating over the range of possible propagation times, we note 
that there is a negligible contribution from times prior to the arrival time 
of the diffusion front, 
\begin{equation}
\tau_D = \frac{r^2}{4D} \,. 
\label{steve}
\end{equation}
For a bursting source, i.e, 
$t_{\rm off} - t_{\rm on} \sim \delta t$, Eq.~(\ref{ndif}) 
leads to~\cite{Farrar:2000nw}
\begin{eqnarray}
n(\vr, t) & = & \frac{n_0}{\tau} \int_{t_{{\rm min}}}^{t_{{\rm max}}} dt' 
\frac{e^{-r^2/4(t-t')}}{[4\, \pi\,D (t-t')]^{3/2}}  
 =  \frac{n_0}{\tau} \frac{1}{4 \,\pi \,D \,r}\,\Phi 
\left[\frac{r}{\sqrt{4 D (t-t')}} \right] \nonumber \\
 & \approx & \frac{n_0}{\tau} \frac{2}{[4 \pi D]^{3/2}} 
[t_{{\rm min}}^{-1/2} - 
t_{{\rm max}}^{-1/2}  + {\cal O} (t^{-3/2})],
\end{eqnarray}
where $\Phi(x)$ is the error function, and we have set 
present observation time at $t=0$. Note that for Kolmogorov diffusion
$D \sim E^{1/3}$. Now, shock acceleration results in a spectrum at the 
source $Q \propto E^{-\gamma}$ with $\gamma$ slightly grater than 2. 
\begin{figure}
\postscript{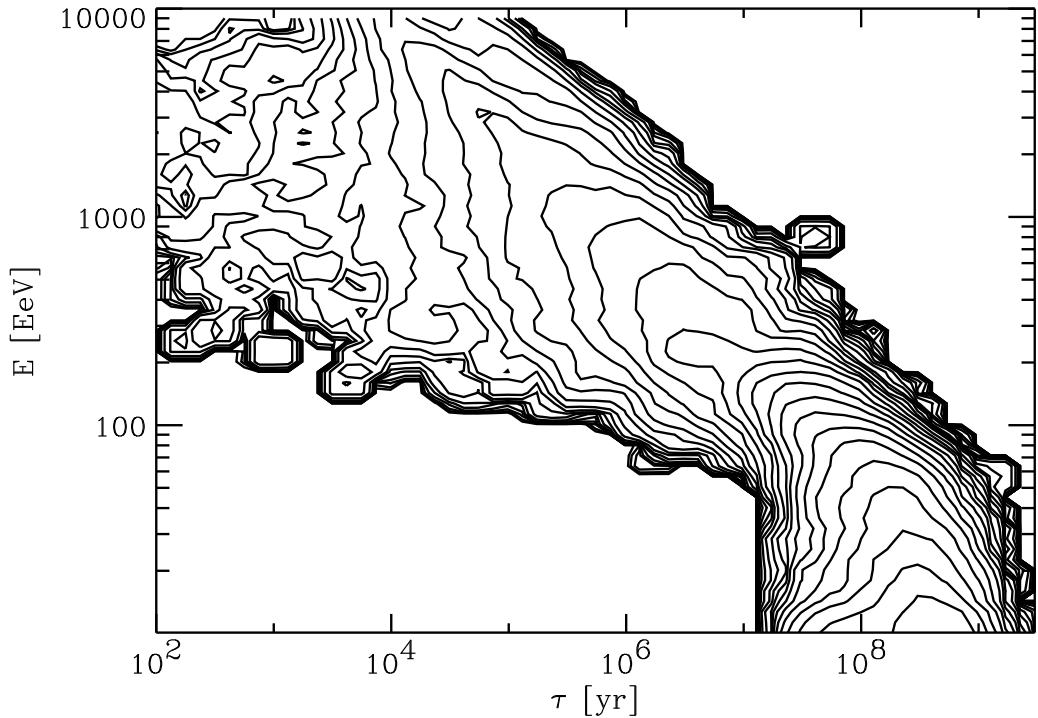}{0.90}
\caption{The distribution of time delays $\tau(E)$ and energies $E$ for a 
burst with spectral index $\gamma = 2.4$ at a distance $d = 10$~Mpc and 
r.m.s. strength $B = 3 \times 10^{-7}$~G. The inter-contour interval is 
0.25 in the logarithm to base 10 of the distribution per logarithmic energy 
and time interval. Published in Ref.~\cite{Sigl:1998dd}.}
\label{SF7}
\end{figure}
Therefore, the purely diffusive 
solution results in a spectrum $n(E) \propto E^{-\gamma + 1/3}$, close to  
the observed spectrum $E^{-2.7}$~\cite{Blasi:1998xp}. The 
condition of diffuse propagation is 
\begin{equation}
\tau_D \gg \tau_s \,,
\label{taud}
\end{equation}
where $\tau_s = r/c$ is
the  time required for a straight line propagation. 

If photopion production becomes important, one has to resort to numerical 
Monte Carlo simulations. Detailed studies of CR propagation in turbulent 
plasmas have been reported~\cite{Sigl:1998dd,Lemoine:1999ys}. In these 
analyses the magnetic field is 
characterized by a Gaussian random 
field with zero mean and a power spectrum with $\langle B^2(k)\rangle 
\propto k^{n_B}$ for $k<k_0$ and $\langle B^2(k) \rangle$ = 0 otherwise, where 
$k_0 = 2\pi/\ell$ represents the numerical cutoff scale and the r.m.s. 
strength is $B^2 = \int_0^\infty dk\,k^2 \,\langle B^2(k) \rangle$. 
The field is then calculated on a grid in real space via Fourier transform. 
Figure~\ref{SF7} shows an example of the 
distribution of arrival time and energies obtained via simulation of a 
bursting source located at $d= 10$~Mpc and assuming $n_B = -11/3$, i.e., a 
turbulent Kolmogorov type spectrum~\cite{Sigl:1998dd,Lemoine:1999ys}. 
Note that for $B=0.3~\mu$G, $\ell=1$~Mpc, 
$r=10$~Mpc, and  $E=10^{19}$~eV, Eq.~(\ref{dg}) leads to $\tau_D=750$~Myr, 
in the midregion of the time-delay given in Fig.~\ref{SF7}. 
For $E=10^{20}$~eV (where proton energy losses are still small), 
$\tau_D =350$~Myr, not inconsistent with the results of Fig.~\ref{SF7}, 
although at the margin of the distribution of time-delays for this energy.
This marginal agreement is indicative of the transition to Bohm diffusion, 
where the validity of Eq.~(\ref{dg}) begins to flag.

\section{Our cosmic backyard}

In order to analyze the effect of energy losses on the observed
spectrum, it is convenient to introduce the accumulation factor
$f_{\rm acc}$, defined as the ratio of energy-weighted fluxes for
``low'' ($10^{18.7}$ eV -- $10^{19.5}$ eV) and high ($> 10^{20}$ eV)
energy CRs above the ankle. For ordinary baryonic CRs, if the Earth is
located in a typical environment and all CR-sources have smooth
emission spectra, the observed spectrum above the ankle should
have an offset in normalization between low and high energy given
by $f_{\rm acc}$. In the case where the cosmic rays are
protons from a uniform distribution of sources active over
cosmological times, the cutoff due to photopion processes
relates the accumulation factor to a ratio of  GZK 
distances~\cite{Farrar:2000nw} and leads to $f_{\rm acc} \sim 100$. 
A similar value
for $f_{\rm acc}$ is obtained for nuclei due to photodisintegration.
The smoothness of the
observed CR spectrum~\cite{Hayashida:2000zr}, i.e., $f_{\rm acc} \sim 1$, 
suggests that the power of nearby sources 
must be comparable to that of all other sources (redshift $z>0.5$) 
added together. Otherwise, one would have to
invoke an apparently miraculous matching of spectra to account for the
smoothness of the CR energy spectrum. Of course, the GZK cutoff can be 
overcome 
if nearby sources are significantly more concentrated.
However, this does not seem to be the case: if one simply assumes that 
the distribution of CR sources follows the distribution of galaxies, the 
local overdensity is only a factor of two above the mean, and thus 
insufficient to explain the measured flux above 
$10^{20}$~eV~\cite{Blanton:2000dr}.\footnote{We note that 
an earlier analysis~\cite{Medina-Tanco:1998uh} overestimated the local 
overdensity by an order of magnitude. This large discrepancy was caused by 
neglecting the necessary galaxy selection functions which account for the 
fact that nearby galaxies are easier to detect than far away 
galaxies~\cite{Blanton:2000dr}.} Several proposals with a smooth high energy
spectrum due to nearby sources have been put forward. In this section 
we discuss these models.  

\subsection{Galactic sources} 

Galactic sources would satisfy the power condition trivially. 
However, the observed approximate isotropy would be difficult to 
reconcile with a galactic explanation, unless CRs diffuse in 
the magnetic field.  Specks of dust could constitute diffuse galactic
CRs, though the depth of shower maxima 
recorded so far are not consistent with the exact picture of showers 
initiated by dust grains~\cite{Linsley:80,Anchordoqui:np}. Heavy nuclei 
will also diffuse in the Galactic magnetic field~\cite{Zirakashvili:xj}. 
Indeed, it was suggested by Blasi, Epstein, and Olinto~\cite{Blasi:1999xm} 
that iron ions from the surfaces of newly formed strongly magnetized pulsars 
may be accelerated to super-GZK energies through relativistic MHD winds. 
The iron ejected with energies $\sim 10^{20}$~eV will reach the Earth after some diffusion through the Galactic magnetic field. Note that the Larmor radius 
($Z_{26} \equiv Z/26$)
\begin{equation}
r_{_{\rm L}} \sim 1.4 \, \frac{E_{20}}{Z_{26}}\, 
\left(\frac{B}{3\,\mu{\rm G}}\right)^{-1}\,\,{\rm kpc}\,,
\end{equation}
is a few times the typical distance to a young neutron star, which is of order $d \sim 8$~kpc.

Pulsars begin their life with very fast rotation, $\omega \sim 3000$~s$^{-1}$, 
and very large magnetic fields, $B_s \agt 10^{13}$~G. Inside the light 
cylinder, i.e. $r \alt c/\omega$, a magnetosphere 
of density~\cite{Goldreich:1969} 
\begin{equation}
n_{_{\rm GJ}} (r) \sim \frac{B(r)\, \omega}{4\,\pi\,Ze\,c}
\end{equation}
corotates with a dipole magnetic field component that scales as $B(r) = B_s (r_{\rm ns}/r)^3$. As the distance from the star increases, the dipole field 
structure cannot be maintained and beyond the light cylinder the field is 
mostly 
azimuthal. From the light cylinder a relativistic plasma with Alfv\'en 
speed close to the speed of light expands as a MHD wind. 

The surface of a young neutron star is composed of elements near to the 
iron peak formed during the supernova event. Iron ions can be stripped off 
the hot surface of the neutron star by strong electric fields and can 
be present throughout much of the magnetosphere~\cite{Ruderman,Arons}. The typical CR 
energy can be estimated by considering the magnetic energy per ion at the 
light cylinder~\cite{Blasi:1999xm} 
\begin{equation} 
E_{\rm CR} = \frac{B^2_{\rm lc}}{8\pi\,n_{_{\rm GJ}}}\,,
\label{pasquale}
\end{equation}
where $B_{\rm lc} \sim 10^{10}\, B_{13}\, \omega_{3k}^3$~G 
is the magnetic field strength at the light cylinder, 
$n_{\rm GJ} \sim 1.7\times 10^{11}\, B_{13}\, \omega_{3k}^4/Z$~cm$^{-3}$, 
$\omega_{3k} \equiv \omega/3000$~s$^{-1}$ and $B_{13} \equiv B_s/ 10^{13}$~G.
Replacing by fiducial values, Eq.~(\ref{pasquale}) leads 
to~\cite{Blasi:1999xm} 
\begin{equation}
E_{\rm CR}  = 4 \times 10^{20} Z_{26} \,B_{13}\, \omega_{3k}^2\, {\rm eV}\,,
\end{equation}
less than the maximum energy of particles that can be contained in the wind 
near the light cylinder~\cite{Gallant,Begelman:94}.

The injection spectrum is determined by the evolution of the rotational frequency. As the pulsar spin slows down because of electromagnetic and gravitational radiation~\cite{Lindblom:1998wf,Andersson}, the energy available for CR production 
decreases. For $B_s \agt 10^{13}$~G, the rotation speed decreases mainly 
by magnetic dipole radiation~\cite{Rezzolla},
\begin{equation}
\omega^2_{3k} (t) = \frac{\omega_{i3k}^2}{1 + t_8\,B_{13}^2\,\omega_{i3k}^2}\,,
\end{equation}
where $\omega_{i3k}$ is the initial spin period and $t_8 = t/10^{8}$~s.
With this in mind, the particle spectrum from each neutron star is given by~\cite{Blasi:1999xm}
\begin{equation}
N(E) = \epsilon \frac{5.5 \times 10^{31}}{B_{13}\,E_{20}\,Z_{26}}\,\,{\rm GeV}^{-1}\,,
\end{equation} 
where $\epsilon < 1$ is the efficiency for accelerating particles at the light cylinder.

Even though young neutron stars are surrounded by remnants of the presupernova star, the accelerated particles can easily escape the supernova remnant without significant degradation for a wide range of initial magnetic fields and spinning rates. Specifically, consider a supernova that 
imparts $E_{\rm SN} \sim 10^{51} E_{51}$~erg on the stellar envelope 
of mass $M_{\rm env} \sim 10 M_1$ solar masses, where 
$E_{51} \equiv E/10^{51}$~erg, and $M_1$ is the mass of the envelope in units 
of $10~M_\odot$ . The column density of the envelope 
surrounding the neutron star is given by
\begin{equation}
\Sigma \sim \frac{M_{\rm env}}{4\,\pi\,(r_0 + v_e \,t)^2} 
\end{equation}
where $v_e \sim \sqrt{2 E_{\rm SN}/M_{\rm env}}$ is the envelope dispersion 
velocity, $t$ is in seconds, and $r_0 \sim 10^{13}\, r_{13}$~cm. The condition for iron nuclei 
to traverse the supernova envelope without significant losses is  
$\Sigma < 100$~g cm$^{-2}$. At late times compared with $t_e = r_0/v_e$, this transparency condition gives $t>t_{\rm tr} \sim 1.3 \times 10^{7} 
M_1 E_{51}^{-1/2}$~s.   

The evolution of the maximum energy is thus given by
\begin{equation}
E_{\rm CR} (t) \sim 4 \times 10^{20} 
\frac{Z_{26}\,B_{13}\,\omega_{i3k}^2}{1 + t_8\, B_{13}^2\, \omega_{i3k}^2}\,\,{\rm eV}\,.
\end{equation}
The maximum energy decreases as the source evolves, and so the condition to produce ultrahigh energy CRs is that $E_{\rm CR}$ exceeds the needed energy when the envelope becomes transparent~\cite{Blasi:1999xm}
\begin{equation}
E_{\rm CR} (t_{\rm tr}) > E_{20},
\end{equation}  
or equivalently,
\begin{equation}
\omega_i > \frac{3000}{B_{13}^{1/2}\, [4 Z_{26} E_{20}^{-1} - 0.13 \,M_1\, 
B_{13}\,E_{51}^{-1/2}]^{1/2}}\,\,{\rm s}^{-1}\,.
\label{tanto}
\end{equation}
Eq.~(\ref{tanto}) translates into upper bounds on the surface magnetic field strength and the star initial spin period $P_i = 2\pi/\omega_i$,
\begin{equation}
B_{13} < \frac{31 \,Z_{26}\,E_{51}^{1/2}}{M_1\,E_{20}}\,,
\label{nhg}
\end{equation}
and
\begin{equation}
P_i < 8 \pi B_{13}^{1/2}\,Z_{26}\, E_{20}^{-1}\,.
\label{nhg2}
\end{equation}
For $M_1 = 2$ and $E_{20} = E_{51} = Z_{26} = 1$, Eq~(\ref{nhg}) gives 
$B_{13} < 15.4$, whereas Eq.~(\ref{nhg2}) leads to $P_i \alt 10$~ms, not very restrictive values for a young neutron star. Moreover, for a reasonable pulsar 
production rate, the CR-flux on Earth is easily consistent with 
observation, i.e., $\epsilon \agt 4 \times 10^{-7}$~\cite{Blasi:1999xm}.

\subsection{Centaurus A}

Centaurus A (Cen A) is the nearest active galaxy, at a distance of 
$\sim 3.4$~Mpc~\cite{Israel}. It is a complex FRI radio-loud 
source with galactic coordinates $l \approx 310^\circ$, $b \approx 20^\circ$, 
and identified
at optical frequencies with the galaxy NGC 5128.\footnote{Galactic 
longitude, $l$, is measured eastward along the galactic equator from $0^\circ$ (in Sagittarius) to $360^\circ$. The galactic equator is defined by the central plane of the Milky Way. Galactic latitude, $b$, is the angular distance from the galactic equator ($b=0^\circ$)  toward either north ($b = 90^\circ$) or 
south ($b = 
- 90^\circ$) galactic pole. Note that before August 1958 a different, now obsolete, system was used.}
Different multi-wavelength studies have revealed a rather complex morphology:
it comprises a compact core, a jet also visible at $X$-ray
frequencies, a weak counterjet, two inner lobes, a kpc-scale
middle lobe, and two giant outer lobes. The jet would be
responsible for the formation of the northern inner and middle
lobes when interacting with the interstellar and intergalactic
media, respectively. There appears to be a compact structure in
the northern  lobe, at the extrapolated end of the jet. This
structure resembles the
hot spots such as those existing at the extremities of FRII
galaxies. However, at Cen A it lies at the
side of the lobe rather than at the most distant northern edge,
and the brightness contrast (hot spot to lobe) is not as extreme~\cite{Burns}.

\begin{figure}
\postscript{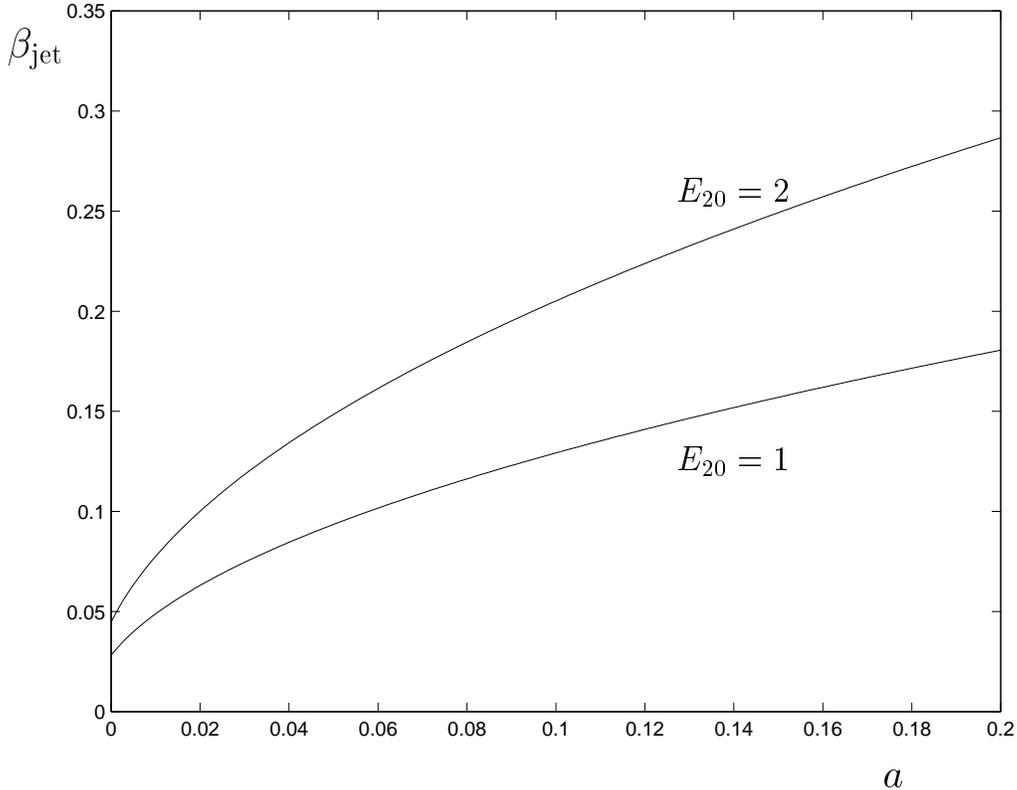}{0.90}
\caption{Jet velocity as a function of the parameter $a$ (defined in the text),
for different proton energies. Published in Ref.~\cite{Anchordoqui:2001bs}.}
\label{cena}
\end{figure}

In order to ascertain the capability of Cen A to accelerate
particles to ultrahigh energies, one first applies the Hillas
criterion \cite{Hillas:1984} (see Fig.~\ref{hillasf}).  Low resolution polarization measurements 
in the region of the suspected hot spot give fields as high as $25\
\mu{\rm G}$ \cite{Burns}. However, in certain regions where
measurements at both high and low resolution are available, the
$B$-field amplitude at high resolution can be seen to be twice that at 
low resolution. The higher resolution can reveal amplification in the 
post-shock
region \cite{Landau}, yielding $B$-fields possibly as high as $50-60\
\mu{\rm G}$~\cite{Romero:ed,Romero:1995tn}. The radio-visible size of the
hot spot can be directly measured from the large scale map~\cite{Junkes}, 
giving $R_{\rm HS}\simeq 2$ kpc. The actual
size can be larger by a factor $\sim 2$ because of uncertainties
in the angular projection of this region along the line of 
sight.\footnote{For example, an explanation of  the apparent
absence of a counterjet in Cen A via relativistic beaming
suggests that the angle of the visible jet axis with respect to
the line of sight is at most 36$^{\circ}$~\cite{Burns}, which
could lead to a doubling of the hot spot radius. It should be
remarked that for a distance of 3.4 Mpc, the extent of the entire
source has a reasonable size even with this small angle.}
All in all, if the magnetic field of the
hot spot is confined to the visible region, the limiting
energy imposed by Eq.~(\ref{rg}) is $\sim 2\times 10^{20}$ eV. 
Estimates of the radio spectral index of synchrotron emission in the 
hot spot and the observed degree of linear polarization in the same 
region suggests that the ratio of turbulent to ambient magnetic energy density 
in the region of the shock is $u \sim 0.4$ \cite{Combi}.
The jet velocity is model dependent: possible values  range from
$\sim 500$ km s$^{-1}$ to $0.99c$ \cite{Burns}. For FRI galaxies, 
the ratio of photon to magnetic 
energy densities, $a$, is expected to be $\ll 1$. Figure~\ref{cena} shows 
curves of constant ultrahigh energy in the $\beta_{\rm jet}$ - $a$ plane, 
for $B=60\,\mu{\rm G},\ R=4$ kpc, according to
Eq.~(\ref{ab}). Since the range of values for  $a$, and the jet 
velocity in units of $c$ conform to expected values, it is plausible that 
Cen A can accelerate particles to energies $\agt 10^{20}$~eV. 

Recent observations of the gamma ray flux for energies $>100$ MeV
by EGRET~\cite{Sreekumar:1999xw}\footnote{The Energetic Gamma Ray 
Experiment Telescope (EGRET) was on board the CGRO mission.} allow an 
estimate $L_{\gamma} \sim 10^{41}\ \es$
for the source.\footnote{Note that the received
radiation is negligibly affected by
interactions with the various radiation
backgrounds~\cite{Aharonian:2000pv}.} This value of $L_{\gamma}$ is
consistent with an earlier observation of photons in the TeV-range
during a period of elevated activity \cite{Grindlay}, and is
considerably smaller than the estimated bolometric luminosity
$L_{\rm bol}\sim 10^{43}\es$\cite{Israel}. Data across the entire
gamma ray bandwidth of Cen A is given in Ref.\cite{Steinle},
reaching energies as high as 150~TeV~\cite{Clay:uy}, though data at 
this energy 
await confirmation. For  values of $B$ in the $\mu$G range,
substantial proton synchrotron
cooling is suppressed, allowing the production of high energy
electrons through photomeson processes. The average energy of
synchrotron photons scales as $\overline{E}_\gamma \simeq 0.29
E_\gamma$ \cite{Ginzburg}. With this in mind, it is straightforward to
see that to account for TeV photons, Cen A should harbor a
population of ultra-relativistic electrons with $E \sim 6 \times
10^{18}$ eV. We further note that this would require the presence
of protons with energies between one and two orders of magnitude
larger, since the electrons are produced as 
secondaries.\footnote{Consecutive factors of $\sim 2$ energy loss occur in the
processes $p\gamma\rightarrow N\pi^0,\ \pi^0\rightarrow
\gamma\gamma,\ \gamma\rightarrow e^+e^-.$ Eq.(\ref{synch}) then
implies proton energies of $\sim 10^{20}$ eV for 100 TeV photons.}

There are plausible physical arguments~\cite{Mannheim:jg,Waxman:1995vg}
as well as some observational reasons \cite{Rawlings} to believe
that when proton acceleration is being limited by energy losses,
the CR luminosity $L_{\rm CR}\approx L_{\gamma}$.
Defining $\epsilon$, the efficiency of ultra
high energy CR production compared to high energy $\gamma$
production -- from the above, $\epsilon\simeq 1$ -- and using
equal power per decade over the interval $(\emin,\ \emax)$,
the source luminosity is found to be~\cite{Farrar:2000nw}
\begin{equation}
\frac{E^2 \,dN^{p+n}_0}{dE\,dt} \, \approx \frac{6.3\,\epsilon\lf
\,10^{52} {\rm eV/s}}{\ln(\emax/\emin)}\,, \label{2}
\end{equation}
where $\lf
\equiv$ luminosity of Cen A$/10^{41}\es$ and the subscript ``0'' refers
to quantities at the source.

It is straightforward to see, using Eqs.~(\ref{dg}) and (\ref{steve}) 
that for $E=10^{19}$\ eV, $B=0.5~\mu$G, $\ell = 0.5$~Mpc, the diffusive 
distance traveled $c \tau_D=50$~Mpc $\gg d =3.4$~Mpc is in very good 
agreement with the diffusion condition given in Eq.~(\ref{taud}). For higher 
energies, the 
validity of the diffusive approach must be checked on a case-by case basis. 
For these purposes, in the case of a continuously emitting source, the 
definition of a diffusion time is somewhat arbitrary. In the following 
discussion we adopt as the diffusion time $\tau_D,$ a choice partially 
consistent with simulations (see Sec.IV-E). Moreover, one can easily check 
that for 3.4~Mpc the diffusion 
time of any proton with energy above the photopion production threshold is 
always less than the GZK-time, and consequently energy losses can be safely 
neglected.
The density of protons at the present time $t$
of energy $E$
at a distance $r$ from Cen A, which is assumed to be continuously emitting at
a constant spectral rate $dN^{p+n}_0/dE\,dt$ from time $t_{\rm on}$ until 
the present, is~\cite{Anchordoqui:2001nt}
\begin{equation}
\frac{dn(r,t)}{dE}  =  \frac{dN^{p+n}_0}{dE\,dt}
\frac{1}{[4\pi D(E)]^{3/2}} \int_{t_{\rm on}}^t dt'\,
\frac{e^{-r^2/4D(t-t')}}{(t-t')^{3/2}}
  =  \frac{dN^{p+n}_0}{dE\,dt}
\frac{1}{4\pi D(E)r} \,\,I(x)\,,
\label{22}
\end{equation}
where $D(E)$ stands for the diffusion coefficient,
$x = 4D\ton/r^2 \equiv \ton/\tau_D$, $\ton=t-t_{\rm on},$ and
\begin{equation}
I(x) = \frac{1}{\sqrt{\pi}} \int_{1/x}^\infty 
\frac{du}{\sqrt{u}} \,\, e^{-u}\ \ . \label{I}
\end{equation}
For $\ton\rightarrow \infty$, the density approaches its time-independent
equilibrium value $n_{\rm eq}$, while for $\ton= \tau_D$,
$n/n_{\rm eq} = 0.16$.

To estimate the power of Cen A, one can  evaluate the
energy-weighted approximately isotropic proton flux at  
$1.5\times 10^{19}$ eV, which lies in the center of the flat
``low'' energy region of the spectrum,
\begin{equation}
E^3 J_p(E)  =
 \frac{Ec}{(4\pi)^2d\,D(E)} \frac{E^2 \,dN^{p+n}_0}{dE\,dt}\,
I(t/\tau_D)  \approx  7.6 \times 10^{24} \, \epsilon \lf\, I
\,\, {\rm eV}^2 \, {\rm
m}^{-2} \, {\rm s}^{-1} \, {\rm sr}^{-1}. \label{jpt}
\end{equation}
In Eq.~(\ref{jpt}) we have used the fiducial values of
$B\ \mbox{and}\ \ell$ as given in the previous paragraph, and set 
$\emin = 1 \times 10^{19}$~eV, $\emax = 4 \times 10^{20}$~eV, 
assuming (as discussed in Sec.III-B.2) that the shock structure can be 
more extended than the visible region in the non-thermal radioemission. 
As noted by Farrar and Piran~\cite{Farrar:2000nw}, one can see, by stretching 
at most the source parameters, that the ``low'' energy flux from 
Cen A could be comparable to that of all other sources in the Universe.
To this end, first fix $\epsilon\, \lf\, I =0.40$, after 
comparing Eq.~(\ref{jpt}) to the observed CR-flux: 
$E^3 J_{\rm obs}(E)  = 10^{24.5}$ eV$^2$ m$^{-2}$ s$^{-1}$ sr$^{-1}$~\cite{Hayashida:2000zr}. Next, $\epsilon
\lf\simeq 1,$ determines $I\simeq 0.40,$ and consequently
the required age of the source $\ton$ to be about 400~Myr. There is still a 
great uncertainty on jet
evolutionary behavior. Experimental approaches and theoretical
studies suggest ages that run from a few tens of Myr, up to half
the Hubble time in some extreme cases~\cite{Begelman}. Moreover,
$\ton \sim 400$~Myr is consistent with FRI lobe ages~\cite{Rawlings}.
To maintain flux at the ``ankle'' for the same $\ton$,
one requires an approximate doubling of $L_{\rm CR}$ at $5\times
10^{18}$ eV. Because of the larger diffusive time delay at this energy, this
translates into an increased luminosity in the early phase of Cen A.
From Eq.~(\ref{synch}), the associated synchrotron photons are  emitted at 
energies $< 30$ MeV. The increase in radiation luminosity in this 
region is not inconsistent with the flattening of the spectrum observed  
at lower energies\cite{Kinzer,Steinle:98}.

Let us turn now to the discussion of anisotropy. This can be found by 
computing the incoming current flux density $D\nabla n$ as viewed by an 
observer on Earth, and one finds for a continuously-emitting source
a distribution $\sim(1+\alpha \cos\theta)$
about the direction of the source, where $\theta$ is the angle to 
the zenith and
\begin{equation}
\alpha = \frac{2D(E)}{cr}\cdot \frac{I\pri}{I}.
\label{anisotropy}
\end{equation}
Here,
\begin{equation}
I\pri(x) = \frac{1}{\sqrt{\pi}} \int_{1/x}^\infty du\ \sqrt{u}
\,\, e^{-u}, \ee with $x=\ton/\tau_D$, and $I$ as defined in
Eq.~(\ref{I})~\cite{Anchordoqui:2001nt}. For our choices of $B$ and 
$\ell,$ and $T_{\rm on}=400$~Myr, we 
find for $E=10^{19}$~eV $(E=10^{20}$~eV) that $\alpha = 0.04\ (\alpha = 0.07).$

Neutrons at the highest energies could survive decay and produce a spike in 
the direction of the source (see Appendix E). However,
those that are able to decay will beget secondary proton diffusion fronts 
with asymmetry parameters given by  
\begin{equation}
\alpha = \frac{2D(E)}{cr}\cdot \frac{I\prii}{I}\  \ \ ,
\label{anisotropya}
\end{equation}
where
\begin{equation}
I\prii(x) = \frac{1}{4\sqrt{\pi}\kappa} \int_{1/x}^\infty \frac{du}{u^{3/2}}
\left[  \left((1-\kappa)u + \thalf\right)\ e^{-(1-\kappa)^2u}
 - \left((1+\kappa)u + \thalf\right)\ e^{-(1+\kappa)^2u}\right]
\label{anisa}
\end{equation}
and $\kappa=\lambda(E)/r,$ where $\lambda(E)=0.9\ E_{20}\, \mbox{Mpc}$ is 
the neutron decay length. In spite of the complicated nature of 
Eq.(\ref{anisa}), the results for
$\alpha$ are very
similar to the ones for the primary diffusion front given above.

\begin{figure}
\centering\leavevmode \epsfxsize=5.5in \epsfbox{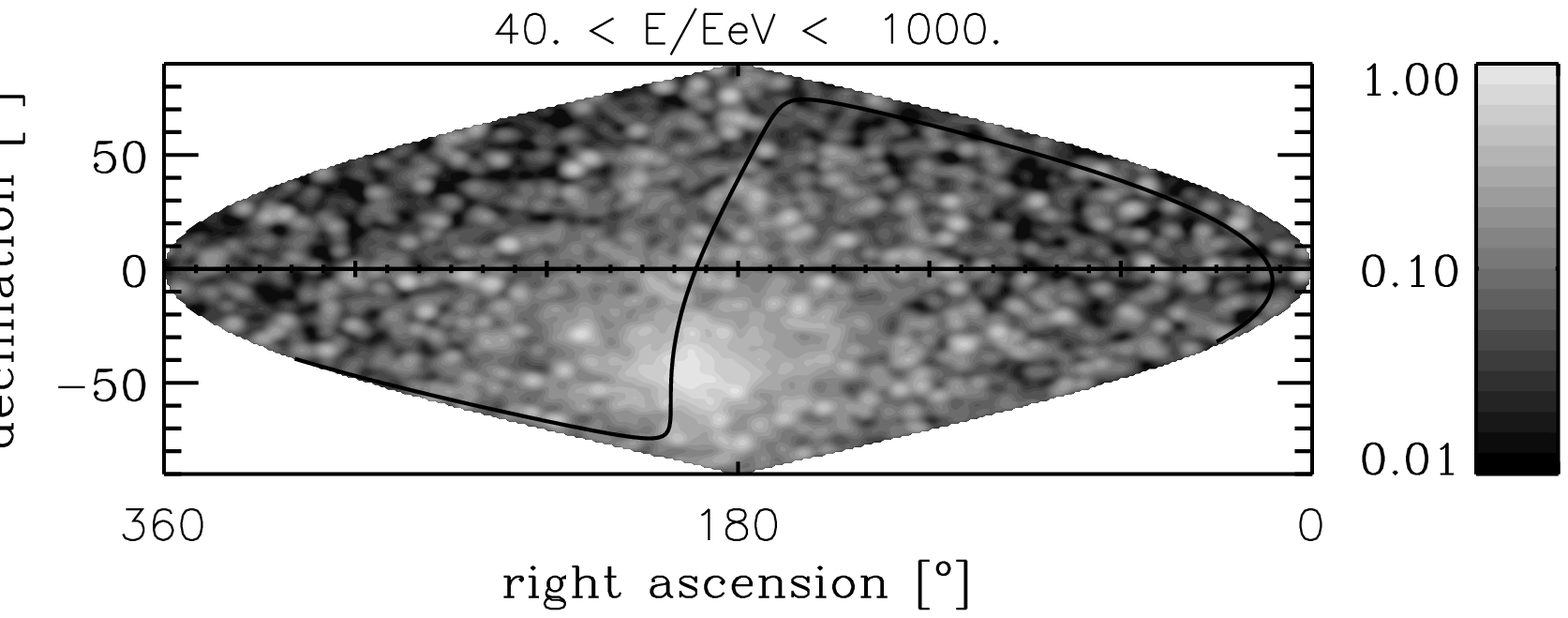}
\epsfxsize=5.5in \epsfbox{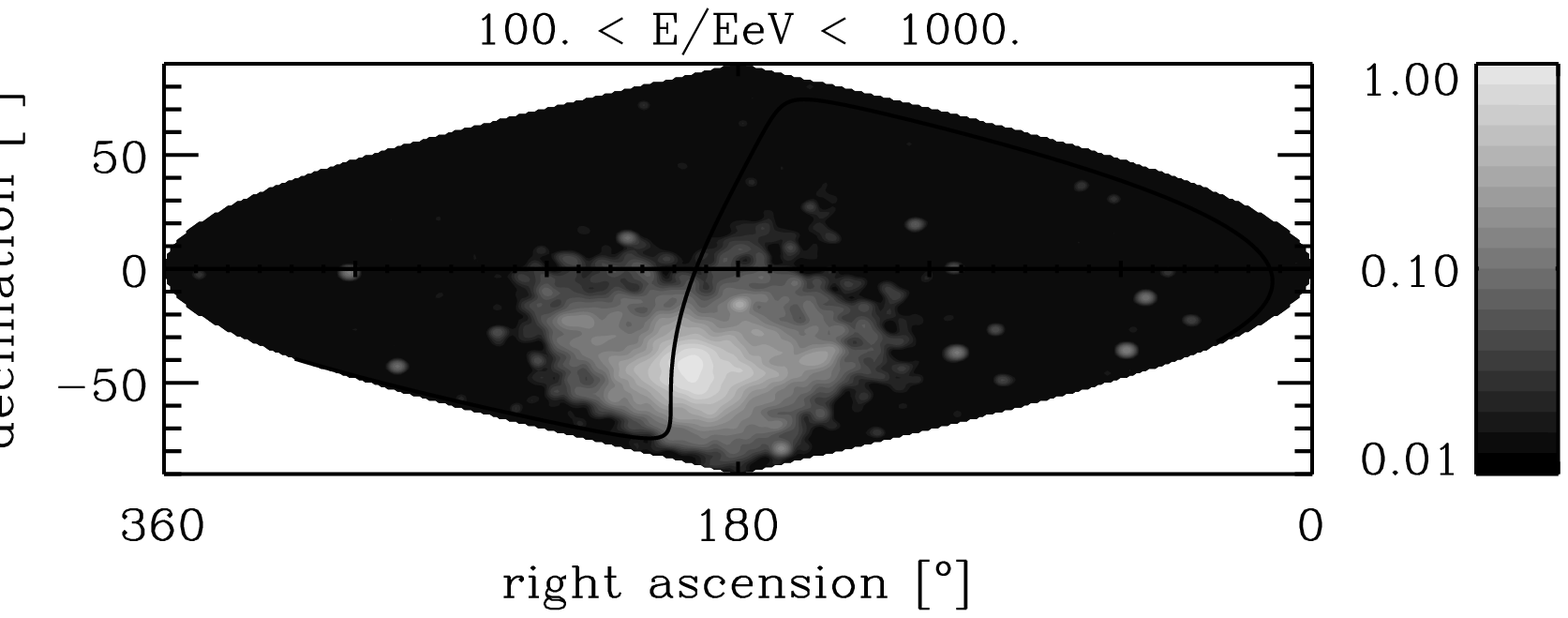}
\caption{The angular image in equatorial coordinates, averaged over
all 20 magnetic field realizations of 5000 trajectories each, for
events above $40\,$EeV (upper panel) and above $100\,$EeV (lower
panel), as seen by a detector covering the whole sky, for the case
suggested in Ref.~\cite{Farrar:2000nw} corresponding to $B=0.3\,\mu$G, 
and the source Cen A. The grey scale represents the integral flux per 
solid angle. The solid line marks the Super-Galactic
plane.  The pixel size is $1^\circ$; the image has
been convolved to an angular resolution of 2.4$^\circ$ corresponding to
AGASA. Published in Ref.~\cite{Isola:2001ng}.}
\label{cena2}
\end{figure}

Detailed Monte Carlo simulations~\cite{Isola:2001ng,Isola:2002ei}, however, 
reveal that the predicted distributions of arrival directions are more
anisotropic than current data even at $10^{19}$~eV (see Fig.~\ref{cena2}.). In particular, 
for a spectral index 
$\gamma = 2.4$ and a maximum injection energy of $10^{21}$~eV, the angular 
power spectrum shows a 3~$\sigma$ quadrupole deviation from observation, 
and the auto-correlation function is not consistent with the clustering at 
small scale observed by AGASA~\cite{Isola:2002ei}. The discrepancy between 
phenomenology and Monte Carlo simulations could arise from a number of 
factors. The simulations are considerably more detailed than
the phenomenological arguments; for instance effects
of photopion production are taken into account.  Furthermore, they
simulate the subtleties in transitions from
Kolmogorov to Bohm diffusion.  On the other hand,
the Monte Carlo simulations rely on a statistical
description of the magnetic field, in which the field is 
arranged into domains or ``cells'' with a characteristic coherence length.
There is some uncertainty in the sensitivity to the geometry of the cell, 
which is reflected in the presence or absence of resonant modes in the 
spectrum. For instance, if one considers coherence domains generated in 
some Gaussian manner (say with spatial spread $R$) rather than as small cubes, 
then $k_0 = 1/R$ instead of $k_0 = 2\pi/\ell$, and for $R \sim 0.7$~Mpc, 
the diffusive resonant condition is satisfied even for the highest observed 
energies. Whichever point of view one may find more convincing, it seems most 
conservative at this point to depend on experiment (if possible) to resolve 
the issue. In particular, PAO will be in a unique position to search the sky 
in the direction of Cen A.

\subsection{M87}

At a distance of $\sim 20$~Mpc, M87 is the dominant radio galaxy in the 
Virgo cluster ($l = 282^\circ, \, b=74^\circ$)~\cite{Blandford:1999}. 
This powerful FRI has been under 
suspicion to be the primary 
source of ultrahigh energy cosmic rays for a long 
time~\cite{Ginzburg:63,Wdowczyk:sg}. The emission of 
synchrotron radiation with a steep cutoff at frequencies about $3 \times 10^{14}$~Hz from its radiojets and hot spots~\cite{Heinz,Bicknell} implies an 
initial turbulence injection scale having the Larmor radius of 
protons at $10^{21}$~eV (see Sec.III-B.2). Hence, because of its relative
proximity to Earth, M87 becomes a potential candidate to account for the 
observed events above $10^{20}$~eV. The major difficulty with
this idea is the observation of the nearly isotropic distribution of the 
CR arrival directions. One can again argue that the orbits are bent. 
However, the bending cannot add substantially to the travel time,  
otherwise the energy would be GZK-degraded. An interesting explanation 
to overcome this difficulty relies on a Galactic wind, akin the solar wind, 
that would bend all the orbits of the highest energy CRs towards 
M87~\cite{Ahn:1999jd,Biermann:fd}. Indeed, it has long been expected 
that such a kind of wind 
is active in our Galaxy~\cite{Burke,Johnson,Mathews}. In the analysis 
of~\cite{Ahn:1999jd}, it was assumed that the magnetic field in the Galactic 
wind has a dominant azimuthal component, with the same sign 
everywhere. This 
is because in a spherical wind the polar component of the magnetic field 
becomes negligible rather quickly, decaying like $1/r^2$, and thus the 
azimuthal part of the magnetic field quickly becomes dominant, with 
$B_\phi \sim \sin \theta/r$ in polar coordinates~\cite{Parker}. Under 
these considerations 
one is left with two degrees of freedom: the strength of the azimuthal 
component at the location 
of the Sun, and the distance to which this wind extends. Recent estimates 
suggest that the magnetic field strength near the Sun is 
$\sim 7~\mu$G~\cite{Beck}. The 
second parameter is more uncertain. Our Galaxy dominates its near environment 
well past our neighbor, M31, the Andromeda galaxy, and might well extend its 
sphere of influence to half way to M81. This implies an outer halo wind of $\sim 1.5$~Mpc. With this in mind, the mean flight time of the protons in the 
Galaxy is $\sim 5.05 \times 10^{6}$~yr $\ll \tau_s$, the time for straight 
line propagation from M87~\cite{Gustavo}. Figure~\ref{gw} shows the directions where the 13 highest energy CR events 
point towards when they leave the halo wind of our Galaxy. Note that 
except for the two highest energy events, all other events 
can be traced back to within less than about 20 degrees from Virgo. If one 
takes into account the uncertainty of the magnetic field distribution, all events are consistent with arising originally from Virgo. Moreover, if one assumes that the two highest event are helium nuclei, all 13 events point within 
$20^\circ$ of Virgo. Note that a bending of $20^\circ$ could be easily 
accommodated with a fine tuning of the magnetic field strength within the 
super-Galactic sheet from here to 
Virgo (more on this below).  Criticisms of this 
model~\cite{Billoir:2000wi} have been addressed in~\cite{Biermann}.

\begin{figure}
\postscript{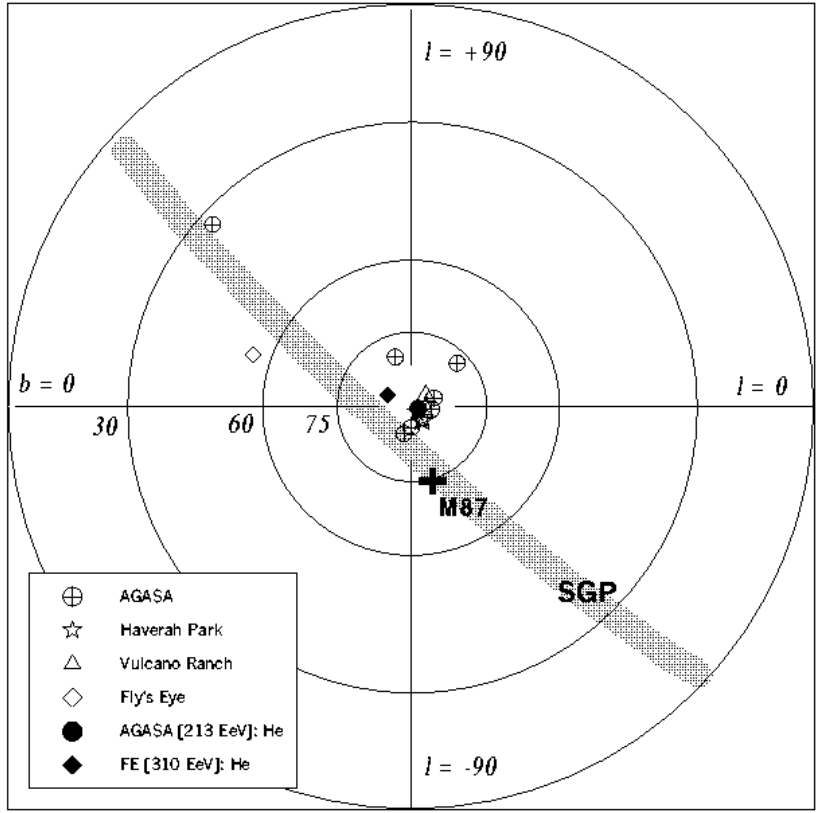}{0.7}
\caption{Directions in polar projection of the highest energy CR events when they enter the halo of our Galaxy. The two highest energy events have two 
entries, under the assumptions that they are either protons or 
helium nuclei (filled black symbols). The cross indicates the position of M87 and the grey band stands for the super-Galactic plane. Published in Ref.~\cite{Ahn:1999jd}.}
\label{gw}
\end{figure}

We now focus attention on the power of M87. To this end, we analyze another 
interesting limit of Eq.~(\ref{spectrum}) in which diffusion is 
found to be extremely small. In such a 
case, the second term in the r.h.s. can be neglected, and the solution for 
a single point source reads~\cite{Stecker:1989ti}
\begin{equation}
n(E, t) = \frac{1}{b(E)} \, \int_E^{\infty} \, Q(E_0, t') \,dE_0, 
\label{phdsspectrum}
\end{equation}
where
\begin{equation}
t'=  t - \int_E^{E_0} \frac{d\tilde{E}}{b(\tilde{E})},
\label{phdsespectro}
\end{equation}
and $E_0$ is the energy of the nucleon when emitted by the source.
Using Eq.~(\ref{Ae}) we model the injection spectrum of a single source 
located at $t_0$ from the observer
by 
\begin{equation}
Q(E, t) \,= \,\kappa \, E^{- \gamma} \,
\delta(t - t_0)\,, 
\end{equation}
where $\kappa$ is a normalization constant and for simplicity we consider 
the distance as measured from the source, i.e., $t_0 = 0 $. 
At very high energies, the total number of particles at a given distance 
from the source is 
\begin{equation}
n(E, t) \approx \frac{\kappa}{b(E)} \int_E^{\infty}\, E_0^{-\gamma} \,
\, \, \delta\left( t - \frac{1}{C} \ln 
\frac{E_0}{E} \right) dE_0\,,
\end{equation}
or equivalently,
\begin{equation}
n(E, t) \approx \kappa \, E^{-\gamma} e^{- \, (\gamma -1) \, C \, t }\,,
\end{equation}
where $C$ is given in Eq.~(\ref{ce}). Thus, the injection spectrum is 
uniformly 
damped by a factor that depends on the proximity of the source.

At low energies, in the region dominated by baryon resonances, the
parameterization of $b(E)$ does not allow a complete analytical solution.
However using the change of variables, 
\begin{equation}
{\tilde t} =
\int_E^{E_g} \frac{d\tilde{E}}{ b(\tilde{E})},
\end{equation}
with $E_0 = \xi (E, \tilde{t})$ and $d{\tilde t} =
dE_0/b(E_0)$, one easily obtains,
\begin{equation}
n(E, t) =  \frac{\kappa}{b(E)} \int_0^{\infty}\, \xi(E, \tilde{t})^
{-\gamma}\,
\,\delta({\tilde t} - t) \,\, b [ \xi(E, {\tilde t}) ] \,\,d{\tilde t},
\end{equation}
and then, the compact form,
\begin{equation}
n(E, t) = \frac{\kappa}{b(E)} E_0^{-\gamma} b(E_0)\,,
\label{hojo}
\end{equation}
where the relation between $E_0$ and  $E$ is 
given in Eq.~(\ref{degradacion}).
Now, to take into account the extension of the cluster, we assume that the 
concentration of potential sources at the center is higher than that 
in the periphery, and adopt a spatial gaussian distribution.
With this hypothesis, the particle injection rate into the
intergalactic medium is given by  
\begin{equation}
Q(E,t) = \kappa \int_{-\infty}^{\infty} \frac{E^{-\gamma}}{\sqrt{2 \, \pi}\, 
\sigma} \, \delta(t-T) \; 
{\rm exp}\left\{ \frac{-(T-t_0)^2}{2\,\sigma^2} \right\} dT
\end{equation}
A delta function expansion around  $t_0$, with derivatives denoted
by lower case Roman superscripts,
\begin{equation}
\delta(t-T) = \delta(t-t_0) + \delta^{(i)}(t-t_0) (T-t_0)+\frac{1}{2!}
\delta^{(ii)}(t-t_0) (T-t_0)^2 + \dots
\end{equation}
leads to a convenient form for the injection spectrum, which is given
by,
\begin{equation}
Q(E,t^\prime) =  \kappa E^{-\gamma} \,[\,\delta(t^{\prime}-t_0) + 
\frac{\sigma^2}{2!} \, 
\delta^{(ii)}(t^{\prime}-t_0) + \frac{\sigma^4}{4!}\, \delta^{(iv)}(t^{\prime}-t_0)
+\dots] \, .
\end{equation}
By replacing the above expression into Eq.~(\ref{phdsspectrum}) 
one obtains~\cite{Anchordoqui:1996ru}
\begin{equation}
n = \kappa \frac{E_0^{-\gamma}\, b(E_0)}{b(E)} \, 
 \left\{ 1+\frac{ \sigma^2 A^2 e^{-2B/E_0}}{2!} F_1(E_0) + 
\frac{\sigma^4 A^4 e^{-4B/E_0}}
{4!} F_2(E_0)  + {\cal O}(6) \right\}\,,
\label{hoj}
\end{equation}
where
\begin{equation}
F_1(E_0)= 2 B^2 E_0^{-2} + 
(2-3\gamma) B E_0^{-1} +(1-\gamma)^2,
\end{equation}
\begin{eqnarray}
F_2(E_0) & = & 24 B^4 E_0^{-4} + (4 - 50
\gamma) B^3 E_0^{-3} + (35 \gamma^2 -25 \gamma +8) B^2 E_0^{-2}
\nonumber \\
& + & (-10
\gamma^3 +20 \gamma^2 - 15 \gamma +4) B E_0^{-1} + (1- \gamma)^4\,, 
\end{eqnarray} 
and $A$ and $B$ as given in Eq.~(\ref{parame}).
\begin{figure}
\postscript{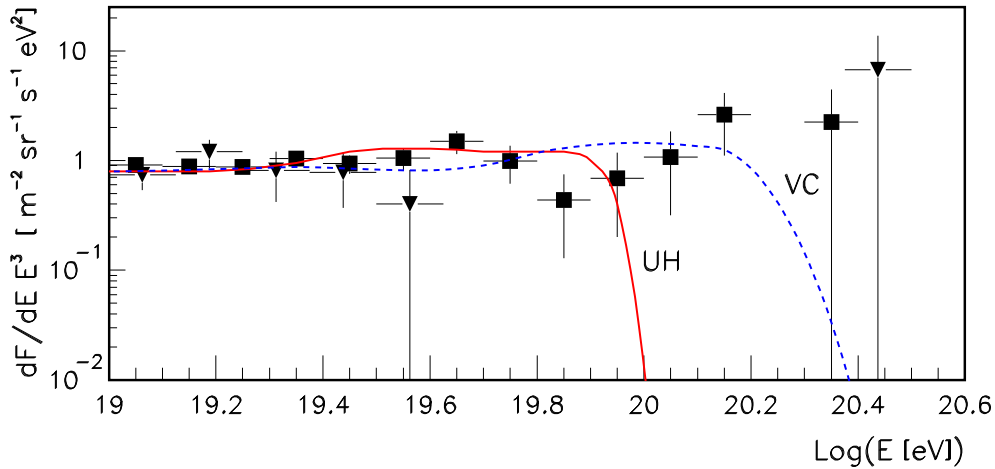}{0.90}
\caption{The CR flux as observed by AGASA (square) and Fly's Eye (triangle) 
experiments. The solid line indicates the CR spectrum assuming the universal 
hypothesis (UH), i.e., homogeneous source distribution, whereas the dotted 
line indicates the evolved spectrum from the Virgo cluster (VC). Published in 
Ref.~\cite{Anchordoqui:1999aj}.} 
\label{M87}
\end{figure}
Figure~\ref{M87} shows the evolved energy spectrum of nucleons assuming a 
cosmologically homogeneous population of sources 
(universal hypothesis)~\cite{Yoshida:pt} 
together with the modified 
spectrum as given by Eq.~(\ref{hoj}) for an extended source modeled by 
a Gaussian 
distribution of width 2~Mpc at a distance of 18.3~Mpc. The extended source 
spectrum was normalized in the ``low'' energy region with
the spectral index, $\gamma = 3.27$, obtained from a maximum likelihood 
fit to the Fly's Eye data~\cite{Dova}. From Fig.~\ref{M87} one can see that 
the proton component of the Virgo cluster partially reproduces AGASA data, but 
apparently cannot account for the super-GZK Fly's Eye event. However, lensing 
effects may come to the rescue~\cite{Harari:2000he}. The magnetic field in 
the Galactic wind strongly magnifies the flux, reducing the power 
requirements of the source. Specifically, with  fine-tuning of the source 
direction relative to  the symmetry axis of the wind, M87 could be as high as 
$> 10^{2}$ times more powerful than if unlensed at energies below 
$E/Z \sim 1.3 \times 10^{20}$~eV~\cite{Harari:2000he}. Moreover, as noted 
in~\cite{Stanev:2001rr}, the super-Galactic plane sheet can focus ultrahigh 
energy CRs along the sheet. Hence, the particles would arrive at the
boundary of our Galactic wind with the arrival directions described by an 
elongated ellipse along the super-Galactic plane sheet. This in turn would 
allow magnetic lensing to produce many hot spots, just as sunlight dapples the 
bottom of a swimming pool, and so easily explain the apparent CR-clustering.

\subsection{Starbursts}

If the super-GZK particles are heavy nuclei from outside our Galaxy, 
then the nearby ($\sim 3$ Mpc \cite{Heckman}) starburst 
galaxies M82 ($l=141^\circ,
b=41^\circ$) and NGC 253 ($l=89^\circ, b= -88^\circ$) would probably
be the sources of most ultrahigh energy CRs observed
on Earth. Starbursts
are galaxies undergoing a massive and large-scale star formation episode.
Their characteristic signatures are strong infrared emission (originating in
the high levels of
interstellar extinction), a very strong HII-region-type emission-line
spectrum (due to a large number of O and B-type stars), and  considerable
radio emission produced by recent supernova remnants (SNRs). Typically,
the starburst region is confined to the
central few hundreds of parsecs of the galaxy, a region that can be
easily 10 or more times brighter than the center of normal spiral galaxies.
In the light of such a concentrated activity, the
existence of  galactic superwinds is not surprising~\cite{Heckman}.

Galactic-scale superwinds
are driven by the collective effect of supernovae and massive star winds. The
high supernovae rate creates a cavity of hot gas ($\sim10^8$ K) whose
cooling time is much greater than the expansion time scale. Since the wind
is sufficiently powerful, it can blow out the interstellar medium of the
galaxy preventing it from remaining trapped as a hot bubble. As the cavity
expands a strong shock front is formed on the contact surface with the
cool interstellar medium. The shock velocity can reach several thousands
of kilometers per second and ions like iron nuclei can be  efficiently
accelerated in this scenario, up to ultrahigh energies,
by Fermi's mechanism~\cite{Anchordoqui:1999cu}.

In a first stage, ions are diffusively accelerated  at
single supernova shock waves within the nuclear region of the galaxy.
Energies up to $\sim 10^{15}$ eV can be achieved in this step~\cite{Lagage}.
Heavy nuclei are not photodissociated in the process despite
the large photon energy densities (mostly in the far infrared) measured
in the central region of the starburst. The escape of the CR outflow is
convection dominated. In fact, the
presence of several tens of young SNRs with very high expansion velocities
and thousands of massive O stars,
with stellar winds of terminal velocities up to 3000 km s$^{-1}$, must
generate collective plasma motions of several thousands of km per second.
Then, due to the coupling of the magnetic field to the hot plasma, the
magnetic field is also lifted outwards and forces the CR gas to
stream along from the starburst region. Most of the nuclei escape
in opposite directions along the symmetry axis of the system, as the total
path traveled is substantially shorter than the mean free 
path~\cite{Anchordoqui:1999cu}.

Once the nuclei escape from the central region of the galaxy, with energies
of $\sim 10^{15}$~eV, they are injected into the galactic-scale wind and
experience further acceleration at its terminal shock up to $10^{20}$~eV.
For this second step in the acceleration process, the photon field energy
density drops to values of the order of the microwave background radiation,
as we are now far from the starburst region, and consequently, iron nuclei
are safe from photodissociation. In terms of parameters that can be determined
from observations, the nucleus maximum energy is given 
by~\cite{Anchordoqui:1999cu}
\begin{equation}
E_{\rm max}\approx\frac{1}{2}\, Ze\,B \,
\frac{\dot{E}_{\rm sw}}{\dot{M}}\, \ton\,\,,
\label{14}
\end{equation}
where  $\dot{E}_{\rm sw} \sim 2.7 \times 10^{42}$~erg~s$^{-1}$ is the
superwind kinetic energy flux and  $\dot{M} = 1.2
{\rm M}_{\odot}$~yr$^{-1}$ is the mass flux generated by the
starburst~\cite{Heckman}.
The  age $\ton$ can be estimated from numerical models that use theoretical
evolutionary tracks for individual stars and make sums over the entire stellar
population at each time in order to produce the galaxy luminosity as a
function of time \cite{Rieke}. Fitting the observational data, these
models provide a range of suitable ages for the starburst phase that
goes from $50$~Myr to $160$ Myr \cite{Rieke}. These models
must assume a given initial
mass function (IMF), which usually is taken to be a power-law with a variety of
slopes. Recent studies have shown that the same IMF can account for the
properties of both NGC 253 and M82~\cite{Engelbracht}.
Besides, a region referred to as M82 ``B'' near the galactic center of
M82, has been under suspicion  as a fossil starburst site in which an
intense episode of star formation occurred over 100~Myr 
ago~\cite{deGrijs:1999hc,deGrijs}.
The derived age distribution suggests steady, continuing cluster formation at
a modest rate at early times ($> 2$ Gyr ago), followed by a concentrated
formation episode 600~Myr ago and more recent suppression of cluster
formation. In order to get some
estimates on the maximum energy, let us assume $B\sim 50\,\mu$G,
a choice consistent with observation~\cite{Paglione}.
Inserting all of these figures in Eq.~(\ref{14}), already for
$\ton = 50$ Myr one obtains for iron nuclei
\begin{equation}
E_{\rm max}^{\rm Fe} > 10^{20}\;\;\;\;{\rm eV}.
\end{equation}

Now, we can use the rates at which starbursts inject mass, metals and
energy into superwinds to get an estimate on the CR-injection spectra.
We once again use $\epsilon$ to denote the efficiency of ultrahigh energy CR
production by the superwind kinetic energy flux.
Using equal power per decade over the interval $10^{18.5}\,{\rm eV}<
E< 10^{20.6}\,{\rm eV}$, we obtain a source CR-luminosity
\begin{equation}
\frac{E^2 \,dN_0}{dE\,dt} \, \approx 3.5 \,\epsilon \,10^{53}
{\rm eV/s}\,. 
\end{equation}

The assumption that the giant air showers with $E > 10^{20}$ eV were triggered by
heavy nuclei implies ordered extragalactic magnetic 
fields $B_{\rm nG} < 15$ (at least in the outskirts of the Galaxy), or else 
nuclei would be captured in magnetic subdomains suffering catastrophic
spallations. Moreover, even for fields ${\cal O}$ (nG), CR nuclei with
energies $E_c < 10^{18}\, \ell_{\rm Mpc}\, Z B_{{\rm nG}}$~eV remain
trapped inside cells of size $\ell_{\rm Mpc}$, attaining efficient
diffusion, with
\begin{equation}
D(E) \approx 0.048 \left( \frac{E_{18} \,\ell^2_{{\rm
Mpc}}}{Z\,B_{{\rm nG}}} \right)^{1/3} \,{\rm Mpc}^2/{\rm Myr}.
\end{equation}
The power of the starbursts can then be estimated by straightforward 
generalization of 
the procedure discussed in Sec.V-B. Specifically, first evaluate the 
energy-weighted approximately isotropic nucleus flux at  $10^{19}$ eV, 
\begin{equation}
E^3 J(E)  =
 \frac{Ec}{(4\pi)^2d\,D(E)} \frac{E^2 \,dN_0}{dE\,dt}\,
I_\star   \approx  2.3 \times 10^{26} \, \epsilon \,
I_\star
\, {\rm eV}^2 \, {\rm
m}^{-2} \, {\rm s}^{-1} \, {\rm sr}^{-1}, \label{jp}
\end{equation}
where $I_\star = I_{\rm M82} + I_{\rm NGC\ 253}$,
$B_{\rm nG} = 15$, $\ell_{\rm Mpc} = 0.5$, and $\langle Z \rangle =20$.
Then, fix
\begin{equation}
\epsilon\, I_\star =0.013,
\label{mimi}
\end{equation}
after comparing Eq.(\ref{jp}) to the observed CR-flux. 
Note that the contribution of $I_{\rm M82}$ and
$I_{\rm NGC\ 253}$ to $I_\star$ critically depends on the age of
the starburst. Figure~\ref{starburst} shows the relation
``starburst-age/superwind-efficiency'' derived from Eq.~(\ref{mimi}), 
assuming that both M82 and NGC 253 were
active for $115$~Myr ($\epsilon \approx 10\%$). Beyond this epoch,
CR-emission must be associated with M82 ``B''.

\begin{figure}
\postscript{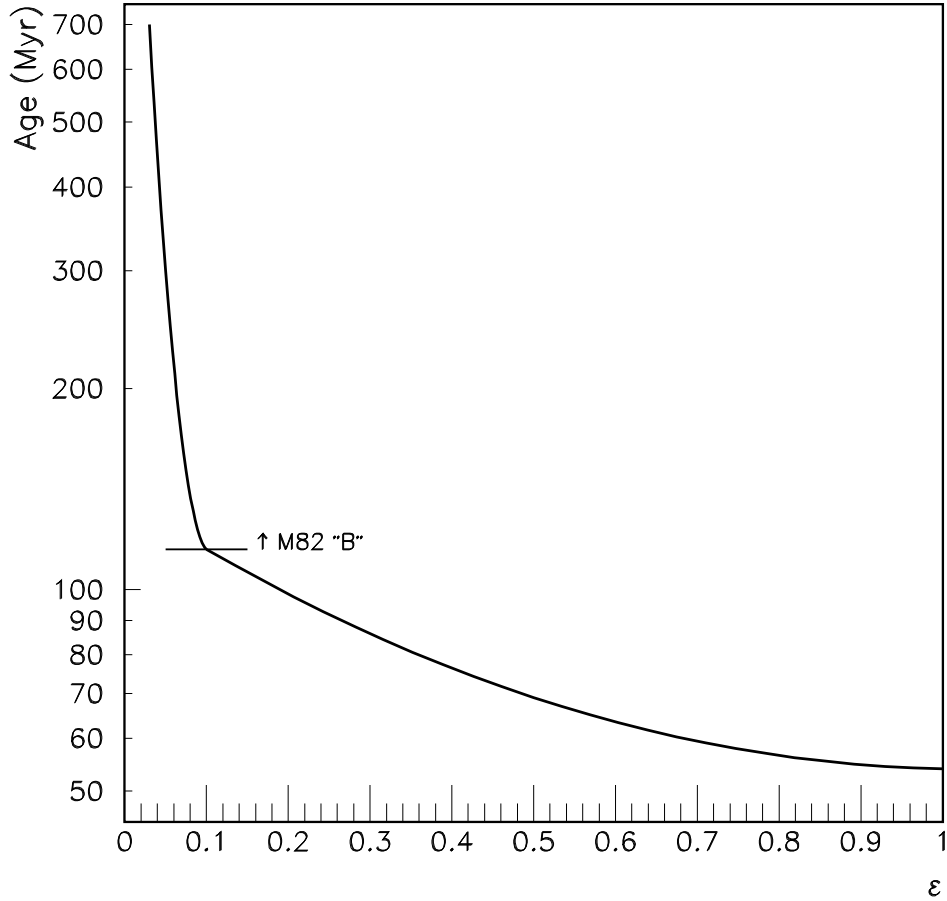}{0.90}
\caption{Age of the starbursts as a function of the efficiency
of CR-production, $\epsilon$. Published in~\cite{Anchordoqui:2001ss}.}
\label{starburst}
\end{figure}

Above $> 10^{20.2}$ eV iron nuclei do not propagate diffusively.
Moreover, the CR-energies get attenuated
by photodisintegration  on the microwave background radiation and the
intergalactic infrared background photons.
In the non-diffusive regime, the accumulated deflection angle $\theta(E)$
from the direction of the source can be estimated using Eq.~(\ref{random}).
Therefore, if $B \sim 15$~nG all directionality is lost. The resulting time 
delay with respect to linear propagation is given by Eq.~(\ref{tdelay}), 
and the total travel time is
\begin{equation}
t\approx \frac{d}{c}\ \left(1+\tfrac{1}{4}\theta^2\right)\ \ .
\label{time}
\end{equation}

As an interesting exercise one can apply these considerations to the 
highest energy Fly's Eye event. Including statistical and systematic
uncertainties, the energy of this event is $3.2\pm 0.9\times
10^{20}$ eV. Eqs.~(\ref{oop}) and (\ref{pats}) relate the
uncertainty in energy to the uncertainty in the attenuation time
\begin{equation}
\frac{\delta t}{t}\simeq \left(\frac{11.3}{E_{20}}\right)\,\,
\left(\frac{\delta E}{E}\right)\ \ .
\end{equation}
From these considerations, we find that the upper limit on the transit time 
for a nuclear candidate for the highest energy Fly's Eye event is 
$\sim 6\times 10^{14}$~s.
The arrival direction of the highest energy
Fly's Eye event is  $37^{\circ}$ from M82~\cite{Elbert:1994zv}.
With $d\simeq 3$ Mpc and $\theta=37^{\circ},$ one finds from 
Eq.~(\ref{time}) a transit time $t\simeq 3.4\times 10^{14}$ s, well
within the stated upper limit.

For average deflections of $60 \deg$, the time of flight is 
$\sim 3.9 \times 10^{14}$ s,
and consequently the CR-spectrum falls off sharply in the neighborhood
of the GZK limit.
A slight anisotropy should arise just before this cutoff.
As the candidate sources in the northern and southern hemispheres are
presumably at
different distances,\footnote{The location of NGC 253 is still subject to
large uncertainties~\cite{Davidge}.} a north-south asymmetry should also 
eventually emerge.  It is rather difficult to assess whether events with
energies $> 10^{20.5}$~eV are plausible, because the maximum energy depends 
strongly on $\tau_{\rm delay}$.

The photodisintegration process results in the production of nucleons of 
ultrahigh energies with the same Lorentz factor as the parent nucleus. 
As a consequence, the total number of particles is not conserved during 
propagation. However, to a very good approximation one 
can treat separately the evolution of the heaviest fragment and those 
fragments corresponding to nucleons emitted from the traveling nuclei.
With this in mind, the evolution of the differential spectrum of the 
surviving fragments is given by Eq.~(\ref{spectrum}). Now, generalizing the formalism 
discussed in the previous section for the case of a single source located at 
$t_0$ from the observer, one obtains a slightly modified expression for the 
evolution of the spectrum~\cite{Anchordoqui:1997rn}
\begin{equation}
n(E, t)  \sim \frac{\kappa E_0^{-\gamma+1}}{E} \,, 
\label{phdsespectros}
\end{equation}
where the relation between $E_0$ and $E$ is given in Eq.~(\ref{pats}). 
Putting all this together, 
the energy-weighted flux beyond the GZK-energy 
due to a single M82 flare 
\begin{equation}
E^3 J(E)  =
 \frac{E}{(4 \pi d)^2} \frac{E_0^2 \,dN_0}{dE_0\,dt}\,e^{-R\,t/56}
 \approx  2.7 \times 10^{25} E_{20}
\epsilon \,e^{-R\,t/56}\,
{\rm eV}^2 \, {\rm m}^{-2} \, {\rm s}^{-1} \, {\rm sr}^{-1},
\end{equation}
is easily consistent with observation~\cite{Hayashida:2000zr}. It is important to stress that 
these phenomenological arguments are in agreement with Monte Carlo 
simulations~\cite{Bertone:2002ks}. In 5 years of operation PAO will provide enough statistics 
to test the starburst anisotropy predictions~\cite{Anchordoqui:2002dj}.

\section{GZK--evading messengers}

Several attempts have been made to explain the high end of the spectrum as 
a manifestation of some kind of physics beyond the SM. These suggestions 
became really interesting after Farrar and 
Biermann~\cite{Farrar:1998we} reported an intriguing correlation between the 
arrival directions 
of 5 extremely high energy CRs and compact radio quasars at high redshifts
($z = 0.2 - 0.3$). However, with the present data, such 
``evidence'' for directional correlation is still a subject of 
debate~\cite{Hoffman:1999ev,Farrar:1999fw,Sigl:2000sn,Virmani:2000xk}.
The most economical proposal to sneak away the GZK-cutoff involves a familiar 
extension of the SM, namely, neutrino masses. As noted by 
Weiler~\cite{Weiler:1982qy,Weiler:apj}, neutrinos can travel over 
cosmological distances 
with negligible energy loss and could, in principle, produce $Z$ bosons on 
resonance through annihilation on the relic neutrino background within a GZK 
distance of Earth. If this were the 
case, the highly boosted decay products of the $Z$ could be observed as 
super-GZK primaries, since they do not have to travel cosmological 
distances to reach us, and would be pointing directly back to the source. 
This proposal requires very luminous sources of 
extremely high energy neutrinos throughout the Universe. Another possibility 
in which the ultrahigh energy CR should point directly to its source assumes 
that the CR is a new, neutral, stable or very long lived 
supersymmetric hadron of mass a few GeV~\cite{Farrar:1996rg}. From all the 
candidates {\it en vogue}, the light glueballino seems the most plausible, 
because it can be efficiently produced in $pp$ collisions, it is not strongly 
absorbed by the CMB, and produces extensive air showers with longitudinal 
development very similar to those observed~\cite{Berezinsky:2001fy}. An even 
more radical proposal postulates a tiny violation of Lorentz invariance, 
such that 
some processes become kinematically 
forbidden~\cite{Sato,Kirzhnits:sg,Gonzalez-Mestres:1997dr,Coleman:1998en,Coleman:1998ti}. In particular, photon-photon pair production and photopion 
production may be 
affected by Lorentz invariance violation. Hence, the absence of the 
GZK-cutoff would result from the fact that the threshold for photopion 
production ``disappears'' and the process becomes kinematically not allowed. 

For models that rely on GZK-evading messengers, the accumulation factor 
$f_{\rm acc}$ depends
on the details of the model. In the case of messengers which can
induce showers across the entire energy spectrum, one expects
enhancement on the low energy side only from the baryonic
component, and $f_{\rm acc}$ depends on the interaction length in
the CMB and on the energy spectra at the different sources. For messengers 
whose attenuation length is comparable to the horizon, and which do not 
shower at the lower energies, $f_{\rm acc}\ll 1.$ In what follows, we give 
a somewhat more detailed account of these general considerations.

\subsection{Super-GZK CRs from the edge of the Universe}

The neutrino is the only known stable particle that can propagate through 
the CMB essentially uninhibited even at the highest CR energies. 
Specifically, the corresponding $\nu\bar{\nu}$ annihilation mean 
free path on the cosmic neutrino background, 
\begin{equation} 
\lambda_\nu
= (n_\nu \, \sigma_{\nu \bar{\nu}})^{-1} \approx  4
\times 10^{28}\,\,{\rm cm}\,\,,
\end{equation}
is just above the present size of the horizon 
(recall that $H_0^{-1} \sim 10^{28}$~cm)~\cite{Roulet:1992pz}. One may then 
entertain the notion that neutrinos are indeed the super-GZK primaries. 
However, in the SM model a neutrino incident 
vertically on the atmosphere would pass through it uninhibited, never 
initiating an extensive air shower. Consequently one has to postulate new 
interactions so that these neutrinos acquire a strong cross section above 
$10^{20}$~eV. 

The idea that neutrinos are the super-GZK primaries was introduced 
by Beresinsky and Zatsepin in 1969~\cite{Beresinsky}. Since then,
explicit models have been devised in which the  neutrino-nucleon 
cross section is enhanced by some new physics beyond the electroweak scale. 
One interesting proposal is that leptons are bound states of dual QCD gluons, 
and therefore can interact ``strongly'' with  all partons in the nucleon 
attaining an effective cross 
section comparable to the geometric nucleon cross 
section~\cite{Domokos:1986qy,Domokos:1988qh,Bordes:1997bt,Bordes:1997rx}. 
More recently, low scale gravity models have provided new impetus to the 
hypothesis of strongly interacting neutrinos. In particular, the neutrino 
nucleon cross section can be 
made large enough due to a quick rise of the density of states, possibly 
increasing exponentially, $\rho \sim e^{\sqrt{s/s_0}}$, above a 
characteristic energy scale $s_0$~\cite{Domokos:1998ry}. A specific 
implementation of this idea is given in theories with $n$ additional large 
compact dimensions and TeV scale quantum 
gravity~\cite{Nussinov:1999jt}. A detailed 
discussion of this last proposal is left until Sec.VII-A. For now, we will 
focus on the $Z$-burst model, in which everyday weekly interacting neutrinos 
explain the super-GZK primaries.

\subsection{${\bm Z}$-burst}

To ascertain whether the $Z$-burst hypothesis is viable, we would like
to know what neutrino energy and flux would actually be required to generate
observed highest energy CRs.  This in turn depends on the
neutrino mass, the distribution of thermal neutrinos and the
kinematics of the $Z$ decay products.  These three points are
discussed in the following paragraphs.  At the end of the section,
we comment on the compatibility of the $Z$-burst hypothesis with current
EAS data and plausible astrophysical sources.
                                              
In recent years, stronger and stronger experimental evidence for neutrino 
oscillations has been accumulating. Certainly, this evidence -- which is
compelling for atmospheric neutrinos~\cite{Fukuda:1998mi,Fukuda:1998tw,Fukuda:1998ub,Fukuda:2000np,Ambrosio:2001je}, strong for solar neutrinos~\cite{Hampel:1998xg,Abdurashitov:1999zd,Fukuda:2001nj,Fukuda:2001nk,
Altmann:2000ft,Ahmad:2001an}, and so 
far unconfirmed for neutrinos produced in the laboratory and studied by 
various groups~\cite{Athanassopoulos:1995iw,Athanassopoulos:1996wc,Athanassopoulos:1996jb,Athanassopoulos:1997pv,Athanassopoulos:1997er,Aguilar:2001ty,Apollonio:1999ae,Eitel:2000by,Boehm:1999gk} -- would extend the SM
by requiring non-vanishing neutrino masses and mixing. A precise estimate of 
the $\nu$-masses and mixing parameters would be an important clue to any 
physics beyond the SM. However, neutrino oscillations are sensitive to 
the mass (squared) splittings $\Delta m_{ij}^2 = m_{\nu_i}^2 - m_{\nu_j}^2$, 
and so only a lower bound on the mass of the heaviest neutrino 
can be obtained.\footnote{$i$, $j$ run over the possible neutrino species.} 
Strictly speaking, using the atmospheric mass splitting 
in a 3 neutrino flavor scenario one obtains
\begin{equation}
m_{\nu_3}\, \geq  \,\sqrt{\Delta m_{\rm atm}^2}\,\, \agt 0.04 \, {\rm eV}\,,
\end{equation}
whereas by using the LSND mass splitting in a 
4 flavor scenario the limit is shifted  an order of magnitude, i.e.,  
\begin{equation}
m_{\nu_4} \,\geq \, \sqrt{\Delta m_{\rm LSND}^2}\,\, \agt 0.4 \, {\rm eV}\,.
\end{equation}
For recent surveys 
see~\cite{Fogli:2001vr,Bahcall:2001zu,Gonzalez-Garcia:2002dz}. The search for mass imprints in the endpoint spectrum of 
tritium $\beta$ decay combined with experimental constraints from oscillations yield upper bounds on the mass of the heaviest neutrino,
\begin{equation}
m_{\nu_3} \,\leq \, \sqrt{m_\beta^2 + \Delta m_{\rm atm}^2}\,\, \alt 2.5 \, {\rm eV}\,,
\end{equation}
in a 3 flavor, and
\begin{equation}
m_{\nu_4} \,\leq\,  \sqrt{m_\beta^2 + \Delta m_{\rm LSND}^2}\,\, \alt 3.8 \, {\rm eV}\,,
\end{equation}
in a 4 flavor scenario~\cite{Barger:1998kz}. 
Additional information on the neutrino mass scale can be derived through cosmological and astrophysical observations~\cite{Dolgov:2002wy}. For instance, 
analyses of galaxy clustering, including recent CMB measurements and other cosmological constraints, 
give an upper bound
\begin{equation}
\sum_i m_{\nu_i} < 1.8 - 4.4 \,\, {\rm eV} 
\end{equation}
on the sum of the neutrino 
masses~\cite{Croft:1999mm,Fukugita:1999as,Gawiser:2000yg,Wang:091,Hannestad:2002xv,Elgaroy:2002bi}.

Big bang cosmology predicts that the Universe is filled with a shallow 
degenerate Fermi sea of neutrinos~\cite{Weinberg}. Neutrino decoupling 
occurred at a 
temperature of  1.9~K, just one second after the bang. This 
thermal neutrino background (TNB) has an average number density of
\begin{equation}
\label{standard_number_dens}
\langle n_{\nu_i}\rangle_0 = \langle n_{\bar\nu_i}\rangle_0
= \frac{3}{22}\, 
\underbrace{\langle n_{\gamma}\rangle_0}_{\rm CMB}
\simeq 56\ {\rm cm}^{-3}\,,
\end{equation}
per neutrino flavor. The dominant interaction mode of extremely high energy 
neutrinos with the TNB is the exchange of a $W^\pm$ boson 
in the $t$-channel $(\nu_i \bar{\nu_j} \rightarrow l_i \bar{l}_j)$, or of a 
$Z^0$ boson in either 
the $s$-channel $(\nu_i \bar{\nu}_j \rightarrow f \bar{f}$) 
or the $t$-channel $(\nu_i  \bar{\nu}_j  \rightarrow \nu_i  \bar{\nu}_j)$.
Here, $l$ denotes a charged lepton, $f$ a charged fermion, and $i \neq j$ 
for the first reaction~\cite{Roulet:1992pz}. Therefore, one expects
neutrinos within a few $Z$ widths of the right energy,
\begin{equation}
E^{R_Z}_\nu = \frac{M_Z^2}{2 m_{\nu_i}} = 4 \,\, \left(\frac{{\rm eV}}{m_{\nu_i}}
\right) \times 10^{21}\,\,{\rm eV}\,,
\end{equation} 
to annihilate with the TNB into hadrons at the $Z$-pole 
with large cross section,
\begin{equation}
\langle \sigma_{\rm ann} \rangle^Z \equiv \int \frac{ds}{M_Z^2}\, 
\sigma_{\rm ann} (s) = 2 \,\pi\,\sqrt{2}\,G_F  \sim 40.4\, {\rm nb}\,\,,
\end{equation}
producing a ``local'' flux of nucleons and 
photons~\cite{Weiler:1997sh,Fargion:1997ft}.\footnote{$G_F = 1.16639(1) 
\times 10^{-5}$~GeV$^{-2}$ is the Fermi coupling constant.}
Remarkably, the energy of the 
neutrino annihilating at the peak of the $Z$-pole has to be well above the 
GZK limit. At this stage, it is worthwhile to point out that in view of 
the expected rapid decrease of the ultrahigh energy $\nu$ flux with 
rising energy, the  
$t$-channel $W$- and $Z$-exchange annihilation processes can be safely 
neglected. On resonance, the $s$-channel $Z$-exchange interactions completely 
overwhelm them.

The mean energies of the $\sim $ 2 nucleons and $\sim$ 20 $\gamma$-rays 
in each process can be estimated by distributing the resonant energy
among the mean multiplicity of 30 secondaries. The proton energy is given by 
\begin{equation}
\langle E_p \rangle\, \sim \frac{M_Z^2}{60\, m_{\nu_j}} \sim 1.3 \,\,\left(\frac{{\rm eV}}{m_{\nu_j}}
\right) \times 10^{20} \,\,{\rm eV},
\end{equation}
whereas the $\gamma$-ray energy is given by
\begin{equation}
\langle E_\gamma\rangle\, \sim  \frac{M_Z^2}{120\,m_{\nu_j}} \sim 0.7 \,\,\left(\frac{{\rm eV}}{m_{\nu_j}}
\right) \times 10^{20} \,\,{\rm eV}.
\end{equation}
The latter is a factor of 2 smaller to account for the photon origin in two 
body $\pi^0$ decay. 

If the neutrino sources are randomly distributed in space, then the total 
rate of super-GZK events induced by $\nu \bar{\nu}$ annihilation at 
the $Z$ pole within a distance $D$ of the Earth is~\cite{Weiler:1997sh}
\begin{equation}
F_Z \sim E_\nu^{R_Z} \, F_\nu \, \langle\sigma_{\rm ann}\rangle^Z \, B_h^Z\, 
\int d^3x \, \frac{n_\nu}{4 \pi r^2}\, = E_\nu^{R_Z} \, F_\nu\, 
\langle\sigma_{\rm ann}\rangle^Z \, B_h^Z\, \int_0^D n_\nu\, dr \,,
\label{far}
\end{equation} 
where $F_\nu$ is the incident neutrino flux evaluated at the resonant energy, $B_h^Z \sim 0.70$ is the hadronic branching fraction of the $Z$, and $n_\nu$ is the column number density of  relic neutrinos. In deriving 
Eq.~(\ref{far}) one assumes that $
\langle \sigma_{\rm ann} \rangle \int_0^D\, dr \,\,n_\nu \ll 1$.
The $Z$-burst rate can be amplified if neutrinos are clustered rather than 
distributed uniformly throughout the 
Universe~\cite{Weiler:1997sh,Fargion:1997ft}. In such 
a case the probability of neutrinos to annihilate within the 
GZK zone is generally on the order of 1\%.

An exhaustive analysis of the $Z$-burst parameter space has  recently  been
reported by Fodor, Katz and 
Ringwald (FKR)~\cite{Fodor:2002hy,Fodor:2002prl}. The analysis includes 
two possibilities for a diffuse background of protons, as distinct from 
protons resulting from the $Z$-burst itself. On the one hand 
they analyze the case 
where all events above 
$4 \times 10^{19}$~eV consist of protons which are produced within our 
Galactic halo or at least within the GZK zone. For this case, 
hereafter referred to as halo background (Halo bk'd), no GZK attenuation is 
considered. On the other hand, they analyze the situation where CRs are 
protons which originate from uniformly distributed extragalactic sources, or  
so-called ``extragalactic background'' (EG bk'd). In view of the observed 
distributions of arrival directions, the EG bk'd seems to be 
phenomenologically more realistic. In such a case, a FKR maximum likelihood 
fit yields $0.08~{\rm eV} \leq m_{\nu_i} \leq 1.3~{\rm eV}$, while 
for a Halo bk'd $2.1~{\rm eV} \leq m_{\nu_i} \leq 6.7~{\rm eV}$, both 
at the 68 \% CL.

The FKR analysis seems to be 
pretty insensitive to the precise values of the cosmological parameters:
$h$, $\Omega_M$, and $\Omega_\Lambda$, the normalized Hubble expansion 
rate, and the matter and vacuum energy densities, respectively. 
The high energy 
$\nu$-flux is 
described by
\begin{equation}
F_\nu (E_\nu, z) = F_\nu (E_\nu) \,\, (1 +z)^{\alpha}\,, 
\end{equation}
where $z$ is the redshift and $\alpha$ characterizes the source evolution.
Since neutrinos are produced as secondaries in hadronic astrophysical 
sources, the flux at zero redshift is expected to fall off like, 
$F_\nu (E_\nu) \propto E_\nu^{-\gamma}$, with 
$\gamma \agt 1$. The FKR analysis takes into account two extreme 
scenarii: (i) 
strong $\gamma$-ray attenuation, where ultrahigh energy photons from the 
$Z$-bursts do not contribute to the observed flux [this is certainly the 
case if the radio background is sufficiently large 
and/or the EGMF is $\cal{O}($nG)],  and (ii) minimal universal radio background with 
vanishing EGMF. 

\begin{figure}
\postscript{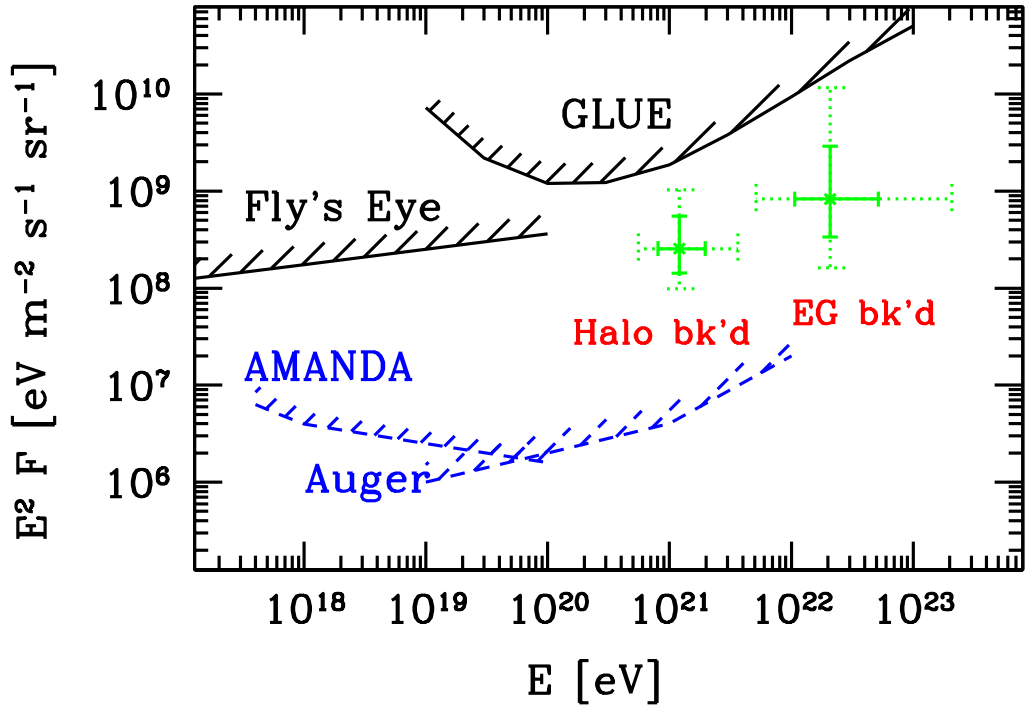}{0.90}
\caption{Neutrino fluxes, 
$F = \frac{1}{3} \sum_{i=1}^3 ( F_{\nu_i}+F_{\bar\nu_i})$, required 
by the $Z$-burst hypothesis for the case of a halo and an extragalactic 
background of 
ordinary cosmic rays, respectively ($\alpha =0, h=0.71, \Omega_M= 0.3,
\Omega_\Lambda =0.7,z_{\rm max}=2$).
Shown, as two crosses, are the necessary fluxes obtained 
for the case of a strong ultrahigh energy $\gamma$ attenuation.
The horizontal errors indicate the 1\,$\sigma$ (solid) and 
2\,$\sigma$ (dotted) uncertainties of the
mass determination and the vertical errors also include the uncertainty
of the Hubble expansion rate.
Also shown are upper limits from Fly's Eye~\cite{Baltrusaitis:mt} on 
$F_{\nu_e}+F_{\bar\nu_e}$ and the 
Gold\-stone Lunar Ultrahigh energy neutrino Ex\-pe\-ri\-ment 
GLUE~\cite{Gorham:2001aj} on $\sum_{\alpha = e,\mu} ( F_{\nu_\alpha}+F_{\bar\nu_\alpha})$, 
as well as projected sen\-si\-tivi\-ties of AMAN\-DA~\cite{Hundertmark:2001} 
on $F_{\nu_\mu}+F_{\bar\nu_\mu}$ and  PAO 
on $F_{\nu_e}+F_{\bar\nu_e}$. Published in 
Ref.~\cite{Fodor:2002hy}.}
\label{z}
\end{figure}
 
EGRET~\cite{Sreekumar:apj} measurements of the diffuse $\gamma$-background 
in the energy range between 30~MeV and 100~GeV strongly constrain the 
source evolution, yielding $\alpha \alt 0$ quite independently of different 
assumptions about the radio background. It should be noted that a correlation 
between ultrahigh energy CRs and BL Lac objects at redshifts $z>0.1$ has been 
claimed~\cite{Tinyakov:2001nr}. There is evidence that such a sub-class of 
AGN has zero or negative 
cosmological evolution. 

The required neutrino fluxes for the $Z$-burst 
hypothesis are given in Fig.~\ref{z}, together with existing upper limits 
and projected sensitivities of present and near future experiments. Power-law 
extrapolation of the fluxes $F_\nu \propto E^{-\gamma}$ below the resonance 
energy with spectral indices $\gamma \agt 1.5$ are excluded by the 
Fly's Eye data. 
Furthermore, in the $Z$-burst hypothesis,
the flux required for power-law   
extrapolations with indices $\gamma \agt 1$ to lower energies 
is larger than the theoretical upper limit from ``hidden'' hadronic 
astrophysical sources, 
$E^2 F_\nu \sim 2 \times 10^{7}$ eV 
m$^{-2}$ s$^{-1}$ sr$^{-1}$~\cite{Mannheim:1998wp}.
Hidden hadronic astrophysical sources are those from which only photons 
and neutrinos can escape. Therefore, 
one has to invoke sources that
have to be opaque to their primary 
protons, and should emit secondary photons (unavoidably produced together 
with the neutrinos) only in the sub-MeV region to avoid conflict with the 
diffuse $\gamma$-ray background measured by the EGRET 
experiment~\cite{Ringwald:2001mx,Kalashev:2001sh}.

It is fair to say that nowadays no convincing astrophysical sources are known 
which can accelerate protons at least up to $10^{23}$~eV, have zero or  
negative cosmological evolution, are opaque to nucleons, and emit photons 
only in the sub-MeV region.\footnote{The authors of~\cite{Davidson:2001ji} 
have pointed out that an extra $U(1)$ boson (the $Z^{\prime}$) with 
sufficiently low mass could reduce the extreme proton energies required in 
the $Z$-burst hypothesis. Speculations about a $Z^{\prime}$ of a few GeV have 
been entertained as one possible explanation of the NuTeV
anomaly~\cite{Davidson:2001ji,Zeller:2001hh}.} It is still an open question 
whether such challenging conditions can be 
realized in Nature. As one can see in Fig~\ref{z}, the PAO  
should give  a definite answer to this question.     

\subsection{SUSY ${\bm U}$}

Certainly, a novel beyond--SM--model explanation to the high end of the 
spectrum is to assume that ultrahigh energy CRs are not known particles 
but a new species of particle, generally referred to as the uhecron, 
$U$~\cite{Chung:1997rz,Albuquerque:1998va}. The meager information we have 
about super-GZK particles allows a na\"{\i}ve description of the properties 
of the $U$. The muonic content of EASs suggests $U$ should interact strongly. 
At the same time, if uhecrons are produced at cosmological distances, they 
must be stable, or at least remarkably long lived, with mean-lifetime 
\begin{equation}
\tau \sim 10^6  \left(\frac{m_U}{3 {\rm GeV}}\right)\, \left(\frac{d}{
{\rm Gpc}} \right)\,{\rm s}\, ,
\end{equation}  
where $d$ is the distance to the source and $m_U$, the uhecron's mass.
In addition, to avoid photopion production on the CMB, 
$m_U \agt 1.5$~GeV, because the 
threshold energy for the reaction $U \gamma \rightarrow U \pi$,
increases linearly with $m_U$,
\begin{equation}
E_{\rm th} = m_\pi \frac{(m_U + m_\pi/2)}{w}.
\end{equation} 
Moreover, to avoid deflections on EGMFs and the consequent energy loss due to 
pair production and other mechanisms, the hadron has to be neutral. Note that 
the latter may not be an essential requirement, depending on the distance 
to the accelerator. 

It has been known since the early days of Supersymmetry (SUSY) that the 
gluino $\tilde g$, the spin $1 \over 2$ SUSY partner of a gluon, could be the 
lightest supersymmetric particle (LSP), with the possible exception of the 
gravitino 
$g_{3/2}$~\cite{Fayet:1974pd,Farrar:xj,Farrar:te,Farrar:ee}.\footnote{If the LSP is the gravitino, the gluino decays gravitationally into a gluon and 
gravitino $\tilde g \rightarrow g g_{3/2}$, and its lifetime can be long 
enough.} Therefore, in a SUSY background the plausible candidates satisfying 
the $U$ requirements are gluino containing hadrons 
($\tilde{g}$-hadrons).\footnote{We note that an alternative proposal 
(where CRs above 
the GZK-energy originate from sources at cosmological distances) that relies 
on a supersymmetric extension of the SM, assumes the existence 
of a new exotic pseudo-scalar axion-like particle~\cite{Gorbunov:2001gc}.}
QCD sum rules suggest that the glueballino $\tilde G = \tilde g g $ is the 
lightest hadron. In the 1980s, $\tilde G$ was suggested as the possible 
primary particle in attempts to explain the hadronic nature of CRs from 
Cygnus X-3~\cite{Auriemma:sv,Berezinsky:xt}. 

There are, however, arguments against a light quasi-stable 
gluino~\cite{Okun:tn}. The main problem follows from the fact that along with 
a neutral hadron $\tilde G$ there must exist a charged hadron of the type 
$\tilde g qqq$ having baryon number (gluebarino). The lightest gluebarino must be a hadron $(\tilde g uud)$ in which the triplet of light quarks has total 
spin $1 \over 2$ and is in a color octet state. The gluebarino must be stable 
to 
the same extent as the gluino, since the lightest two particle state with 
the same quantum numbers $(\tilde g g) + (uud)$ must be heavier than 
$(\tilde g uud)$ by the mass of a constituent gluon ($> 0.6$~GeV).
Now, production 
of charged gluebarinos in the Earth atmosphere by CRs and their accumulation 
in the oceans would result in a too high abundance of anomalous heavy isotopes of 
hydrogen and oxygen in contradiction with observational data~\cite{Okun:tn}. 
However, it may well be that the lightest baryonic state is the neutral 
flavor singlet $S^0 = \tilde g uds$~\cite{Farrar:1996rg,Buccella:cs}, 
due to strong quark attraction in this 
state. We stress that even in this case conflict with observational 
data emerges if $\tilde g uds$-gluebarino and proton are bound into anomalous 
deuterium.    

In addition, direct searches for glueballino decays~\cite{Adams:1997ht,Fanti:1998kx,Alavi-Harati:1999gp} as well as for 
decays of other unstable $\tilde g$-hadrons~\cite{Albuquerque:1994xi} have 
severely eroded the attractiveness of the light gluino scenario.
A practically model-independent analysis, which takes into account 
contributions of the light gluino to the running of $\alpha_s$ and to QCD 
colour coefficients excludes light gluinos with 
$m_{\tilde g} = 3 (5)$~GeV with 93\% (91\%) CL~\cite{Csikor:1996vz}. 
By combining these analysis with the determination of QCD colour coefficients from the analysis of multi-jets events~\cite{Barate:1997ha}, the conclusion of~\cite{Csikor:1996vz} becomes much stronger: light gluinos with mass 
$\alt 5$~GeV are excluded with at least 99.89\% CL. Recent analyses 
based on currently available OPAL and CDF data conclude that the range 
$3~{\rm GeV} \alt m_{\tilde g} \alt 130 -150~{\rm GeV}$ can be excluded at 
the 95\% CL~\cite{Baer:1998pg}. For certain choices of the parameters, 
a window in the 
intermediate mass region $25~{\rm GeV} \alt m_{\tilde g} \alt 35~{\rm GeV}$ 
remains open~\cite{Raby:1998xr,Mafi:1999dg}. This leaves two narrow 
windows for allowed masses of 
$\tilde g$-hadrons $1.5~{\rm GeV} \alt m_{\tilde g-{\rm h}} \alt 3~{\rm GeV}$
and $25~{\rm GeV} \alt m_{\tilde g-{\rm h}} \alt 35~{\rm GeV}$. 
Gluino-containing hadrons corresponding to the second window would produce 
EASs 
very different from those detected by current 
CR-experiments~\cite{Berezinsky:2001fy}. However, light $\tilde g$-hadrons 
corresponding to the first gluino window produce EASs 
very similar to those initiated by protons. Furthermore, as shown in 
Fig.~\ref{gluino}, light glueballinos, 
$m_{\tilde G} \agt 1.5~{\rm GeV}$, have a spectrum with GZK-cutoff beyond the 
currently observed energy range~\cite{Berezinsky:2001fy}. 

\begin{figure}
\postscript{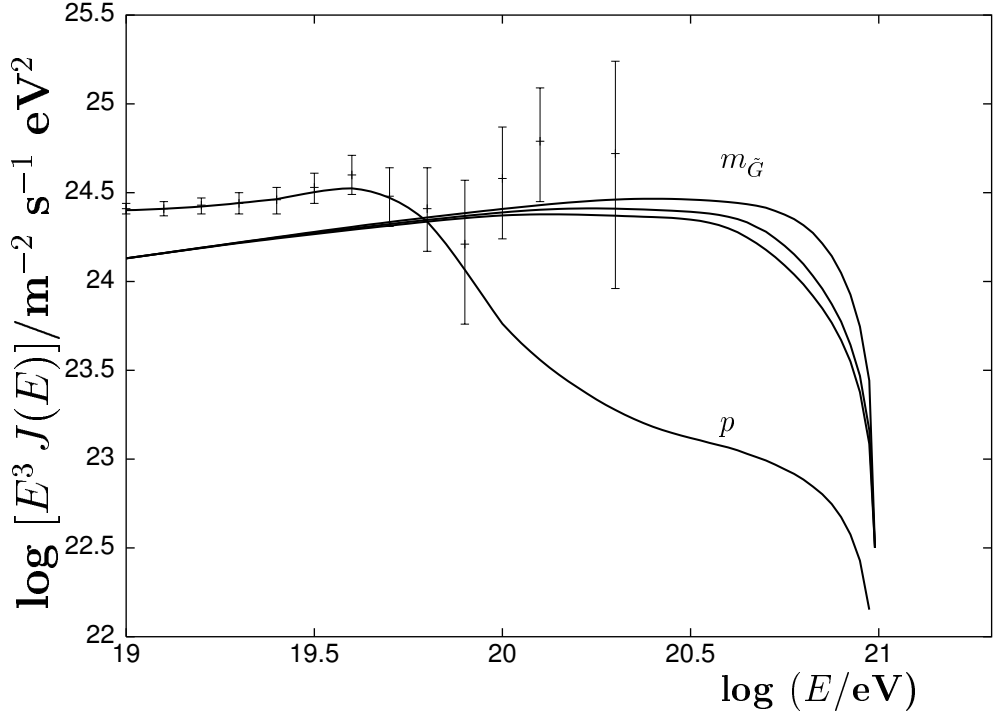}{0.90}
\caption{The three solid curves that accurately fit the observational data 
from AGASA above $10^{20}$~eV indicate the diffuse glueballino flux from 
uniformly distributed sources with injection spectra 
$dN_0/dE \propto E^{-2.7}$ and intrinsic cutoff $E = 10^{21}$~eV 
for $m_{\tilde{G}}$ = 1.5, 2 and 3~GeV (from below). The  
proton flux is also shown for comparison. 
Published in~\cite{Berezinsky:2001fy}.}
\label{gluino}
\end{figure}

\subsection{Lorentz symmetry violations with anomalous kinematics}

At present, there is no reason to anticipate the existence of a universal 
scale below which our present notion of flat spacetime geometry is not 
valid. However, local Lorentz invariance (LLI) should not be accepted on 
faith but rather as a plausible hypothesis subject to experimental test. 
It is possible to introduce the notion of LLI violation either with or 
without accompanying anomalous kinematics. 
If no anomalous kinematics is involved, any search for LLI non-conservation 
effects will require testing length scales  below $10^{-16}$~cm or 
less~\cite{Anchordoqui:1995im}. However, introducing anomalous 
kinematical constraints allows tiny departures from LLI, which would be 
undetectable at the electroweak scale, to be magnified rapidly with rising 
energy. LLI-violation with anomalous kinematics could then help to break the 
GZK barrier~\cite{Sato,Kirzhnits:sg,Gonzalez-Mestres:1997dr,Coleman:1998en,Coleman:1998ti}.

For instance, it was noted by Coleman and Glashow~\cite{Coleman:1998en} that 
renormalizable and gauge invariant perturbations to the SM Lagrangian that 
are rotationally invariant in a preferred frame, but not Lorentz invariant, 
lead to species-specific maximum attainable velocities $c_i$ for different 
particles. In such a case, the possible departure from LLI can be phrased 
in terms of the difference between the particle maximum attainable velocities
\begin{equation}
\delta_{ij} = c_i - c_j \,.
\end{equation}
Within this framework the energy conservation can be expressed 
by~\cite{Coleman:1998ti}
\begin{equation}
4 \, w \geq \delta_{\Delta p} \,\,E + \frac{m_\Delta^2 - m_p^2}{E} \,.
\label{con}
\end{equation}
If Lorentz symmetry is unbroken, $\delta_{\Delta p} =0$ and 
Eq.~(\ref{con}) leads to the conventional threshold for a head-on collision.
Otherwise, Eq.~(\ref{con}) 
is a quadratic form in $E$ with discriminant $4\,w^2 -( m_\Delta^2 - m_p^2)\, 
\delta_{\Delta p}$. For
\begin{equation}
\delta_{\Delta p} > \hat{\delta}(w) \equiv \frac{4 w^2}{m_\Delta^2 -m_p^2} \approx 3.5 \times 10^{-25}\,\left(\frac{w}{w_0}\right)^2\,
\end{equation}
the discriminant $<0$, and therefore the reaction $p  \gamma \rightarrow 
\Delta \rightarrow p  \pi$ is kinematically forbidden for all $E$. 
Now, recalling that the relic photons follow a Planck distribution, it is 
easily 
seen that for $\delta_{\Delta p}$ comparable to $\hat{\delta}(w_0)$, the GZK 
effect would be relaxed. 

There are no direct experimental constraints on the parameter 
$\delta_{\Delta p}$. However, if we compare the speed of photons to that of 
high energy CRs, it is possible to obtain a very stringent bound on 
violations of LLI. This bound follows from the emission of \v{C}erenkov 
radiation and consequent loss of energy by charged CRs, a process 
which is allowed if $c_\gamma<c_{_{\rm CR}}$. Primary protons (electrons) 
with energies up to $10^{20}$~eV (1~TeV) have been seen, thus 
$\delta_{p\gamma} < 3.0 \times 10^{-23}$~\cite{Coleman:1997xq} 
($\delta_{\gamma e} <1.3 \times 10^{-13}$~\cite{Stecker:2001vb}). 
Moreover, if LLI is broken and $c_e > c_\gamma$ the threshold energy for pair 
production is altered, yielding $\delta_{\gamma e} \leq 2 m_e^2/E_\gamma^2$.
Multi-TeV $\gamma$-ray observations from Markarian 501 then 
lead to $\delta_{\gamma e} <1.3 \times 10^{-15}$~\cite{Stecker:2001vb}. 
On the other hand, if $c_\gamma> c_e$, the decay of a photon into an electron 
positron pair is kinematically allowed for photons with energies exceeding 
$E_{\rm max} = m_e \sqrt{2/|\delta_{e \gamma}|}$. The decay would take place rapidly, so that photons with energies above $E_{\rm max}$ would not be 
observable. 
Detection of primary photons with energies $E_\gamma \geq 50$~TeV leads to
$\delta_{e\gamma} <1.3 \times 10^{-15}$~\cite{Stecker:2001vb}.
The failure to detect velocity oscillations of neutrinos translates into 
$|\delta_{\nu\nu'}| < 2 \times 10^{-21}$, whereas the detection of neutrinos from SN 1987a yields $|\delta_{\nu\gamma}| < 10^{-8}$~\cite{Stodolsky:1987vd}.
An additional constraint $|\delta_{m\gamma}| < 6 \times 10^{-22}$ results
from a Hughes-Drever type experiment~\cite{Lamoreaux:1986xz}. In the latter, 
the maximum attainable velocity of material matter $c_m$ was taken to be the 
same for all massive particles.
It is interesting to remark that none of these constraints reaches the level 
of sensitivity to Lorentz violation needed to overcome the GZK paradox.
However, the future PAO observations of faraway sources could 
provide constraints on, or even a measurement of, the violation of Lorentz
symmetry, yielding essential insights into 
the nature of gravity-induced wave dispersion in the 
vacuum~\cite{Amelino-Camelia:1996pj,Amelino-Camelia:1997gz,Bertolami:1999da,Aloisio:2000cm,Bertolami:2000qa,Aloisio:2002ed}.

\section{Spacetime's unseen dimensions}

In recent years, it has become evident that a promising route towards 
reconciling the apparent mismatch of the fundamental scales of particle 
physics and gravity is to modify the short distance behavior of gravity 
at scales much larger than the Planck length. Such modification can be most 
simply achieved by introducing extra dimensions, generally thought to be 
curled-up, in the sub-millimeter range~\cite{Arkani-Hamed:1998rs,Antoniadis:1998ig,Arkani-Hamed:1998nn,Randall:1999ee}. Within this 
framework the fundamental scale of gravity  can be lowered all the way 
to ${\cal O}$ (TeV), and the observed Planck scale turns out to be just an 
effective scale valid for energies below the mass of Kaluza--Klein (KK) 
excitations.\footnote{Theories with small periodic internal dimensions where 
the low energy degrees of freedom are restricted to zero modes and 
separated by a large mass gap from the massive modes were first 
proposed by Kaluza~\cite{Kaluza:tu} and Klein~\cite{Klein:tv,Klein:fj}.} 
Clearly, while the 
gravitational 
force has not been directly measured far below the millimeter 
range~\cite{Hoyle:2000cv}, SM interactions have been investigated well 
below this scale. Therefore, if large extra dimensions really exist, one 
needs some mechanism 
to prevent 
SM particles from feeling them. Remarkably, there 
are several possibilities to confine SM fields to a 4 dimensional 
subspace (referred to as a brane-world) within the $(4+n)$ dimensional 
spacetime~\cite{Dvali:1996xe,Bajc:1999mh}.

Theories with TeV scale gravity can be classified into two broad categories 
according to whether they do or do not assume that the full spacetime 
manifold is of factorized form. In these theories, in addition to 
the 4 dimensions that we see (with coordinates $x^\mu$) there are $n$ unseen 
dimensions with coordinates $y^m$. In the case of non-factorizable geometries
the metric takes the form
\begin{equation}
ds^2 = dx^\mu \,dx_\nu + g_{mn} (y)\, dy^m dy^n,
\end{equation}
where the characteristic size of the {\it large} extra dimensions is of the 
order of the fundamental Planck length, explaining their invisibility. In 
the case of {\it warped} extra dimensions, the scale of 
the 4-dimensional metric may vary depending on the location in the extra 
dimension, due to the warp factor $e^{2 A(y)}$ that can be thought of as 
the carrier of a position-dependent redshift. The metric in this case reads, 
\begin{equation}
ds^2 = e^{2 A(y)} \,dx^\mu\,dx_\nu + g_{mn} (y)\, dy^m dy^n \,.
\end{equation}
 To illustrate the two cases we treat the 
following representative examples.

\underline{\it Large extra dimensions:} Arkani-Hamed, Dimopoulos, 
and Dvali (ADD)~\cite{Arkani-Hamed:1998rs} imagine the spacetime as a 
direct product of ordinary four-dimensional spacetime 
and a (flat) spatial $n$-torus
with circumferences $L_i = 2 \pi r_i$ ($i=1,\dots,n$), generally of 
common linear size $r_i=r_c$.
As mentioned above, SM fields cannot propagate freely in the extra dimensions
and remain consistent with observations. This is avoided by trapping the 
fields 
to a thin shell of thickness $\delta \sim M_s^{-1}$. Assuming that the higher 
dimensional theory at short distance is a string theory, one expects that 
the fundamental string scale $M_s$ and the higher dimensional Planck scale 
$M_*$ are not too different. 
 The only 
particles propagating in the (4+n) dimensional bulk are the (4+n) gravitons.
Because of the compactification, the extra $n$ components of the 
graviton momenta are quantized
 \begin{equation}
k_i = \frac{2\pi\ell_i}{L_c} = \frac{\ell_i}{r_c}, \,\,\,\, i=1,\dots,n.  
\end{equation}
Thus, taking into account the degeneracy on $\ell_i \in \mathbb{Z}$, 
the graviton looks like a massive KK state with mass 
\begin{equation}
m_{\ell_1, \dots, \ell_n} = \left(\sum_{i=1}^n 
\ell_i^2\right)^{1/2} \, r_c^{-1}.
\label{3}
\end{equation}
It is important to stress that the graviton's self-interactions 
must conserve both ordinary 4-momenta and KK momentum components, whereas 
SM fields (that break 
translational invariance) do not have well 
defined KK momenta in the bulk  
for $\ell/r_c \leq M_s$.\footnote{For $\ell/r_c > M_s$ we would expect 
higher order quantum gravity and string effects to become important,  
providing some sort of a natural cut--off in the field theory.}
Therefore, interactions of gravitons with SM particles do not conserve KK 
momentum components. Applying Gauss' 
law at $r \ll r_c$ and $r \gg r_c$, it is easily seen that the Planck scale 
of the four dimensional world is 
related to that of a higher dimensional space-time simply by a volume factor,
\begin{equation}
r_c \sim \left( \frac{M_{\rm Pl}}{M_*} \right)^{2/n} \, \frac{1}{M_*} 
\sim 2.0 \times 10^{-19} \left(\frac{{\rm TeV}}{M_*}\right)  \left( \frac{M_{\rm Pl}}{M_*} \right)^{2/n} \,\,{\rm cm},
\label{rcomp}
\end{equation}
so that $M_*$ can range from $\sim$ TeV to $M_{\rm Pl} \sim 10^{19}$ GeV, 
for $r_c \alt  1$ mm and $n \geq 2$.
For $n\leq 6$, the mass splitting, 
\begin{equation}
\Delta m \sim \frac{1}{r_c} \sim M_* \left(\frac{M_*}{M_{\rm Pl}} \right)^{2/n} \sim
\left(\frac{M_*}{{\rm TeV}} \right)^{n+2/2} 10^{(12\,n-31)/n} \,\,{\rm eV},
\end{equation}
is so small that the sum over the tower of KK states can be replaced by a 
continuous integration. Then the number of modes between $|\ell|$ and 
$|\ell| + d\ell$ reads,
\begin{equation}
dN = d\ell_1 d\ell_2 \dots d\ell_n = S_{n-1} \,|\ell|^{n-1} d\ell, 
\label{states}
\end{equation}
where 
\begin{equation}
S_{n-1} = \frac{2\,\pi^{n/2}}{\Gamma(n/2)}
\end{equation}
is the surface of a unit-radius sphere in $n$ dimensions.
Now using Eqs. (\ref{3}) and (\ref{rcomp}), Eq. (\ref{states}) can be 
re-written as
\begin{equation}
dN  = S_{n-1} \left(\frac{M_{\rm Pl}}{M_*}\right)^2 \, \frac{1}{M_*^n} \, m^{n-1}\, dm.
\end{equation}
From the 4-dimensional viewpoint the graviton 
interaction vertex is suppressed by $M_{\rm Pl}$. Roughly speaking, 
$\sigma_m \propto  M_{\rm Pl}^{-2}$. 
Now, introducing $d\sigma_m/dt$, the differential cross section 
for producing a single mode of mass $m$, one can write down 
the differential cross section for inclusive graviton production 
\begin{equation}
\frac{d^2\sigma}{dt\,dm} = S_{n-1} \left(\frac{M_{\rm Pl}}{M_*}\right)^2 \, 
\frac{1}{M_*^n} \, m^{n-1}\, \frac{d\sigma_m}{dt},
\label{zero}
\end{equation}
or else, the branching ratio for emitting any one of  
the available gravitons  
\begin{equation}
\Gamma_g  \sim \frac{s^{n/2}}{M_*^{2+n}},
\label{silence}
\end{equation}
where $s^{1/2}$ is the c.m. energy available for graviton-KK emission.
As one can see by inspection of Eq. (\ref{silence}), the enormous 
number of accessible KK-states can compensate for the $M_{\rm Pl}^2$ factor 
in the scattering amplitude.

\underline{\it Warped extra dimensions:} 
Randall and Sundrum (RS)~\cite{Randall:1999ee} have proposed 
a simple and attractive scenario to ameliorate the hierarchy problem
in which our Universe plus a hidden brane are 
embedded in a 5-dimensional bulk with a negative cosmological constant.
The set-up is in the shape of a gravitational 
capacitor:  two 3-branes with equal and opposite tensions 
rigidly reside at $S_1/Z_2$ orbifold fixed points at the
boundaries ($y = 0$ and $y = \pi r_c$) of a slab of anti-de Sitter (AdS)
space of radius $\ell$. 
The line element satisfying this Ansatz in horospherical coordinates reads
\begin{equation}
ds^2 = e^{-2 |y|/\ell} \,\,\eta_{\mu\nu} \,dx^\mu dx^\nu + dy^2\,,
\label{lisa-metric}
\end{equation}
where $\eta_{\mu \nu}$ is the metric of Minkowski space.
Examination of the action in the 4-dimensional 
effective theory leads to~\cite{Randall:1999ee}
\begin{equation}
\overline{M}_{\rm Pl}^2 = M^3\,\ell\,\, \left( 1 - e^{-2\,\pi\,r_c/\ell}\right)\,,
\label{rel}
\end{equation}
where $\overline{M}_{\rm Pl}$ is the reduced effective 4-dimensional
Planck scale, and $M$ the fundamental scale of gravity. 
The AdS warp factor is exponential in the $y$-coordinate, and 
so the energy scales on the negative tension brane become 
exponentially redshifted. In other words, a field with the fundamental 
mass parameter $m_0$ will appear to have the 
physical mass $m = e^{-\pi\,r_c/\ell} \,m_0$ on the 
the 3-brane living at $y = \pi\,r_c$. Therefore, the weak scale is 
generated from fundamental scales of order $M_{\rm Pl}$ through the 
exponential hierarchy, requiring $r_c/\ell \approx 12$. 
Additionally, in the ``single brane'' limit 
($r_c\rightarrow \infty$)~\cite{Randall:1999vf}, the bulk continuum can be 
described by a Conformal Field Theory 
(CFT)~\cite{Hawking:2000kj,Duff:2000mt,Anchordoqui:2000du}, reproducing the 
duals discussed in~\cite{Maldacena:1997re}. For finite but large $r_c,$ 
deviations from conformality are exponentially damped in the 
infrared~\cite{Rattazzi:2000hs,Perez-Victoria:2001pa}.
In the RS model there exists a KK tower of gravitons 
with essentially electroweak couplings and masses 
$m_n = \ell^{-1}\,x_n\, e^{-\pi r_c/\ell}$, where 
$x_n$ is the $n^{\rm th}$ root of the Bessel 
function $J_1$~\cite{Randall:1999vf}. This implies 
that the tower mass spectrum, $m_n = m_1 x_n/x_1$, is completely determined 
by the lowest lying excitation. Note that while the zero 
mode graviton couples with a strength $\overline{M}_{\rm Pl}^{-1}$, all the 
remaining states couple as $\left(\overline{M}_{\rm Pl}\,e^{-\pi r_c/\ell} \right) ^{-1}$. The phenomenology of this scenario (recall that only gravity 
spills into the extra dimension) is 
governed by two parameters: 
$m_1$, and $c =(\ell \overline{M}_{\rm Pl})^{-1}$ which
is expected to be near, though somewhat less than, 
unity~\cite{Davoudiasl:1999jd}. The analysis~\cite{Davoudiasl:1999jd} of 
Tevatron data~\cite{Abachi:1996ud,Abachi:1995yi,Abbott:rs,Abe:1997fd,Abe:1996mj,Abe:1994zg} for anomalous Drell-Yan and dijet
production as well as the calculation of indirect contributions
to electroweak observables~\cite{Davoudiasl:2000wi} yield 
$m_1 \agt 500$~GeV, whereas AdS/CFT considerations suggest 
$c\alt 0.1$~\cite{Anchordoqui:2002fc}.

Clearly, the ADD and the RS model would have different manifestations   
in scattering processes. In the case of the product spacetime, 
each excited state couples with gravitational strength, and the key to 
observing KK-modes in particle collisions is the large multiplicity of 
states, due to their fine splittings. However, in the RS geometry, 
instead of gravitational strength coupling $\sim $ 
energy$/M_{\rm Pl}$, each excited state coupling is of order energy$/$TeV, 
and hence each resonance can be individually detected via its decay products. 
In the first part of this section we discuss how virtual graviton exchange 
would disturb high energy neutrino interactions~\cite{Nussinov:1999jt}. 
Under some extremely speculative hypotheses this 
phenomenon may allow  neutrinos to 
interact strongly in the atmosphere~\cite{Jain:2000vg,Jain:td}. More realistic assumptions would lead to a lower cross section which would predict an  
increase in the event rate of quasi-horizontal deeply developing showers. 
In the last part of the section, we discuss the 
possibility that ultrahigh 
energy neutrino interactions on the TNB produce gravitons of weak scale mass 
and coupling, resulting in ``gravi-burst'' fragmentation jets that can 
contribute to the super-GZK spectrum in a way that is similar to the 
$Z$-burst mechanism~\cite{Davoudiasl:2000hv}.

\subsection{Influence of KK-modes on the development of extensive air showers}

A novel feature of the contribution of KK graviton exchange to high energy 
scattering cross sections is the fact that 
it is projectile independent, and thus the same for scattering of 
$pp$, $\gamma p$, $\nu p$, etc. We focus here on $\nu p$ interactions where 
the contribution from KK-modes leads to a radical departure from SM particle 
physics.

A simple Born approximation to the elastic
$\nu$-parton cross section~\cite{Jain:2000pu} (which underlies the
total $\nu$-proton cross section) leads, without modification, to
$\snp\sim s^2.$ Unmodified, this behavior by itself eventually
violates unitarity. This  may be seen either by examining the
partial waves of this amplitude, or by noting the high energy
Regge behavior of an amplitude  with exchange of the graviton
spin-2 Regge pole: with  intercept $\alpha(0)=2$, the elastic
cross section
\begin{equation}
\frac{d\sigma_{el}}{dt}\, \sim\, \frac{|A_R(s,t)|^2}{s^2}\, \sim 
s^{2\alpha(0)-2}\,\sim s^2,
\end{equation} 
whereas
\begin{equation}
\sigma_{tot}\, \sim \frac{{\rm Im}[A_R(0)]}{s}\,\sim s^{\alpha(0)-1}\,\sim s,
\end{equation}
so that eventually $\sigma_{el}>\sigma_{tot}.$ Eikonal
unitarization schemes modify these behaviors: in the case of the
tree amplitudes~\cite{Nussinov:1999jt} the resulting 
(unitarized) cross section ($n=2$) $\snp\sim s,$ the all-order loop 
resummation 
yields $\snp\sim s^{2/n}$~\cite{Emparan:2001kf}, whereas for the single Regge 
pole exchange amplitude, $\snp\sim \ln^2(s/s_0)$~\cite{Kachelriess:2000cb}.\footnote{Note that the KK contribution to the cross section at 
$s \sim 2 \times 10^4$~TeV$^2$ gives $\sigma_{\rm KK} \alt 10^{-34}$~cm$^{2}$, 
whereas the usual hadronic $pp$ cross section at this energy is 
$\sigma_{pp} \sim 1.5 \times 10^{-25}$~cm$^2$, so the impact 
on the $pp$ scattering process is almost 
innocuous~\cite{Anchordoqui:2001ez}.}

The other relevant parameter which has direct influence on the shower 
profile is the inelasticity. The KK-modes couple to neutral currents, and 
thus the scattered neutrino carries away 90\% of the incident energy per 
interaction. Therefore, the cross section has to be at least a few $10^{-26}$~cm$^{2}$ to be consistent with observed showers which start within the first 50 g/cm$^2$ of the atmosphere. Specifically, the survival probability 
$N$ at atmospheric depth $X$ of a particle 
$a$ with mean free path
\begin{equation}
\lambda_a = \frac{m_{\rm air}}{\sigma_{a-{\rm air}}},
\label{po}
\end{equation}
is given by
\begin{equation}
N(X) = e^{-X/\lambda_a},
\label{po2}
\end{equation}
where $m_{\rm air} \sim 2.43 \times 10^{-23}$~g is the mass of an average 
atom of air, 
and $\sigma_{a-{\rm air}}$ the cross section on air. For a proton 
energy $\sim 10^{20}$~eV, the mean free path is 
$\lambda_p \sim 40$~g/cm$^2$. Therefore, a proton air shower is initiated 
at the top of the atmosphere. The key feature in the evolution of the shower 
is the ratio of decay to interaction of secondary hadrons
along their path in the atmosphere. The latter strongly depends 
both on particle energy and target density. For protons of this energy, 
$\langle X_{\rm max} \rangle \sim 800$~g/cm$^2$, corresponding 
to 20 hadronic interaction lengths. The neutrino nucleon cross section 
estimates discussed above indicate that $\lambda_\nu \agt 10 \lambda_p$. 
Due to the larger mean free path, 
$X_{\rm max}$ is shifted  360~g/cm$^2$ downwards in the atmosphere.  
Moreover, in a neutral 
current interaction the neutrino only transfers 10\% of its energy to the 
shower and consequently elongates the longitudinal development even 
further~\cite{Kachelriess:2000cb,Anchordoqui:2000uh}. 
Events consistent with these features have not been observed.
However, the model may still be salvaged if the neutrino-nucleon
cross section can be enlarged enough ($\sigma_{\nu N} \agt 20$~mb), 
then multiple scattering within the nucleus may provide a sufficiently 
large energy transfer to reproduce the 
super-GZK events~\cite{Jain:2000vg,Jain:td}. 

The question of whether the interaction cross section of neutrinos with matter
could reach typical hadronic values at high energies is yet undecided.
The Regge picture of graviton exchange is not 
yet entirely established, the apparently increasing dominance assumed
by successive Regge cuts due to multiple Regge pole exchange, as well as the 
presence of the zero mass graviton can introduce considerable uncertainty 
in the eventual energy behavior of the cross section~\cite{Muzinich:in,Nussinov:2001pu}. However, it is fair to say that recent 
calculations for the rate 
of rise of the cross section in the context of string 
theory tend not to support hadronic-like cross sections in 
$\nu$-interactions~\cite{Cornet:2001gy,Kachelriess:2001jq}.

Now we turn to the possibility of experimental input into the issue.
In the following we discuss a method to test scenarii in which 
neutrinos are super-GZK primaries which relies directly on EAS observables 
and is independent of the type of interaction enhancing physics.
Any physics beyond the SM that increases the neutrino-nucleon cross
section to typical hadronic values at $E_\nu \agt 10^{20}$~eV   
should also affect standard shower observables at lower energies, where 
the cross section 
attains sub-hadronic-sizes.  In particular, 
for $\sigma_{\nu N} \agt 10^{-27}$~cm$^2$, one expects neutrinos to trigger 
moderately penetrating showers, where $1000~{\rm g/cm}^2 \alt X_{\rm max} 
\alt 2500~{\rm g/cm}^2$. For the specific case discuss 
in Ref.~\cite{Jain:2000vg,Jain:td} the cross section is likely to be 
sub-hadronic near the energy at which the cosmogenic neutrino flux peaks 
(see Fig.~\ref{nuflux}), 
and so this kind of 
shower should be copiously produced~\cite{Anchordoqui:2000uh}.\footnote{There is no significant signal of showers with $X_{\rm max} \agt 1000$~g/cm$^2$ in the Fly's Eye data~\cite{Gaisser:ix}.} Clearly, 
the absence of moderately 
penetrating showers in the CR data sample should be understood as a serious 
objection to the hypothesis of neutrino progenitors of the super-GZK 
events. Even though neutrino-nucleon interactions in TeV-gravity models 
seems insufficient to explain the showers above the GZK-limit, virtual graviton exchanged can still lead to interesting new phenomena which may be 
observed in upcoming CR and neutrino experiments~\cite{Tyler:2001gt,Alvarez-Muniz:2001mk}. Recall that the atmosphere provides a detector medium with a column depth $\sim 36000$ g/cm$^2$ for horizontal arrival directions, and so can 
probe cross sections in the range $\sim 10^{-29} - 10^{-27}$~cm$^2$. Due to the increase column depth, water/ice detectors would probe cross sections in the range $\sim 10^{-31} - 10^{-29}$~cm$^2$.

\subsection{Gravi-burst}

The RS~\cite{Randall:1999ee} model of localized gravity would open up new 
channels for high energy neutrinos to annihilate with the TNB to produce a 
single graviton KK state on resonance which subsequently decays 
hadronically~\cite{Davoudiasl:2000hv}. For neutrino masses $m_{\nu_j} 
\sim 10^{-2} - 10^{1}$~eV, and graviton resonance of order a TeV, super-GZK 
events can be produced. Therefore, if one assumes that the incoming neutrino 
spectrum extends in energy with a reasonably slow fall-off, the existence of a series of 
$s$-channel KK graviton resonances may lighten the 
requirements on neutrino fluxes given in the $Z$-burst model shown in 
Fig.~\ref{z}.

\begin{figure}
\centerline{\psfig{figure=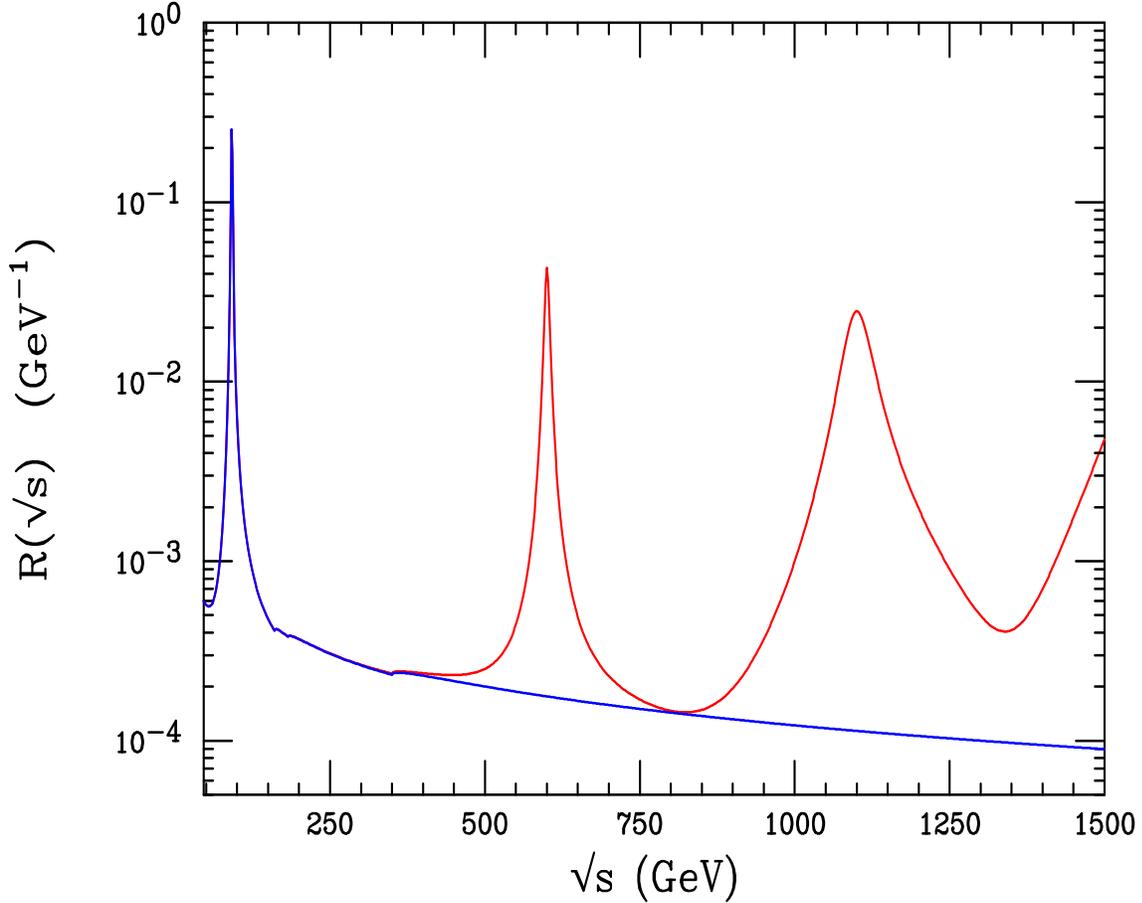,height=15cm,width=12cm,angle=90}}
\caption{Energy weighted total cross section for hadron production in units 
of that for the $Z$ pole in the Weiler $Z$-burst model for $\gamma = 0$ as a 
function of c.m. energy. The relatively flat lower 
curve shows the SM, and the upper curve shows the RS model with 
$c= 0.1$ and $m_1=600$ GeV. Published in Ref.~\cite{Davoudiasl:2000hv}.}
\label{graviburst}
\end{figure}

To estimate the event rate of a $\nu \bar\nu$ graviton mediated process, 
consider a point in the parameter space allowed by all current 
constraints: $m_1 = 600$~GeV and $c =0.1$. 
For these parameters, Davoudiasl, Hewett, and Rizzo~\cite{Davoudiasl:2000hv}
have estimated the cross section for $\nu \bar\nu$ annihilation into hadrons. 
Such a cross section has a number of distinct contributions, because 
gravitons not only lead to pairs of 
quarks, and $W$ and $Z$ final states, but also to pairs of gluons and 
Higgs bosons.\footnote{For numerical purposes the mass of the Higgs 
is set to 120~GeV.}  By combining all these individual process cross sections 
and assuming the neutrino spectrum above the $Z$ pole energy falls  
like  $\sim E_\nu^{-\gamma}$, it is straightforward to compute 
the full energy dependence of the total $\nu \bar\nu \rightarrow$ 
hadrons cross section, and afterwards the ratio of expected CR rates in 
units of the $Z$-pole induced rate given in Eq.~(\ref{far}),
\begin{equation}
R(\sqrt{s}) = \frac{F_{\rm SM + GRAV} (\sqrt{s})}{F_Z} = 
\frac{2 \,\,\sqrt{s}\,\,\, 
\sigma_{\rm ann}^{\rm SM + GRAV} (s)\,\,\, 
(M_Z/\sqrt{s})^{2\,\gamma}}{M^2_Z \langle \sigma_{\rm ann}\rangle^Z\,B_h^Z}\,.
\label{DHR}
\end{equation}
To get an idea of what this ratio looks like, Fig.~\ref{graviburst} shows the 
case of $\gamma = 0$. Integration of $R$ over a range of $\sqrt{s}$ leads to 
the relative rate of events expected in the RS model to those originating 
from the $Z$-burst model. Note that the integral of $R$ under the $Z$ pole gives the value unity as it should to reproduce the $Z$-burst results.
If the ultrahigh energy neutrino 
spectrum above the $Z$-pole falls with spectral index $\gamma \sim 0.5$, 
it is easily seen from Eq.~(\ref{DHR}) that  correction from the first 
RS excitation to the $\nu$-fluxes required to fit the observed spectrum 
(see Fig.~\ref{z}) is $\alt 5\%$, quickly 
falling below $1\%$ for $\gamma \agt 1$.

\section{Exotica}

The difficulties so far encountered in modeling the production of ultrahigh 
energy CRs arise from the need to identify a source capable of launching 
particles to extreme energy. In contrast to the ``bottom-up'' 
acceleration of charged particles, the ``top-down'' scenario avoids 
the acceleration problem by assuming that charged and neutral primaries simply 
arise in the quantum mechanical decay of supermassive elementary $X$ 
particles. Sources of these exotic particles could be:
\begin{itemize}
\item Topological 
defects (TDs) left over from early Universe phase transitions associated with 
the spontaneous symmetry breaking that underlies unified models of high energy 
interactions~\cite{Hill:1982iq,Hill:1986mn,Bhattacharjee:vu,Bhattacharjee:1990js,Bhattacharjee:1991zm,Bhattacharjee:1994pk}. 
\item Some long-lived metastable 
super-heavy ($m_X~\agt~10^{12}$~GeV) relic particles produced through vacuum 
fluctuations during the inflationary stage of the 
Universe~\cite{Gondolo:1991rn,Berezinsky:1997hy,Kuzmin:1997cm,Kuzmin:1998uv}. 
\end{itemize}
Due to their topological stability, the TDs (magnetic monopoles, 
cosmic strings, domain walls, etc.) can survive forever with $X$ particles 
($m_X \sim 10^{16} - 10^{19}$~GeV) trapped inside them. Nevertheless, from 
time to time, TDs can be destroyed through collapse, annihilation, or 
other processes, and the energy stored would be released in the form of 
massive quanta that would typically decay into quarks and leptons.
Similarly, super-heavy relics would also have quarks and leptons as the 
ultimate decay products. The strongly 
interacting quarks fragment into jets of hadrons resulting in 
typically 10$^4$ - 10$^5$ mesons and baryons. In this way, very energetic CRs,
with energies up to $m_X$, can be produced directly without any acceleration 
mechanism. Another exotic explanation of the highest 
energy CRs postulates that relic TDs themselves 
constitute the primaries~\cite{Kephart:1995bi,Bonazzola:1997tk}. 
General features of the exotic 
scenario have been discussed in several comprehensive 
reviews~\cite{Bhattacharjee:1998qc,Bhattacharjee:1997ps,Kuzmin:1999zk,
Berezinsky:mw,Olinto:2000sa,Bhattacharjee:2000vh}. 
Here we give just a general overview with emphasis on recent developments.

\subsection{Top-down origin}

In an epic paper, well ahead of its time, Lema\^{\i}tre~\cite{Lemaitre} 
introduced what may well be the ultimate top-down model.  
According to Lema\^{\i}tre the entire material filling the Universe, as well 
as the Universe's
expansion, originated in the super-radioactive disintegration of 
a ``Primeval Atom'', which progressively decayed into atoms of smaller and 
smaller atomic weight. The CRs were introduced as the energetic particles 
emitted in intermediate stages of the decay-chain. As a 
matter of fact, Lema\^{\i}tre regarded the CRs as  
vestigial evidence of the primeval fireworks. Of course, we now know that the existence 
of the CMB precludes the origin of super-GZK CRs in a very early cosmological 
epoch. However, it is amusing that one of the earliest explanations 
considered for the CR origin was a top-down mechanism.

\underline{\it Topological defects:} A more contemporary top down model proposes the vestiges of phase transitions in the early Universe as responsible 
for the highest energy CRs.
According to current unified models of high energy interactions, the Universe 
may have experienced several spontaneous symmetry breakings, where some 
scalar field, generally referred to as the Higgs field, 
acquired a non-vanishing expectation value in the new vacuum (ground) 
state. Quanta associated with these 
fields are typically of the order of the symmetry-breaking scale, which in 
Grand Unified Theories (GUTs) can be $\sim 10^{16} - 10^{19}$~GeV. During 
a phase transition, non-causal regions may evolve 
towards different states, so that in different domain borders the Higgs field 
may keep a null expectation value. Energy is then stored in a TD whose 
characteristics depend on the topology of the manifold where the Higgs 
potential reaches its minimum~\cite{Kibble:1976sj,Vilenkin:ib,Hindmarsh:1994re,Brandenberger:1993by}. The relic defects such as magnetic 
monopoles~\cite{Hill:1982iq,Bhattacharjee:1994pk}, 
cosmic strings~\cite{Bhattacharjee:vu,Bhattacharjee:1990js}, superconducting 
cosmic strings~\cite{Hill:1986mn}, 
vortons (superconducting  
string loops stabilized by the angular momentum of the charge 
carriers)~\cite{Masperi:1997qf,Masperi:2000sp}, and cosmic 
necklaces~\cite{Berezinsky:1997td}\footnote{A cosmic necklace is a possible hybrid TD consisting of a closed loop of cosmic string with monopole ``beads'' on it~\cite{Hindmarsh:xc}.} are all 
relatively topologically stable, but can release part of their energy 
(through radiation, annihilation, or collapse) in the form of $X$ particles that typically decay to 
quarks and leptons. The quarks hadronize producing jets of hadrons
containing mainly pions together with a 3\% admixture of nucleons. 
Note that, in contrast to the case of  bottom-up acceleration models, 
top-down models 
predict a flux above $10^{20}$~eV which is dominated by gamma rays and 
neutrinos produced via pion decay. Therefore, the photon/proton ratio can be used as 
a diagnostic tool in determining the CR origin~\cite{Aharonian:1992qf}.
In light of the mounting evidence that ultrahigh energy CRs are not gamma 
rays~\cite{Ave:2000nd,Ave:2001xn}, one may try to force a proton dominance 
at ultrahigh energies by postulating efficient absorption of the dominant ultrahigh energy 
photon flux on the universal and/or galactic radio background. However, the 
neutrino flux accompanying a normalized proton flux is inevitably increased 
to a level where it should be within reach of operating experiments such as 
AGASA~\cite{Barbot:2002kh}.

Certainly, the precise decay modes of the $X$ particles and the detailed 
dynamics of the first secondary particles depend on the exact nature of the 
particles under consideration. However, one expects the bulk flow of outgoing 
particles to be almost independent of such details. Moreover, the gross 
features of the hadronic jet systems can be reasonably well described 
by using  Local Parton-Hadron Duality 
(LPHD)~\cite{Azimov:1984np,Azimov:1985by}. 
In this approach, the primary 
hadron spectrum is taken to be the same, up to an overall normalization 
constant, as the spectrum of partons in the parton cascade after evolving the latter all the way down to a cutoff transverse momentum $\langle 
k_\perp^2 \rangle^{1/2}_{\rm cutoff} \sim $ few hundred MeV. A common 
picture for the parton cascade evolution is provided by the so-called 
Modified Leading Logarithmic Approximation (MLLA) of 
QCD~\cite{Mueller:cq,Mueller:cq2}. Within this approximation 
the energy spectrum can be expressed analytically, and simplifies 
considerably when the QCD scale $\Lambda_{\rm eff}$ is equal to the 
transverse momentum cutoff, $\tilde {Q}_0$. In such a limiting case the 
energy distribution 
of the partons is given by~\cite{Azimov:1984np,Azimov:1985by,Khoze:1996dn}
\begin{eqnarray}
x \frac{dN_{\rm part}}{dx} & = & \frac{4 C_F}{b}\, \Gamma(B)\, 
\int_{-\pi/2}^{\pi/2} \frac{d\ell}{\pi}\,\, e^{-B\alpha}\,\, \left[\frac{\cosh \alpha + (2 \xi / Y -1) \sinh \alpha}{(4N_c/b)\,\,Y\,\,(\alpha / \sinh \alpha)}\right]^{B/2} \nonumber \\
 & \times & I_B \left\{ \left[\frac{16 N_c}{b}\, Y\, 
\frac{\alpha}{\sinh \alpha}\, [ \cosh \alpha + (2 \xi/Y -1) \, \sinh \alpha] 
\right]^{1/2} \right \}\,,   
\end{eqnarray}
where $dN_{\rm part}$ is the number of partons with a fraction $(x, x+dx)$ of 
the energy $E_{\rm jet}$ of the original jet-initiating quark $q$, 
$\xi = \ln(1/x)$, $Y = \ln (E_{\rm jet}/\Lambda_{\rm eff})$, and $\alpha = 
[\tanh^{-1} (1 - 2\xi/Y) + i\ell]$. $I_B$ is the modified Bessel function 
of order $B$, where $B = a/b$ with $a = [11N_c/3 + 2n_f/(3N_c^2)]$ and 
$b = (11 N_c - 2n_f)/3$, $n_f$ is the number of flavors of 
quarks, $N_c = 3$ is the number 
of colors, and $C_F = (N_c^2 - 1)/2 N_c = 4/3$.  
Now, using LPHD one obtains the fragmentation function due to the 
hadronization of a quark $q$,
\begin{equation}
x \,\frac{dN_h}{dx} = K(Y)\,\, x \,\,\frac{dN_{\rm part}}{dx}\,\,,
\end{equation}
where $x = E_h/E_{\rm jet}$. The overall normalization constant $K(Y)$, which 
takes into account the effect of conversion of partons into hadrons, is 
fixed from the conservation of energy
\begin{equation}
\int_0^1 x \, \frac{dN_h(Y,x)}{dx} \, dx =1\, .
\end{equation}
Of course, there is a great uncertainty in the 
extrapolation of the QCD (MLLA + LPHD) spectra, which has been tested so far 
only at collider energies, up to super-ultrahigh energies $\agt 10^{23}$~eV. 
In particular, new processes may alter energy thresholds as well as the content of particles in the jets. For instance, if SUSY turns on at an energy scale 
$M_{\rm SUSY}$ of ${\cal O}$~(TeV), the shower development is expected to 
tie up not only quarks and gluons but also their supersymmetric partners 
with equal probability, provided the 4 momentum transfer 
$\tilde Q$ is above the SUSY scale. Once $\tilde Q < M_{\rm SUSY}$, the 
SUSY particles in the cascade would decouple from the cascade, and eventually decay into the stable LSPs. In such a case, the final state may contain ultrahigh energy 
LSPs~\cite{Berezinsky:1997sb}, thus modifying the shape of the fragmentation 
spectrum~\cite{Berezinsky:1998ed}.

\begin{figure}
\postscript{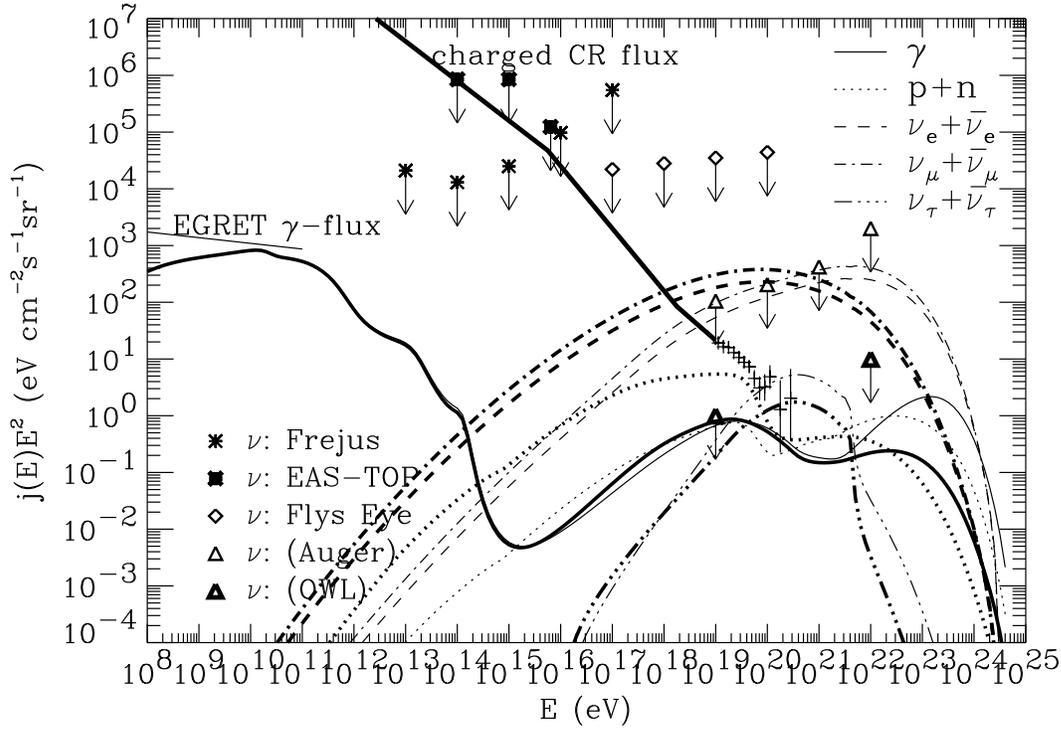}{0.90}
\caption{Energy-weighted spectra of nucleons, $\gamma$-rays and 
neutrinos for the TD model with $m_X~=~10^{16}$~GeV, $p=1$, and decay 
mode $X \rightarrow q +q$, 
assuming an EGMF of $10^{-10}$~G. Thick and thin lines represent the 
SUSY and no-SUSY fragmentation functions, respectively. The 1 sigma error bar 
crosses represent combined data from the Haverah Park, Fly's Eye, and AGASA 
experiments above $10^{19}$~eV. Also 
shown are piecewise power law fits to the observed charged CR flux below 
$10^{19}$~eV, the EGRET measurement of the diffuse $\gamma$-ray flux 
between 30 MeV and 100 MeV, experimental neutrino flux limits from Frejus 
and Fly's Eye, as well as projected neutrino sensitivities of PAO and NASA's OWL project. Published in 
Ref.~\cite{Sigl:1998vz}.}
\label{top-down}
\end{figure}

It is clear that the  wide variety of TDs produce $X$ 
particles at different rates. However, as noted in~\cite{Bhattacharjee:1991zm},
the $X$ particle production rate, $\dot {n}_X$, may be parametrized on 
dimensional grounds in a very general way,
\begin{equation}
\dot {n}_X (t) = \frac{Q_0}{m_X} \,\, \left(\frac{t}{t_0}\right)^{-4+p}\,,
\label{td}
\end{equation}
where $t$ is the Hubble time, $t_0$ denotes the present age of the Universe, 
and $Q_0 \equiv \dot{n}_X(t_0)\, m_X$ is the rate of energy injected in the 
form of $X$ particles of mass $m_X$ per unit volume in the present epoch. 
In most cases $p=1$. Exceptions are superconducting string models where 
$p \leq 0$, or decaying vortons  which have $p=2$. Certainly, the evolutionary 
properties of the TD system are unknown, and so one is not 
able to infer the value $Q_0$ {\it a priori} in a parameter-free manner. 
However, if a TD scenario is to explain the origin of ultrahigh energy CRs, 
Eq.~(\ref{td}) can be normalized to account for the super-GZK events without 
violating any observational flux measurements or limits at higher or lower 
energies, as shown in Fig.~\ref{top-down}. The top-down neutrino and gamma ray 
fluxes depend on the energy released integrated 
over redshift, and thus on the specific TD model. Note that the 
electromagnetic energy injected into the Universe above the pair production 
threshold on the CMB is recycled into a generic cascade spectrum below this 
threshold on a time scale short compared with the Hubble time. Therefore, 
it can have several potential observable effects, such as modified light 
element abundances due to $^4$He photodisintegration, or induce spectral 
distortions of universal gamma-ray and neutrino 
backgrounds~\cite{Sigl:1995kk,Sigl:1996gm}. In particular, measurements of 
the diffuse gamma ray background in the 100~MeV region, to which the generic 
cascade spectrum would contribute directly, limit significantly the parameter 
space in which TDs can generate the flux of the highest energy 
CRs~\cite{Sigl:1996im,Protheroe:1995ft,Protheroe:1996pd,Protheroe:1996zg}.

\underline{\it Super-heavy relics:} The highest energy CRs may also be 
produced from the decay of some metastable superheavy relic particle with 
mass $\agt 10^{12}$~GeV and lifetime exceeding the age of the 
Universe~\cite{Gondolo:1991rn,Berezinsky:1997hy,Kuzmin:1997cm,Kuzmin:1998uv}.  
Of course, there are no metastable superheavy relics within the SM. 
However, a number of candidate metastable relics are predicted in theories 
beyond the SM~\cite{Frampton:1980rs,Ellis:1980ap,Ellis:1990iu,Ellis:1990nb,Chang:1996vw,Frampton:1997gf,Chung:1998zb,Benakli:1998ut,Chung:1998ua,Chung:1998rq,Chung:2001cb,Hamaguchi:1998wm,Hamaguchi:1998nj,Hamaguchi:1999cv}. Here, the predicted CR flux is driven by the 
ratio of the density of the relics to their lifetime. 
Since the decay should occur within the GZK-sphere, a reasonable 
parametrization of 
the decay rate is
\begin{equation}
\dot{n}_X = \frac{n_X}{\tau_{_X}}
\end{equation}
where $\tau_{_X}$ is the relic's lifetime, 
$n_X = \rho_c \,\Omega_X\, h^2/m_X$, is the relic density,
$\Omega_X$ is the cosmic average mass density 
contributed by the superheavy relics in units of the critical 
density $\rho_c \approx 1.05 \times 10^{-4}\, h^2$ GeV cm$^{-3}$, 
and $h$ is the present value of the Hubble constant in units of 100 km 
sec$^{-1}$ Mpc$^{-1}$. Clearly, neither $\Omega_X$ nor $\tau_{_X}$ is 
known with any degree of confidence. Additionally, as in the case of TDs, 
the details of the CR spectrum depend on the fragmentation function. Several 
models have been used to approximate the 
fragmentation in superheavy relic decay, including the 
HERWIG Monte Carlo program~\cite{Birkel:1998nx}, 
MLLA~\cite{Blasi:2001hr,Dick:2002kp}, numerical integration of the 
Dokshitzer-Gribov-Lipatov-Altarelli-Parisi (DGLAP) 
equations~\cite{Sarkar:2001se,Barbot:2002ep,Barbot:2002gt}, 
as well as a combination of MLLA
(for small $x$) and DGLAP (for large $x$)~\cite{Fodor:2000za}.  
In general, these approaches all yield similar results in the region 
$10^{19}$--$10^{20}$~eV~\cite{Blasi:2001hr}. 
As an example, Fig.~\ref{susyfit} shows the 
resulting ultrahigh energy CR flux from the decay of relics  
($m_X = 5 \times 10^{12}$~GeV) clustered in the halo of our Galaxy 
(more on this below) obtained by numerical solution of the DGLAP 
equations, including effects of SUSY~\cite{Sarkar:2001se}.
The hard injection spectra in top-down scenarii ($\propto E^{-1}$) have 
roughly the inverse structure to that of the attenuation length of 
baryonic CRs and $\gamma$-rays. Therefore, over the limited energy range 
$\sim 10^{19.5 - 20}$~eV, where the attenuation length drops quickly, 
the attenuated spectra can be easily fitted to the observed spectrum. However, 
the spectra of such models 
have entirely the wrong shape below the photopion production threshold, and 
at $10^{18.7}$~eV, where the flux is well measured, the predicted flux is 
more than an order of magnitude too small (see Figs.~\ref{top-down} and \ref{susyfit}). 
Thus, to be consistent with observations, top-down 
models need to be smoothly matched to a bottom-up model which describes the 
spectrum below the GZK energy~\cite{Sigl:1995st}. An elegant explanation which can account for all the events with 
energy $\agt 10^{19.0}$~eV  can be concocted by mixing the neutrino fluxes of 
top-down models with the $Z$-burst mechanism 
$(m_{\nu_j} = 0.07$~eV)~\cite{Gelmini:1999ds,Gelmini:2002xy}. This hybrid 
scenario predicts that most primaries above the ankle should be nucleons up 
to $10^{20}$~eV (with a slight accumulation at the GZK energy) and photons 
at higher energies. In addition, the model predicts a new break in the 
observed spectrum above $10^{20}$~eV, reflecting the hard top-down spectra. 

\begin{figure}
\postscript{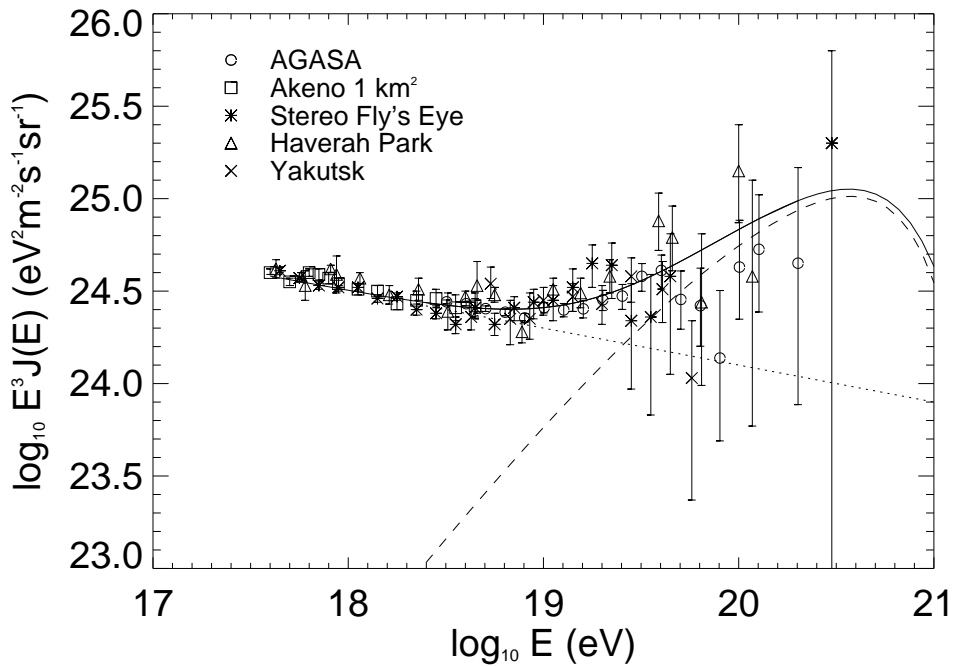}{0.80}
\caption{The best SUSY evolution fit to the CR data with a decaying particle 
mass of $5 \times 10^{12}$~GeV ({\it cf.} Fig.~\ref{top-down}). The dotted 
line indicates the extrapolation of the power law component from lower 
energies, the dashed line shows the 
decay spectrum, and the solid line is their sum. It is important to stress 
that this model would have an anomalous accumulation factor (see Sec.V). 
Published in~\cite{Sarkar:2001se}.}
\label{susyfit}
\end{figure}

Under certain circumstances the superheavy relics could also constitute
the dark matter (DM) of the Universe.\footnote{Nowadays there is  mounting 
evidence that about 90\% of the matter in our Universe may be dark. Current 
observations supporting the DM hypothesis include
gravitational lensing~\cite{Falco:1997ry,Cooray:1998jn}, peculiar velocities 
of large scale structures~\cite{Sigad,Branchini}, CMB 
anisotropy~\cite{Dodelson:1999am,Hu:2000ti}, and recession velocities of 
high redshift supernovae~\cite{Perlmutter:1998np}.} Even though the nature of 
the DM is still unknown, the DM hunt traditionally concentrates
on particles with mass of the order of the weak scale and with 
interaction with ordinary matter on the scale of the weak force. 
Importantly, big bang nucleosynthesis constraints imply that such 
weakly interacting massive particles (WIMPs) cannot be baryonic, and must 
therefore comprise non-SM particles. If the WIMP is a thermal 
relic, then it was once in local thermodynamic equilibrium in the early 
Universe, and its present abundance is determined by its self-annihilation 
cross section.
The largest annihilation cross section at early times is expected to be 
$\propto m_{X}^{-2}$. This implies that heavy WIMPs would have such a small 
annihilation cross section that their present abundance would be too large. 
Therefore, the mass of a thermal WIMP is found to be less than about 
500~TeV~\cite{Griest:1989wd}. Nevertheless, it was recently put 
forward that DM particles may have never experienced local thermal 
equilibrium, and so their masses may be as high as 
$10^{12}$--$10^{19}$~GeV~\cite{Chung:1998zb,Benakli:1998ut,Chung:1998ua,Chung:1998rq,Chung:2001cb,Hamaguchi:1998wm,Hamaguchi:1998nj}. Such monsters have 
been christened ``wimpzillas''~\cite{Kolb:1998ki}. One of 
the more promising candidates is the ``crypton'', the analog of the hadron in 
the hidden sector of supersymmetry breaking string theories~\cite{Benakli:1998ut}. Cryptons naturally emerge from the theory with about the right mass, 
they are cosmologically stable, and their interaction rate is sufficiently 
weak such that thermal equilibrium with the primordial plasma was 
never obtained. Furthermore, a 
sufficient abundance of these particles could have been produced near the end 
of the inflationary epoch.  

If superheavy relics play the role of DM, then irrespective of the abundance 
$\Omega_X$ the 
ratio of the mass density contribution from $X$ 
particles to that from DM particles should be roughly the same everywhere 
in the Universe, because both $X$ and DM particles respond to 
gravity in the same way. Therefore, since DM particles (by definition)  
cluster on galactic halos, so will  the $X$ particles. This suggests
that a clean signature of the superheavy relic $X$ hypothesis is the 
anisotropy imposed by the asymmetric position of the sun in the Galactic 
halo~\cite{Dubovsky:1998pu,Berezinsky:1998ft,Berezinsky:1998rp}. The 
exact value of this asymmetry is, however, model dependent, because
the distribution of dark matter inside the halo is by no means certain. 
The study in~\cite{Benson:ie}, following the {\it cusped} 
Navarro-Frenk-White (NFW) DM density distribution~\cite{Navarro:1995iw}, 
gives a substantial anisotropy signal in the arrival direction of ultrahigh energy CRs. 
Indeed, the analysis seems to indicate that less than 10\% of the 
ultrahigh energy CRs could come from relic particles in the Galactic halo. 
However, recent observations suggest that the NFW profile, for which the 
inner regions are dominated by DM, does not give an accurate description of 
the Galaxy~\cite{Evans:DM}. Specifically, the mass density implied by the 
luminous disk is already sufficient to account for the rotation curve in the inner Galaxy 
without any contribution from DM~\cite{Englmaier}.  
In addition, the microlensing optical depth 
to the red clump stars already shows that almost all the density in the 
inner Galaxy must be in the form of compact objects that yield 
microlensing, and therefore cannot be DM~\cite{Binney:62,Binney:99}. 

Recently, Evans, Ferrer and Sarkar (EFS)~\cite{Evans:2001rv} have analyzed in 
detail the dependence of the expected CR anisotropy with the DM density 
distribution in the Galaxy. 
They studied 4 typical models of the dark matter halo: {\it cusped}, 
{\it isothermal}, {\it triaxial}, and {\it tilted} (see Appendix~F). In these 
models, the amplitude of the anisotropy is controlled by the extent of the 
halo, whereas the phase is controlled by its shape. The results of EFS, 
given in Fig.~\ref{halo}, show that the amplitude, which is 
$\sim 0.5$ for a {\it cusped} halo, falls to 0.3 for an {\it isothermal} halo 
with realistic core radii. The phase points in the direction of the Galactic 
center, with deviation of up to $30^\circ$ when considering {\it triaxial} and {\it tilted} 
haloes. These amplitudes and phases are very similar to those obtained 
by Medina-Tanco and Watson in a  separate 
analysis~\cite{MedinaTanco:1999gw}. Unfortunately, the current CR data are too 
sparse above $10^{19}$~eV to have statistically significant discriminators 
between any dark halo model density profiles, 
or to either confirm or refute any correlation with the 
Galactic halo. Moreover, due to the limited size of the present ultrahigh 
energy CR sample, nothing can yet be said about the hypothesized existence 
of an ultrahigh 
energy CR contribution originating in the dark halo of Andromeda 
(M31)~\cite{Evans:2001rv,MedinaTanco:1999gw}. Certainly these issues will be 
resolved by PAO after a few years of operation.

\begin{figure}
\postscript{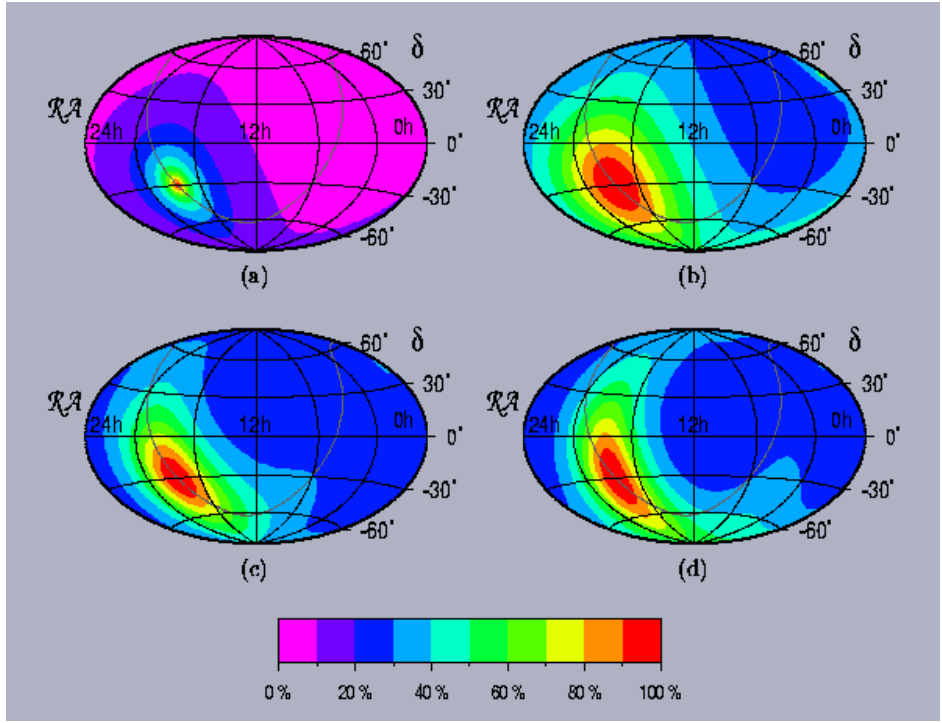}{0.80}
\caption{Contour plots in equatorial coordinates (${\cal RA},\ \delta$) 
of the predicted ultrahigh energy cosmic ray sky for 4 different dark halo 
models (from left to right, downwards: {\it cusped}, {\it isothermal}, {\it 
triaxial} and {\it tilted}). 
The effect of the halo of M31 is seen in the upper right of each plot: 
a hotspot at ${\cal RA} \approx 00^{\rm h}\ 43^{\rm m},\, 
\delta \approx 41^\circ$ (referred to the J2000.0 epoch).
Published in~\cite{Sarkar:2002ip}.}
\label{halo}
\end{figure}

\subsection{Monopoles and other topological defects as primaries}

It has long been known that any early Universe phase transition occurring 
after inflation  which leaves an 
unbroken  $U(1)$ symmetry group may produce magnetic 
monopoles~\cite{'tHooft:1974qc,Polyakov:ek}.
For instance, minimal $SU(5)$ breaking may lead to ``baryonic monopoles'' 
of mass $M \sim T_c/ \alpha$, with magnetic charge $U(1)_{\rm EM}$ and 
chromomagnetic   
$SU(3)_{\rm C}$~\cite{Goldhaber:1999sj}. Here 
$\alpha$ stands for the fine structure constant at 
symmetry breaking temperature $T_c$. These monopoles easily pick up energy from the magnetic fields 
permeating the Universe and can traverse unscathed through the primeval 
radiation. Thus, they are likely to generate extensive 
air showers~\cite{Kephart:1995bi,Weiler:1996mu,Swain:2000yd}.\footnote{The idea of 
monopoles as 
constituents of primary cosmic radiation is 
actually quite old, it  can be traced back at least as far as 
1960 \cite{Porter}.} 
Before proceeding further, it is important to point out 
that if the monopoles are formed at the usual 
GUT scale  $\sim 10^{15}$~GeV, the energy density 
overcloses the Universe.
Thus, to avoid this effect, the symmetry breaking scale associated with the 
production of monopoles has to be shifted to lower energies. Remarkably, 
if the GUT scale is at $\sim 10^9$~GeV, one would end up with 
an abundance of relativistic monopoles well below the closure limit, and yet 
potentially able to explain the tail of the CR-spectrum. In addition, 
for such a critical temperature the observed flux of ultrahigh energy CRs 
is below the flux allowed by the Parker 
limit.\footnote{This bound requires that there not be 
so many monopoles around as to effectively ``short out'' the Galactic 
magnetic field~\cite{Turner:ag}.} Unfortunately, 
contrary to the observed CR arrival directions, the expected 
flux of relativistic monopoles would be highly anisotropic, roughly aligned 
with the magnetic lines near the Earth~\cite{Escobar:1997mr}. 

In models with large extra dimensions, the low-scale unification 
enables the production of light-mass monopoles, say $M\sim100$~TeV. 
Furthermore, the physical embodiment of these theories would allow a natural 
generalization of 
the 't Hooft-Polyakov monopole providing a convenient set of 
representations for D1-branes ending on  D3-branes, and consequently 
even lighter monopoles~\cite{Roessl:2002rv}. 
The light-mass monopoles could lose and gain energy as 
they random-walk towards the Earth. The maximum energy attainable before 
hitting the atmosphere is roughly $10^{25}$~eV~\cite{Wick:2000yc}.
Therefore, these ``particles''  would be ultra-relativistic, and the expected 
flux would have no imprint of correlation with the local magnetic field. 
Note, however,  that direct searches at 
accelerators pretty much exclude masses below a few hundreds of GeV, whereas 
bounds stemming from quantum effects on current observables turn out to be 
$\sim 1$ TeV~\cite{DeRujula:1994nf}. The value for the lower limit 
of the mass of the monopole is still under debate~\cite{Gamberg:1999tv}.

To mimic a shower initiated by a proton the monopole must transfer nearly
all of its energy to the atmospheric cascade in a very small distance.
The large inertia of a massive monopole makes this impossible
if the cross-section is typically strong, say $\sim$ 100 mb.
Wick, Kephart, Weiler and Biermann (WKWB)~\cite{Wick:2000yc}
have recently pointed out that this problem can be avoided in models in which 
the baryonic
monopole consists of $q$-monopoles confined by strings of
chromomagnetic flux. To describe the interactions of such
a monopole in air, WKWB have developed a model based on the four following 
axioms: i) before hitting the atmosphere the monopole-nucleus cross 
section (unstretched state) is roughly hadronic 
($\sigma_0 \sim \Lambda_{\rm QCD}^{-2}$), attaining a geometric growth after 
the impact; ii) in 
each interaction almost all of the exchanged energy goes into 
stretching the  chromomagnetic strings of the monopole; iii) the 
chromomagnetic strings (of tension $T\sim \Lambda_{\rm QCD}^{-1}$) can only 
be broken to create monopole-antimonopole pairs, a process 
highly suppressed and consequently ignored; iv) the average fraction of energy 
transferred to the shower in each interaction is soft 
$\Delta E/E \equiv \eta  \approx \Lambda_{\rm QCD}/M$.  
Generally speaking, in this set-up the monopole will penetrate into the 
atmosphere, because the cross section is comparable to that of a high 
energy proton. 
However, since the geometrical cross-section grows 
proportionally with the Lorentz factor $\Gamma$, the interaction length after 
the impact shrinks to a small fraction of the depth of the first 
interaction. Stated 
mathematically, the unstretched monopole's string length, 
$L \sim \Lambda_{\rm QCD}^{-1}$, increases by $\delta L = \Delta E/T$. Recalling that 
nearly all  of the exchanged energy goes into stretching  the color magnetic strings, the fractional increase in the length is $\delta L/L = \Gamma$, yielding $\sigma_1 \sim (1+\Gamma)/\Lambda_{\rm QCD}^2$. Now, 
 the total mean free path after the $N$-th interaction reads,  
\begin{equation}
\lambda_N \sim \frac{1}{\sigma_N \,n_{\rm nuc}} \sim \frac{\Lambda^2_{\rm QCD}}
{(1 + \sum_{j=1}^N \Gamma_j) \,n_{\rm nuc}} \sim \frac{\Lambda^2_{\rm QCD}}
{N\,\Gamma \,n_{\rm nuc}},
\end{equation}
where we have 
assumed a constant density of nucleons 
($n_{\rm nucl} \approx (4/3)\, \pi\, A \,R_0^3$)
and we have used the approximation 
$\Gamma_N \sim (1 - \Lambda_{\rm QCD}/M)^N 
\Gamma \sim \Gamma$. Here $A$ stands for the mass number of an atmospheric 
nucleus, and $R_0 \approx 1.2 - 1.5$ fm. It should also be stressed 
that for $N = \eta^{-1}$ the approximation has an error  
bounded by $\lim_{N \rightarrow \infty} (1- N^{-1})^N = e^{-1}$. 
For $\eta^{-1} \gg 1$, the total 
distance traveled between the first interaction and 
the $\eta^{-1}$-th interaction is then
\begin{equation}
\Delta X \sim \frac{\Lambda_{\rm QCD}^2}{\Gamma \,n_{\rm nuc}} 
\sum_{N=1}^{\eta^{-1}} \frac{1}{N} \sim \frac{\Lambda_{\rm QCD}^2}{\Gamma \,n_{\rm nuc}} \ln \eta^{-1}.
\end{equation}
Note that the mean free path for all secondary interactions is 
${\cal O} (1/\Gamma)$ of the first one. Thus, a baryonic 
monopole encountering the atmosphere will diffuse like a proton, producing a 
composite heavy-particle-like cascade after the first interaction. 
A distinctive feature of the monopole shower would be the great number of 
muons 
among all the charged particles~\cite{Anchordoqui:2000mk}. 
Although this feature was observed in a poorly understood 
super-GZK event~\cite{Efimov},
it seems unlikely that a complete explanation for the ultrahigh energy CR 
data sample would be in terms of magnetic monopoles alone. Moreover, 
any confirmed 
directional pairing of events would appear difficult to achieve with the 
monopole hypothesis.

An alternative TD that can be easily accelerated to ultrahigh energies is 
the vorton~\cite{Bonazzola:1997tk}. However, the interaction properties of 
vortons with ordinary matter are completely uncertain, and so no clear 
predictions can be made.

\section{The atmosphere as a black hole factory}

In this chapter we shift the focus of our discussion
from what CRs tell us about astrophysical phenomena
to what they tell us about fundamental interactions.
In particular, ultrahigh energy CR neutrinos provide
a means to probe the fundamental Planck scale in a regime
not currently accessible to man-made accelerators.  If
Planck scale is of $\cal O( \mathrm{TeV} )$, then
cosmic neutrinos could produce microscopic BHs when they 
interact in the atmosphere.
The BHs would decay promptly, initiating 
deeply developing showers far above the SM rate~\cite{Feng:2001ib}, and 
with very distinctive characteristics when the BH entropy 
$\gg 1$~\cite{Anchordoqui:2001ei}.  
In the subsequent sections, we discuss first the 
phenomenology of BH production, and then
experimental signatures for BHs created
in CR events~\cite{Anchordoqui:2001cg}. For dessert,
we examine the parameter space left for $p$-brane 
production~\cite{Ahn:2002mj,Anchordoqui:2002it}.

\subsection{Black hole production in particle collisions}

Black holes (BHs) are among the most fascinating and inaccessible 
phenomena in nature. It has been known for quite a long time that 
microscopic BHs can be produced in particle collisions with c.m. 
energies above the fundamental scale of 
gravity~\cite{Amati:1987wq,Amati:1992zb,'tHooft:rb,'tHooft:yr}, where 
they should be well described semi-classically and 
thermodynamically~\cite{Hawking:1975sw}. In the ordinary 
4-dimensional scenario, where the fundamental scale of gravity is 
$\sim~10^{19}$~GeV, the study of such BHs is hopelessly beyond the realm of 
experimental particle physics. However, if TeV scale gravity is realized 
in nature, production of BHs should be observed in particle collisions 
with $\sqrt{s} \gg 1$~TeV and sufficiently small impact 
parameter~\cite{Banks:1999gd,Emparan:2000rs,Giddings:2000ay,Giddings:2001bu,Dimopoulos:2001hw}.

The ensuing discussion will be framed in the context of large extra 
dimensions~\cite{Arkani-Hamed:1998rs}, and will be valid at distances that 
are small 
compared to the compactification radius $r_c$. The requirements for validity 
of the picture in the warped scenario~\cite{Randall:1999ee,Giddings:2001yu} 
are discussed in Appendix~G. For  purposes of 
normalization we use the notation of Ref.~\cite{Giudice:1998ck}, where
the fundamental mass scale reads
\begin{equation}
\md = [(2\pi)^n/8\pi]^{1/(n+2)}\, \mstar \ , \quad D=4+n\ . 
\label{mdmstar}
\end{equation}
The radius of a Schwarzschild BH of mass $\mbh=\sqrt{\hat{s}}$ in $(4+n)$
dimensions is~\cite{Myers:un}
\begin{equation}
\label{schwarz}
r_s(\mbh) =
\frac{1}{\md}
\left[ \frac{\mbh}{\md} \right]^{\frac{1}{1+n}}
\left[ \frac{2^n \pi^{(n-3)/2}\Gamma({n+3\over 2})}{n+2}
\right]^{\frac{1}{1+n}}\,.
\end{equation}
If one envisions a head-on collision involving partons $i$ and $j$ with c.m. energy $\sqrt{\hat{s}} = \mbh$ and impact parameter less than $r_s$, semiclassical reasoning suggests that a BH would be 
formed~\cite{Thorne:ji}.\footnote{Here, $\hat{s} \equiv x s$, where $x$ is the parton momentum 
fraction and $s$ is 
the square of the c.m. energy.} The total cross section of the process can be 
estimated 
from geometrical arguments~\cite{Giddings:2001bu,Dimopoulos:2001hw} and is 
of order 
\begin{equation}
\label{sigma}
\hat{\sigma} \sim \pi r_s^2 \, .
\end{equation}
Criticisms~\cite{Voloshin:2001vs,Voloshin:2001fe} of this cross section, 
which center on the exponential suppression of transitions involving 
a (few-particle) quantum state to 
a (many-particle) semiclassical state, have been addressed 
in~\cite{Dimopoulos:2001qe,Giddings:2001ih,Eardley:2002re,Solodukhin:2002ui,Hsu:2002bd}. However, it is worthwhile to point out that 
the heuristic arguments supporting Eq.~(\ref{sigma}) only determine 
$\hat{\sigma}$ 
up to an overall factor of order one~\cite{Anchordoqui:2001cg}. Uncertainties 
in this factor are due, for instance, to the inclusion of angular momentum 
when considering BH creation for collisions with non-zero impact parameter.

Note that even though the colliding particles are confined to the 
brane, if $r_s \ll r_c$ then the BH should be treated as a fully $(4+n)$ 
dimensional object in an asymptotically Minkowskian spacetime. In addition, 
for $\sqrt{s} \gg \md$, BH production should dominate over all other SM 
processes because of the rapidly increasing cross section 
$\hat\sigma \propto \hat{s}^{1/(n+1)}$. To calculate the total production 
cross section, we have to take into account 
that only a fraction of the total c.m. energy in a collision is 
available to the parton participating. In our discussion of BH production 
in the 
atmosphere, we are most interested in collisions of ultrahigh energy neutrinos 
with the nucleons in the air molecules, where~\cite{Feng:2001ib}
\begin{equation}
\label{partonsigma}
\sigma_{\nu N \to \text{BH}} (E_\nu) = \sum_i \int_{(\mbhmin{})^2/s}^1 dx\,
\hat{\sigma}_i ( \sqrt{xs} ) \, f_i (x, Q) \ .
\end{equation}
Here, $s = 2 m_N E_{\nu}$, $f_i$ are parton distribution functions, 
$\mbhmin$ is the minimum BH mass for which the parton cross section into 
BHs is applicable, and the 
sum is carried out over all partons in the nucleon. The momentum
transfer $Q$ is set to $\min \{ \mbh, 10~\tev \}$, where the upper limit 
comes from the CTEQ5M1 distribution functions~\cite{Lai:2000wy}.  The cross
section $\sigma_{\nu N \to \text{BH}}$ is highly insensitive to the
details of this choice~\cite{Feng:2001ib}. 
Once produced, the BH will Hawking evaporate with a temperature proportional 
to the inverse radius 
\begin{equation}
T_H={n+1\over 4 \pi r_s}\, .
\end{equation}
Note that the wavelength $\lambda = 2\pi/T_H$
corresponding to this temperature is larger than the BH size. Hence,
to first approximation the BH behaves like a point-radiator with 
entropy~\cite{Anchordoqui:2001cg}
\begin{equation}
\label{entropy}
S = {4\pi \, \mbh\ r_s\over n+2} \ .
\end{equation}
The BH lifetime estimated from~\cite{Argyres:1998qn}
\begin{equation}
\frac{dE}{dt} \sim A \,\, T_H^{4+n} 
\end{equation}
is found to be
\begin{equation}
\tau_{_{\rm BH}} \sim {1\over \md} \left({\mbh\over \md}\right)^{3+n \over 1+n}\ ,\end{equation}
where $A$ is the horizon area.
For $M_{\rm BH} \gg M_D$, the BH is a well defined 
resonance and may be thought of as an 
intermediate 
state in the $s$ channel. This means that the decay of a BH
is only sensitive to the radial coordinate and does not make use of the 
angular modes available in the bulk. Therefore, BHs decay 
with equal probability to a particle on the brane and in the bulk~\cite{Emparan:2000rs,Giddings:2001bu,Dimopoulos:2001hw}.\footnote{One caveat is that BHs may jump off the brane, as suggested in a recent article~\cite{Frolov:2002as}.} 
Since there are many more particles on the brane, we expect the BH to decay 
visibly into SM particles 
giving rise to events with large multiplicity, $\langle N  
\rangle \approx M_{\rm BH}/(2T_H)$, large total transverse energy, 
and a characteristic ratio of hadronic to 
leptonic activity of roughly 5:1. As $M_{\rm BH}$ approaches $\md$,  
semiclassical arguments are no longer valid, because the BHs 
become ``stringy'' and their properties rather complex.

\begin{figure}
\postscript{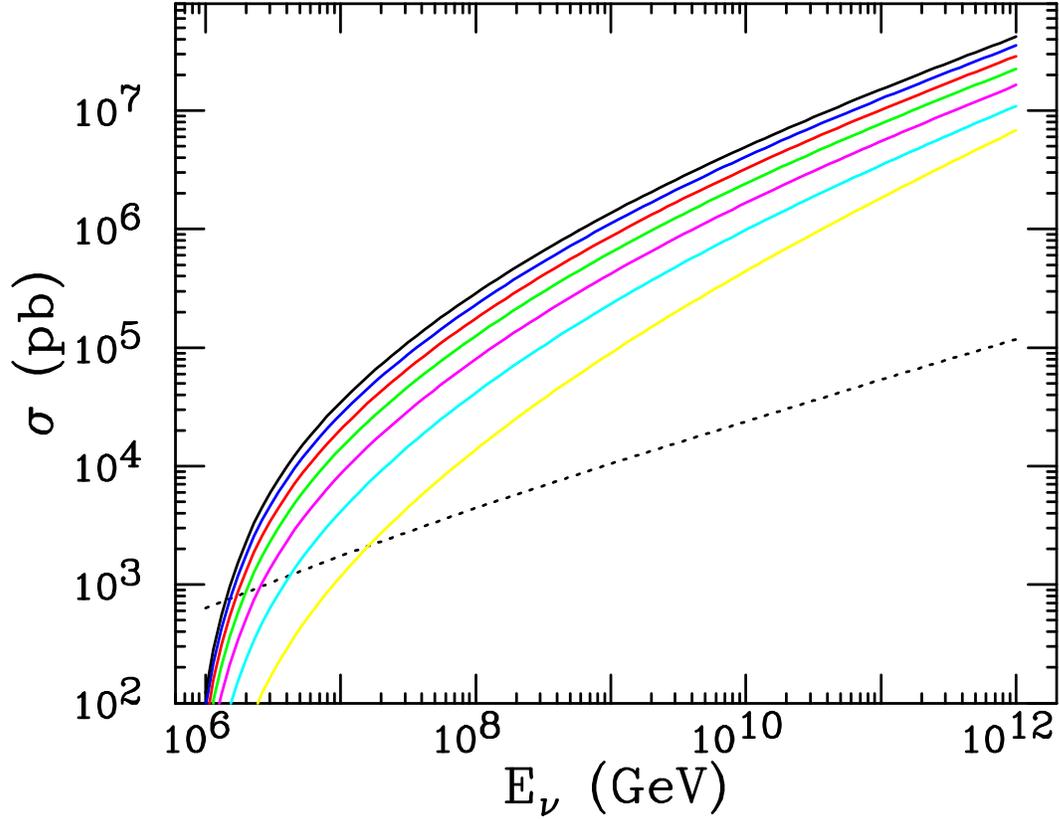}{0.90}
\caption{Cross sections $\sigma_{\nu N \to {\rm BH}}$ for $n=1,\ldots,
7$ (bottom to top) for $\md = 1~\tev$, $x_{\rm min} =1$. 
The SM cross section~\cite{Gandhi:1998ri}
is indicated with a dotted line. Published in Ref.~\cite{Anchordoqui:2001cg}.}
\label{sigmabh}
\end{figure}

The magnitude of the entropy determines the validity of the semiclassical 
approximation. Thermal fluctuations due to particle emission are small
when $S \gg 1$~\cite{Preskill:1991tb}, and statistical
fluctuations in the microcanonical ensemble are small for 
$\sqrt{S} \gg 1$~\cite{Giddings:2001bu}. Searches for BH
production at colliders are viable when  
$x_{\rm min} \equiv M_{\rm BH}^{\rm
min}/M_D$ is high enough that the decay branching ratios
predicted by the semiclassical picture of BH evaporation are
reliable. The QCD background is large, and therefore the
extraction of signal from background at hadron colliders depends
on knowing the BH decay branching 
ratios reliably~\cite{Giddings:2001bu,Dimopoulos:2001hw,Landsberg:2001sj,Giudice:2001ce,Rizzo:2002kb,Giddings:2002av,Giddings:2002kt,Uehara:2002gv,Rattazzi:2002rz,Park:2001xc,Chamblin:2002ad,Han:2002yy,Anchordoqui:2002cp}. Therefore, in searches at colliders a cutoff of $x_{\rm min}  = 5.5$ or more may
be appropriate. In contrast, the search for deeply penetrating
quasi-horizontal showers initiated by BH decay products can afford to be
much less concerned with the details of the final state, since
the background from hadronic showers is filtered out by the atmosphere.  
As a result, the signal
relies only on the existence of visible decay products, which, in
this context, includes all particles other than neutrinos, muons,
and gravitons. Indeed, there is very little about the final
state, other than its total energy and to some degree its
multiplicity and electromagnetic component~\cite{Anchordoqui:2001ei}, 
that we can reasonably expect to observe, since detailed reconstruction 
of the primary BH decay process is not possible at cosmic ray detectors.  
With this in mind, one can choose a significantly lower value of $\mbhmin$
than the one needed for collider searches.  Although BHs of mass around 
$\md$ will
be outside the semiclassical regime, it seems quite reasonable to
expect that they, or their stringy progenitors, will nevertheless
decay visibly, whatever stringy or quantum gravitational description
applies.

Cross sections for BH production by cosmic neutrinos are given in
Fig.~\ref{sigmabh}.\footnote{It should be stressed that the 
neutrino-gluon scattering may dominate neutrino-nucleon collisions 
below $10^{8}$~GeV~\cite{Friess:2002cc}.} They scale as
\begin{equation}
\sigma_{\nu N \to {\rm BH}} \propto \left[ \frac{1}{\md^2} \right]
^{\frac{2+n}{1+n}} \ .
\label{scaling}
\end{equation}
Despite the fact that the BH production cross section reduces the neutrino  
interaction length to
\begin{equation}
\lambda_\nu= 1.7 \times 10^7~{\rm km} \,{\rm w.e.} \left( \frac{{\rm pb}}{\sigma} \right)\,,
\end{equation}
the Earth's atmospheric depth, which is only
$0.36$~km w.e. even when traversed horizontally, is still smaller than $\lambda_\nu$. Neutrinos therefore produce BHs uniformly at all atmospheric depths.

\subsection{Probes of TeV scale gravity in CR experiments}

Given the provocative new features of models with large--compact--dimensions 
(LCDs), there is a strong motivation for phenomenological studies to assess 
their experimental viability. The most obvious consequence of the 
existence of LCDs is the violation of Newton's Gravitational Law at distances of order 
$r_c$. Applying Eq.~(\ref{rcomp}) for $n=1$ one obtains 
$r_c \sim 10^{13}$~cm, which immediately suggests this case is excluded 
because Newtonian gravity would be modified at the scale of our solar system.
For $n \geq 2$, $r_c$ is sufficiently small so that these scenarii are not 
yet excluded. For $n= 2$, sub-millimeter tests of the gravitational 
inverse-square law constrain $M_D > 1.6$~TeV~\cite{Hoyle:2000cv}. For 
$n \geq 3$ LCDs become 
microscopic and therefore elude the search for deviations in gravitational
measurements. In the presence of LCDs, however, the effects of gravity are 
enhanced at high energies due to the large multiplicity of KK-modes, yielding 
astrophysical bounds on $r_c$. For instance, the requirement that the neutrino 
signal of SN 1987A not be unduly shortened by the emission of KK modes into 
the bulk leads to $\md \agt 50$~TeV, for $n=2$, and $\md \agt 5$~TeV, for 
$n=3$~\cite{Cullen:1999hc,Barger:1999jf}. These limits could be even more 
restrictive when taking into account KK graviton decay in typical 
astrophysical environments, yielding $M_D \agt 600 - 1800$~TeV for $n=2$ 
and $\md \agt 10 - 100$~TeV for 
$n=3$~\cite{Hanhart:2001fx,Hall:1999mk,Hannestad:2001jv,Hannestad:2001xi}.

For $n \geq 4$, only high energy colliders and CR-experiments provide 
reasonably sensitive probes of LCDs. The effects of direct graviton emission, 
including production of single photons or $Z$'s, were sought at
LEP~\cite{Acciarri:1999jy,Acciarri:1999bz}. The resulting bounds are fairly
model-independent, as the relatively low energies at LEP imply a
negligible dependence on any soft-brane damping factor. 
For $n=4\ (6)$, these null results imply $M_D \agt 870\
(610)~\gev$~\cite{Pagliarone:2001ff}. The effects of low-scale gravity can 
also be seen through virtual graviton effects. These are most stringently 
bounded by the D\O\ Collaboration, which recently 
reported~\cite{Abbott:2000zb} the first results for virtual graviton effects 
at a hadron collider.  The data
collected at $\sqrt{s} = 1.8~\tev$ for dielectron and diphoton
production at the Tevatron agree well with the SM predictions and
provide  restrictive limits for $n \geq 4$. Specifically, 
if one adopts a Gaussian cutoff on the transverse momentum of the graviton 
due to brane excitations, $\md \agt 1~\tev$~\cite{Anchordoqui:2001cg}.

\begin{figure}[tbp]
\postscript{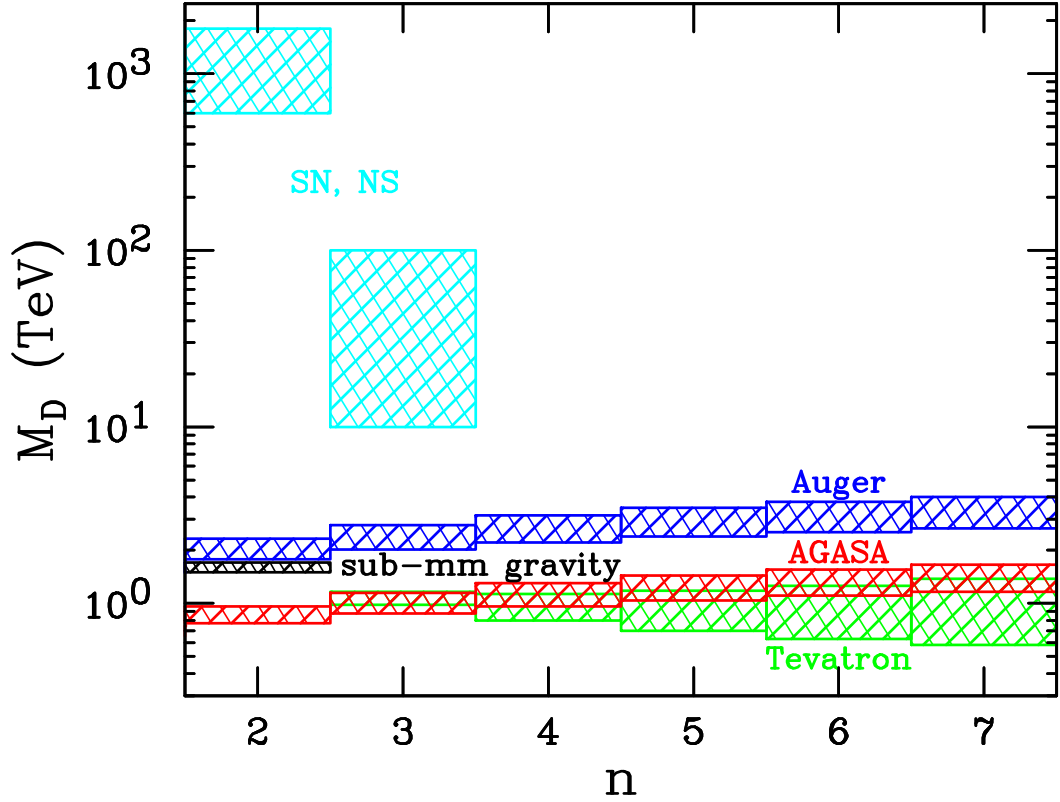}{0.90}
\caption{Bounds on the fundamental Planck scale $\md$ from tests of
Newton's law on sub-millimeter scales, bounds on supernova cooling and
neutron star heating, dielectron and diphoton production at the
Tevatron, and non-observation of BH production at AGASA.  Future
limits from PAO ground array, assuming 5 years of data and no
excess above the SM neutrino background, are also shown. 
The range in Tevatron bounds corresponds to the range of a
brane-softening parameter which sets the
Gaussian cutoff (see text). The range in cosmic ray bounds is for
$\xmin=1-3$. Published in 
Ref.~\cite{Anchordoqui:2001cg}.}
\label{summary}
\end{figure}

Limits on the fundamental Planck scale derived from CR-observations are 
mainly due to searches for BH production in the 
atmosphere by ultrahigh energy 
neutrinos~\cite{Anchordoqui:2001cg,Ringwald:2001vk,Sigl:2002bb}. The number 
of BHs to be detected by a given CR-experiment is 
\begin{equation}
N = \int dE_{\nu}\, N_A \, \frac{d\Phi}{dE_{\nu}} \,
\sigma_{\nu N \to {\rm BH}} (E_{\nu}) \, A(E_{\nu}) \, T \ ,
\label{numevents}
\end{equation}
where $A(E_{\nu})$ is the experiment's acceptance in cm$^3$ w.e. sr, $N_A
= 6.022 \times 10^{23}$ is Avogadro's number, $d\Phi/dE_{\nu}$ is the
source flux of neutrinos, and $T$ is the running time of the detector.
As discussed earlier, for neutrinos with large incident zenith angles 
the likelihood of interaction is maximized.
Observed EASs with  $\theta > 70^\circ$ typically must traverse 
$> 2000$ g/cm$^2$~of atmosphere before interacting. Since the interaction 
length of the atmosphere for protons is $\sim 45$~g/cm$^2$ one expects the 
background 
from hadronic cosmic rays to be eliminated. In 1710.5 days of data collected 
from December 1995 to November 
2000, the AGASA Collaboration found 6 candidate events with 
$\xmax \ge 2500~\g/\cm^2$~\cite{Yoshida:h}.  At AGASA, the location of the 
shower maximum is determined through correlation to two measurable quantities:
$\eta$, which parameterizes the lateral distribution of charged
particles at ground level, and $\delta$, which parameterizes the
curvature of the shower front. The candidate neutrino events must
satisfy $\xmax^{\eta}, \xmax^{\delta} \ge 2500~\g/\cm^2$. The expected 
background from
hadronic showers is $1.72{}^{+0.14}_{-0.07}{}^{+0.65}_{-0.41}$, where
the first uncertainty is from Monte Carlo statistics, and the second
is systematic.  Of the 6 candidate events, however, 5 have values of
$\xmax$ that barely exceed $2500~\g/\cm^2$, and are well within 
$\Delta \xmax$ of this value, where $\Delta \xmax$ is the estimated 
precision with which $\xmax$ can be reconstructed. The AGASA Collaboration 
thus concludes that there is no significant enhancement of deeply 
penetrating shower rates given the detector's resolution.
The AGASA results imply lower bounds on the scale of low-scale
gravity. Of course, the bounds are subject to both
small uncertainties inherent in the parton level cross
section and uncertainty in the cosmogenic neutrino flux.   
For the conservative cosmogenic neutrino fluxes given in Fig.~\ref{nuflux}
the expected rate for deeply penetrating showers at AGASA from SM neutrino 
interactions is about
0.02 events per year, which is negligible.  Given 1 event that
unambiguously passes all cuts, and the central value of 1.72
background events, the AGASA results imply an upper bound of 3.5 black
hole events at 95\% CL~\cite{Feldman:1997qc}. Therefore, the absence of deeply
penetrating showers in the AGASA data implies limits on the size 
of LCDs which are summarized in Fig.~\ref{summary}~\cite{Anchordoqui:2001cg}. 
These bounds are 
conservative. Note that larger cosmogenic neutrino
fluxes, as predicted by some models~\cite{Kalashev:2002kx}, will strengthen 
them, possibly dramatically. Note also that cosmic 
neutrino interactions from sub-Planckian extra-dimensional physics 
can only serve to strengthen these bounds.
The absence of a deeply-penetrating signal in the Fly's Eye
data also implies lower bounds on $M_D$. However, these
are consistently weaker. The Fly's Eye group searched for deeply developing 
showers in the sample recorded with the Fly's Eye I during an observation time 
of $6 \times 10^6$~s~\cite{Baltrusaitis:mt}. Although they found some events 
with large incident 
zenith angle $(\theta >75^\circ)$~\cite{Baltrusaitis:apj}, the 
distribution of $X_0$, the depth of 
the first observed interaction, is consistent with $X_0 < 2000$~g/cm$^2$, 
and there are no events with $\xmax \ge 2500~\g/\cm^2$~\cite{Baltrusaitis:mt}.
This result leads to  limits on the size of LCDs. For example, if $n=6$,  
$M_D > 900~\gev$~\cite{Ringwald:2001vk}. The bounds~\cite{Anchordoqui:2002vb} 
derived with the combined 
exposure of the AGASA and Fly's Eye experiments 
(the latter integrated over all its operating 
epochs) extend up to 2~TeV for $n=7$ and $x_{\rm min} = 1$. 
Moreover, assuming the conservative value $\xmin =3$,
for which the entropy $S> 10$, the bounds derived with the
combined exposure, for $n = 5, 6 ,7$, are $\md > 1.26~{\rm TeV},
1.30~{\rm TeV}, 1.40$~TeV, respectively. All of these exceed the
Tevatron bounds~\cite{Abbott:2000zb}, and  represent 
the best existing limits on the scale of TeV-gravity for $n\ge 5$ extra 
spatial dimensions.

We now discuss the sensitivity of future CR-experiments.
PAO is expected to become fully operational in 2003, and thus would 
have a running time of roughly 5 years before the LHC begins operation. 
If no enhancement of quasi-horizontal showers is seen during the pre-LHC 
epoch, PAO will probe fundamental Planck scales as large as 
$\md = 4~\tev$~\cite{Anchordoqui:2001cg,Feng:2001ib,GAP}. Moreover, 
given the prospects for fairly high statistics in the region of high 
entropy, detailed studies of BH shower profiles~\cite{Anchordoqui:2001ei}
are in principle possible. In addition, the projected sensitivity 
of the OWL satellite project will substantially extend the region 
of $\md$ probed in CR-observations before the first collisions at 
LHC~\cite{Dutta:2002ca}. It should also be stressed that neutrinos that 
traverse the atmosphere unscathed may produce BHs through interactions in 
the ice or water. Detailed 
simulations~\cite{Kowalski:2002gb,Alvarez-Muniz:2002ga,Uehara:2001yk,Jain:2002kz} also find observable BH rates at neutrino telescopes.

In summary, in searches for evidence of TeV scale quantum gravity, 
CR-experiments are competitive with colliders, and in fact, until the 
LHC-era, they will provide the most stringent limits.

\subsection{Prospects for distinguishing BH production in neutrino showers}

Up to now we have only discussed how to set bounds on physics beyond the
SM.  An actual discovery of new physics in cosmic rays is a tall order
because of large uncertainties associated with the depth of the first 
interaction in the atmosphere, and experimental challenges of reconstructing 
cosmic air showers from partial information.  
However, in the following paragraphs we will discuss an observable 
which skirts these uncertainties, and may therefore provide a technique for 
actually discovering BHs in cosmic air showers if TeV-scale quantum gravity 
exists. If an anomalously large quasi-horizontal shower rate is found, it 
may be ascribed to either an enhancement on the incoming neutrino flux, or an 
enhancement in the neutrino-nucleon cross section. However, these two possibilities may be distinguished by separately binning events which arrive at very
small angles to the horizontal, the so-called ``Earth-skimming'' 
events~\cite{Domokos:1997ve,Domokos:1998hz,Fargion:2000iz,Fargion:2000rn,Bertou:2001vm,Feng:2001ue,Kusenko:2001gj}.  An enhanced flux will increase 
both quasi-horizontal and Earth-skimming event rates, whereas a large BH 
cross section suppresses the latter, because the hadronic decay products of 
BH evaporation do not escape the Earth's crust~\cite{Feng:2001ib}.

Consider a flux of $\tau$-neutrinos in the energy decade $10^{18}$ to  
$10^{19}~{\rm eV} $. During propagation inside the Earth, 
a $\tau$-neutrino can produce a tau lepton via a charged current 
interaction.  The interaction length for a neutrino near the Earth's 
surface, which is roughly homogeneous with density $\rho_s = 2.15$ g/cm$^3$, 
is given by
\begin{equation}
\lambda_\nu^{\rm CC} = \left[ N_A \rho_s \sigma_{\nu N}^{\rm CC} 
\right] ^{-1} \ ,
\end{equation}
where 
\begin{equation}
\sigma_{\nu N}^{\rm CC} = 1.0 \times 10^{-32}\,\, 
\left( \frac{E_\nu}{10^{18}~{\rm eV}} \right)^{0.363} \,\, {\rm cm}^2 
\end{equation}
is the charged current cross section accurate to within 10\% as given by 
CTEQ4-DIS parton distribution~\cite{Gandhi:1998ri}. 
Here we neglect neutral current interactions, which at these energies serve 
only to reduce the neutrino energy by approximately 20\%.  
Therefore, for $E_\nu \sim 10^{19}~\ev$, we have $\lambda_\nu^{\rm CC} \sim 250~\km$.  
However, if one takes into account the possibility of BH production, 
the mean free path is reduced to
\begin{equation}
\lambda_\nu^{\rm tot} = \left[ N_A \rho_s (\sigma_{\nu N}^{\rm CC} + 
\sigma_{\nu N \rightarrow {\rm BH}}) \right] ^{-1} \ .
\end{equation}

The energy of the outgoing $\tau$ is degraded in Earth by
Bremsstrahlung, pair production and deep inelastic scattering, and 
can be parametrized by
\begin{equation}
\frac{dE_{\tau}}{dz} = -(\alpha_{\tau} + \beta_{\tau} E_{\tau}) 
\rho_s \ ,
\end{equation}
where $\alpha_{\tau}$ is negligible at energies of interest, and 
$\beta_{\tau} \approx 0.8\times 10^{-6}~\cm^2/\g$~\cite{Dutta:2000hh}.
The maximal path length for a detectable $\tau$ is, then,
\begin{equation}
L^{\tau} = \frac{1}{\beta_{\tau} \rho_s} \ln \left( E_{\rm max} / 
E_{\rm min} \right) \ ,
\label{ltau}
\end{equation}
where $E_{\rm max} \approx E_\nu$ is the energy at which the tau is
created, and $E_{\rm min}$ is the minimal energy at which a $\tau$ can
be detected.  For cosmogenic neutrino fluxes and other reasonable
sources, and the acceptances of typical cosmic ray detectors, taus
cannot lose much energy and still be detected. 
For $E_{\rm max} / E_{\rm min} = 10$, $L^{\tau} = 11~\km$.

For a given isotropic $\nu_{\tau} + \bar{\nu}_{\tau}$ flux, the number of
taus that emerge from the Earth with enough energy to be detected
is proportional to the ``effective solid angle''~\cite{Feng:2001ue}
\begin{equation}
\Omega_{\rm eff} \equiv \int d\cos\theta\, d\phi\, \cos\theta\,
P(\theta,\phi) \ ,
\end{equation}
where 
\begin{equation}
P(\theta,\phi) = \int_0^{\ell} \frac{dz}{\lambda_\nu^{\rm CC}}
e^{-z/\lambda_\nu^{\rm tot}} \ 
\Theta \left[ z - (\ell - L^{\tau} ) \right]
\label{P}
\end{equation}
is the probability for a neutrino with incident nadir angle $\theta$
and azimuthal angle $\phi$ to emerge as a detectable $\tau$~\footnote{
In \eqref{P} we have assumed that  BHs produced in the Earth 
will range out before they can produce a detectable signal.}.   
Here $\ell = 2 R_{\oplus} \cos\theta$ is the chord length
of the intersection of the neutrino's trajectory with the Earth, where
$R_{\oplus} \approx 6371~\km$ is the Earth's radius.  Evaluating the
integrals, one obtains ~\cite{Kusenko:2001gj}
\begin{equation}
\Omega_{\rm eff} = 2 \pi 
\frac{\lambda_\nu^{\rm tot}}{\lambda_\nu^{\rm CC}}
\left[ e^{L^{\tau} / \lambda_\nu^{\rm tot}} - 1 \right]
\left[ \left( \frac{\lambda_\nu^{\rm tot}}{2 R_{\oplus}} \right)^2
- \left( \frac{\lambda_\nu^{\rm tot}}{2 R_{\oplus}} +
\left( \frac{\lambda_\nu^{\rm tot}}{2 R_{\oplus}} \right)^2 \right)
e^{-2R_{\oplus} / \lambda_\nu^{\rm tot}} \right] \ .
\label{Omegaeff}
\end{equation}
For $E_\nu \in (10^{18}~{\rm eV}, 10^{19}~{\rm eV})$, 
$\lambda_\nu^{\rm tot} \ll R_{\oplus}$. Additionally, if the BH cross section 
is not very
large, $\lambda^{\rm
tot}_{\nu} \gg L^{\tau}$, and thus \eqref{Omegaeff} simplifies to
\begin{equation}
\Omega_{\rm eff} \approx 
2\pi \frac{\lambda_\nu^{{\rm tot} \, 2} L^{\tau}}{4 R_{\oplus}^2 
\lambda_\nu^{\rm CC}} \ .
\label{omegaeffapprox}
\end{equation}
Now, since the $\tau$ production is proportional to the incoming neutrino 
flux $\Phi^{\nu}$, the number of Earth-skimming events detected in
5 years reads
\begin{equation}
N_{\rm ES} \approx 
C_{\rm ES}\, \frac{\Phi^{\nu}}{\Phi^{\nu}_0}
\frac{\sigma_{\nu N}^{{\rm CC} ^2}}
{\left( \sigma_{\nu N}^{\rm CC} + \sigma_{\nu N \to {\rm BH}} \right)^2} \ ,
\label{ES}
\end{equation}
where $C_{\rm ES}$ is the number of Earth-skimming events expected for
the standard cosmogenic flux $\Phi^{\nu}_0$ in the absence of BH
production. Assuming maximal neutrino mixing and the $\beta_{\tau}$
value given above,  one obtains $C_{\rm ES}
\approx 3.0$  for the PAO ground array~\cite{Bertou:2001vm}.  The HiRes fluorescence detectors
provide additional sensitivity~\cite{Feng:2001ue}.  In the following 
discussion we make a 
conservative assumption of $C_{\rm ES} = 3.0$ for the SM combined rate 
over 5 years. The presence of a
significant cross section for BH production, as well as their
prompt decay and the rapid absorption of the decay products in
the Earth, gives rise to a substantial depletion of the ES event
rate.

In contrast to \eqref{ES}, the rate for quasi-horizontal showers
follows simply from \eqref{numevents}, and has the form
\begin{equation}
N_{\rm QH} = C_{\rm QH} \frac{\Phi^{\nu}}{\Phi^{\nu}_0}
\frac{\sigma_{\nu N}^{\rm CC} + \sigma_{\nu N \to {\rm BH}}}
{\sigma_{\nu N}^{\rm CC} } \ ,
\end{equation}
where $C_{\rm QH} = 2.5$ for the PAO ground array~\cite{Capelle:1998zz}.

Given a flux $\Phi^{\nu}$ and BH cross section $\sigma_{\nu N \to {\rm BH}}$, 
both $N_{\rm ES}$ and $N_{\rm QH}$ are determined.  Contours
for these two rates are shown in Fig.~\ref{skimbh}. 
As can be seen from the figure, it is impossible to differentiate between
an enhancement from large BH cross
section and large flux, given a quasi-horizontal event rate $N_{\rm QH}$.
However, in the region where significant
event rates are expected, the $N_{\rm QH}$ and $N_{\rm ES}$ contours
are more or less orthogonal, and thus provide complementary information.
With measurements of $N_{\rm QH}$ and $N_{\rm ES}$, both
$\sigma_{\nu N \to {\rm BH}}$ and $\Phi^{\nu}$ may be determined
independently, and neutrino interactions beyond the SM may be
unambiguously identified~\cite{Anchordoqui:2001cg}.

\begin{figure}
\postscript{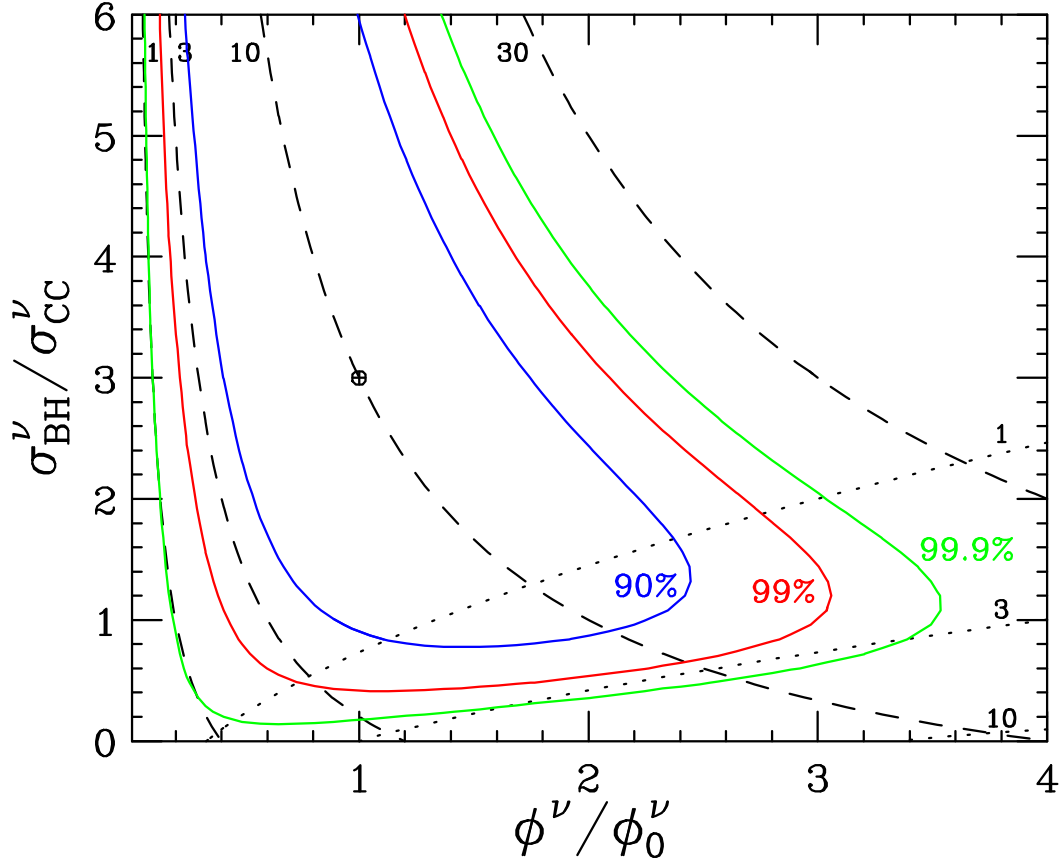}{0.90}
\caption{Contours of constant number of quasi-horizontal showers
$N_{\rm QH}$ (dashed) and Earth-skimming neutrino events $N_{\rm ES}$
(dotted) as functions of source flux $\Phi^{\nu}$ and BH production
cross section $\sigma_{\nu N \to {\rm BH}}$.  5 year running times for PAO
and HiRes are assumed. The figure also shows confidence level contours, 
assuming $\Phi^{\nu} = \Phi^{\nu}_0$ and $\sigma_{\nu N \to {\rm BH}} = 3
\sigma_{\nu N}^{\rm CC} $, corresponding to $(N_{\rm QH}, N_{\rm ES})
\approx (10, 0.2)$. Published in Ref.~\cite{Anchordoqui:2001cg}.}
\label{skimbh}
\end{figure}

As an example, consider the case in which $\sigma_{\nu N \to {\rm
BH}}/\sigma_{\nu N}^{\rm CC} = 3$, and $\Phi^{\nu}/\Phi^{\nu}_0 = 1$, shown as a dot in Fig.~\ref{skimbh}. On
average, one would then observe a total of $N_{\rm QH} = 10$ deep
quasi-horizontal showers, an excess of 8 above SM expectations.  On
average, one also expects $N_{\rm ES} \approx 0.2$ Earth-skimming
events.  A SM explanation (i.e., $\sigma_{\nu N \to {\rm BH}} = 0$) of the
deeply penetrating event rate would require $\Phi^{\nu}/\Phi^{\nu}_0 =
4$ and predict 12 Earth-skimming events, a possibility that would be
clearly excluded at high confidence level.
More generally, one might try to salvage a SM explanation by
attributing the observed rates to statistical fluctuations in both
$N_{\rm QH}$ and $N_{\rm ES}$.  Figure~\ref{skimbh} shows contours 
of constant $\chi^2$ using a maximum likelihood method for
Poisson-distributed data~\cite{James:sr}. It is easily seen that 
the possibility of a SM interpretation along the 
$\sigma_{\nu N \to {\rm BH}} = 0$ axis
would be excluded at greater than 99.9\% CL for any assumed flux.

BH production will most likely be accompanied by more
model-independent sub-Planckian effects.  In particular, 
the virtual exchange of bulk gravitons (KK modes) leads 
to extra contribution to the neutrino nucleon cross section. 
As a consequence, the quasi-horizontal event rate will be increased. 
However, the exchange of spin 2 bulk gravitons has very little effect on the
Earth-skimming event rate, because neutrinos suffer very little energy
loss during this process~\cite{Emparan:2001kf}. Then one expects  
KK modes to further enhance the ratio $N_{\rm QH}/N_{\rm ES}$,
making it still easier to distinguish such effects from SM
expectations. 

It is important to stress that during this section we only consider the 
conservative estimates of the cosmogenic neutrino flux. However, some 
models predict fluxes far above these conservative 
values~\cite{Mannheim:1998wp}.\footnote{Clearly, 
any top-down origin of ultrahigh energy CRs is inconsistent with TeV scale 
gravity. The arguments in this section then do not apply to neutrino
fluxes associated with a top-down origin.} In such a case the atmosphere 
would be effectively a BH factory, and 100's of BH events would be observed 
each year at PAO. 

\subsection{${\bm p}$-branes from Heaven}

In this penultimate section we come back to a main topic of this review and 
discuss yet another proposal~\cite{Jain:2002kf} to sneak away from the 
GZK cutoff. This recent 
suggestion 
is a variant of those discussed in Sec.VI-A where neutrinos interact 
strongly beyond the GZK-energy. Therefore, the validity 
of this model is contingent on the observation of moderately penetrating 
showers.  

The basic scaling relation of Eq.~(\ref{rcomp}) is the simplest one possible. 
Once one envisages the notion of asymmetric compactification radii, a 
plethora of new 
phenomena may arise as one can consider the possibility of several compactification scales. Of particular interest here is the next simplest example, 
where there is a hierarchy of two sets of compactified extra 
dimensions, with $m$ small dimensions of length 
$L \alt$~1~TeV$^{-1}$, and $n-m$ large extra 
dimensions of length $L'$. The relation between $\mstar$ and the low
energy 4-dimensional Planck mass $\mplanck \simeq 1.2 \times
10^{19}~\gev$ is~\cite{Lykken:1999ms}
\begin{equation}
\mplanck^2 = \mstar^{2+n}\, L^{m}\, {L'}^{n-m} \ .
\end{equation}
For simple toroidal compactifications, $L=2\pi r_c$, and $L'=2 \pi r_c'$.
Scenarii with low values of $n-m$ are already tightly constrained from 
table-top gravity experiments, as well as from astrophysical and 
cosmological considerations~\cite{Hoyle:2000cv,Cullen:1999hc,Barger:1999jf,Hanhart:2001fx,Hall:1999mk,Hannestad:2001jv,Hannestad:2001xi}.

In spacetimes with asymmetric compactifications, extended higher dimensional 
objects wrapped around small extra dimensions, so-called ``$p$-branes'', 
can be produced in super-Planckian scattering 
processes~\cite{Ahn:2002mj,Cheung:2002aq,Ahn:2002zn}. In general, 
the formation of higher-dimensional branes   
dominates the formation of lower-dimensional branes and spherically symmetric 
BHs (0-branes). The decay of
$p$-branes is not well-understood.  One possibility is that they may
decay into lower dimensional brane-antibrane pairs, leading to a
cascade of branes. In any case, there is no reason
to expect them to decay only to invisible particles, and it is
reasonable to expect their decays, as with BH decays, to be dominated
by visible quanta (those on the brane) that could be detected by cosmic ray 
observatories~\cite{Ahn:2002mj}. 

$p$-brane production is negligible relative to BH production unless the 
$p$-brane wraps only  Planck-sized dimensions. The enhancement in the 
$p$-brane cross section results from wrapping on small dimensions and is 
a consequence 
of the dependence of the $p$-brane radius~\cite{Ahn:2002mj}, 
\begin{equation}
r_p = \frac{1}{\sqrt{\pi} \mstar}\, \gamma(n,p)\,
\left({M_{p}\over \mstar \ V_p}\right)^{\frac{1}{n+1-p}}\ , \label{rp}
\end{equation}
solely on the density of the $p$-brane. Here,
$V_p = (L/\lstar)^p$ is the volume wrapped by the $p$-brane in fundamental 
Planck units,
\begin{equation}
\gamma(n,p) = \left[ 8\,\Gamma\left( \frac{3+n-p}{2} \right)
\sqrt{\frac{1+p}{(2+n)(2+n-p)}} \ \right] ^ \frac{1}{1+n-p} \ ,
\end{equation}
and $L_*$ is the fundamental Planck length.\footnote{Note that 
for $p=0$, \eqref{rp} reduces to the metric of a (4+$n$)-dimensional
BH and $r_p$ becomes the Schwarzschild radius~\cite{Myers:un}.}
$p$-brane production dominates over BH production only if $L \alt L_*$.
It is worthwhile to point out that in the string-based low
energy Lagrangian, the gauge coupling is inversely proportional to the
compactification volume, and a small volume corresponds to strong
coupling. In certain explicit models, these small volumes can be
removed from the gauge sector via a $T$-duality
transformation~\cite{Shiu:1998pa}.  In what follows, we avoid reference to
specific models, and  for illustrative purposes discuss $p$-brane production 
over an inclusive range $0.1 < L/\lstar < 10$.

The total cross section for brane production is given by~\cite{Anchordoqui:2002it}
\begin{equation}
\sigma_{\nu N \to \text{brane}}  =
 \sum_p \sigma_{\nu N \to p\text{-brane}} \ ,
\label{sam}
\end{equation}
where
\begin{equation}
\label{partonsigmap}
\sigma_{\nu N \to p\text{-brane}} =
\sum_i \int_{\mpmin{}^2/s}^1 dx\,
\hat{\sigma}_i ( \sqrt{xs} ) \, f_i (x, Q) \ ,
\end{equation}
is the cross section for $p$-brane production from neutrino nucleon 
scattering,
$\hat{\sigma}_{ij\to p\text{-brane}} (\sqrt{\hat{s}}) = \pi r_p^2$, 
is the parton-parton cross section, the sum is over all partons in the 
nucleon, the $f_i$ are parton distribution functions, and  
$\mpmin$ is the minimum mass required for $p$-brane production which is set 
equal to $M_D$.

\begin{figure}
\postscript{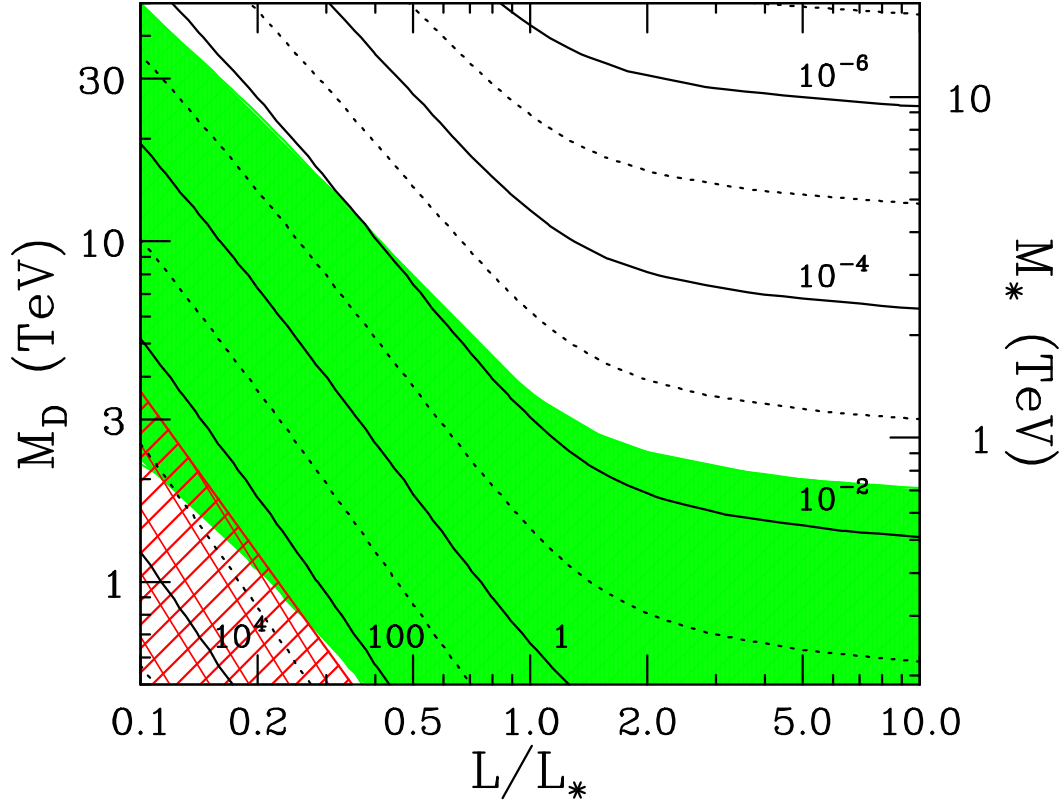}{0.90}
\caption{Contours of $\sigma^{\text{total}}_{\nu N \to \text{brane}}$
(in mb) at $E_{\nu} = 10^{11}~\gev$ in the $(\md,L/L_*)$ plane for
$(n=6,m=5)$.  The right vertical axis shows the corresponding values $M_*$. 
The shaded region is
excluded by the non-observation of deeply penetrating showers at
AGASA. The hatched region is excluded by requiring no large
corrections to standard model physics at lower energies (see text). Published in Ref.~\cite{Anchordoqui:2002it}.}
\label{pbrane}
\end{figure}

It was recently proposed~\cite{Jain:2002kf} that ultrahigh energy neutrinos 
interacting via $p$-brane production may trigger vertical EASs. The 
required cross sections for hadronic-like interactions 
($\sim 100~\mb$ at $10^{20}$~eV), as given by Eq.~(\ref{sam}), are only attained
for $n - m = 1, 2$ and $M_*$ of a few TeV, a region of the parameter space 
excluded by the sub-mm gravity experiments and astrophysical constraints. 
Therefore, cosmic neutrinos with interaction strengths enhanced by $p$-brane 
production cannot initiate super-GZK showers in factorizable spacetimes.
Although no explicit models for warped scenarii are available, in the spirit 
of~\cite{Giddings:2001yu} one may consider the possibility that 
Eq.~(\ref{sam}) is still valid 
for $M_* \sim$ 1~TeV, and $n - m = 1$. Under these speculative assumptions,
hadronic-like neutrino interactions could dominate the scattering to such an 
extent that the non-observation of deeply developing showers is satisfied 
because the cross section is {\em too large} --- the showers would develop 
too high in the atmosphere. 

Figure~\ref{pbrane} displays 
contours in the $(\md,L/L_*)$ plane labeled by the value of the cross
section attained at $10^{20}$~eV (in mb), for $n =6, m=5$ (a similar result 
can be obtained for $n =7, m =5$)~\cite{Anchordoqui:2002it}. The shaded area is
excluded by the absence of a significant signal of
deeply developing showers reported by the AGASA 
Collaboration~\cite{Yoshida:h}. 
It is important to stress that the AGASA
data serve to restrict the parameter space even in the case where the
total cross section at $10^{20}$~eV is of ${\cal O}(100)~\mb$,
because some among the tower of $p$-branes are produced deep in the
atmosphere. We note that the upper boundary of this shaded region is
in agreement with existing limits~\cite{Tyler:2001gt}.  In the lower
left of Fig.~\ref{pbrane} is an unshaded region with large cross
section which ostensibly can evade the AGASA bound because of neutrinos 
showering high in the atmosphere. However, a new consideration
enters in this situation: cross sections which are too large can lead,
via a dispersion relation, to large deviations in SM
predictions at lower energies~\cite{Goldberg:1998pv}. With the cross
section behavior given by Eq.~(\ref{sam}), it is a
straightforward exercise using the result in \cite{Goldberg:1998pv} to
show that low energy corrections of ${\cal O}(100$\%) can arise 
such that
\begin{equation}
\sigma_{\nu N}(10^{20}~{\rm eV}) \ge 300~\mb \ .
\end{equation}
In Fig.~\ref{pbrane} this region is cross-hatched, to indicate that it
is problematic for SM physics at much lower energies
($\sim 100~\gev$). As can be seen from Fig.~\ref{pbrane}, this
leaves very little room for explaining the super-GZK events using 
$p$-brane physics: part of the potential parameter space is ruled out by
AGASA data, most of the rest by requiring small feed-down of these
high energy contributions to low energy neutrino physics. Only a tiny
window is available in a region where the small extra dimension
$L <0.2 L_*$. Furthermore, in this allowed region one has strong coupling 
effects in the underlying stringy regime.

\section{Looking forward}

It is difficult -- and perhaps hazardous -- to speculate where CR physics will
go in the next ten years. One can never foretell the serendipitous
discoveries that will transform our understanding. However,
on-going and future experiments will surely provide us with the statistics to
begin discriminating among the many promising ideas so far proposed to explain
the origin and nature of CRs above the GZK energy limits.
The superior angular and energy resolution of the Pierre Auger Observatory 
will allow the high end of the energy spectrum and the CR arrival directions to be
measured with unprecedented precision.  These
observables provide the most powerful discriminators among candidate 
sources and primary species. 
         
Distinguishing among CR sources that produce anisotropic distributions is 
comparatively straightforward.
For example, CRs from radio galaxies would appear clustered in the
super-Galactic plane,
whereas those from super heavy relics are expected to cluster in the halo.
It is more challenging to distinguish among the various models that 
reproduce the observed isotropy below $10^{20}$~eV.  In this case, 
there are two broad categories: scenarii where
isotropy is realized by diffusive CR propagation, and those where the
distribution of the sources themselves is isotropic.

First we comment on the case of an isotropic CR distribution resulting from
diffusion.  Diffuse propagation of CRs must be associated with relatively
nearby sources, and the most likely candidates are characterized by distinctive
signatures.  Galactic sources, for example, will emit mainly nuclei which
will begin to exhibit a correlation with the Galactic plane for energies
in excess of $10^{20}$ eV.  If Centaurus A is the culprit, one can expect
to observe neutrons above $10^{20}$ eV which of course would point back to the
source. In addition, magnetic fields of $\cal O(\mu \mathrm{G})$ are required to
produce enough diffusion to render an isotropic distribution for the lower
energy particles. A third possibility is that CRs coming from M87 are diffused
by
a Galactic wind, in which case a north-south asymmetry should emerge in the
galactic coordinates.  Finally, if nearby starburst galaxies are the ultrahigh energy
CR generators, one should observe primarily nuclei which, at high energy, 
are correlated with the sources. In this case, fields of 
$\cal O(\mathrm{nG})$ would provide
enough diffusion to render the distribution isotropic below about $10^{20}$~eV.
                                           
An isotropic CR distribution could also arise from isotropic sources,
such as highly red-shifted AGNs, GRB-fireballs, or decaying TDs.
To ascertain whether CRs share a common origin with GRBs, we can make use of
the data recorded by the HETE satellite and directly search for correlations
between CR and GRB samples. To fill in the picture further, we have to be 
able to discriminate between the signatures of AGNs and TDs.  
TD scenarii predict that $\gamma$-rays should dominate the super-GZK energy 
spectrum. In contrast, AGNs could produce GZK-evading messengers like uhecrons
or neutrinos that attain hadronic cross sections, both
of which would generate typical hadronic EASs. These two types of showers 
can be distinguished by the rate of vertical compared to 
inclined showers.
Concerning the uhecron
option, we note that the allowed parameter space is narrowing and collider
experiments will have the final word. Theories which postulate strong 
neutrino interactions
at super-GZK energies also predict that moderately penetrating
showers should be produced at lower energies,
where the neutrino-nucleon cross section reaches a sub-hadronic size. 
In TeV scale gravity models the neutrino-nucleon cross section is likely to be 
sub-hadronic near the energy at which the cosmogenic neutrino flux peaks, and 
so moderately penetrating showers should be copiously produced. Finally,     
the $Z$-burst model eludes a definitive confirmation based on
 the prediction of 
the arrival directions, since the distribution will depend on whether relic 
neutrinos are clustered in the halo or not. However, the sensitivity of the 
Pierre Auger Observatory 
will be sufficiently high to test if the neutrino flux is big 
enough to render the model viable. Interestingly, if the $Z$-burst model 
could be verified it would provide experimental evidence for the thermal 
neutrino background. Moreover, in this case 
the CR arrival directions would encode some information on the distribution 
of relic neutrinos in the Universe.   Table~\ref{PAOsignal} summarizes 
the signatures corresponding to some of the models in the discussion above.

\begin{table}[htb]
\vspace*{-0.1in}
\caption{Possible signatures for different ultrahigh energy cosmic ray sources.}
\label{PAOsignal}
\begin{tabular}{|@{}l|c|c|}
\hline
\hline
\multicolumn{2}{@{}|c|}{Source}   & Signal \\
\hline
\hline
\multicolumn{2}{@{}|c|}{M87} & Asymmetry in flux with respect to Galactic plane\\ 
\hline
\multicolumn{2}{@{}|c|}{Pulsars}  & Arrival directions correlated with  Galactic plane \\
\hline
\multicolumn{2}{@{}|c|}{Starbursts} & Anisotropy towards sources above $2 \times 10^{20}$~eV \\
\hline
\multicolumn{2}{@{}|c|}{Cen A} & Spike in source--direction due to $n$-channel \\
\hline
\multicolumn{2}{@{}|c|}{$Z$-burst} & Can only check for required $\nu$ flux \\  
\hline
  ~~   & Superheavy relics  & Anisotropy towards Galactic halo \\
\cline{2-3}
~~ Top Down $\;\;\;\;\;\;\;$ & $\Longleftrightarrow$   & Flux dominated by $\nu$'s and $\gamma$'s    \\
\cline{2-3}
 ~~ &  Topological defects & Isotropy \\
\hline
\hline
\end{tabular}
\end{table}

Future CR data will not only provide clues to the origin of the
super-GZK events, but could enhance our understanding of fundamental
particle physics. For example, if CR primaries are found to have a significant photon component above $10^{20}$~eV, this could suggest an exotic ingredient 
in CRs,
such as decay products of TDs, and thus could provide       
insight into the description of the early Universe as well as particle
physics beyond the SM.
On the other hand, an absence of ultrahigh energy
photons would imply hadronic primaries interacting at
c.m. energies well above that achievable at any current
or foreseeable collider, thus providing a unique opportunity to
evaluate hadronic interaction models at ultrahigh energies.
Moreover, a similar technique to that employed
in discriminating between photons and hadrons can be applied to search for
signatures of extra-dimensions. In this case, comparison of event rates
of quasi-horizontal deeply developing showers and Earth-skimming
neutrinos will allow better limits to be placed on low scale gravity.
An optimist might even imagine the discovery of microscopic BHs, the telltale
signature of the Universe's unseen dimensions.  The puzzle of ultrahigh energy CRs may even have something to say about issues as fundamental as local
Lorentz invariance.

All in all, after 40 years of careful work by many research groups around 
the world, we are in possession of a tantalizing body of data, more than 
sufficient to stimulate our curiosity but not yet sufficient to unravel the 
mystery of the highest energy CRs. The upcoming high quality observations promise to make the
next 10 years an extremely exciting time for CR physics.

\hfill

{\bf Note added}: After this review was completed, the HiRes 
Collaboration~\cite{:2002ta,Abu-Zayyad:2002sf} re-examined their sample 
and some of the highest energy events where discarded. The small modification 
induced on the energy spectrum is, however, of little help in attempts to 
understand the tail of the CR spectrum, leaving the nature and origin of 
the highest energy events still a mystery. The AGASA Collaboration~\cite{Takeda:2002at} 
have also re-evaluated their highest energy events and conclude that there are ``surely 
events above $10^{20}$~eV and the energy spectrum extends up to a few 
times $10^{20}$~eV.'' Certainly, the conclusion of this
review remains the same: {\it more data is needed to understand the shape of 
the energy spectrum above the GZK-limit}

\begin{acknowledgments}

We are most grateful to Haim Goldberg for
ongoing collaboration and many illuminating discussions. We also want 
to thank Jorge Combi, Tere Dova, Luis Epele, Jonathan Feng, Tom McCauley, 
Santiago Perez Bergliaffa, Gustavo Romero, Sergio Sciutto, Al Shapere, 
Guenter Sigl, Diego Torres, and Tom Weiler for fruitful collaborations. 
We also benefited from discussions with Felix Aharonian, Ileana Andruchow, 
Peter Biermann, Murat Boratav, Anal\'{\i}a Cillis, Jim Cronin, Huner 
Fanchiotti, Carlos Garc\'{\i}a Canal, Diego Harari, Carlos Hojvat,
Masha Kirasirova, Johannes Knapp, Jeremy Lloyd-Evans, Jared MacLeod, 
Gustavo Medina Tanco, Charly Nu\~nez, 
Kasper Olsen, John Ralston, Esteban Roulet, Martin Schvellinger, 
Paul Sommers, Tomasz  
Taylor, el Tum, and  Alan Watson. We acknowledge Rainer Dick, Peter Fisher, 
Zoltan Fodor, Michal Ostrowski, and Subir Sarkar for valuable comments on the manuscript.
We are 
indebted to several collegues and collaborators mentioned above as well as 
 Michael Kachelriess, Andreas Ringwald, 
Tom Rizzo, Simon Swordy, Masahiro Takeda,  
and the AGASA and Fly's Eye Collaborations for allowing 
us to use various figures from their papers in this review. We also wish to 
thank the American Physical Society for permission to reproduce some of 
these figures. Special thanks go to Creedence Clearwater Revival for 
inestimable inspiration. This work has been supported, in part, by the US 
National Science Foundation (under grant No. PHY--0140407).

\end{acknowledgments}

\appendix

\hfill

\section{Hadronic interaction models in the {\em terra incognita}}

The analysis of ultrahigh energy cosmic rays requires the
extrapolation of hadronic 
interaction models more than 2 orders of magnitude in  
energy beyond the highest accelerator energies ($\sqrt{s} = 1.8$ TeV). 
Actually, the required extrapolation is much greater than this because the 
showers involve nuclei as well as single hadrons both as targets and 
projectiles.
 
Soft multi-particle production with small transverse momenta with
respect to the collision axis is a dominant feature of most events
in high energy hadronic collisions.  Though strict calculations based on
ordinary QCD perturbation theory are not feasible, some
phenomenological approaches successfully describe the main properties
of soft diffractive processes (for 
reviews see Refs.~\cite{Capella:yb,Drescher:2000ha}). 

The most theoretically advanced type of model 
(like {\sc venus}~\cite{Werner:1993uh},
{\sc qgsjet}~\cite{Kalmykov:ys} 
and {\sc dpmjet}~\cite{Ranft:fd,Roesler:2000he}) is 
based on Gribov-Regge theory~\cite{Regge:mz,Gribov:fc}. The interactions 
are not described by single
particle exchange, but by the exchange of highly complicated collective modes
known as reggeons. The slow growth of the
cross section with $\sqrt{s}$ for many different hadronic 
reactions measured at colliders can 
be well described by the form 
$\sigma_j = Y_j \,s^{-\eta} + X_j \,s^{\epsilon}$, with  
the universal parameters
$\eta \sim 0.47$ and $\epsilon \sim 0.08$~\cite{Barnett:1996hr}.
Here, $X_j$ and $Y_j$ are relative amplitudes differing for each reaction 
$j$ and they must be determined experimentally. The two terms represent 
the effect of the exchange of reggeons and pomerons,\footnote{This 
leading Regge trajectory with vacuum quantum
numbers was originally proposed by Chew and Frautschi~\cite{Chew:ev}.}
 respectively, between the scattering 
objects, the latter dominating at high energies. Elastic amplitudes and 
total cross-sections can be calculated on the basis of 
multi-pomeron exchange, whereas inelastic reactions are 
introduced by cut pomerons~\cite{Werner:1993uh}. The transition from the 
number of cut pomerons to the secondaries produced is
modeled by string theory \cite{Andersson:ia,Sjostrand:1987xj}.

On a different track, in {\it minijet} models ({\it e.g.} 
{\sc sibyll}~\cite{Fletcher:1994bd}) 
the rise of the cross section with energy is driven by a growth of the 
fraction of hard interactions. The probability 
distribution for obtaining $N$ jet pairs 
(with $p_{\rm T}^{\rm jet}\,>\,p_{\rm T}^{\rm min}$, where
$p_{\rm T}^{\rm min}$ is a sharp threshold in the transverse momentum
below which hard interactions are neglected)
in a collision  at energy $\sqrt{s}$, is computed regarding elastic $pp$ or
$p\bar{p}$ scattering as a diffractive shadow scattering associated
with inelastic processes. The energy of each produced particle is generated 
using Lund techniques~\cite{Bengtsson:1987kr}. The algorithms are
tuned to reproduce the central and fragmentation region data
up to $p\bar{p}$ collider energies,
and with no further adjustments they are then extrapolated several orders of 
magnitude.

The most difficult point in connection with the air shower development
is certainly the treatment of nuclear reactions. According to the 
traditional notion, which treats the projectile simply as a superposition 
of free nucleons~\cite{Engel:vf}, fluctuations in a nucleus initiated 
shower should be smaller. This notion still stands if one uses 
the quark-gluon picture of nucleus-nucleus interaction, but the resulting 
fluctuations in EASs initiated by nuclei become nearly twice as large as 
semi-superposition model predictions~\cite{Kalmykov:br}. 
Such an enhancement is due to the more adequate description of the stochastic 
behavior of projectile nucleons. Moreover, the different approaches used to 
model the underlying physics of $p\bar{p}$ collisions show clear 
differences in multiplicity predictions which increase with rising 
energy~\cite{Knapp,Anchordoqui:1998nq,Ranft:2000mz,Alvarez-Muniz:2002ne}. 
Therefore, distinguishing 
between a proton and a nucleus shower is extremely difficult at the 
highest energies~\cite{Anchordoqui:1999hn}. Recently, a complete analysis on 
the hadronic core of showers around the knee done by 
the KASCADE Collaboration~\cite{Horandel:br} provides important information 
about the quality of the hadronic event generators. The study seems to 
indicate that {\sc qgsjet} is the model which best reproduces the spectrum 
below $5 \times 10^{15}$ eV. However, above the knee deviations in several 
observables are seen, suggesting that further investigations (and data) are 
needed.\footnote{For a more extensive discussion on hadronic interaction 
models the reader is referred to~\cite{Knapp:2002vs}.}

\section{Redshifting}

There is now a substantial body of observations that support, both directly 
and indirectly, the idea that our Universe expanded from a super-dense 
hot phase roughly 
13 billion years ago~\cite{Peebles:1991ch}. In particular, in the late 1920's 
Hubble established that the spectra of galaxies at greater distances are 
systematically shifted to longer wavelengths. The change in wavelength of 
a spectral line is expressed as the redshift of the observed feature,
\begin{equation}
1 + z \equiv \frac{\lambda_{\rm observed}}{\lambda_{\rm emitted}}\,.
\end{equation}
Interpreting the redshift as a Doppler velocity, for $z\ll 1$, Hubble's 
relationship can be re-written as $z \sim H_0 d/c$, where $H_0$ is the 
expansion rate at the present epoch. 

The adiabatic energy losses suffered by cosmic rays due to the expansion 
of the Universe can be associated with the internal energy loss of a 
relativistic gas with particle density $n$ and temperature $T$ doing 
work to expand its volume $V$, 
\begin{equation}
dU=-P\,dV \,, 
\end{equation}
where $P\,V = N\,k\,T$ or $P = n\,k\,T$ (i.e., $n\,V=N$) is the gas pressure and $k$ is the Boltzmann constant.
Now, since the mean energy of each particle is $3kT$, one obtains
$dU = n\, V\, dE = - n\, E\, dV/3$. For a fixed number of 
particles  
\begin{equation}
\left(\frac{dE}{dt}\right)_{\rm adiabatic} = - \frac{1}{3} \frac{nE}{N} \frac{dV}{dt},
\label{notti}
\end{equation}
where $dV/dt$ is the expansion rate of the region with field velocity 
$\vec{v}(\vec{r})$. The change in the volume $dx \ dy \ dz$ moving 
with the flux is given by,
\begin{equation}
\frac{dV}{dt} = (v_{x+dx} - v_{x}) dy\,dz+(v_{y+dy} - v_y) dx\,dz +
(v_{z+dz}-v_z) dx\,dy,
\end{equation}
or, using Taylor's expansion
\begin{equation}
\frac{dV}{dt} = \left( \frac{\partial v_x}{\partial x} +
\frac{\partial v_y}{\partial y} + \frac{\partial v_z}{\partial z}
\right) dx\,dy\,dz = (\nabla . \vec{v}) V.
\label{escobar}
\end{equation}
Now, the adiabatic fractional energy loss is obtained by replacing 
Eq.~(\ref{notti}) into Eq.~(\ref{escobar}) 
\begin{equation}
\left(\frac{dE}{dt}\right)_{\rm adiabatic}  = -\frac{1}{3}\,(\nabla . \vec{v})\,E \,.
\label{pari}
\end{equation}
In standard Friedman-Robertson-Walker cosmology, with vanishing 
cosmological constant and scale factor $R$ which expands at a
velocity $v_0$, $v=v_0(r/R)$, and
\begin{equation}
\nabla . \vec{v} = \frac{1}{r^2}\, \left[
\frac{\partial}{\partial r} (r^2\, v_r ) \right] = 3\,
\frac{v_0}{R}\, . 
\label{pari2}
\end{equation}
Hence,
\begin{equation}
-\left(\frac{dE}{dt}\right)_{{\rm adiabatic}} = \frac{v_0}{R} \,E = 
\frac{1}{R}\frac{dR}{dt}\, E = H_0 \,E\,.
\end{equation}

\section{Kinematics of photomeson production}

The inelasticity $K_\pi$ depends not only on the outgoing particles but also 
on the kinematics of the final state. Nevertheless, averaging over final 
state kinematics leads to a good approximation of $K_\pi$.  The 
c.m. system quantities are determined from the relativistic 
invariance of the square of the total 4-momentum $p_\mu p^\mu$ of 
the photon-proton system. This invariance leads to the relation.
\begin{equation}
s=(w^{c.m.} + E^{c.m.})^2 = m_p^2 + 2m_p w_r.
\end{equation}
The c.m. system energies of the particles are uniquely determined 
by conservation of energy and momentum. 
For reactions mediated by resonances one can assume a decay, which in the 
c.m. frame is symmetric in the forward and backward directions with 
respect to the collision axis (given by the incoming particles). 
For instance, we consider single pion production via the reaction
$p  \gamma \rightarrow \Delta \rightarrow p  \pi.$ Here,
\begin{equation}
E_\Delta^{\rm c.m.} = \frac{(s + m_\Delta^2 - m_\pi^2)}{2\, \sqrt{s}}\,.
\end{equation}
Thus, the mean energy of the outgoing proton is
\begin{equation}
\langle E_p^{\rm final\,\,c.m} \rangle = \frac{(s + m_\Delta^2 - m_\pi^2)}{2
\,\sqrt{s}\,m_\Delta} \frac{(m^2_\Delta + m_p^2 - m_\pi^2)}{2\,m_\Delta},
\end{equation}
or in the lab frame
\begin{equation}
\langle E_p^{\rm final} \rangle =\frac{E}{s} \frac{(s-m_\pi^2+m_\Delta^2)}{ 2\,m_\Delta}
\frac{(m_\Delta^2 - m_\pi^2 + m_p^2)}{2\, m_\Delta}.
\label{6}
\end{equation}
The mean inelasticity $K_\pi = 1 - (\langle E^{\rm final}\rangle/E)$ of 
a reaction
that provides a proton after $n$ resonance decays can be obtained by 
straightforward generalization of 
Eq.~(\ref{6}), and is given in Eq.~(\ref{kj}).

\section{Iron photodisintegration}

The photoabsorption cross section roughly obeys the electric dipole 
sum rule
\begin{equation}
\Sigma_d \equiv \int_0^\infty \sigma(w_r) dw_r = 60\,\frac{n\,Z}{A}\,\, 
{\rm MeV\, mb}\,,
\end{equation}
where $n = A - Z$ is the number of neutrons.\footnote{Indeed, this integral 
is experimentally $\sim 20-30\%$ larger~\cite{Hayward:62}.} 
These cross sections contain essentially 
two regimes. At $w_r < 30$~MeV there is the domain of the giant dipole 
resonance where disintegration proceeds mainly by the emission of one 
or two nucleons. A Gaussian distribution in this energy range is found to 
adequately fit the cross section data~\cite{Puget:nz}. At higher energies, 
the cross section is dominated by multinucleon emission and is approximately 
flat up to $w_r \sim 150$~MeV. Specifically,
\begin{equation}
\sigma_{Ai}= \frac{\xi_{Ai} \Sigma_d \, \Theta(w_r-2) \,
\Theta(30 - w_r
)
\, e^{ -2 ( w_r -\epsilon_{0i})^2/\Delta_i^2}}{W \Delta_i}
 + \frac{f_i \Sigma_d \, \Theta(w_r- 30)}{120}\,,
\label{cs1}
\end{equation}
for $i$=1, 2, and 
\begin{equation}
\sigma_{Ai} = \frac{f_i \Sigma_d \, \Theta(w_r - 30)}{120}\,,
\label{cs2}
\end{equation}
for $i > 2$~\cite{Puget:nz}. 
Here, $W$ is a normalization factor given by
\begin{displaymath}
W = \left(\frac{\pi}{8}\right)^2 \left[ \Phi(\sqrt{2}(30 -
\epsilon_{0i} )/\Delta_i) + \Phi (\sqrt{2}(\epsilon_{0i}-2 )/\Delta_i) \right],
\end{displaymath}
$\Phi(x)$ is the error function, and  $\Theta (x)$
the Heaviside step function. The dependence of the 
width $\Delta_i$, the peak energy $\epsilon_{0i}$, the branching 
ratio $f_i$, and 
the dimensionless integrated cross section $\xi_i$ are given in 
Ref.~\cite{Puget:nz} for isotopes up to $^{56}$Fe. 

The photon background relevant for nucleus disintegration consists 
essentially of photons of the 2.7~K CMB. The background of optical 
radiation turns out to be of no relevance for ultrahigh energy cosmic ray 
propagation. The infrared background radiation~\cite{Salamon:1997ac}
\begin{equation}
\frac{dn}{dw} = 1.1 \times 10^{-4} \,
\left(\frac{w}{{\rm eV}}\right)^{-2.5}\,\, {\rm cm}^{-3} \,{\rm eV}^{-1}\,, 
\end{equation}
only leads to sizeable effects far below $10^{20}$~eV and for time-scales 
${\cal O}$ ($10^{17}$ s)~\cite{Epele:1998mc}. 

By substituting Eqs. (\ref{cs1}) and (\ref{cs2}) into Eq.
(\ref{phdsrate}) the photodisintegration rates on the CMB can be expressed as 
integrals of two basic forms. The first one is
\begin{equation}
I_1=\frac{{\cal A}}{2 \Gamma^2 \pi^2 \hbar^3 c^2}
\left[\int_{1/\Gamma}^{15/\Gamma} dw
(e^{w/kT}-1)^{-1} {\cal J} + \int_{15/\Gamma}^{\infty} dw
(e^{w/kT}-1)^{-1} {\cal J'}\right]\,,
\end{equation}
where the functions ${\cal J}$ and ${\cal J'}$ are given by the expressions,
\begin{eqnarray}
{\cal J} & = & \sqrt{\frac{\pi}{8}} \epsilon_{0i} \Delta_i \left[
 \Phi
(\sqrt{2}(2 \Gamma w - \epsilon_{0i} )/\Delta_i) +\Phi
(\sqrt{2}(\epsilon_{0i}-2) / \Delta_i) \right] \\  \nonumber
 & + & \left(\frac{\Delta_i}{2} \right)^2 \left\{ e^{ -2 (
(\epsilon_{0i}-2)/\Delta_i)^2} - e^{ -2 ((2\Gamma w -\epsilon_{0i})/
\Delta_i)^2} \right\}\,,
\end{eqnarray}
and
\begin{eqnarray}
{\cal J'} & = & \sqrt{\frac{\pi}{8}} \epsilon_{0i} \Delta_i \left[ \Phi
(\sqrt{2}(30 - \epsilon_{0i} )/\Delta_i) +\Phi
(\sqrt{2}(\epsilon_{0i}-2) / \Delta_i) \right] \\ \nonumber
 & + & \left(\frac{\Delta_i}{2} \right)^2 \left\{ e^{ -2 (
(\epsilon_{0i}-2)/\Delta_i)^2} - e^{ -2 ((30 -\epsilon_{0i})/
\Delta_i)^2} \right\}.
\end{eqnarray}
The second basic integral is of the form
\begin{equation}
I_2 = (\pi^2 \hbar^3 c^2)^{-1} \sigma_{Ai}\left[
\int_{15/\Gamma}^{\infty} \frac{w^2
dw}{e^{w/kT} - 1} - \left( \frac{15}{\Gamma}\right)^2
\int_{15/\Gamma}^{\infty} \frac{
dw}{e^{w/kT} - 1}\right].
\end{equation}
With this in mind, Eq.~(\ref{phdsrate}) can be re-written as~\cite{Anchordoqui:ed}
\begin{eqnarray}
R_{Ai} & = & \frac{1}{\pi^2 \hbar^3 c^2 \Gamma^2}
\left\{ \frac{{\cal A}}{2}  \left(  \frac{\pi}{8} \right)^{1/2}
\epsilon_{0i} \Delta_i \left[
e^{-2 \epsilon_{0i}^2/\Delta_i^2}  {\cal S}_1  + {\cal S}_2  \right] -
 \frac{{\cal A }}{2} \, {\cal J'} k T  \ln (1 - e^{-15/\Gamma k T})
\right.   \nonumber \\
&  &  -  \frac{{\cal A}}{8}\,
e^{-2 \epsilon_{0i}^2/\Delta_i^2} \left(\frac{\pi}{32}\right)^{1/2}
\frac{ \Delta_i^3}{ \Gamma}
   {\cal S}_3 +  \frac{{\cal A}}{2} \,
{\cal K} k T \left[
\ln (1 - e^{-15/\Gamma k T}) - \ln (1 - e^{-1/\Gamma
k T}) \right]  \nonumber \\
 & & + \left. \frac{f_i \Sigma_d }{120}  \left[  \Gamma^2 {\cal S}_4  +
15^2 k T \ln (1 - e^{-15/\Gamma k T})
\right] \right\}\,,
\label{R}
\end{eqnarray}
with ${\cal A}$, ${\cal S}_i$, and ${\cal K}$ as given in Table I.

\begin{table}
\caption{Series and functions of Eq.~(\ref{R})}

\hfill

\begin{tabular}{|c|c|}
\hline
\hline
${\cal A}$ & $W^{-1} \xi_{Ai} \Sigma_d \Delta_i^{-1} $\\
\hline
${\cal S}_1$ &  $  \sum_{_{j=1}}^{^{\infty}} \,\, k T j^{^{-1}}\,
\exp [{\cal B}^{^2}] \,
\{ \Phi( {\cal B} +
15 \sqrt{8}/ \Delta_i) - \Phi ({\cal B} + \sqrt{8} /  \Delta_i)
  \}$ \\
\hline
${\cal S}_2$ & $\sum_{_{j=1}}^{^{\infty}} \,\,k Tj^{^{-1}}\,
\exp\{-j/ \Gamma kT \} [
\Phi (\sqrt{2}(2-\epsilon_{0i})/\Delta_i) -
\Phi (\sqrt{2} ( 30-\epsilon_{0i})/\Delta_i ) ]$ \\
\hline
${\cal S}_3$ &  $  \sum_{_{j=1}}^{^{\infty}} \,\,
\exp [{\cal B}^{^2}] \,
\{ \Phi ({\cal B} + 15 \sqrt{8} /  \Delta_i)
 - \Phi( {\cal B} + \sqrt{8}/ \Delta_i)\}$
\\
\hline
${\cal S}_4$ & $\sum_{_{j=1}}^{^{\infty}}
\exp\{-15j/\Gamma k T\} [ (kT/j)
(15 / \Gamma)^2 +(kT/j)^2
(15/\Gamma) + (kT/j)^3 \,]$ \\
\hline
${\cal B}$ & $j \Delta_i / \Gamma k T \sqrt{32} - 2 \epsilon_{0i} /
\sqrt{2} \Delta_i$\\
\hline
${\cal K}$ & $\sqrt{\frac{\pi}{8}}\,\epsilon_{0i}\,\Delta_i
\,\;\Phi (\sqrt{2}\,(\epsilon_{0i}-2)/\Delta_i) + (\Delta_i/2
)^2 \exp\{-2 (\epsilon_{0i}-2)^2/\Delta_i^2 \}$\\
\hline \hline
\end{tabular}
\end{table}

\section{The neutron channel}

As remarked in Sec.III-B.2, if the radiation energy
density of the source is sufficiently high, photopion production
leads to copious neutron flux (that can readily escape the
system) and associated degradation of the proton spectrum. This
occurs only  near the maximum proton energy~\cite{Biermann:ep}. 
It is
reasonable to assume that the ambient photon density of Cen A is
sufficiently high so that near the end of the
spectrum the efficiency of neutron production $\epsilon_n$
becomes comparable to the proton channel $\epsilon_p$. 
Because of the leading particle effect, 
one expects a cutoff in the neutron spectrum at
approximately half the maximum injection energy. In what follows, we 
adopt an energy of
$1\times 10^{20}$ eV as a lower cutoff on the neutron spectrum,
and simplify the discussion by assuming that in the narrow
interval $E_{20} \in [1,2]$ $\epsilon_n\approx \epsilon_p.$ 
The neutron beam observed at Earth is further narrowed because of both
decay {\em en route} and interactions with
the CMB. The interaction length is approximately
6 Mpc, so that after 3.4 Mpc, 45\% of the neutrons interact. Half
go back into neutrons, so this 22\% depletion effect falls  within errors.
However, because of the exponential
depletion,  about 2\% of the neutrons survive the trip at
10$^{20}$ eV, and about 15\% at $2\times 10^{20}$ eV. Note that the 
increasing survival of neutrons at energies above $1.5\times 10^{20}$~eV 
has, as a consequence of the Cen A
model, that the observed diffuse flux $E^3J_{\rm obs}(E)$ should
begin to decrease at these energies (unless other factors
contribute to an increase).

We may now estimate a signal-to-noise ratio for the detection of
neutron CRs in the southern hemisphere.
If we assume circular pixel sizes with  2\deg\ diameters, the
neutron events from Cen A will be collected in a pixel
representing a solid angle $\Delta \Omega({\rm Cen A}) \simeq
10^{-3}$ sr.
For PAO, the event rate of (diffuse) protons coming
from the direction of Cen A (say in a $2^{\circ}$ angular cone)
is found to be~\cite{Anchordoqui:2001nt}
\begin{equation}
\frac{dN_p}{dt}  =  S\, \Delta \Omega({\rm Cen A})\,\,
\int_{E_1}^{E_2} E^3 \, J_p(E) \, \frac{dE}{E^3} \,
  \alt  \frac{0.014}{E_{1,20}^2} \,\,\,\, {\rm events}/{\rm yr},
\label{ruido}
\end{equation}
where we have assumed $E^3 J_p(E)$ to be (approximately) constant
up to at least
$E \approx 3 \times 10^{20}$ eV, in agreement with the observed
isotropic flux in this region, $E^3\,J_{\rm
obs}(E) = 10^{24.5 \pm 0.2} \evmss$~\cite{Hayashida:2000zr}.
The  neutron rate~\cite{Anchordoqui:2001nt}
\begin{equation}
\frac{dN_n}{dt}  =  \frac{S }{4 \pi d^2} \,\,\int_{E_1}^{E_2}
\frac{dN_0^n}{dE dt} \, e^{-d/\lambda(E)} 
    =  116 \, \epsilon_n \lf\int_{E_{1,20}}^{E_{2,20}}
\frac{dE_{20}}{E_{20}^2} \, e^{-d/\lambda(E)}\,\,\,\,  {\rm
events}/{\rm yr}, \label{7}
\end{equation}
is potentially measurable. For $E_{20}\in[1,2]$ one expects
\begin{equation}
\frac{dN_n}{dt} \approx 4 \,\,\epsilon_n \lf \,\,\,\, {\rm events}/{\rm yr}
\label{pipi}
\end{equation}
arriving from the Cen
A direction of the sky. With
$\epsilon_n\lf\approx 1/2,$ this gives about 2 direct events per
year, against the negligible background of
Eq.~(\ref{ruido})~\cite{Anchordoqui:2001nt}. An increase in the maximum 
energy attainable at the
source will shift the end of the spectrum to higher energy. This
will lead to significant enhancement of the neutron flux (at the
higher energy) because of a greater survival probability.

\section{Recipes for a dark matter halo}

There is a great uncertainty in the structure of the Galaxy's dark 
matter halo. Strictly speaking, the HI gas rotation of the Galaxy cannot be 
traced for distances $\agt 20$~kpc, and therefore
the best estimates of the total mass 
($\sim 2 \times 10^{10} M_\odot$) and the extent of the Galaxy 
($\agt 200$~kpc) arise from kinematical analyses of the 
distant satellite galaxies~\cite{Wilkinson:1999hf}.

Numerical simulations of the structure formation in hierarchical merging 
cosmogonies seem to favor a {\it cusped} density profile for dark 
halos~\cite{Navarro:1995iw}
\begin{equation}
n(\vec{r}) \propto \frac{1}{(r + r_\epsilon)\, (r+r_s)^2} \,,
\end{equation}
where $\vec{r}$ is the position with respect to the Galactic center, 
$r = |\vec{r}|$ is the spherical polar radius, $r_s$ is the scale radius 
(which is approximately 10 kpc for the Galaxy), and $r_\epsilon \sim 0.5$ is 
set by the resolution limit of the simulations. This profile implies that the 
inner regions are dominated by DM.

An alternative description, in which the luminous matter dominates the 
central regions and DM the outer parts, assumes the Galaxy's dark halo has an 
{\it isothermal} distribution with large core radius 
($r_c \sim 10$~kpc)~\cite{Evans:1993}
\begin{equation}
n(\vec{r}) \propto \frac{3 r^2_c + r^2}{(r_c^2 + r^2)^2}\,.
\label{profile}
\end{equation} 
Actually, the profile of Eq.~(\ref{profile}) is the spherical limit of a more 
general family of {\it triaxial} distributions~\cite{Evans:2000gr} where
\begin{equation}
n(\vec{r}) \propto \frac{Ax^2 + By^2 + CZ^2 + D}{(r_c^2 + x^2 + y^2\,p^{-2}+ z^2 q^{-2})^2}\,,
\end{equation}
with $\vec{r} = (x,y,z)$, $A=(p^{-2} + q^{-2} -1)$, $B = p^{-2} (1 - p^{-2} + q^{-2})$, $C = q^{-2} (1 + p^{-2}- q^{-2})$, and $D = r_c^2 ( 1 + p^{-2} 
+ q^{-2})$. Here, $p$ and $q$ are axis ratios of the potential. If $p=1$, 
the halo is oblate, whereas if $p=q$, the halo is prolate. In Fig.~\ref{halo}, 
$p = 0.9$ and $q = 0.75$. This means that the density contours have 
semi-axes in the ratio 1 : 0.788 : 0.428, yielding a highly flattened profile 
with an ellipticity\footnote{The ellipticity is $10 \times (1-b/a)$, where $b$ and $a$ are the projected minor and major axes respectively. A spherical galaxy is E0, whilst the most flattened elliptical galaxies are E7.} of roughly E6.

It was also pointed out~\cite{Binney} that in some cases the outer parts of the halo may 
be misaligned with the disk, because of the warping of the neutral gas disk. 
This effect, which is known to be present in the Galaxy, leads to a new family of {\it tilted} halos with density
\begin{equation}
n(R',z') \propto \frac{r_c^2 (2 + q^{-2}) + q^{-2} R'^2 + z'^2 q^{-2} (2 - q^{-2})}{(r_c + R'^2 + z'^2 q^{-2})^2}\,,
\end{equation}
where $R'^2 = x'^2 + y'^2$. The coordinates $(x', y', z')$ are related to 
($x, y, z$) by a rotation through an angle $\theta$ about the $x$-axis, 
on which the Sun lies. The pretext for this transformation is that the Sun 
lies nearly on the line of nodes of the warp. In Fig.~\ref{halo}, to obtain 
the main qualitative features of {\it tilted} haloes, the rotation angle 
was set to an extreme value: $\theta = 30^\circ$.

\section{Tubular black holes}

In what follows, we discuss the domain of the RS~\cite{Randall:1999ee} 
parameter space for which a 
high energy collision, as viewed from the SM brane, can result in the 
formation of a 5 dimensional flat space BH. Two different kinds of BHs 
can be produced in super-Planckian particle collisions within the RS setup. 
On the one hand, there is the AdS/Schwarzschild BH that can propagate freely 
into the bulk, and in general will fall towards the AdS horizon once 
produced. On the other hand, there is the tubular pancake shape BH that is 
bound to the brane~\cite{Giddings:2000ay}. As in the 
ADD~\cite{Arkani-Hamed:1998rs} scenario one expects that 
this type of BH radiates mainly on the brane with interesting 
phenomenological consequences. To characterize the main features of 
tubular BHs it is convenient to re-write Eq.~(\ref{lisa-metric}) by 
introducing the coordinate $z = \ell \, e^{y/\ell}$, so that the metric
\begin{equation}
ds^2 = \frac{\ell^2}{z^2}\,\, \left(dz^2 + \eta_{ij}\, dx^i\, dx^j \right)\,
\end{equation}
is manifestly conformally flat~\cite{Chamblin:1999by}.
Next, to describe the SM brane, define $w = z - z_c$, where 
$|w| \in (0, w_c)$, $z_c = \ell \, e^{r_c/\ell}$, 
$w_c = \ell\, (e^{r_c/\ell}-1)$, and locate the TeV brane at $w=0$.
A calculation  of metric perturbations due to a source on the
$w=0$ brane can be made in the flat 5-dimensional space-time
approximation, {\em i.e.,} ignoring the effects of the AdS curvature, 
if $w\ll z_c$~\cite{Giddings:2000mu}. This implies that the flat space 
BH formulae discussed in Sec.IX-A are also valid in the RS scenario 
if the Schwarzschild radius as measured by an observer in the canonical 
frame $\ll z_c$. The upper limit on the BH horizon translates into an upper 
bound on the BH mass
\begin{equation}
\frac{M_{\rm BH}}{M_D}< 24\ c^{-4/3}\ \ ,
\label{upperc}
\end{equation}
where, $M_D = (4\,\pi)^{1/3}\, M\,\ell/z_c$, and
$c \equiv \left(\ell \overline{M}_{\rm Pl}\right)^{-1}$~\cite{Anchordoqui:2002fc}. When the energy exceeds this bound, the
behaviour of the cross section may be analyzed within the AdS/CFT
dual picture, and may assume the ln$^2E$ behavior conforming to the Froissart 
bound~\cite{Giddings:2002cd}.

\end{document}